\documentclass{aa}  

\usepackage{graphicx}
\usepackage[varg]{txfonts}
\usepackage{lscape}
\usepackage{natbib}
\usepackage{rotating}
\usepackage{array}

\usepackage{amssymb}
\usepackage{indentfirst}
\usepackage{multirow}
\usepackage{hhline}
\usepackage{tabularx}
\usepackage{booktabs}
\usepackage{caption}
\usepackage{pict2e,slashbox}
\usepackage{multicol}
\usepackage{supertabular}
\usepackage{setspace}
\usepackage{threeparttable}
\usepackage{enumerate}
\usepackage{textcomp}

\def\deg{\ifmmode^\circ\else$^\circ$\fi}
\def\alphaTF{\ifmmode{\alpha_{\mathrm{\,{\small TF}}}}\else{$\alpha_{\mathrm{\,{\small TF}}}$}\fi}

\def\imagetop#1{\vtop{\null\hbox{#1}}}

\def\Msun{\ifmmode{\mathrm M_\odot}\else{M$_\odot$}\fi}
\newcommand{\rbreak}{\ensuremath{R_\mathrm{brk III}}}
\newcommand{\mubreak}{\ensuremath{\mu_\mathrm{brk III}}}
\newcommand{\hi}{\ensuremath{h_{\mathrm{i}}}}
\newcommand{\mui}{\ensuremath{\mu_{0,\mathrm{i}}}}
\newcommand{\ho}{\ensuremath{h_{\mathrm{o}}}}
\newcommand{\muo}{\ensuremath{\mu_{0,\mathrm{o}}}}
\newcommand{\risoph}{\ensuremath{R_\mathrm{25}}}
\newcommand{\nir}{3.6\,$\mu$m}

\newcommand{\rbreakrisoph}{\ensuremath{R_\mathrm{brk III}/\risoph}}

\newcommand{\logrbreak}{\ensuremath{\log(R_\mathrm{brk III})}}
\newcommand{\loghi}{\ensuremath{\log(h_\mathrm{i})}}
\newcommand{\logho}{\ensuremath{\log(h_{\mathrm{o}})}}
\newcommand{\logrisoph}{\ensuremath{\log(R_\mathrm{25})}}

\newcommand{\logrbreakrisoph}{\ensuremath{\log(R_\mathrm{brk III}/\risoph)}}

\begin{document}

\title{Photometric scaling relations of antitruncated stellar discs\\in S0-Scd galaxies}

\author{M.~Carmen Eliche-Moral\inst{1}, Alejandro Borlaff\inst{2}, John E.~Beckman\inst{2,3,4}, \and Leonel Guti\'{e}rrez\inst{5}}

\institute{Departamento de Astrof\'{\i}sica y CC.~ de la Atm\'osfera, Universidad Complutense de Madrid, E-28040 Madrid, Spain\\\email{mceliche@ucm.es}
\and 
Instituto de Astrof\'{\i}sica de Canarias, C/ V\'{\i}a L\'actea, E-38200 La Laguna, Tenerife, Spain,
\and
Departamento de Astrof\'{\i}sica, Universidad de La Laguna, E-38200 La Laguna, Tenerife, Spain
\and
Consejo Superior de Investigaciones Cient\'{i}ficas, Spain
\and
Instituto de Astronom\'{i}a, Universidad Nacional Aut\'onoma de M\'exico, Ensenada BC 22860, Mexico 
}

   \date{Received July 28, 2014; accepted May 18, 2015}

\abstract{
It has been recently found that the characteristic photometric parameters of antitruncated discs in S0 galaxies follow tight scaling relations.
}{
We investigate if similar scaling relations are satisfied by galaxies of other morphological types.
}{
We have analysed the trends in several photometric planes relating the characteristic surface brightness and scalelengths of the breaks and the inner and outer discs of local antitruncated S0--Scd galaxies, using published data and fits performed to the surface brightness profiles of two samples of Type-III galaxies in the $R$ and Spitzer \nir\ bands. We have performed linear fits to the correlations followed by different galaxy types in each plane, as well as several statistical tests to determine their significance.
}{
We have found that: 1) the antitruncated discs of all galaxy types from Sa to Scd obey tight scaling relations both in $R$ and \nir, as observed in S0s; 2) the majority of these correlations are significant accounting for the numbers of the available data samples; 3) the trends are clearly linear when the characteristic scalelengths are plotted on a logarithmic scale; and 4) the correlations relating the characteristic surface brightnesses of the inner and outer discs and the breaks with the various characteristic scalelengths significantly improve when the latter are normalized to the optical radius of the galaxy. The observational uncertainties prevent us from discerning robustly whether the trends differ or not between the different types and bands, but we do not find statistical evidence of significant differences between the distributions of S0s and spirals and of barred and unbarred galaxies either. These results suggest that the scaling relations of Type-III discs are independent of the morphological type and the presence (or absence) of bars within the observational uncertainties of the available datasets. However, larger and deeper samples are required to confirm this.
}{
The tight structural coupling implied by these scaling relations impose strong constraints on the mechanisms proposed for explaining the formation of antitruncated stellar discs in the galaxies across the whole Hubble Sequence.
}


\titlerunning{Photometric scaling relations of antitruncated stellar discs in S0-Scd galaxies}	
\authorrunning{Eliche-Moral et al.}

   \maketitle

\section{Introduction}
\label{sec:introduction}

\citet[]{2005ApJ...626L..81E} introduced for the first time a definition of antitruncated or Type-III galaxies, as those in which the surface brightness of the disc does not follow the typical exponentially-decaying profile with the radius \citep{1940BHarO.914....9P,1957AJ.....62...69D,1958ApJ...128..465D,1970ApJ...160..811F}, but presents an up-bending profile, with the outer disc exhibiting a distinct shallower slope than the inner disc outside a given radius (known as the break radius, \rbreak). This nomenclature was an extension of the classification defined by \citet{1970ApJ...160..811F}, who classified Type-I discs as those with single exponentially-decaying profiles and Type-II discs as those with down-bending profiles outside the break radius \citep[see also][]{1979A&AS...38...15V,1987A&A...173...59V}.

In edge-on systems, antitruncations tend to coincide with the superposition of a thin disc and a thick disc \citep[]{2012ApJ...759...98C}. However, while nearly all Type-II profiles are associated with galaxy subcomponents (such as rings, pseudorings, lenses, or strong star formation regions), only $\sim 1/3$ of Type-III profiles are related to distinct morphological substructures \citep[see][L14 henceforth]{2014MNRAS.441.1992L}. The structural properties and frequencies of antitruncations seem to differ in S0 and spiral types. The percentage of antitruncations rises from \mbox{$\sim$10--20\%} in Sc--Sd galaxies to \mbox{$\sim$20--50\%} in S0--Sa types \citep[][G11 hereafter; L14; Maltby et al.~2015]{2008AJ....135...20E,2012A&AT...27..313I,2011AJ....142..145G}. In spirals, antitruncations are basically disc-related phenomena, with less than \mbox{$\sim$15\%} of them associated with the contribution of central spheroidal components to the galaxy outskirts \citep{2012MNRAS.420.2475M,2015MNRAS.447.1506M}. However, this percentage rises to \mbox{$\sim$25\%} in S0--Sb galaxies \citep{2005ApJ...626L..81E} and up to \mbox{$\sim$50\%} if only S0s are considered \citep{2015MNRAS.447.1506M}.

Type-II profiles are known to be related to bars in most cases \citep[see, e.g.,][]{2014ApJ...782...64K}, but the origin of Type-III discs is still poorly constrained. Diverse mechanisms have been proposed to explain the formation of antitruncations. The majority of them are related to gravitational or tidal interactions, such as minor mergers \citep{2001MNRAS.324..685L,2006ApJ...650L..33P,2007ApJ...670..269Y}, major mergers \citep{2014A&A...570A.103B}, interactions of the disc with dark matter subhaloes \citep{2009ApJ...700.1896K}, high-eccentricity fly-by encounters \citep{2008ApJ...676L..21Y}, or harassment \citep{2012ApJ...758...41R}. Other formation scenarios include the existence of different star formation thresholds as a function of the radius in the galaxy \citep{2006ApJ...636..712E}, ram-pressure stripping \citep{2012ApJ...758...41R}, ongoing gas accretion \citep{2012A&A...548A.126M}, and simple fading of stellar discs \citep{2015MNRAS.447.1506M}. \citet{2015MNRAS.448L..99H} have also proposed that the disc profile type of a galaxy may basically depend on the initial spin of its host halo. It seems that bars are unrelated to antitruncations, as derived from the observational fact that the relative frequency of Type-III profiles found in samples of barred and unbarred galaxies is similar \citep[][E08 hereafter; G11; L14]{2009IAUS..254..173S,2008AJ....135...20E}. However, this needs to be confirmed by other means.

Recently, \citet{2014A&A...570A.103B} have found that the structures of the inner and outer discs and the location of the break in Type-III S0 galaxies are strongly coupled, and this coupling seems to be independent of the existence of bars in the galaxies. These authors have shown that the characteristic photometric parameters of the inner and outer discs and the breaks in S0s satisfy several scaling relations, tighter in many cases if the scalelengths are normalized to the optical size of the galaxy. The question is whether or not these scaling relations (or similar ones) are also obeyed by Type-III discs of other morphological types. If not this would imply that antitruncations do form through diverse and independent mechanisms in different Hubble types (which seems reasonable, accounting for the wide variety of possible formation processes). However, if the antitruncated discs of spiral galaxies satisfy scaling relations similar to those observed in S0s, it becomes challenging to understand the physical processes underlying this coupling in galaxies spanning the whole Hubble Sequence. Analogously, if bars are relevant in determining the structure of some antitruncated discs or have triggered their formation in some cases, we should expect to find significant differences between the photometric trends followed by Type-III discs of barred and unbarred galaxies, whereas negligible differences would be expected if both phenomena are structurally unrelated. 

Therefore, we have investigated whether the Type-III discs of spirals obey scaling relations as tight as those observed in antitruncated S0s and, in this case, whether the scaling relations can be considered similar or both galaxy types exhibit significant differences between them. The same analysis has been performed for barred and unbarred galaxies, to find out whether bars and antitruncations are structurally related or not.

For this purpose, we have used the data published by E08 and G11 in the $R$ band, and by L14 in the \nir\ Spitzer band. In Section\,\ref{sec:data}, we briefly comment on the galaxy samples of these authors, their data, and the procedures they followed to obtain and characterize the surface brightness profiles. Section\,\ref{sec:fits} describes our fitting technique to the trends found in the studied photometric planes, as well as the tests performed to identify the correlations that were statistically significant. In Section\,\ref{sec:results} we show the main trends and scaling relations that we have found involving the characteristic scalelengths of the inner and outer discs (\hi\ and \ho, respectively), \rbreak, and \risoph. There we also statistically analyse the differences and similarities of the trends followed by S0 vs.\,spiral galaxies, by barred vs.\,unbarred galaxies, and of $R$ vs.\,\nir\ data. Finally, the discussion and main conclusions are provided in Sections\,\ref{sec:discussion} and \ref{sec:conclusions}.

\begin{table}
\caption{Statistics of Hubble types (S0 and spiral) and barred-unbarred galaxies for the $R$ and \nir\ samples of Type-III galaxies}
\label{tab:samples}
{\footnotesize
\begin{center}
\begin{tabular}{l lll}
\toprule\\\vspace{-0.6cm}\\
\multicolumn{4}{l}{$R$ band$^\mathrm{a}$} \\
 & Barred$^\mathrm{b}$  & Unbarred$^\mathrm{c}$ & Total \\
\midrule
S0--S0$/$a           & 9 (22.5\%) & 12 (30\%) & 21 (52.5\%)\\ 
Sa--Sbc                & 7 (17.5\%) & 12 (30\%) & 19 (47.5\%)\\ 
All Hubble types & 16 (40\%)  & 24 (60\%) & 40 (100\%)\\ 
\bottomrule\\\vspace{-0.6cm}\\
\multicolumn{4}{l}{\nir\ band$^\mathrm{a,d}$} \\
 & Barred  & Unbarred & Total \\
\midrule
S0--S0$/$a           & 13 (21\%)  & 18 (29\%) & 31 (50\%)\\ 
Sa--Scd                & 16 (26\%)  & 15 (24\%) & 31 (50\%)\\ 
All Hubble types & 29 (47\%)  & 33 (53\%) & 62 (100\%)\\ 
\bottomrule

\end{tabular}
\begin{minipage}[t]{0.45\textwidth}{\vspace{0.2cm}
\emph{Notes}: \\
$^\mathrm{a}$ The percentages are given with respect to the total number of galaxies in the sample of each band.\\
$^\mathrm{b}$ All barred galaxies in the $R$-band sample are from E08.\\ 
$^\mathrm{c}$ All unbarred galaxies in the $R$-band sample are from G11.\\ 
$^\mathrm{d}$ The data in the \nir\ band come from L14.\\ 

}
\end{minipage}
\end{center}
} \vspace{-0.6cm}
\end{table}

\section{Data}
\label{sec:data}

We have analysed the possible correlations between the characteristic parameters of the breaks and the inner and outer discs of two samples of local galaxies with Type-III stellar discs, in the $R$ and \nir\ bands. The $R$-band dataset contains the photometric parameters derived for 16 Type-III barred nearby galaxies by E08 and for 24 Type-III unbarred ones by G11 (40, in total), with S0-Sbc types. The \nir\ dataset comprises the 62 Type-III (barred and unbarred) galaxies from the sample analysed by L14, with types spanning from S0 to Scd. Our study is exclusively centered on galaxies with pure Type-III profiles, i.e., the galaxies with hybrid profiles from the original samples (Type II$+$III) have been excluded in our subsamples to avoid a possible additional dispersion in the trends we were looking for (they were 4 galaxies in the E08 sample, 5 from G11, and 7 from L14). The E08 sample overlaps with the L14 sample in 4 galaxies, while G11 sample has 7 galaxies in common with L14.

Table\,\ref{tab:samples} summarizes the statistics of the samples in terms of morphological types and barred/unbarred nature in both bands. The statistics of the two subsamples is not very high (40 S0--Sbc galaxies in $R$ and 62 S0--Scd's in \nir), but the numbers are sufficiently large to allow us to look for photometric scaling relations in S0s and spirals separately, because the galaxies distribute nearly equally among the two types in both bands. A similar argument holds for barred and unbarred galaxies. The original data, reduction, and methodology to characterize the surface brightness profiles are extensively described in the original papers, so we provide just a brief description here. 

The original samples were defined using different selection criteria for the radial velocities, angular sizes, galactic latitudes, and morphologies of the galaxies. The galaxies in the $R$-band sample have distances $<$\,30\,Mpc, while those of the L14 sample lie at $<$\,80\,Mpc, but both datasets present \mbox{-18}\,$<$\,$M_B$\,$<$\,\mbox{-22}\,magnitudes. E08 and G11 used data in the $r$ and $R$ bands taken with different telescopes, with PSF FWHM$\sim$0.7'' and limiting surface brightness $\mu_\mathrm{lim}$\,$\sim$\,26--27\,mag\,arcsec$^{-2}$ in $R$ (Vega system). L14 combined data obtained in the \nir\ IRAC band for Sa-Sd galaxies of the S$^4$G survey \citep[FWHM$\sim$1.7'', see][]{2010PASP..122.1397S} with $K_s$-band images for S0-S0/a galaxies from the NIRS0S survey \citep[FWHM$\sim$0.7'', see][]{2011MNRAS.418.1452L}. In L14, the distances of the galaxies coming from the S$^4$G sample are $<$\,40\,Mpc, and $<$\,80\,Mpc for those coming from NIRS0S. 

L14 converted the $K$-band surface brightness profiles of the galaxies from the NIRS0S sample to AB magnitudes in the \nir\ band accounting for the color differences and magnitude offsets derived for the 93 galaxies that the two surveys have in common. These authors computed total magnitudes in elliptical apertures tracing the $\mu=22.5$\,mag\,arcsec$^{-2}$ isophote in \nir\ in each galaxy. The median difference between the magnitudes obtained in the two surveys was derived, including a linear term to describe the dependence on  colour. This conversion factor was then applied to the surface brightness profiles and total magnitudes of the NIRS0S data to transform them into \nir. L14 data finally presented $\mu_\mathrm{lim}\sim$\,26.4\,mag\,arcsec$^{-2}$ for the Sa-Scd's and $\mu_\mathrm{lim}\sim$\,24.7\,mag\,arcsec$^{-2}$ for the S0-S0/a's in \nir\ (AB magnitudes).

Both data samples are analogous in terms of depth for the spiral types, but the $R$-band sample is at least $\sim 1$\,mag deeper than the \nir\ sample for the S0 galaxies. L14 compared the limiting surface brightness of their sample with that of the $V$-band sample by \citet{2012MNRAS.419..669M}, finding that $V - [3.6] \sim 1.5$\,mag (AB system, see their Section\,6). Considering that $V-R$ ranges $\sim 0.2$--0.5 in the discs of Sa--Sd galaxies \citep{2004A&A...415...63M} and $(V-R) = 0.5$-- 0.65 in those of S0s \citep{1989ApJS...69..217G}, we find that the limiting magnitudes of the \nir\ sample by L14 roughly correspond in the $R$ band to $\mu_\mathrm{lim}\sim 27.5$ for the spirals and $\mu_\mathrm{lim}\sim 25.5$ for the S0s (Vega system). Here, we have considered that $V(\mathrm{AB}) - V(\mathrm{Vega}) = 0.02$ \citep{2007AJ....133..734B}. Assuming that $\mu_\mathrm{lim} \sim 26.5$\,mag\,arcsec$^{-2}$ on average in the E08 and G11 samples, this means that the \nir\ data sample is $\sim 1$\,mag\,arcsec$^{-2}$ deeper than the $R$-band sample for the spirals. On the other hand, the E08 and G11 samples are $\sim 1$\,mag\,arcsec$^{-2}$ deeper than the L14 sample for the S0s. However, some profiles in E08 and G11 achieve $\mu_\mathrm{lim}\sim 28$\,mag\,arcsec$^{-2}$. So, the $R$-band sample may be reaching similar depths to the L14 sample in some specific cases.

E08 and G11 used the morphological types available in the RC3 catalog \citep{1991rc3..book.....D}, based on the optical morphology of the galaxies, whereas the types in L14 were assigned according to the morphology in their $K$ or \nir\ images \citep[][]{2011MNRAS.418.1452L,2015arXiv150100454B}. E08 considered as barred galaxies those exhibiting strong (SB) or weak (SAB) bars according to the RC3 classification, but revised the classes according to their deep $R$ band images and rejected the galaxies without clear bars in them or involved in strong interactions. The barred/unbarred classification in L14 was, however, made on the basis of their deep $K$ and \nir\ images from the NIRS0S and S$^4$G surveys.

The surface brightness profiles were obtained by azimuthally averaging the light within ellipses fitted to the isophotes of the galaxies. The three studies (E08, G11, and L14) held the values of the centre, ellipticity, and position angle of isophotes fixed to the values of the outer discs in the fits. 

E08 and L14 fitted the disc profiles using ''broken-exponential'' functions, which describe the inner and outer discs through two exponentially-decaying profiles joined by a transition region, according to the following expression:

\begin{equation} \label{eq:broken}
 I(r) = S\, I_0\, \exp\left[\frac{-r}{\hi}\right]\,\lbrace 1 + \exp\left[ \alpha\,(r - \rbreak) \right] \rbrace^{\frac{1}{\alpha}\,(\frac{1}{\hi} - \frac{1}{\ho})},
\end{equation}

\noindent where $I_0$ represents the central intensity of the inner exponential section, $\alpha$ parameterizes the sharpness of the break, and $S$ is a scaling factor, given by

\begin{equation} \label{eq:Sbroken}
S = \left[ 1 + \exp(-\alpha\,\rbreak)\right]^{\frac{1}{\alpha}\,(\frac{1}{\hi} - \frac{1}{\ho})}.
\end{equation}

\noindent On the other hand, G11 performed independent exponential fits to the inner and outer discs (''piecewise fits''), defining \rbreak\ as the radius at which the fitted profiles cross. The surface brightness of the profile at $r = \rbreak$ is defined as the break surface brightness (\mubreak). E08 showed that the two fitting procedures provide very similar results (within 5\% for the characteristic scalelengths in case of Type-III profiles). This allows the comparison of the characteristic parameters of the samples by E08 and G11 in the $R$ band. 

Consequently, we have used the characteristic parameters of the breaks and the inner and outer discs of Type-III galaxies derived by E08, G11, and L14 to compare the trends of S0 and spiral types in several photometric planes, in the $R$ and \nir\ bands. We remark that the magnitudes of the $R$-band data are the Vega system and in AB for the \nir\ dataset.


 \begin{figure*}[th!]
\begin{tabular}{cc}
\framebox[0.48\textwidth][c]{Trends with \rbreak\ and \rbreak$/$\risoph\  in $R$ (S0s/spirals)} &   \framebox[0.48\textwidth][c]{Trends with \rbreak\ and \rbreak$/$\risoph\  in \nir\ (S0s/spirals)} \\
   \imagetop{
 \begin{minipage}{.48\textwidth}
\centering
   \includegraphics[width = 0.48\textwidth,bb=-35 -15 472 425, clip]{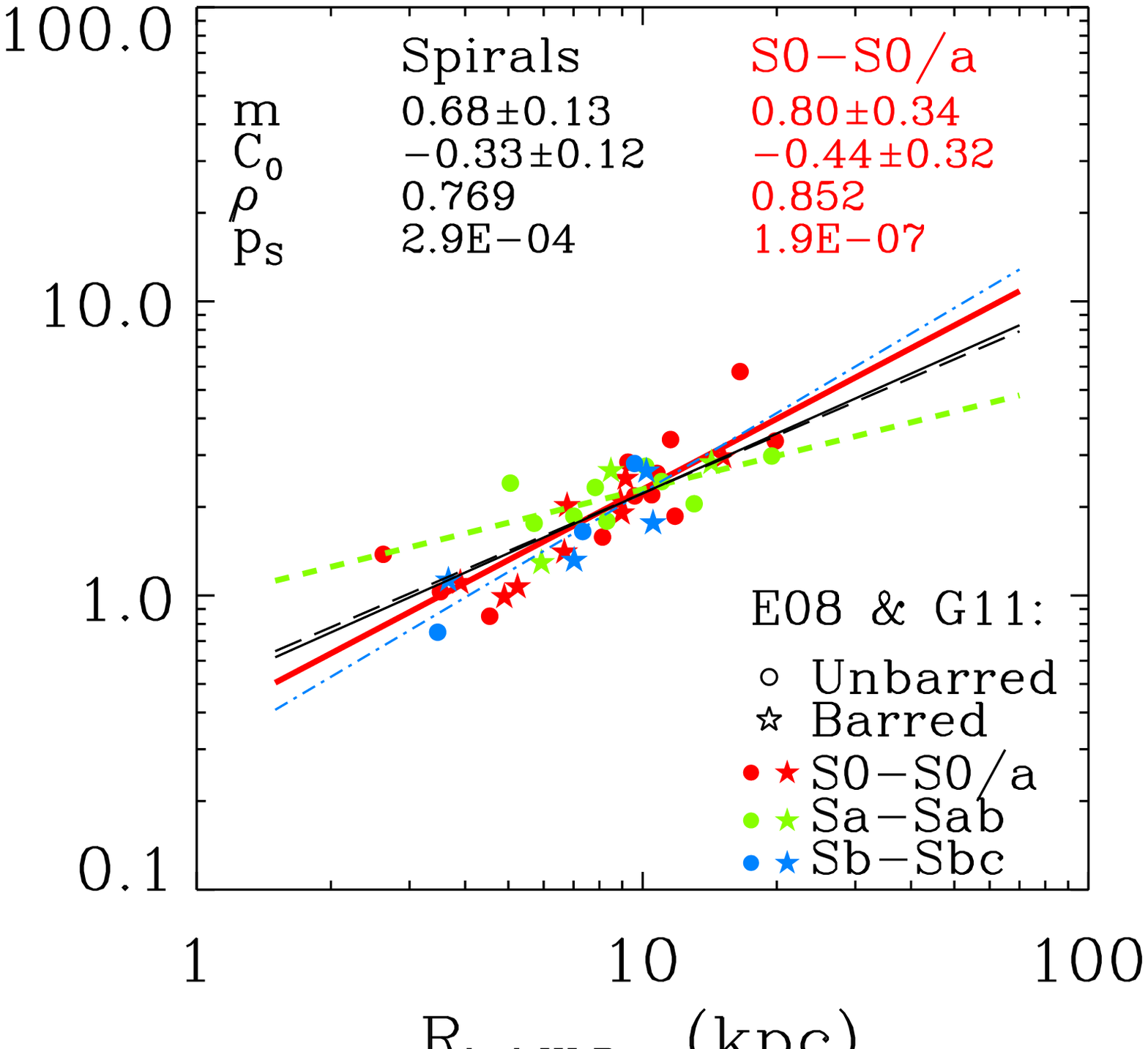}
   \includegraphics[width = 0.48\textwidth,bb=-35 -15 472 425, clip]{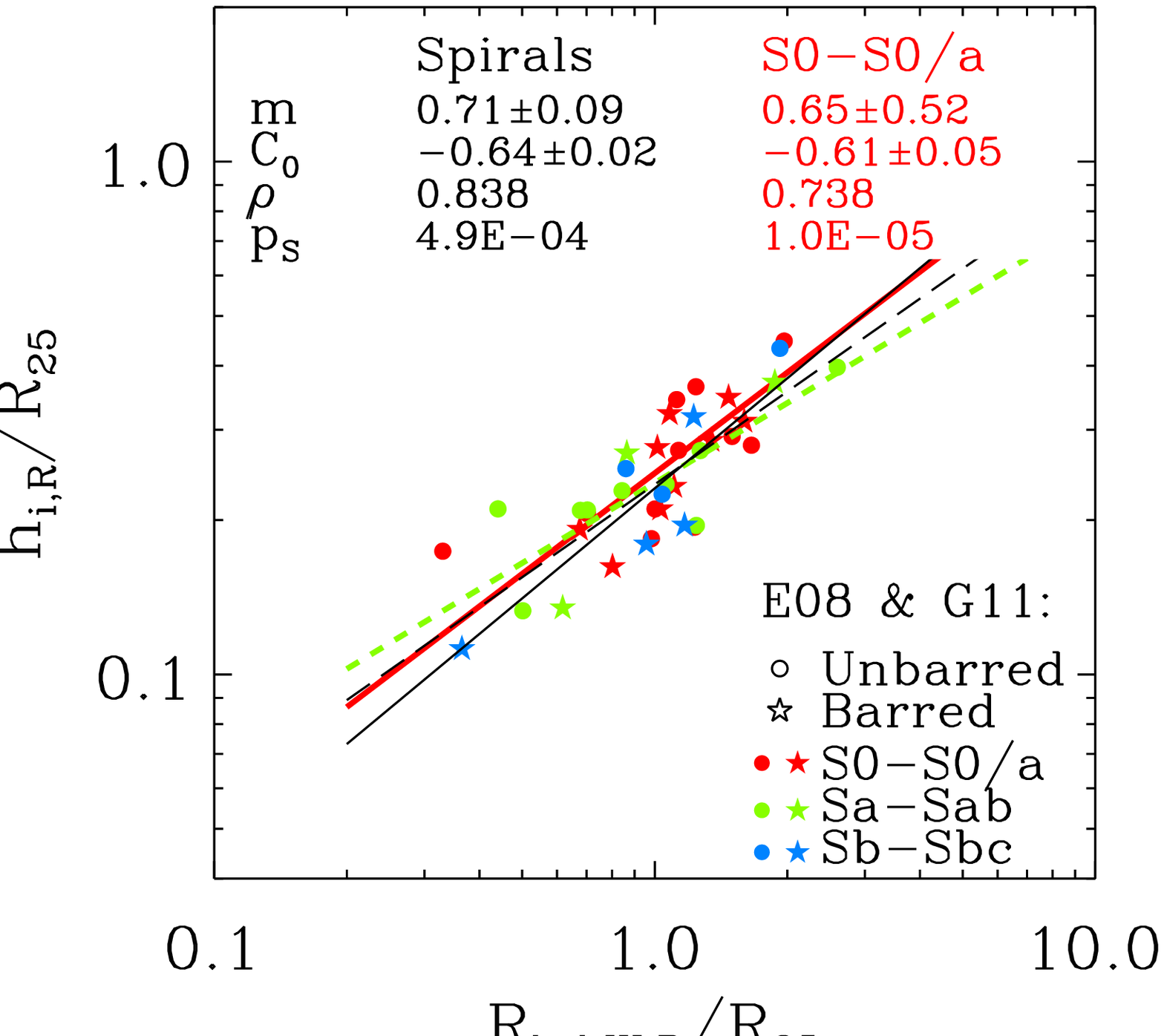}
   \includegraphics[width = 0.48\textwidth,bb=-35 -15 472 425, clip]{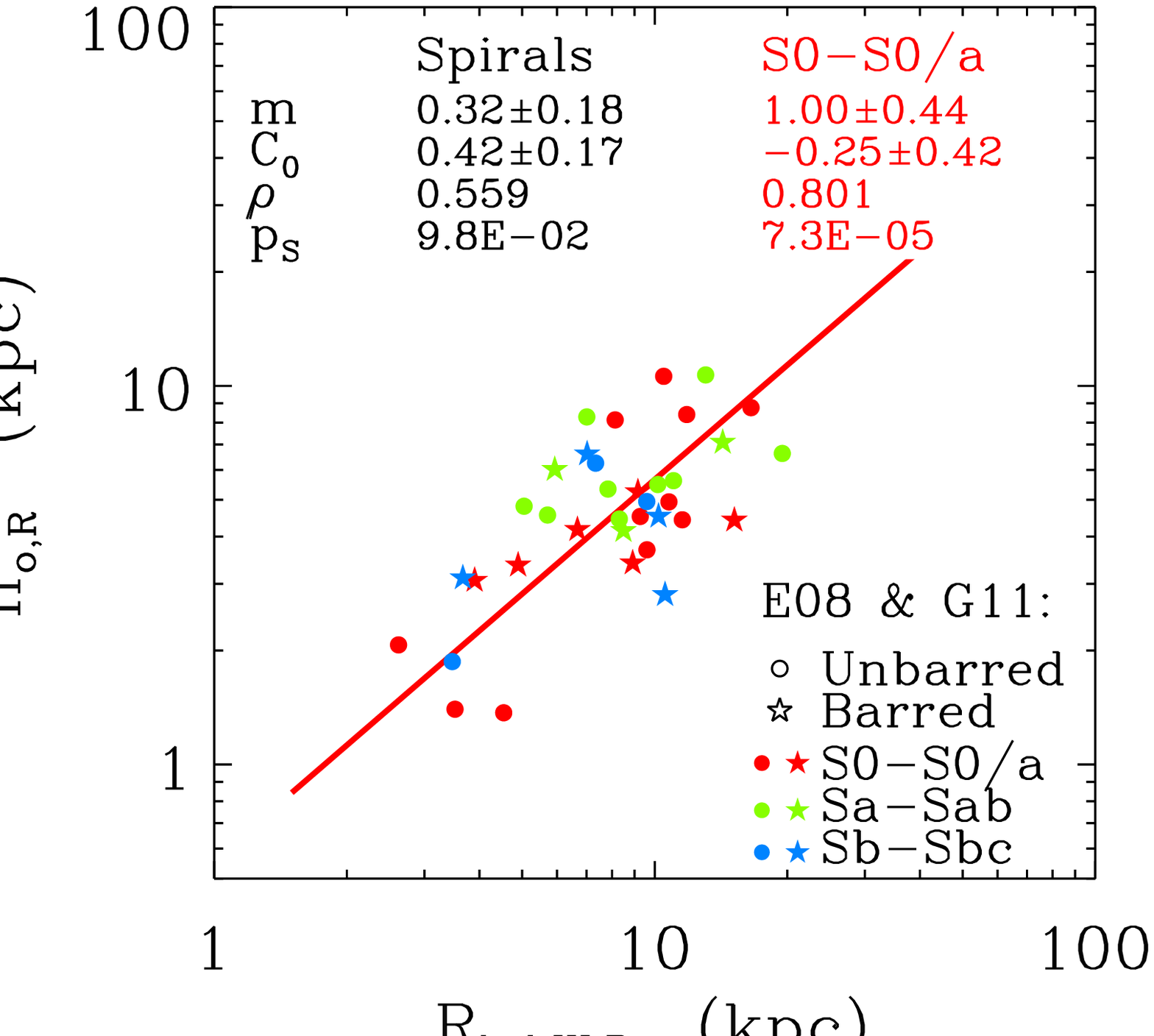}
   \includegraphics[width = 0.48\textwidth,bb=-35 -15 472 425, clip]{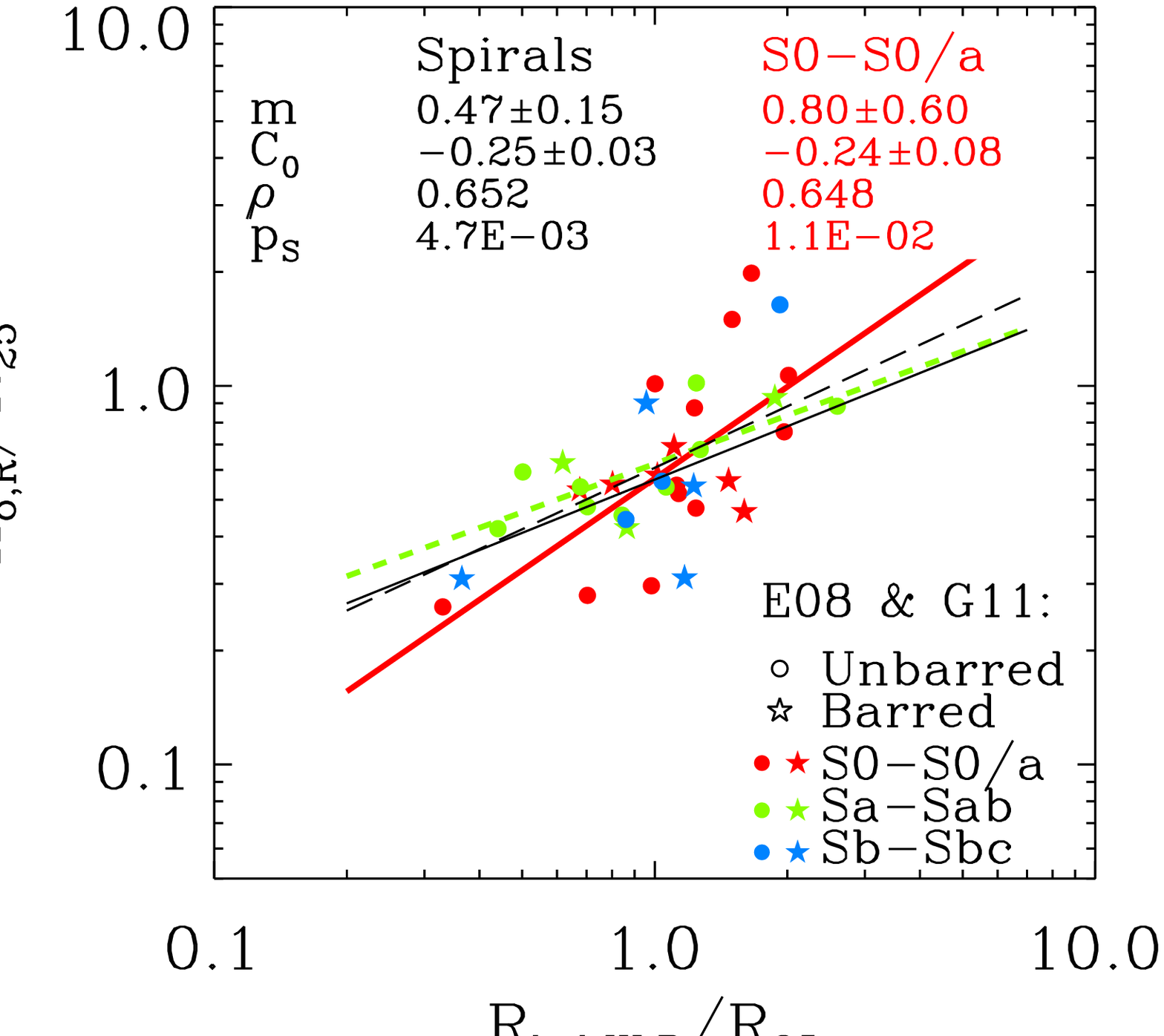}
   \includegraphics[width = 0.48\textwidth,bb=-35 -15 472 425, clip]{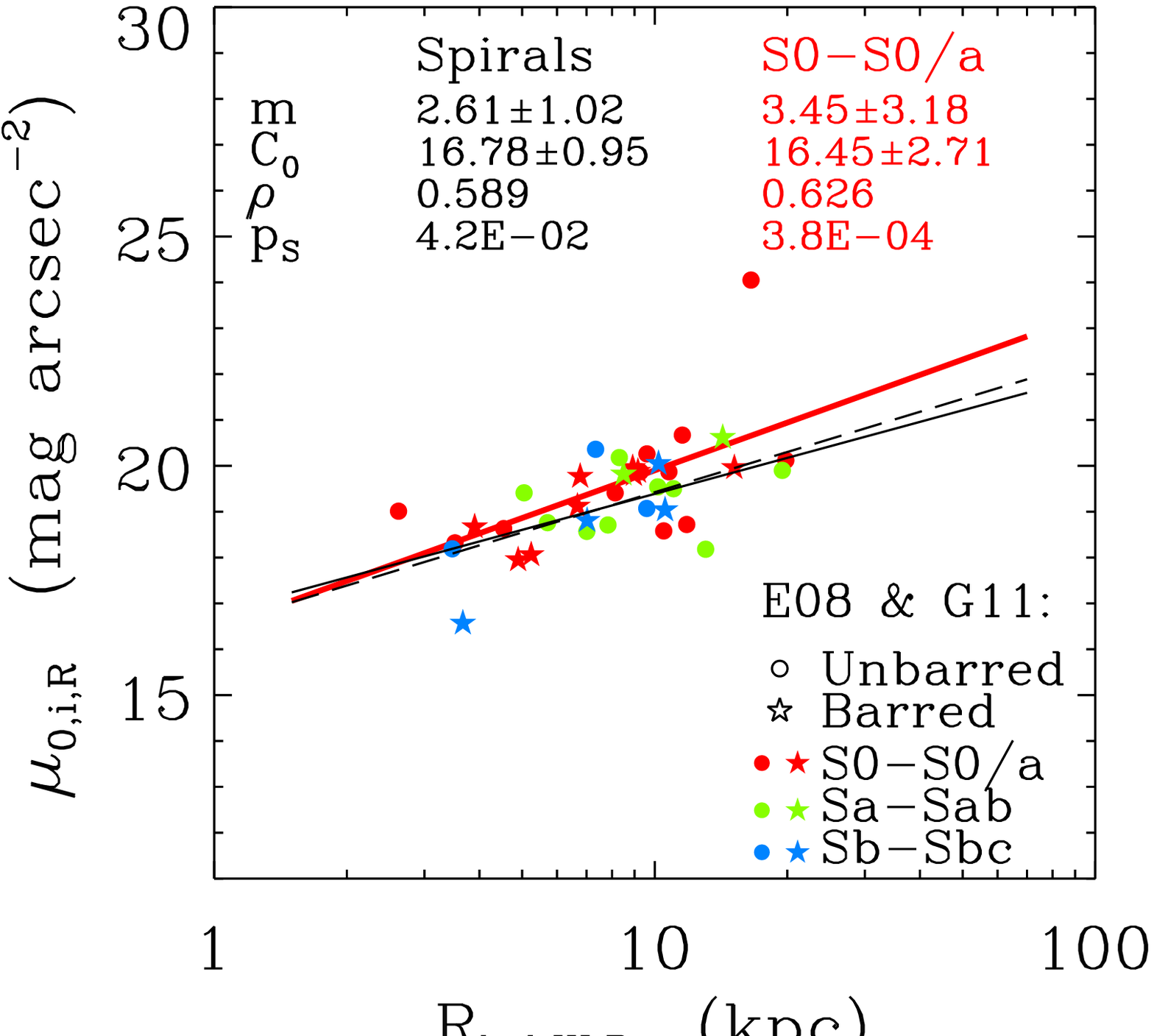}
   \includegraphics[width = 0.48\textwidth,bb=-35 -15 472 425, clip]{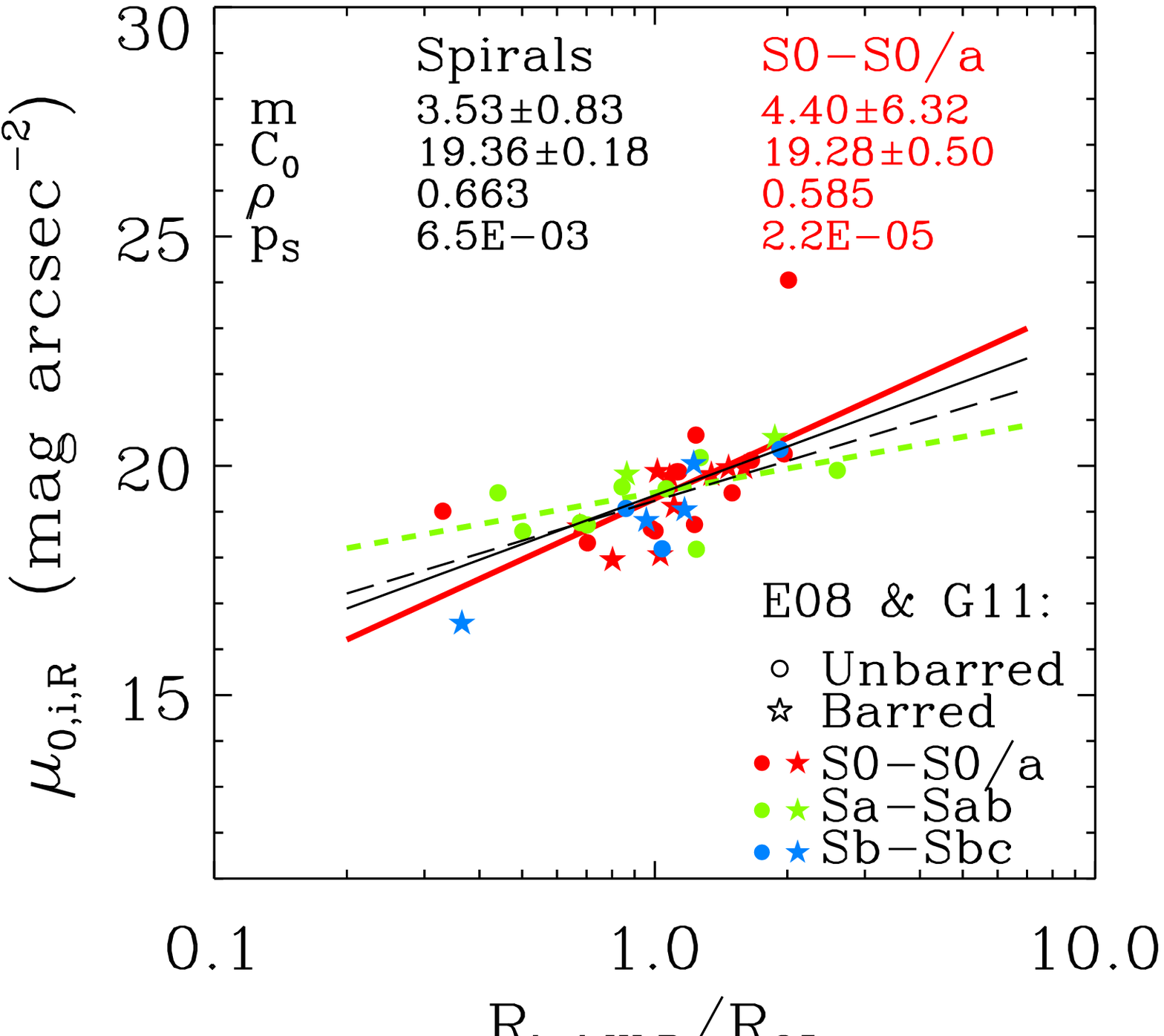}
   \includegraphics[width = 0.48\textwidth,bb=-35 -15 472 425, clip]{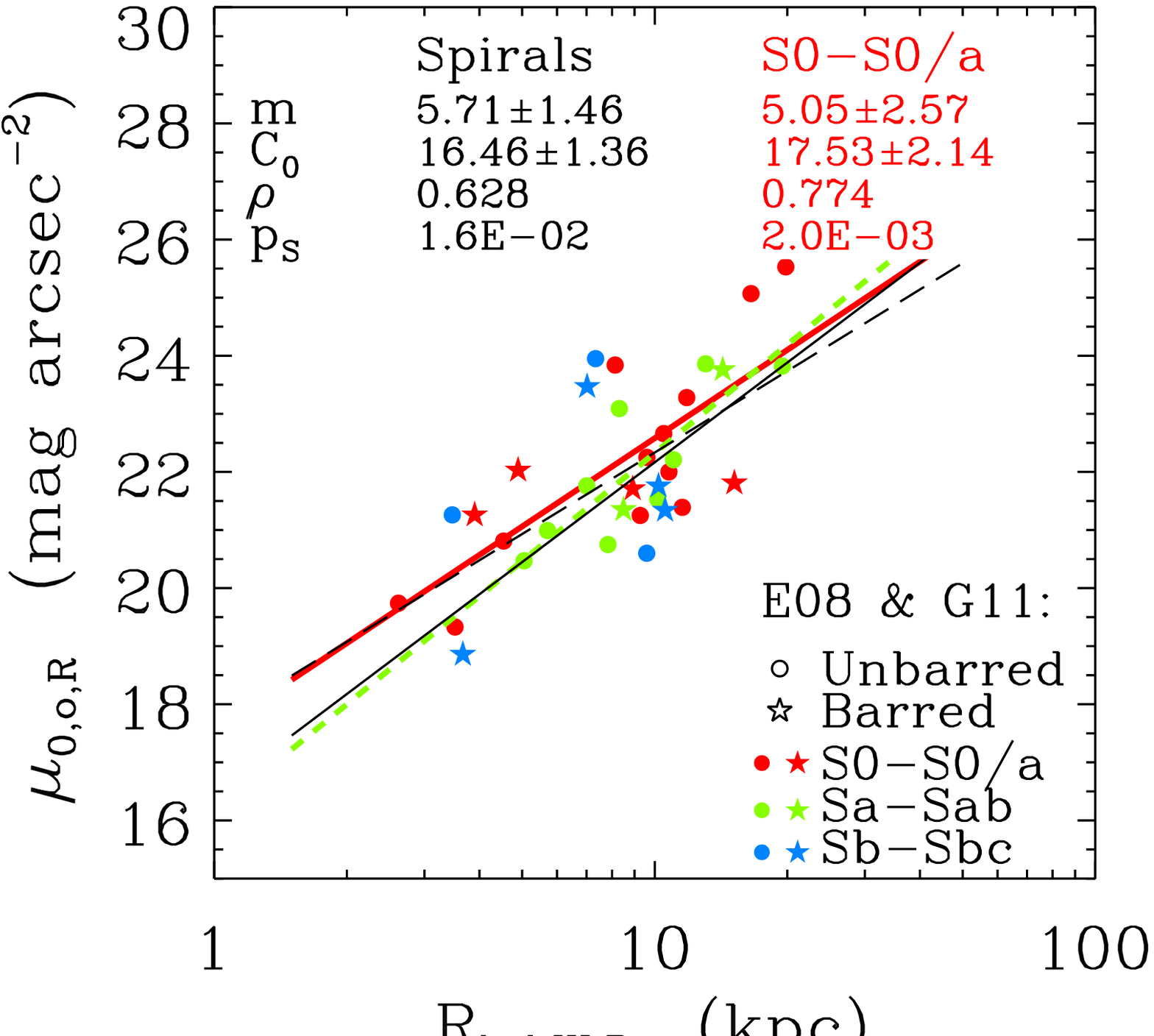}
   \includegraphics[width = 0.48\textwidth,bb=-35 -15 472 425, clip]{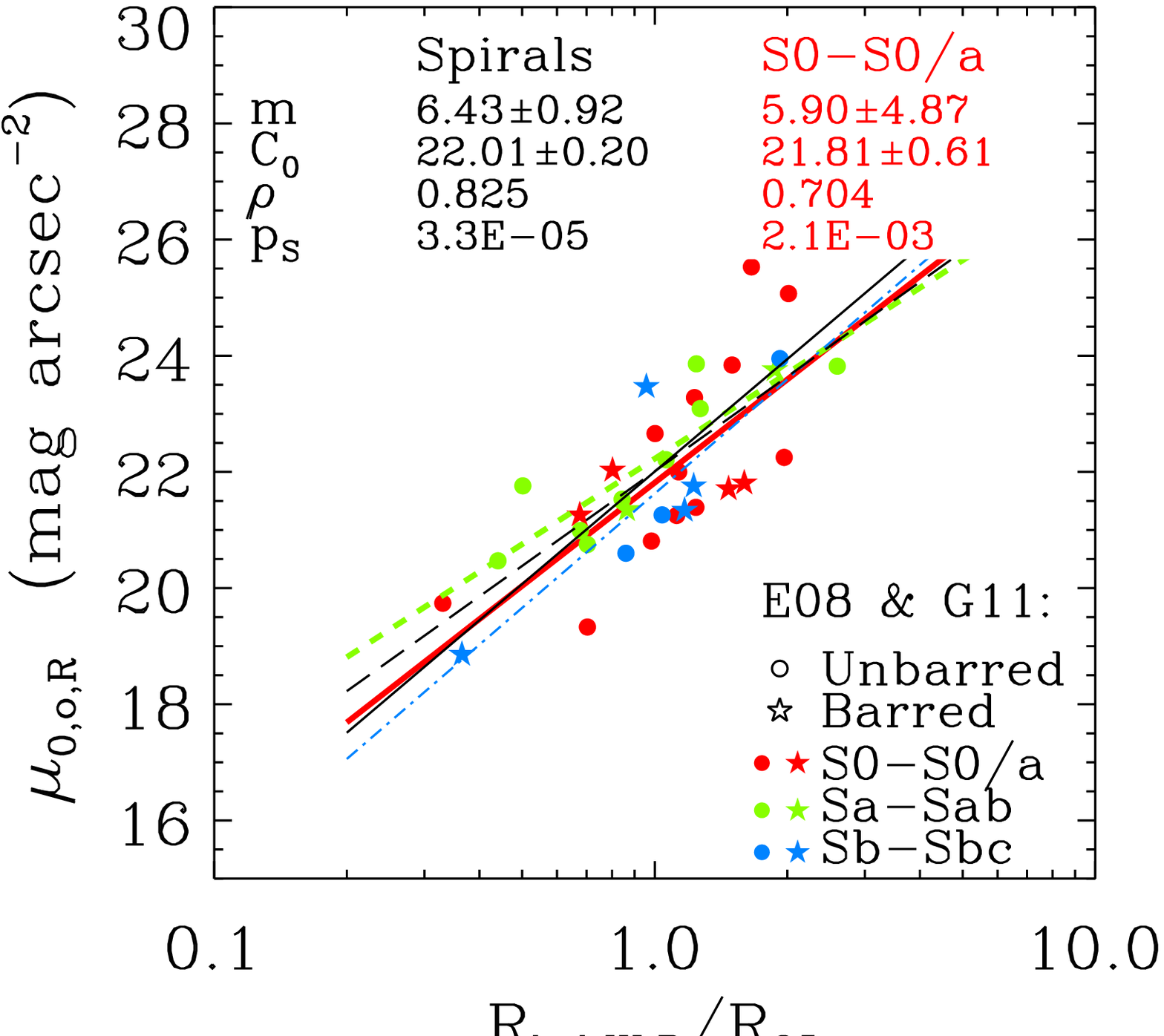}
   \includegraphics[width = 0.48\textwidth,bb=-35 -15 472 425, clip]{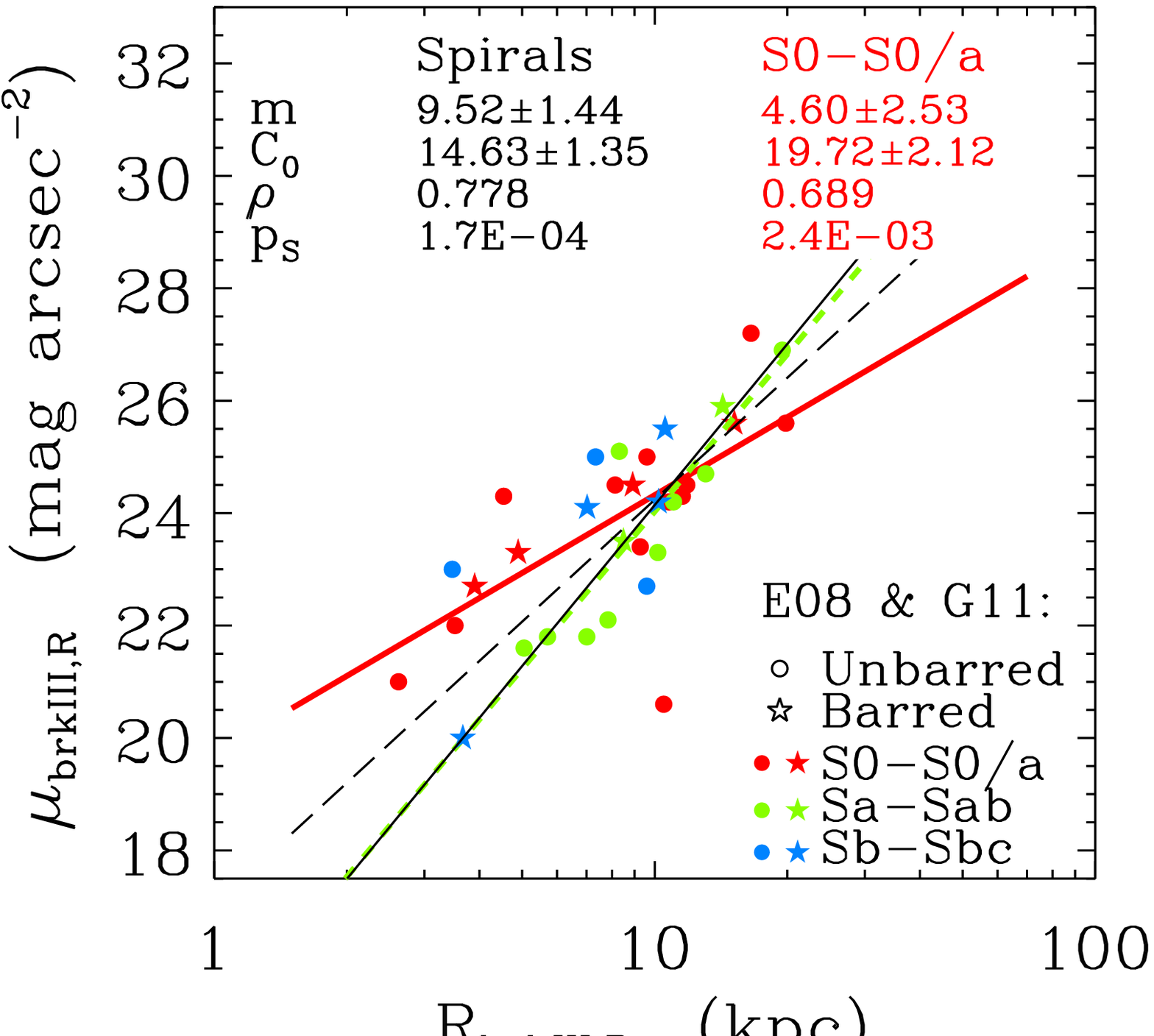}
   \includegraphics[width = 0.48\textwidth,bb=-35 -15 472 425, clip]{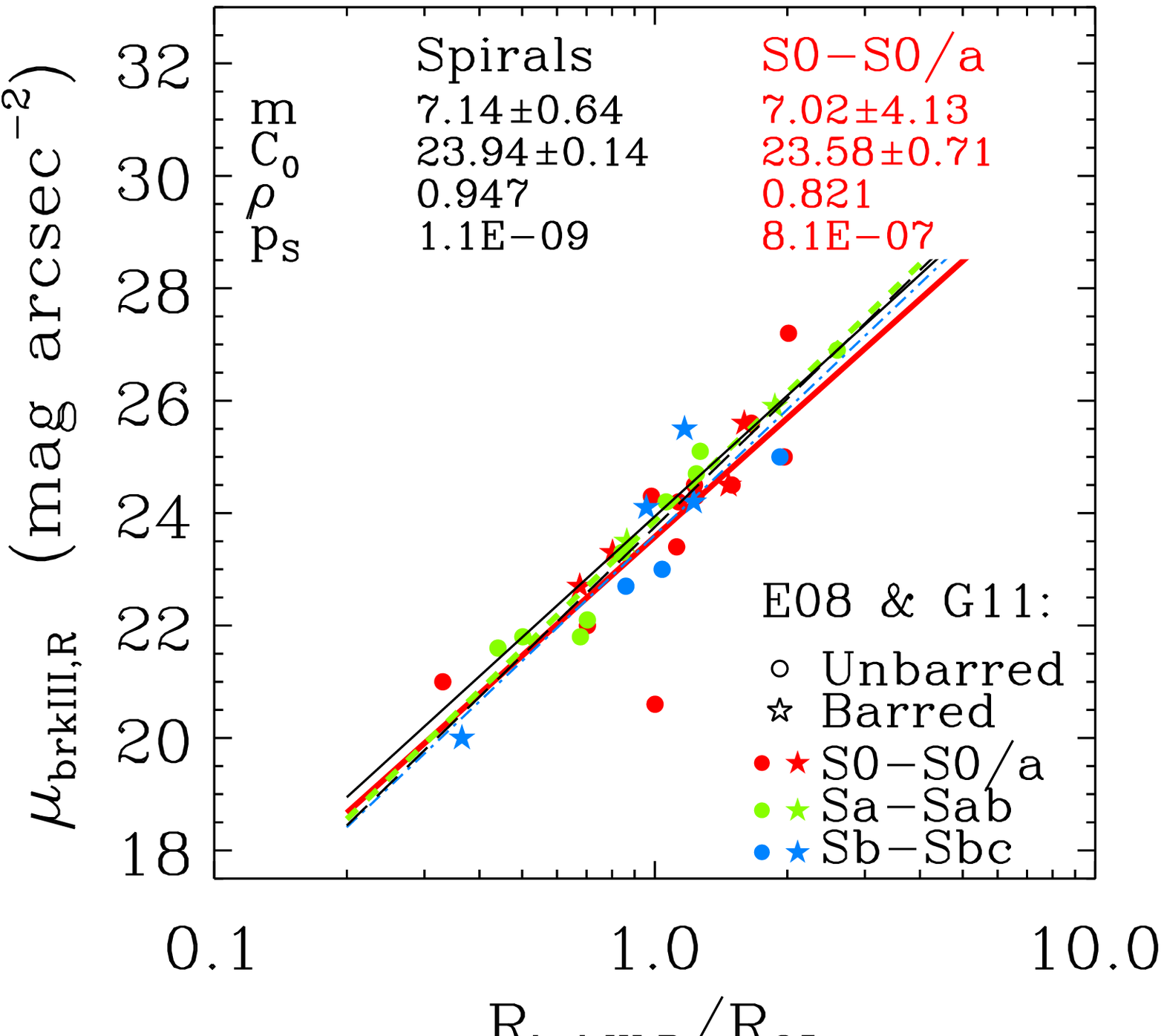}
  \caption{Trends of the photometric parameters of the break and the inner and outer discs with \rbreak\ for the local antitruncated S0--Sbc galaxies in the $R$ band from the E08 and G11 samples (see Tables\,\ref{tab:hihorbreak}-\ref{tab:mubreakrbreak}). \emph{Left}: trends with \rbreak. \emph{Right}: trends with \rbreak$/$\risoph. The linear fits performed to each galaxy type are overplotted only if they are significant (\emph{red thick solid line}: S0--S0/a, \emph{grey thin solid line}: all spirals, \emph{green dashed line}: Sa-Sab, \emph{blue dashed-dotted line}: Sb--Sbc).  The results of the linear fits performed for the spirals and S0s are indicated at the top of each panel. The errors of the fits shown in the panels have been symmetrized for simplicity (the results are available in the corresponding Tables). See the legend in the panels.}
 \label{fig:withRbreak_R}
\end{minipage}
}
& 
\imagetop{
 \begin{minipage}{.48\textwidth}
\centering
   \includegraphics[width = 0.48\textwidth,bb=-35 -15 472 425, clip]{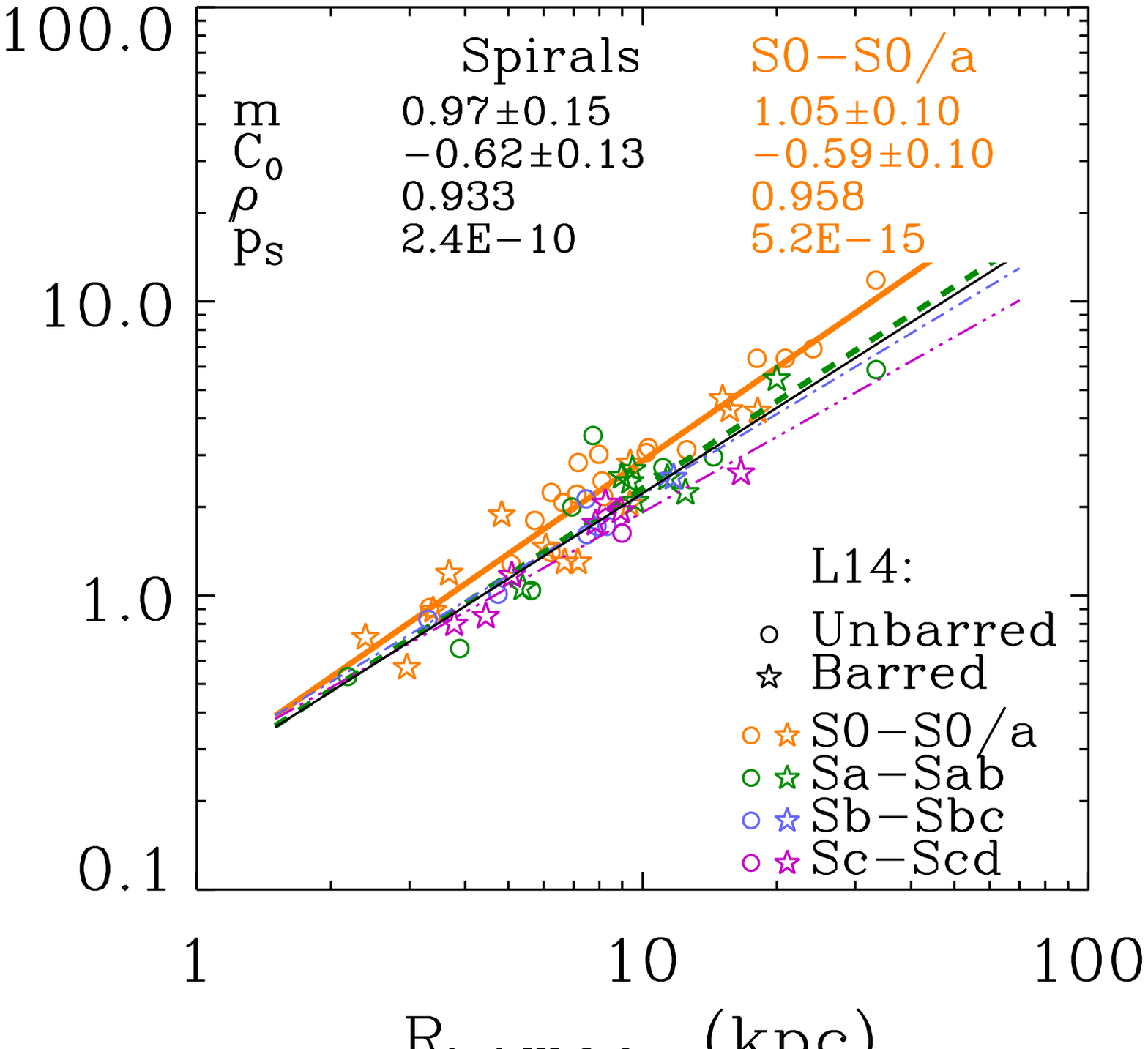}
   \includegraphics[width = 0.48\textwidth,bb=-35 -15 472 425, clip]{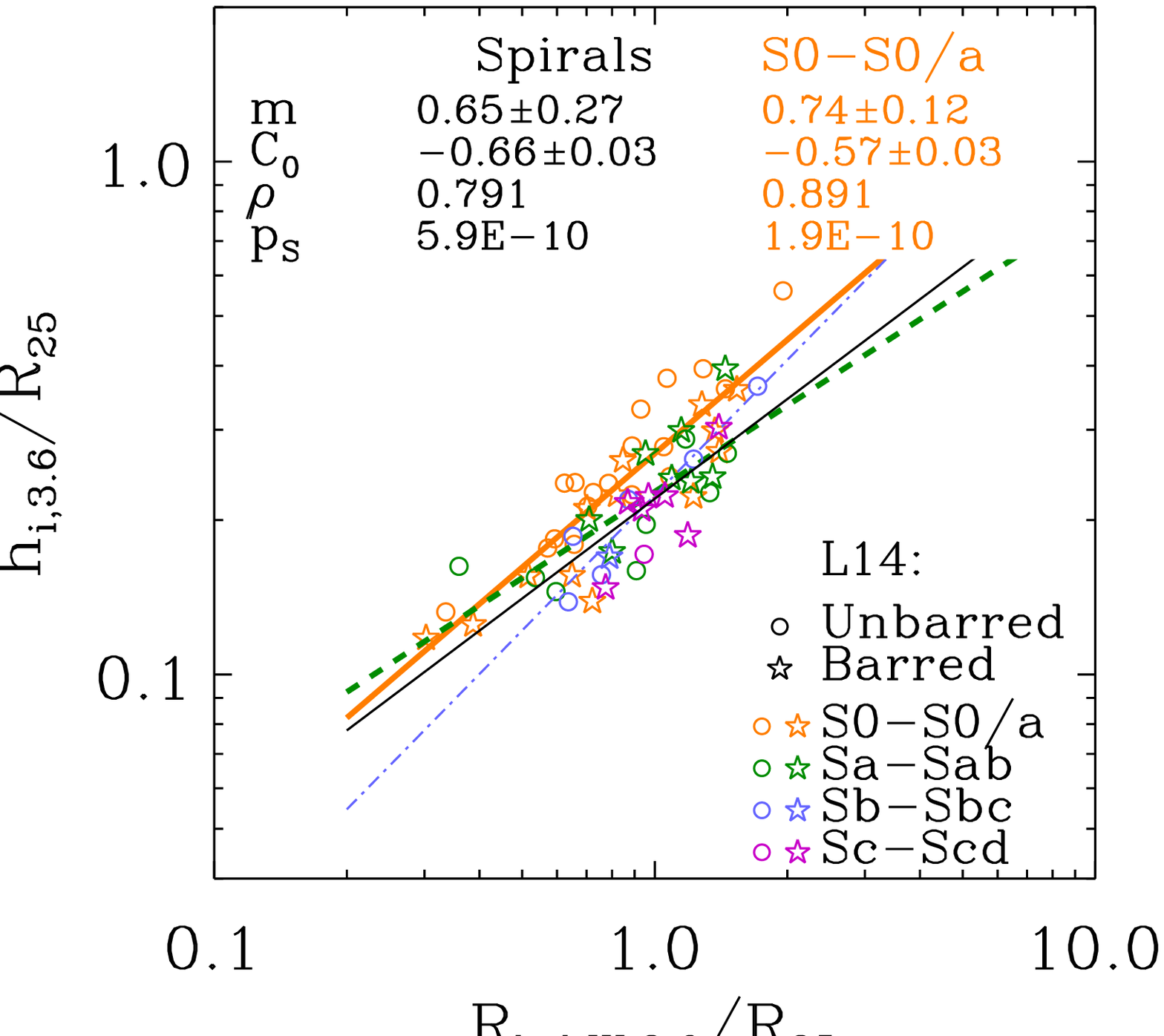}
   \includegraphics[width = 0.48\textwidth,bb=-35 -15 472 425, clip]{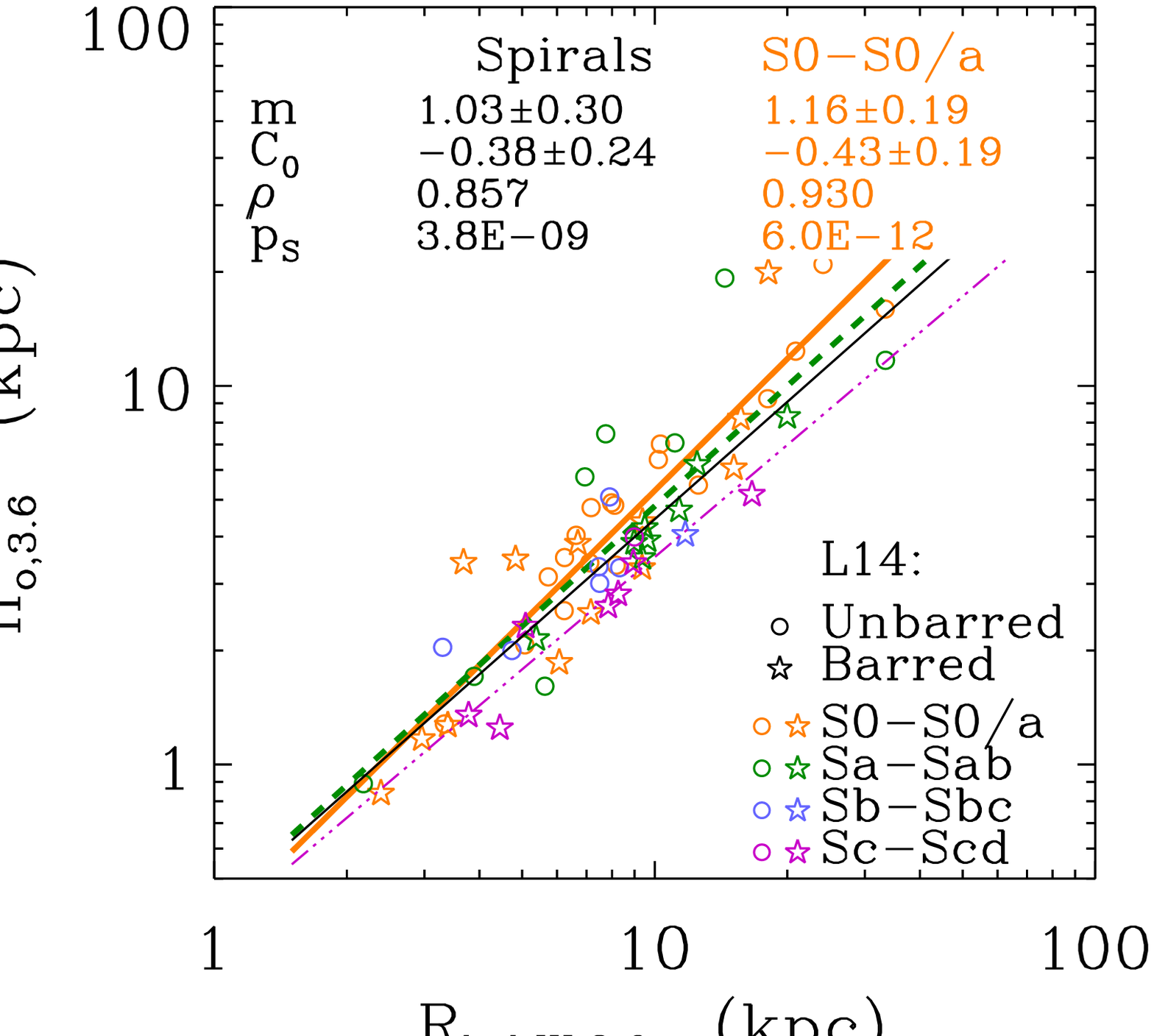}
   \includegraphics[width = 0.48\textwidth,bb=-35 -15 472 425, clip]{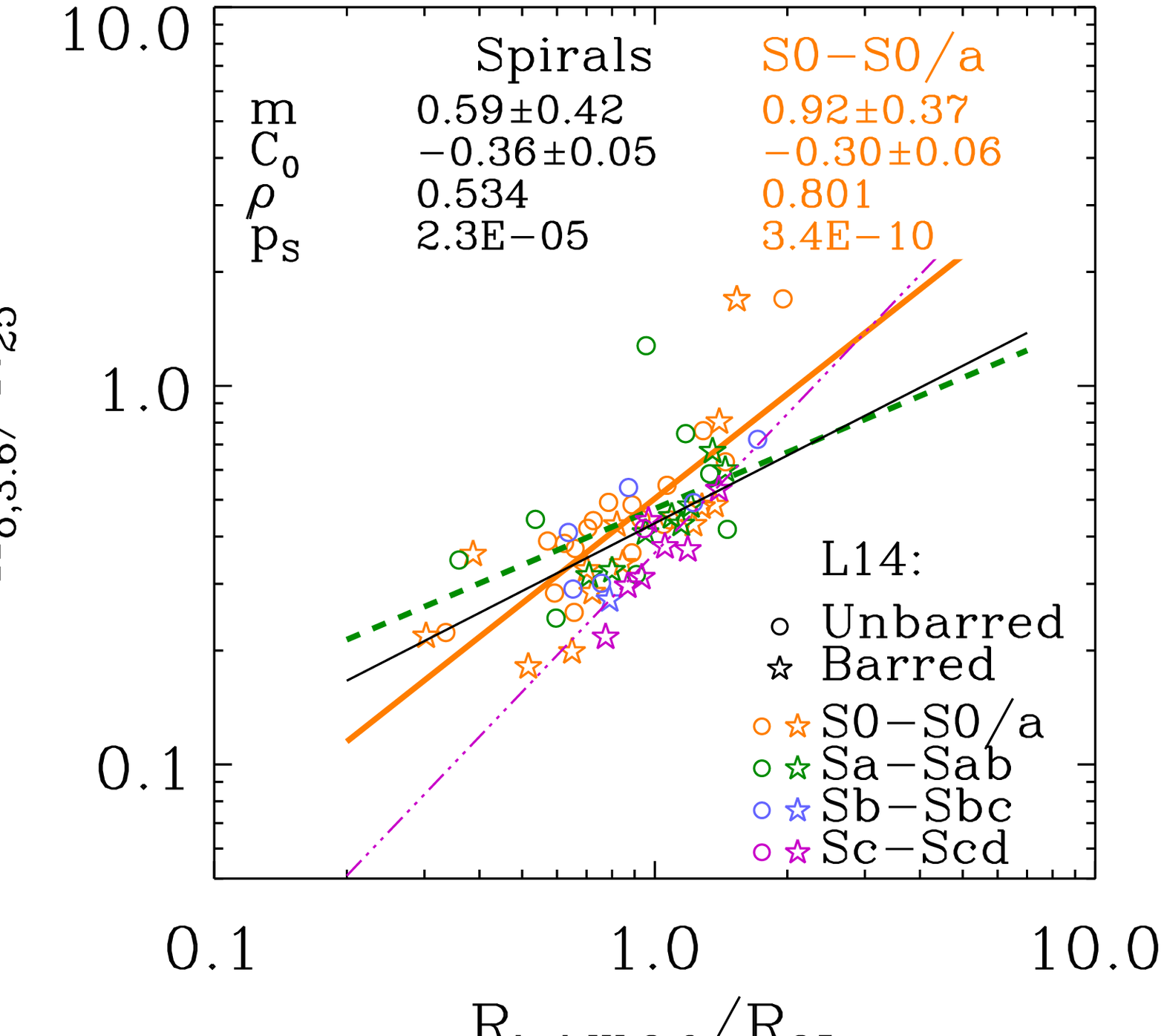}
   \includegraphics[width = 0.48\textwidth,bb=-35 -15 472 425, clip]{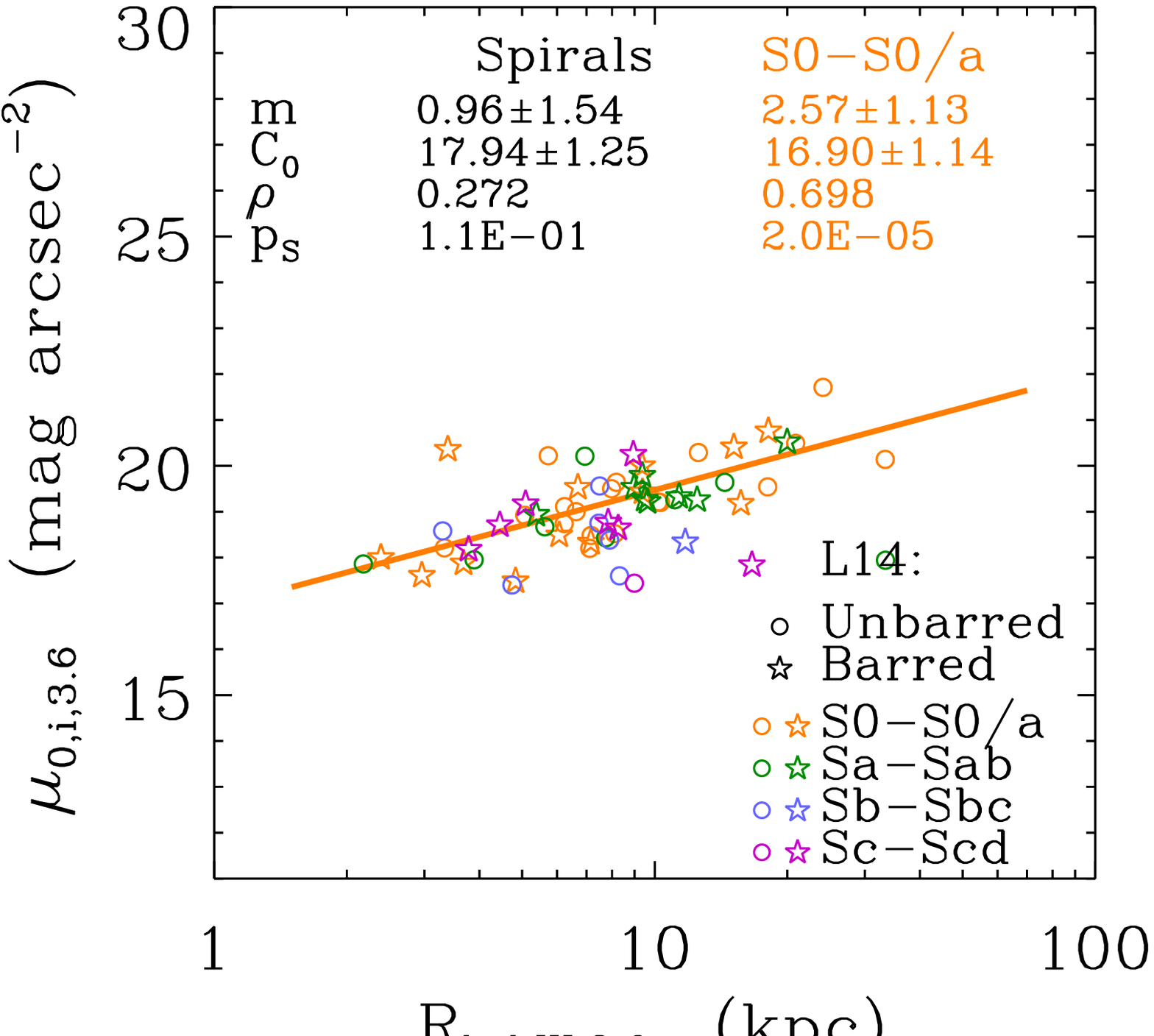}
   \includegraphics[width = 0.48\textwidth,bb=-35 -15 472 425, clip]{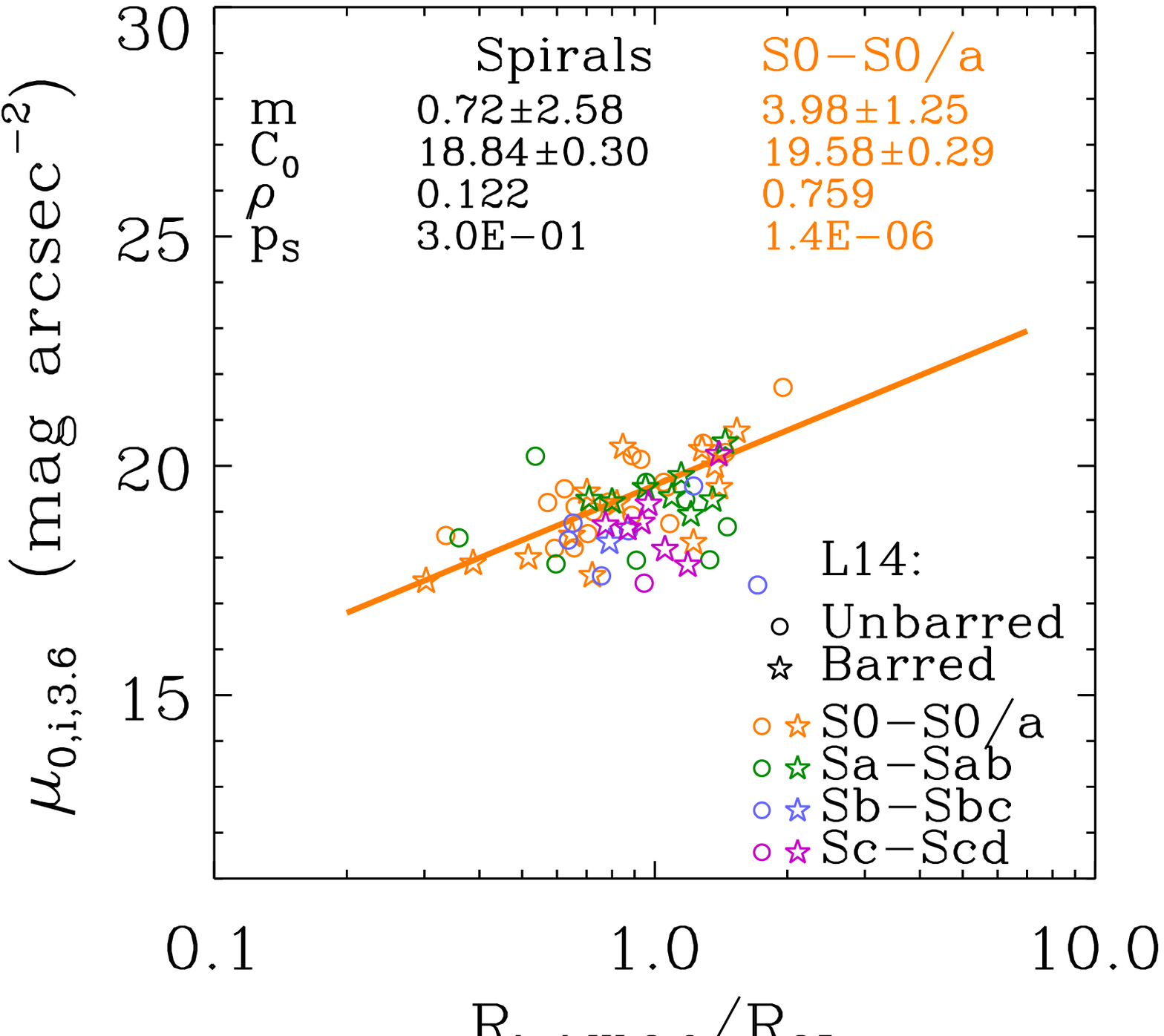}
   \includegraphics[width = 0.48\textwidth,bb=-35 -15 472 425, clip]{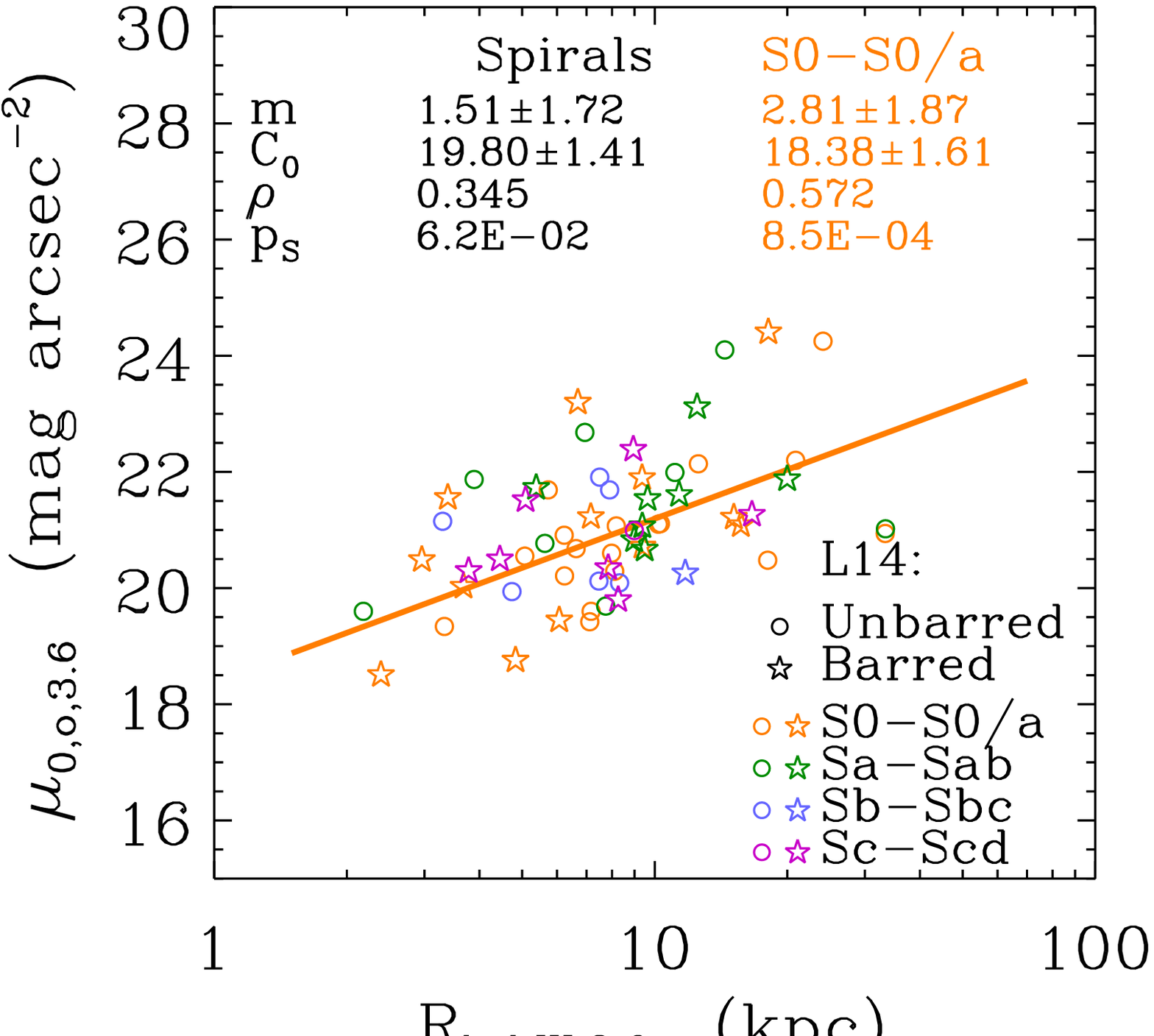}
   \includegraphics[width = 0.48\textwidth,bb=-35 -15 472 425, clip]{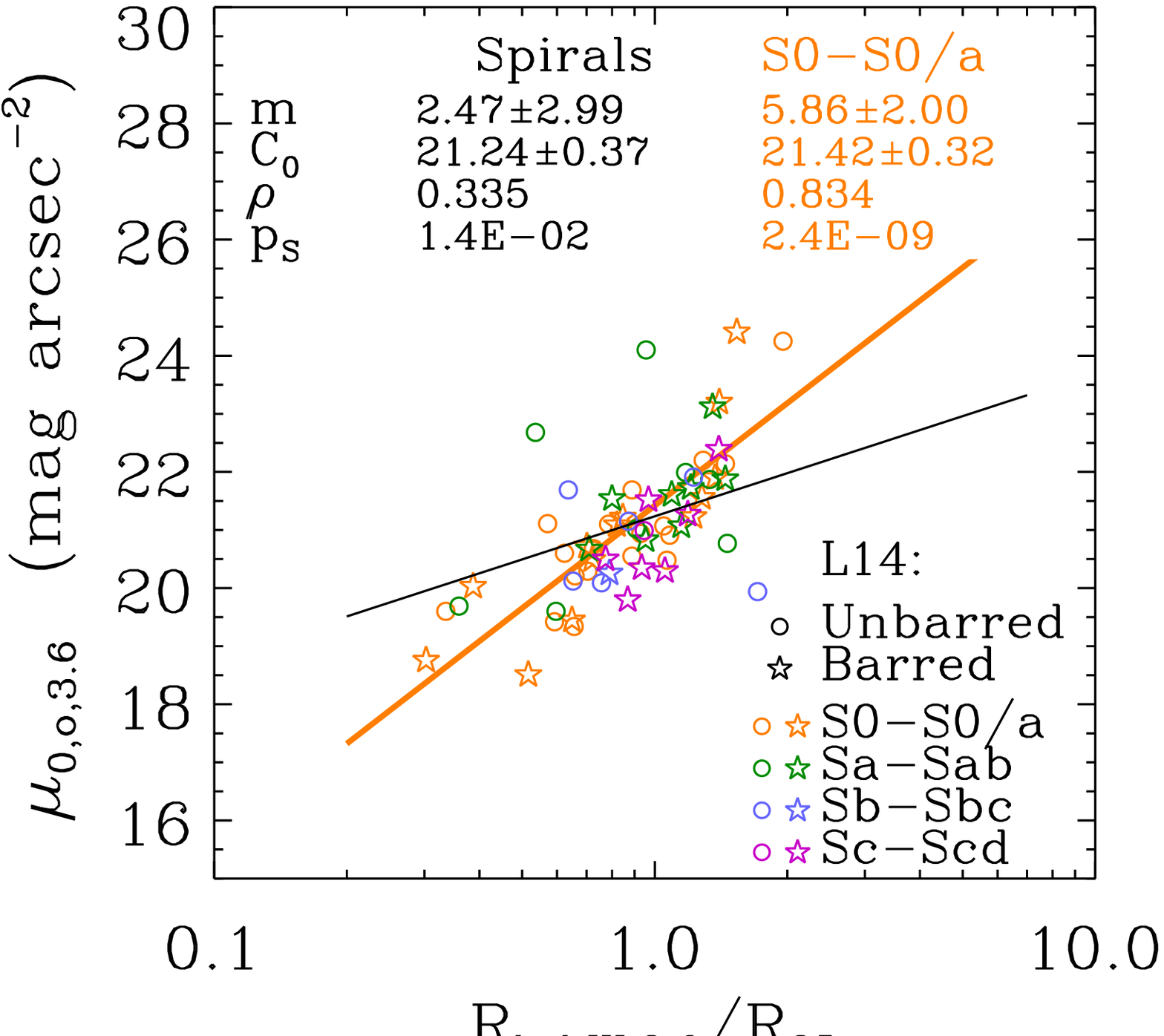}
   \includegraphics[width = 0.48\textwidth,bb=-35 -15 472 425, clip]{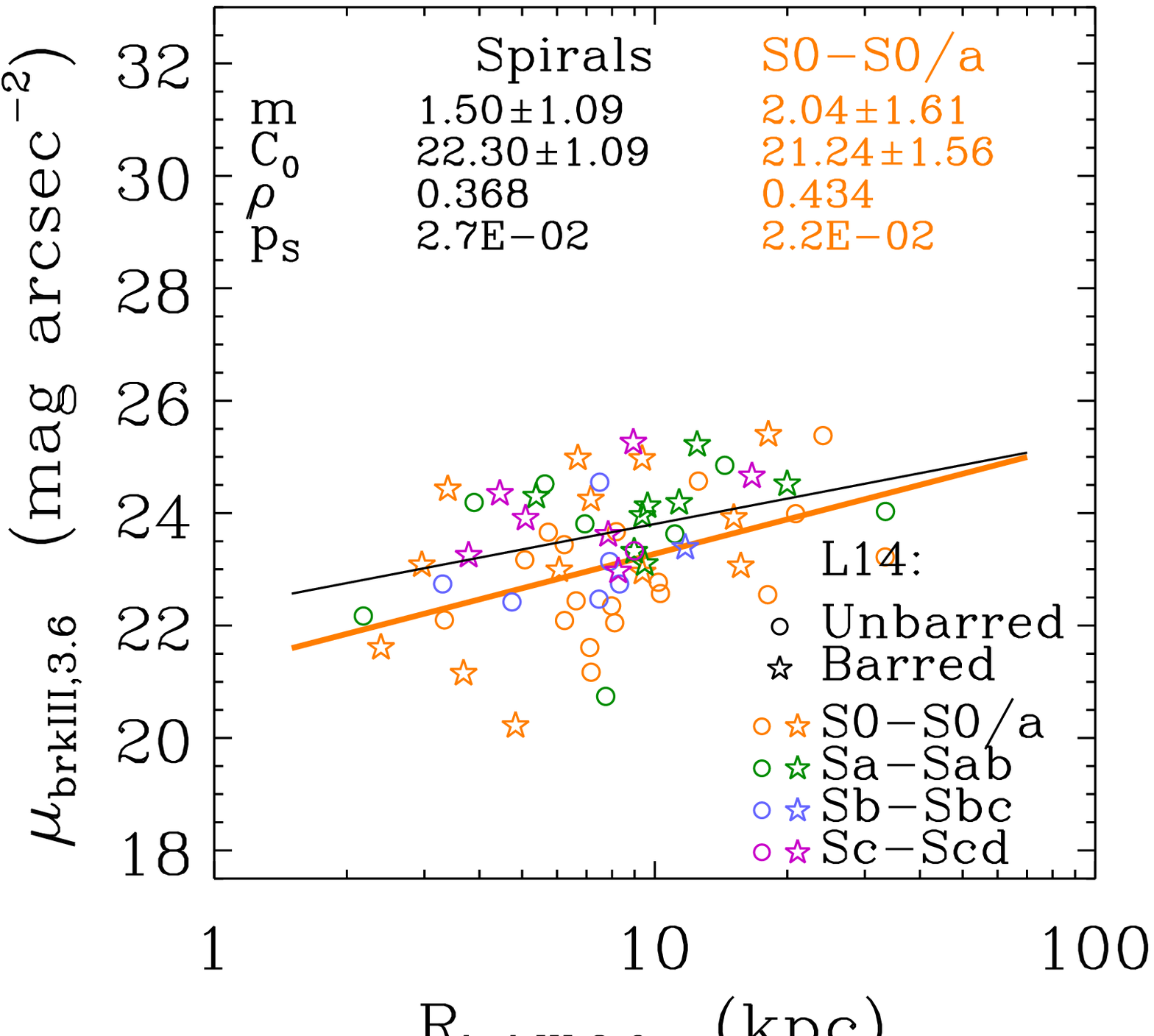}
   \includegraphics[width = 0.48\textwidth,bb=-35 -15 472 425, clip]{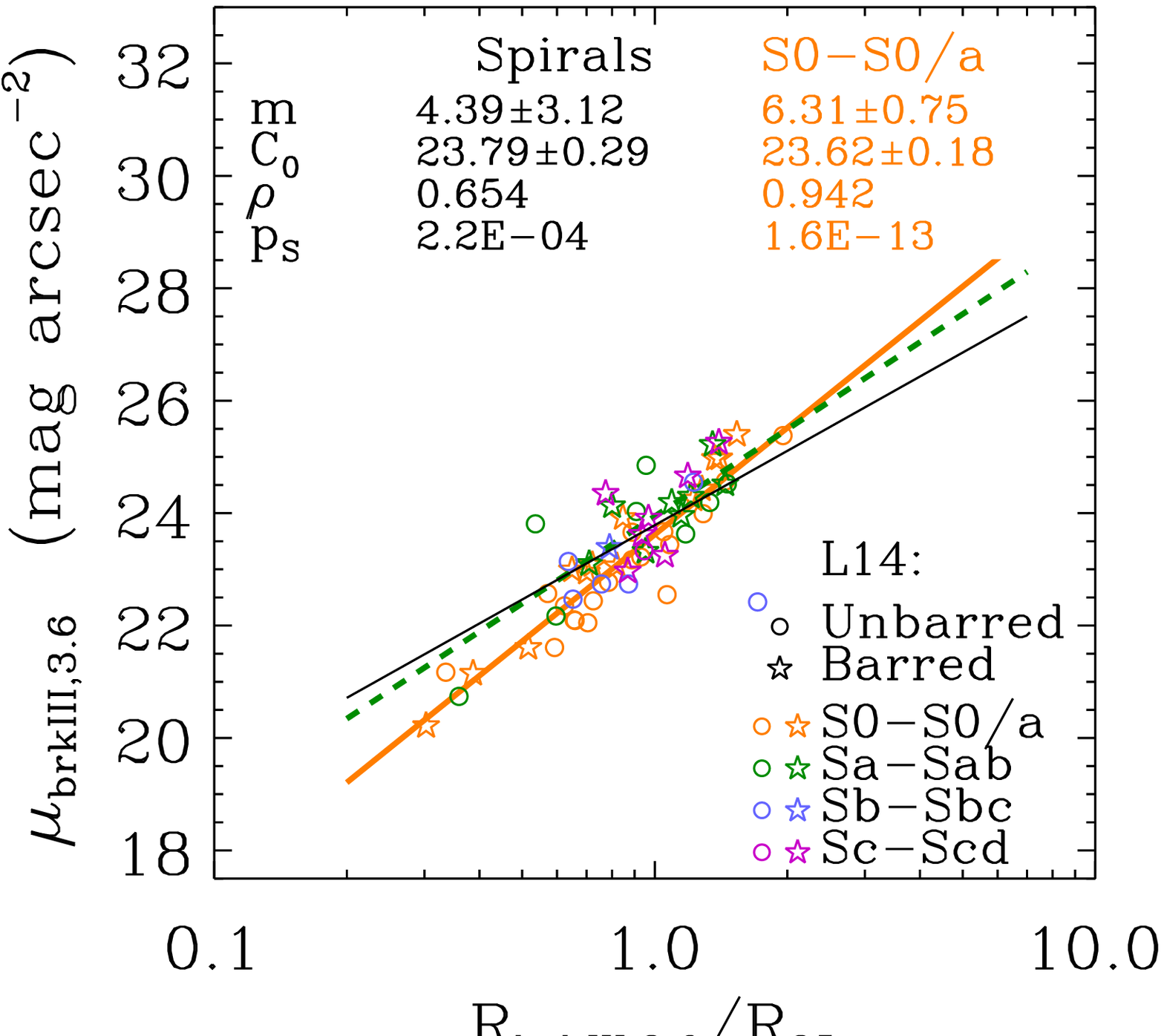}
\caption{The same as Fig.\,\ref{fig:withRbreak_R}, but for local antitruncated S0--Scd galaxies in the \nir\ band from the L14 sample (see Tables\,\ref{tab:hihorbreak}-\ref{tab:mubreakrbreak}). \emph{Left}: trends with \rbreak. \emph{Right}: trends with \rbreak$/$\risoph. The linear fits performed to each galaxy type are overplotted only if they are significant (\emph{orange thick solid line}: S0--S0/a, \emph{grey thin solid line}: all spirals, \emph{green dashed line}: Sa-Sab, \emph{blue dashed-dotted line}: Sb--Sbc, \emph{purple three dotted-dashed line}: Sc-Scd).  The results of the linear fits performed for the spirals and S0s are indicated at the top of each panel. The errors of the fits shown in the panels have been symmetrized for simplicity (see the results in the corresponding Tables). See the legend in the panels.}
 \label{fig:withRbreak_3.6}
\end{minipage}
}
\end{tabular}
\end{figure*}


\begin{figure*}[th!]
\begin{tabular}{cc} 
\framebox[0.48\textwidth][c]{Trends with \rbreak\ and \rbreak$/$\risoph\ in $R$ (barred/unbarred)} &   \framebox[0.48\textwidth][c]{Trends with \rbreak\ and \rbreak$/$\risoph\ in \nir\ (barred/unbarred)} \\
   \imagetop{
 \begin{minipage}{.48\textwidth}
\centering
   \includegraphics[width = 0.48\textwidth,bb=-35 -15 472 425, clip]{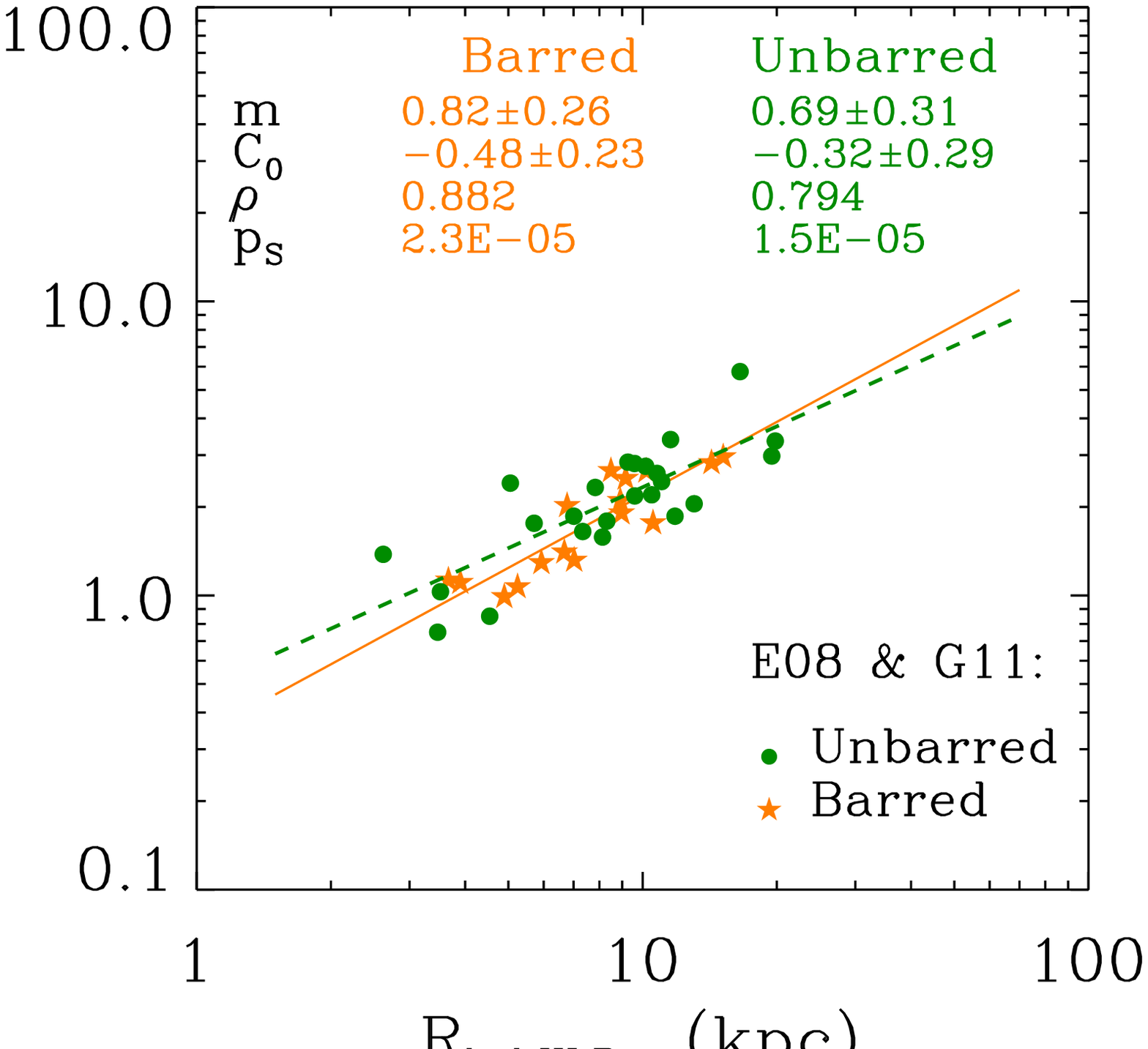}
   \includegraphics[width = 0.48\textwidth,bb=-35 -15 472 425, clip]{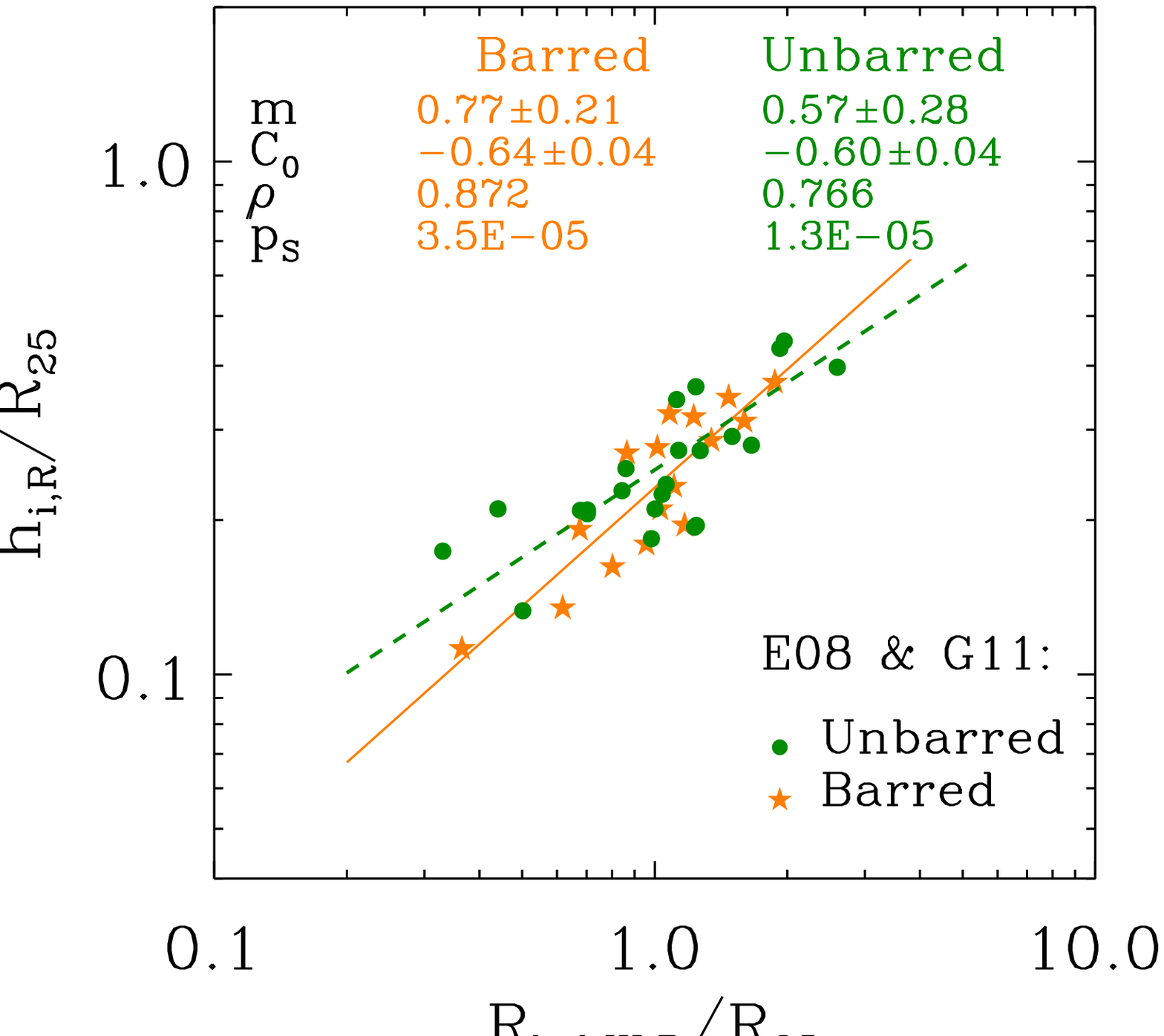}
   \includegraphics[width = 0.48\textwidth,bb=-35 -15 472 425, clip]{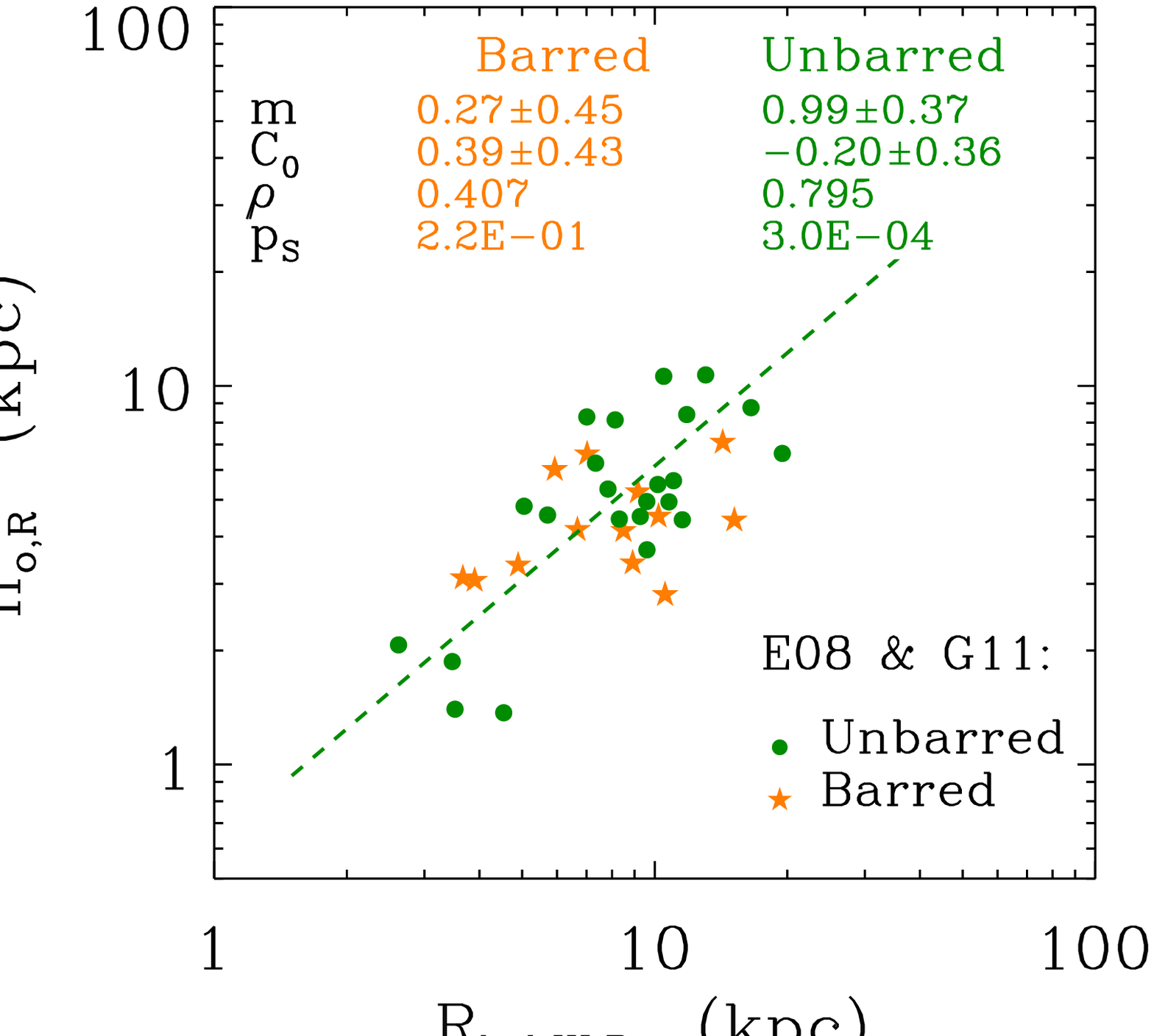}
   \includegraphics[width = 0.48\textwidth,bb=-35 -15 472 425, clip]{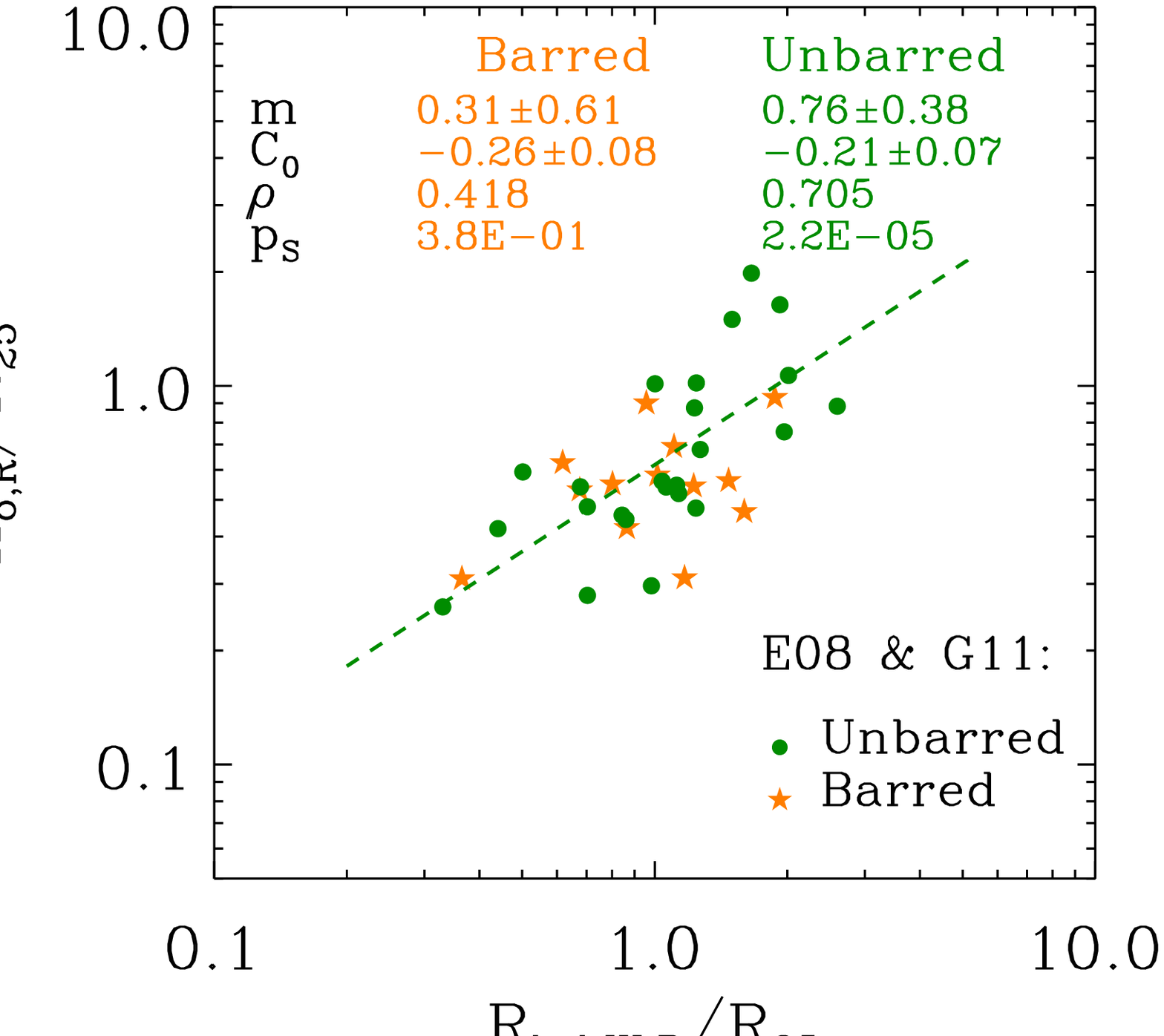}
   \includegraphics[width = 0.48\textwidth,bb=-35 -15 472 425, clip]{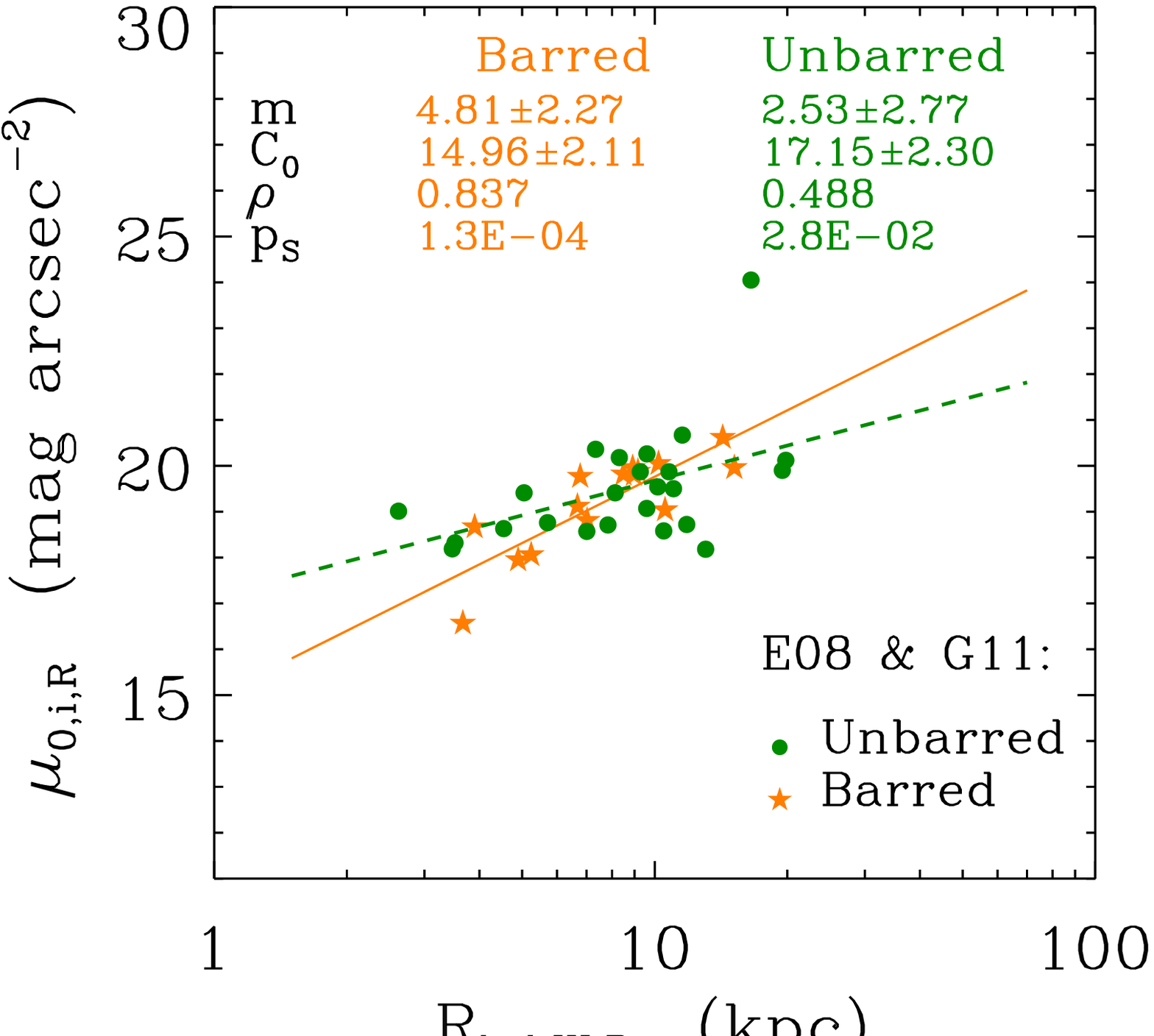}
   \includegraphics[width = 0.48\textwidth,bb=-35 -15 472 425, clip]{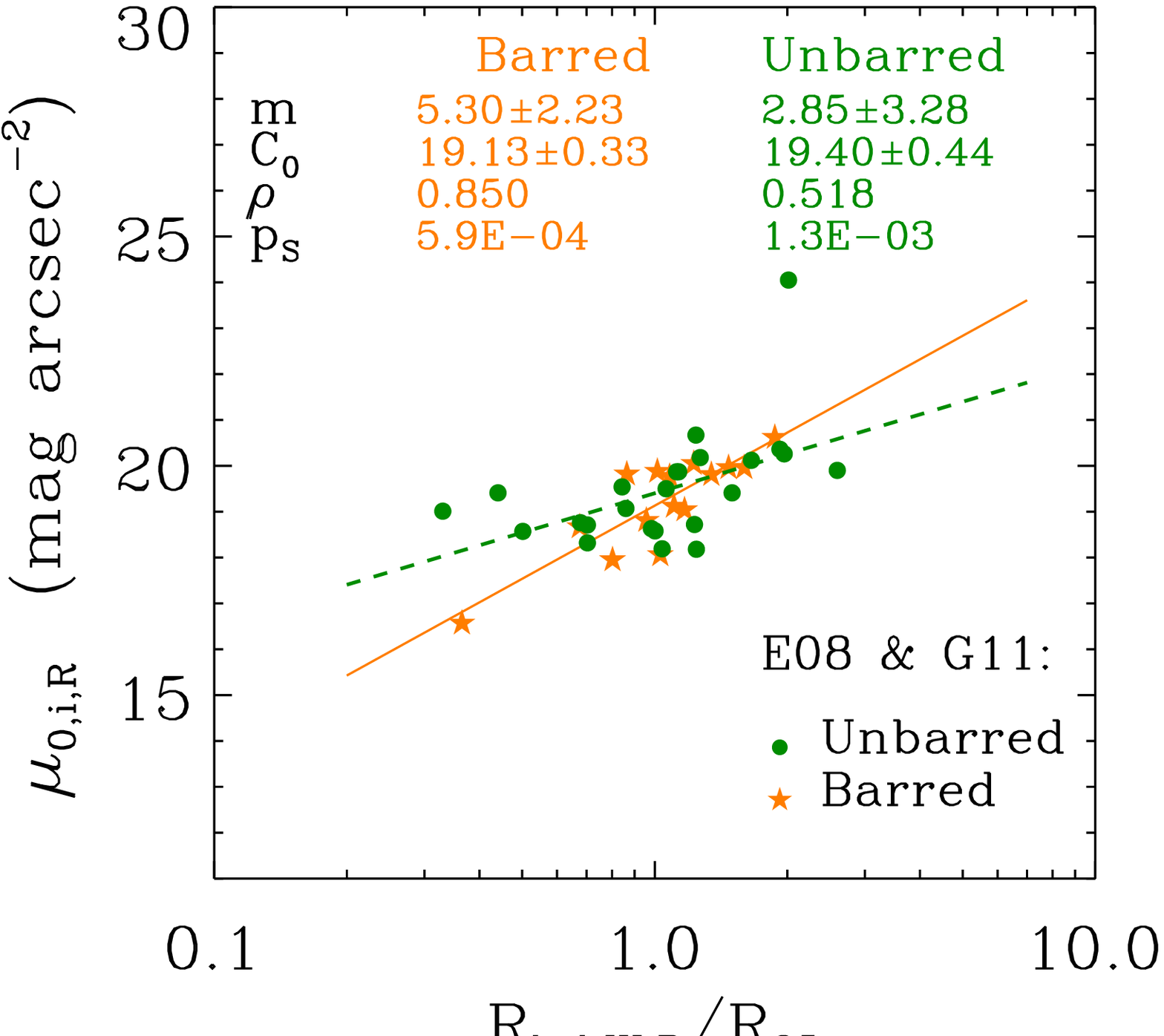}
   \includegraphics[width = 0.48\textwidth,bb=-35 -15 472 425, clip]{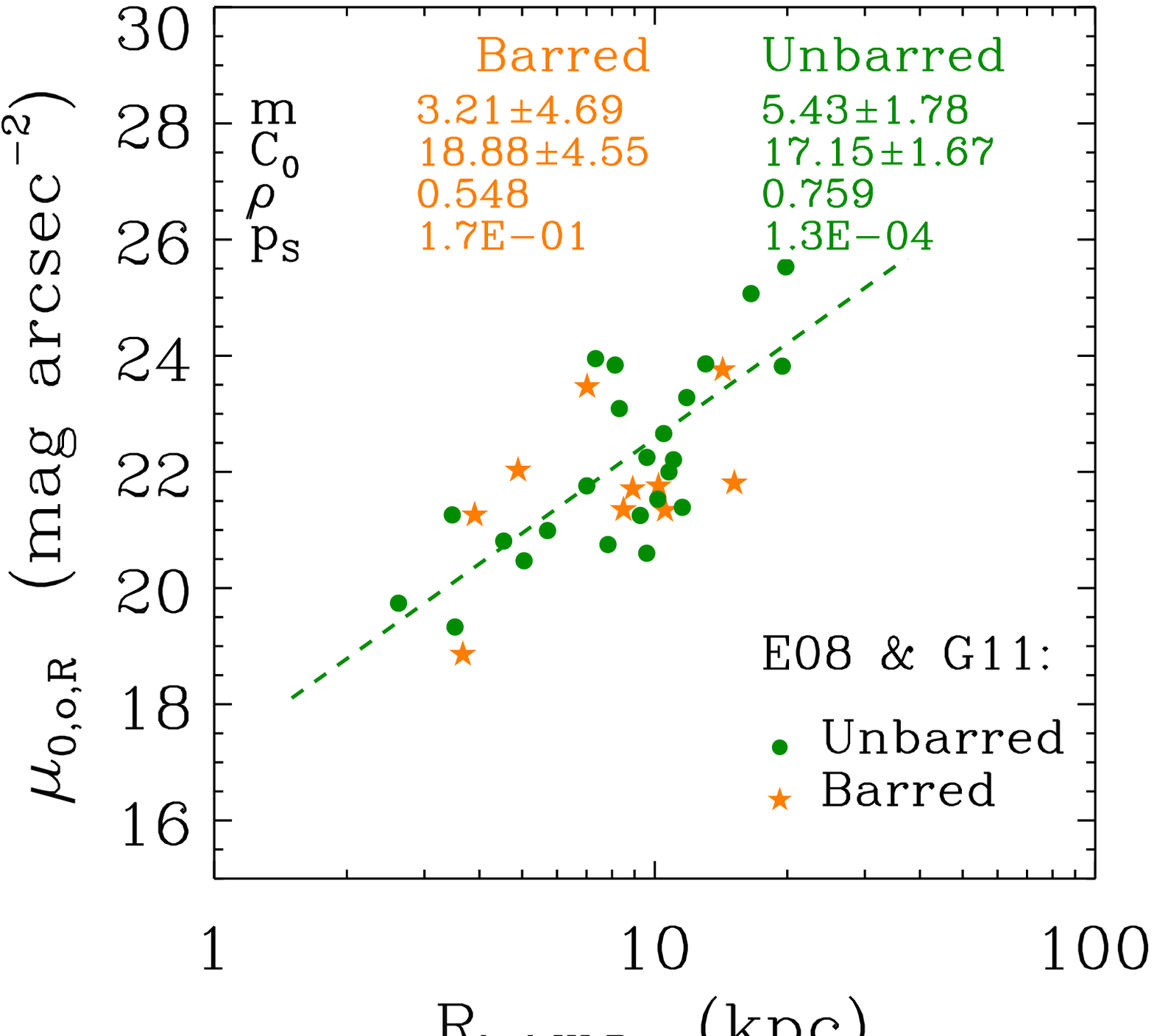}
   \includegraphics[width = 0.48\textwidth,bb=-35 -15 472 425, clip]{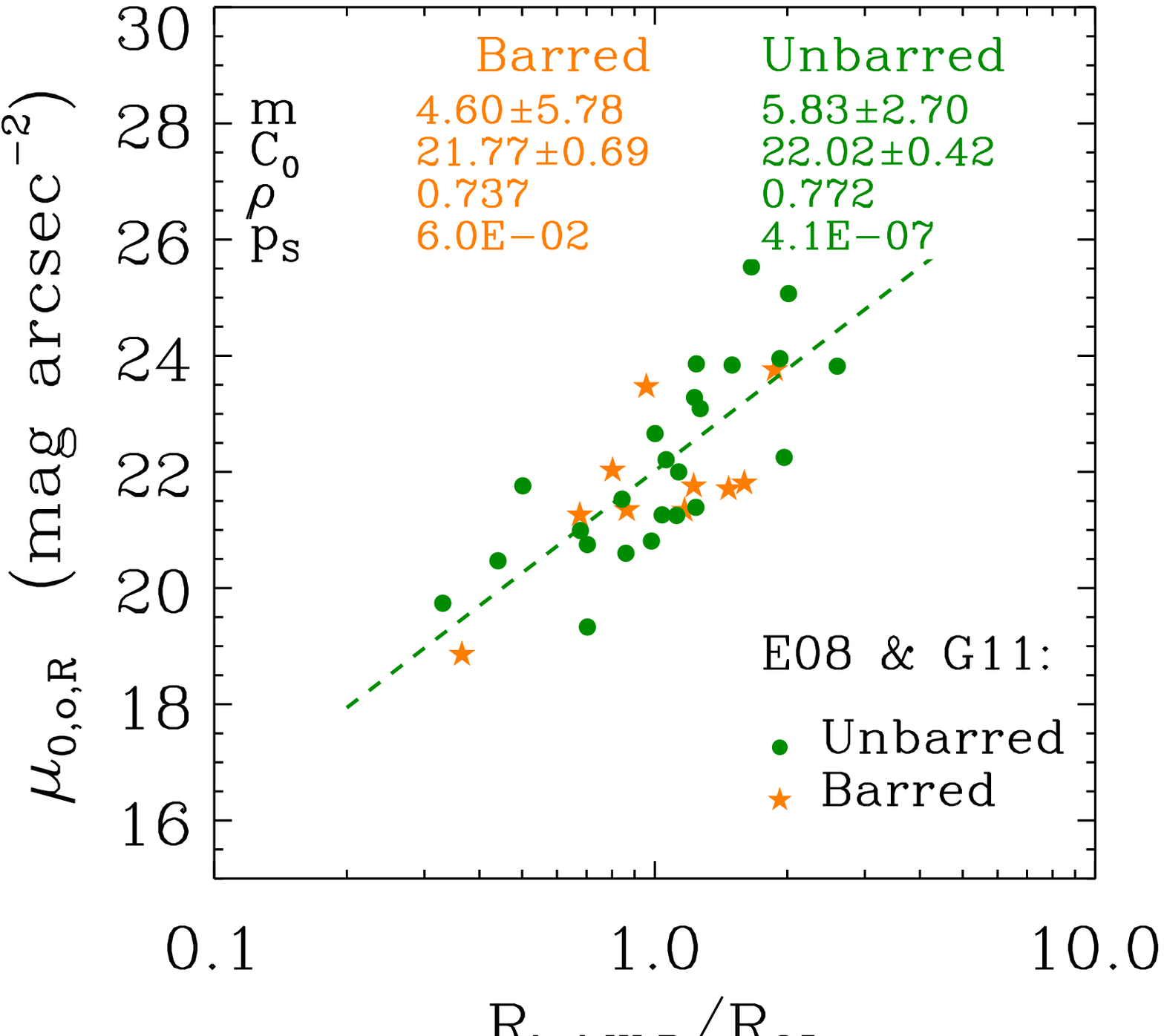}
   \includegraphics[width = 0.48\textwidth,bb=-35 -15 472 425, clip]{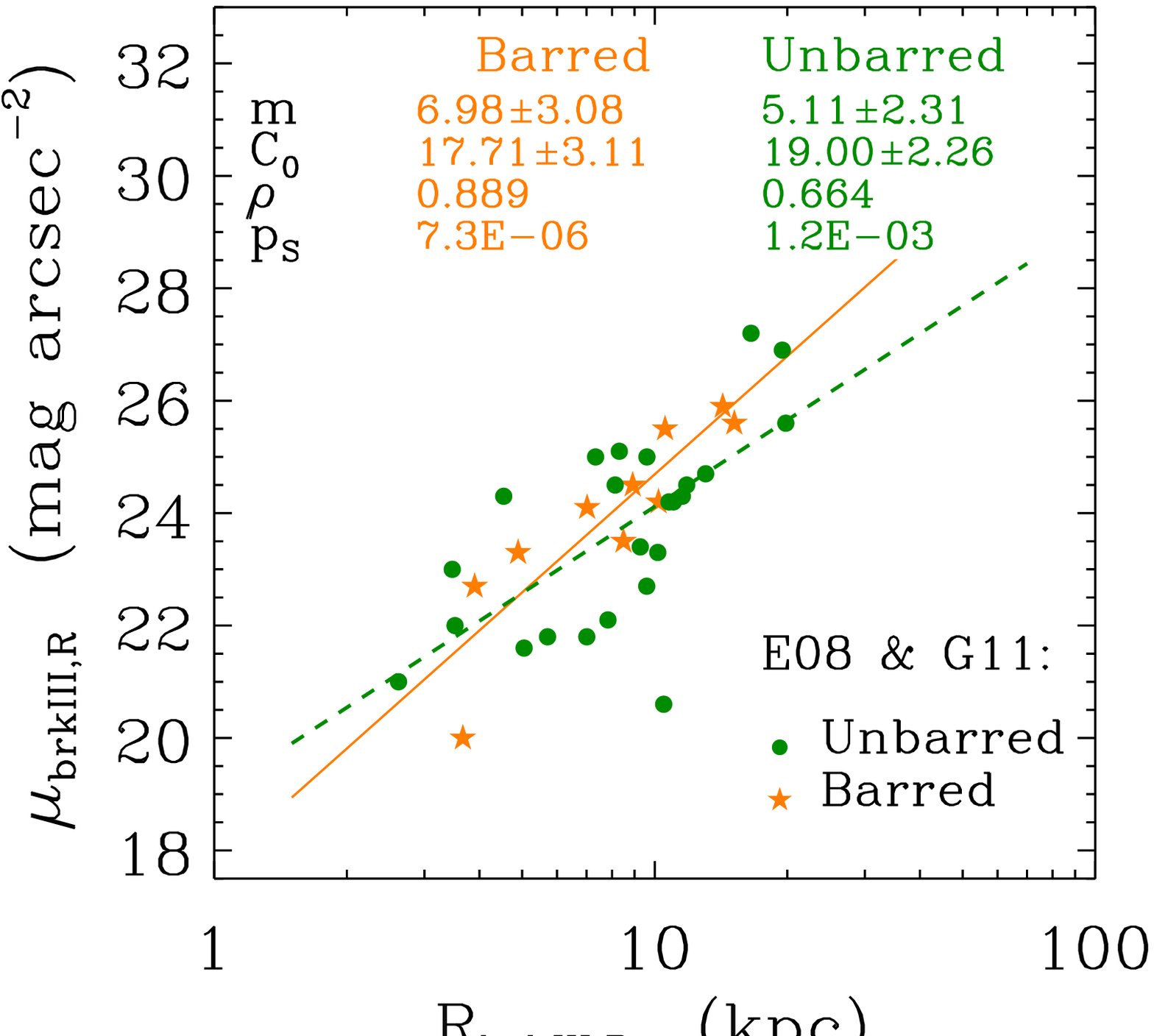}
   \includegraphics[width = 0.48\textwidth,bb=-35 -15 472 425, clip]{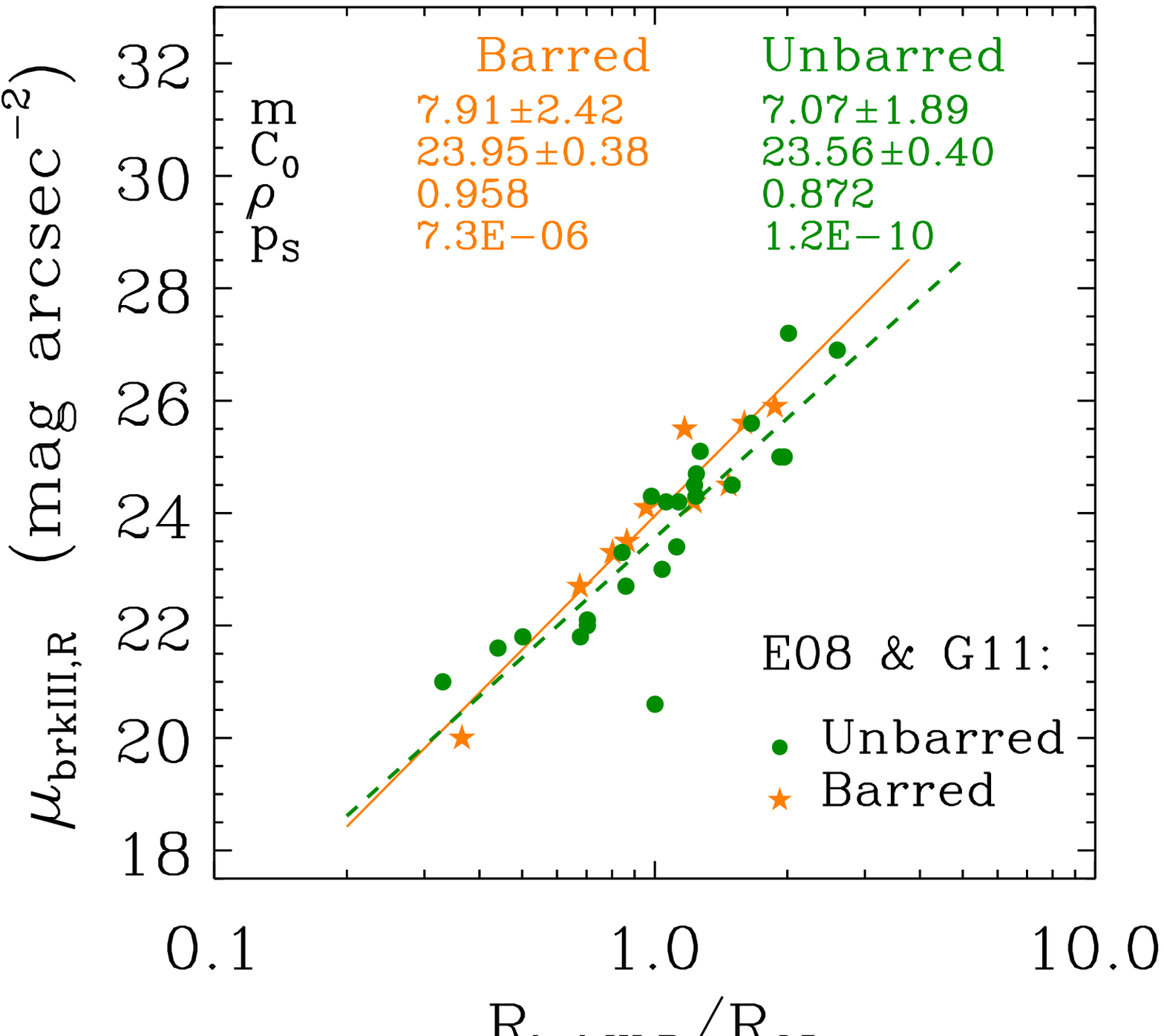}
  \caption{The same as Fig.\,\ref{fig:withRbreak_R}, but for barred and unbarred galaxies in the $R$ band (see Tables\,\ref{tab:hihorbreak}-\ref{tab:mubreakrbreak}). The linear fits performed to barred and unbarred galaxies are overplotted only if they are significant (\emph{yellow solid line}: barred galaxies, \emph{green dashed line}: unbarred galaxies).  The results of the linear fits are indicated at the top of each panel. The errors of the fits shown in the panels have been symmetrized for simplicity (the results are available in the corresponding Tables). See the legend in the panels. 
}
 \label{fig:withRbreak_R_barred}
 \end{minipage}
}
& 
\imagetop{
 \begin{minipage}{.48\textwidth}
\centering
   \includegraphics[width = 0.48\textwidth,bb=-35 -15 472 425, clip]{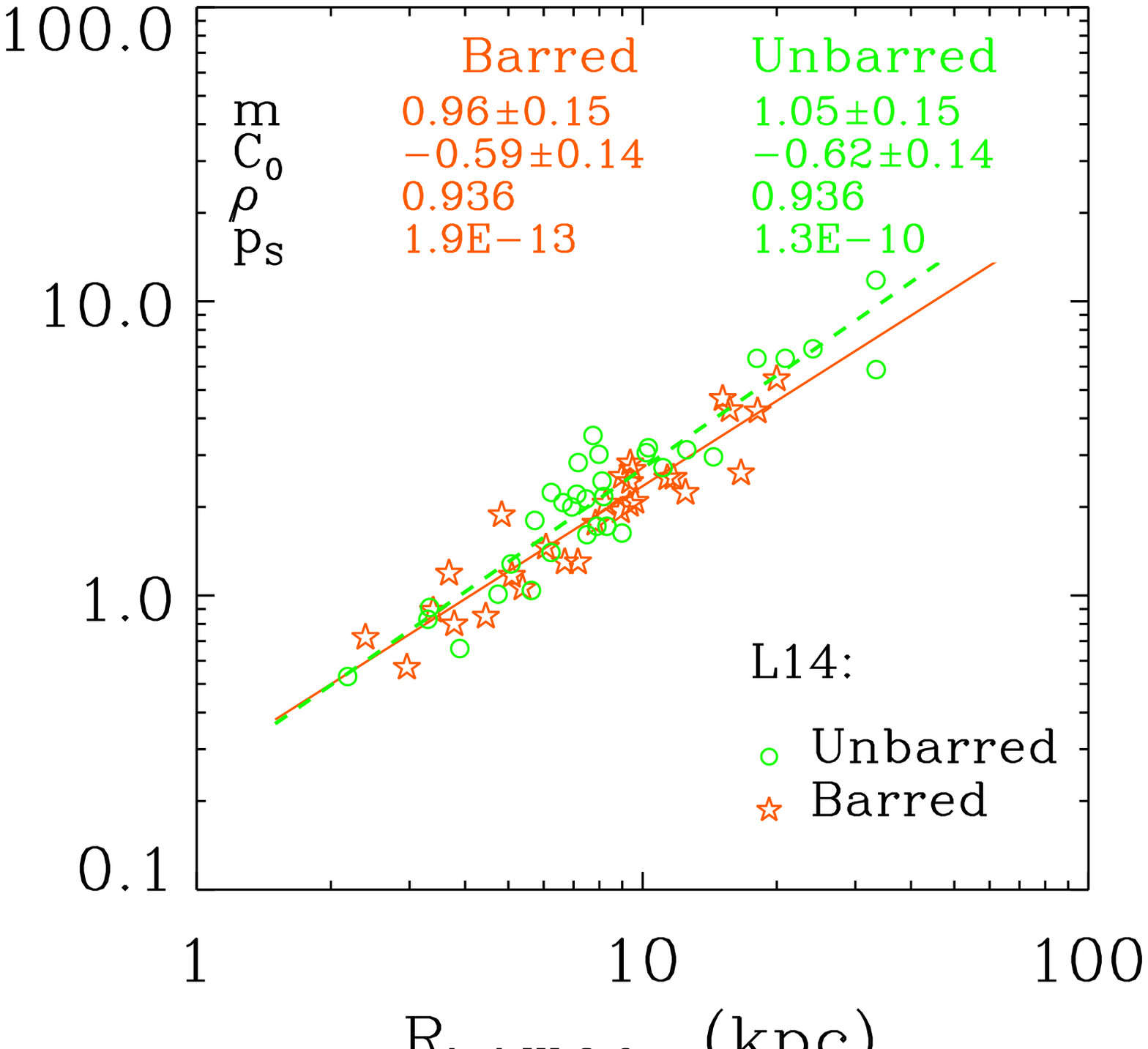}
   \includegraphics[width = 0.48\textwidth,bb=-35 -15 472 425, clip]{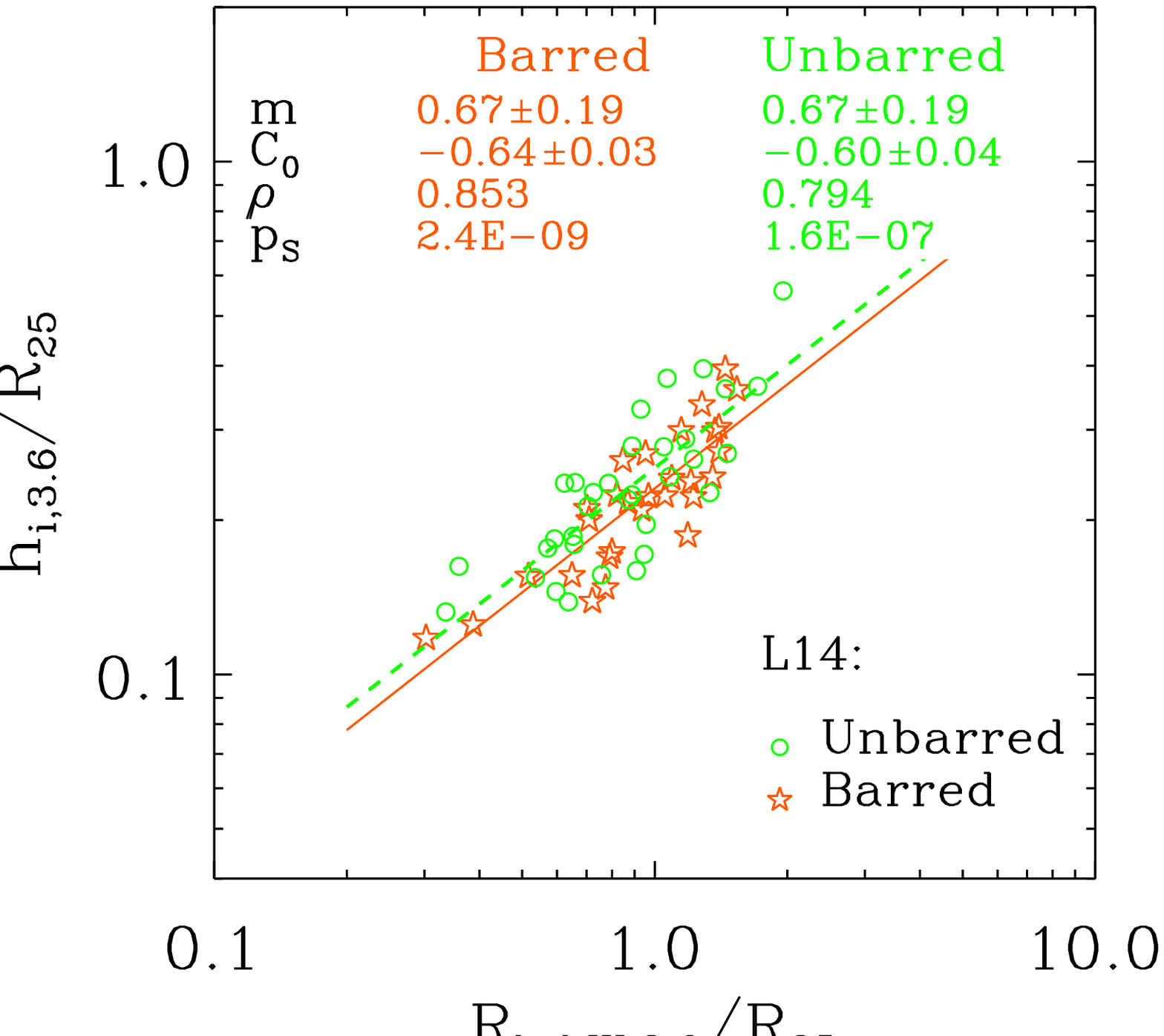}
   \includegraphics[width = 0.48\textwidth,bb=-35 -15 472 425, clip]{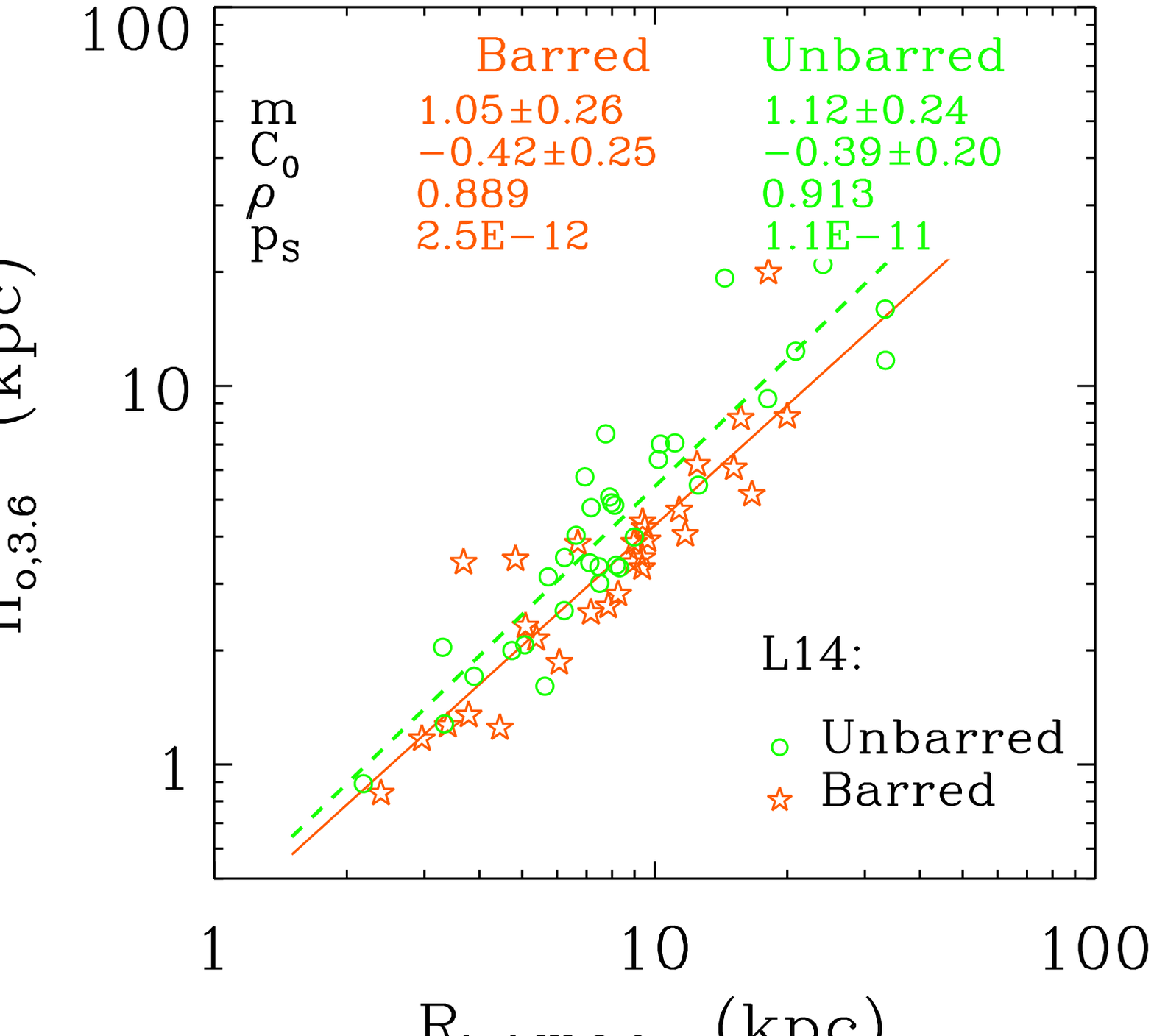}
   \includegraphics[width = 0.48\textwidth,bb=-35 -15 472 425, clip]{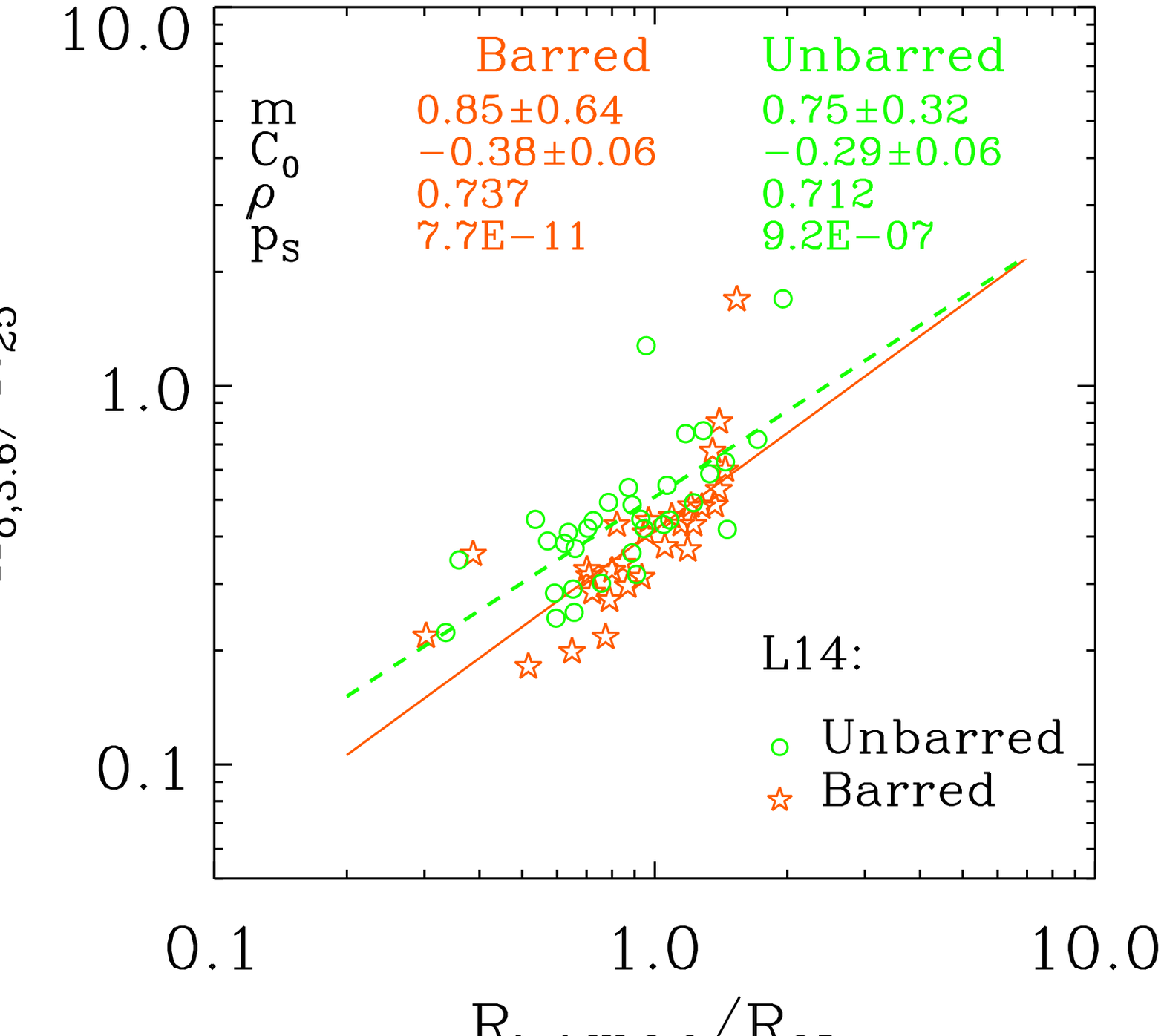}
   \includegraphics[width = 0.48\textwidth,bb=-35 -15 472 425, clip]{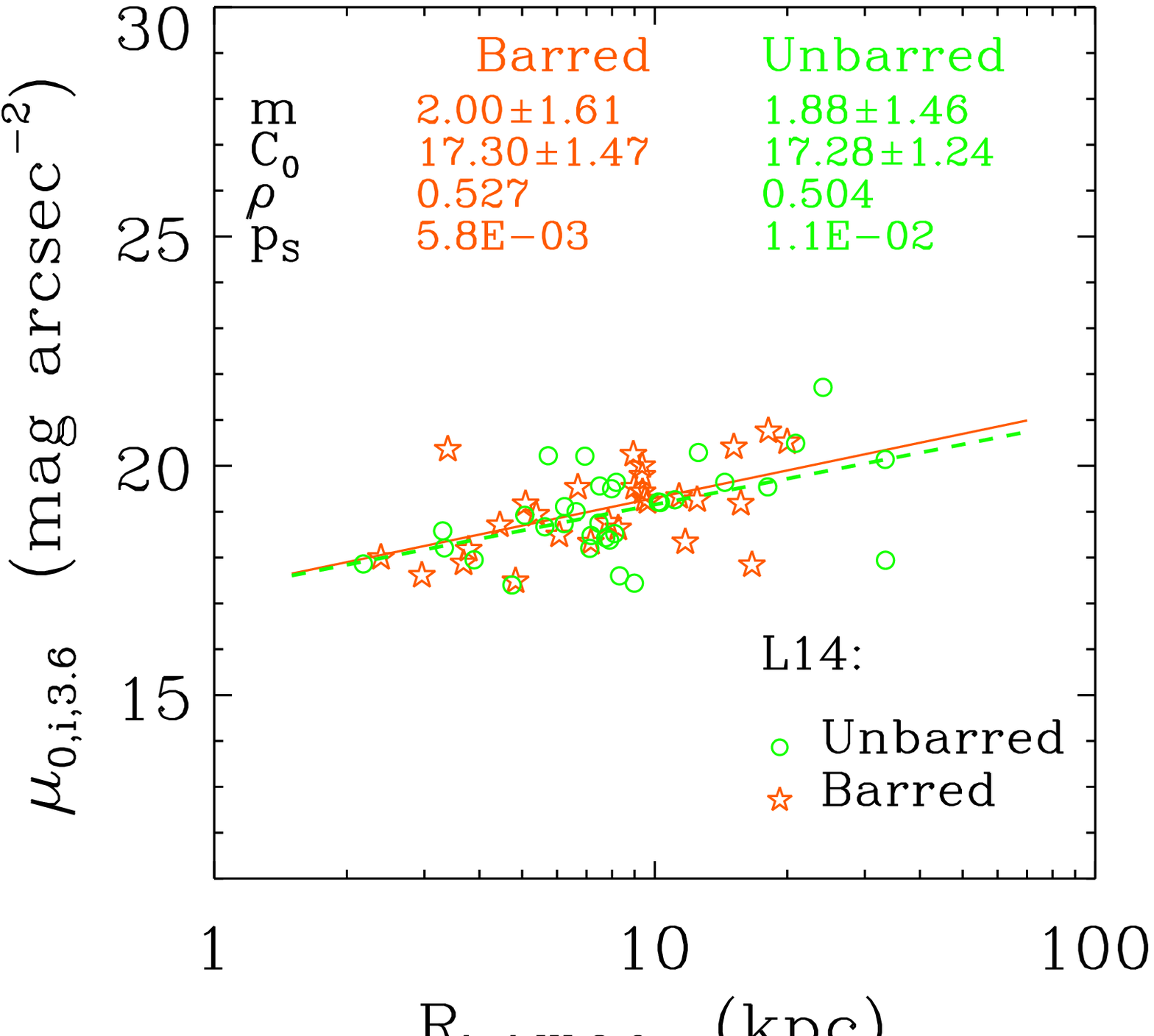}
   \includegraphics[width = 0.48\textwidth,bb=-35 -15 472 425, clip]{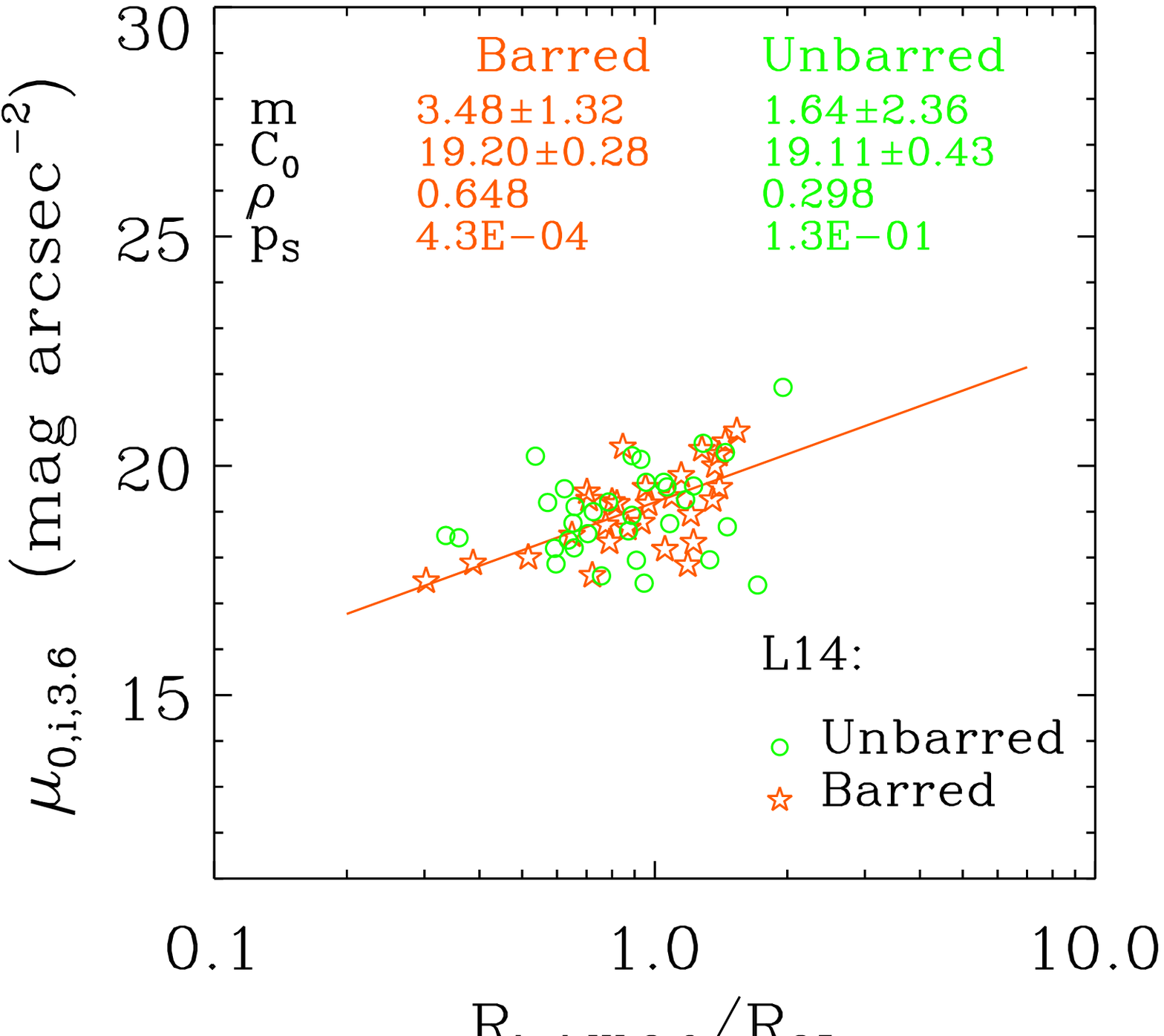}
   \includegraphics[width = 0.48\textwidth,bb=-35 -15 472 425, clip]{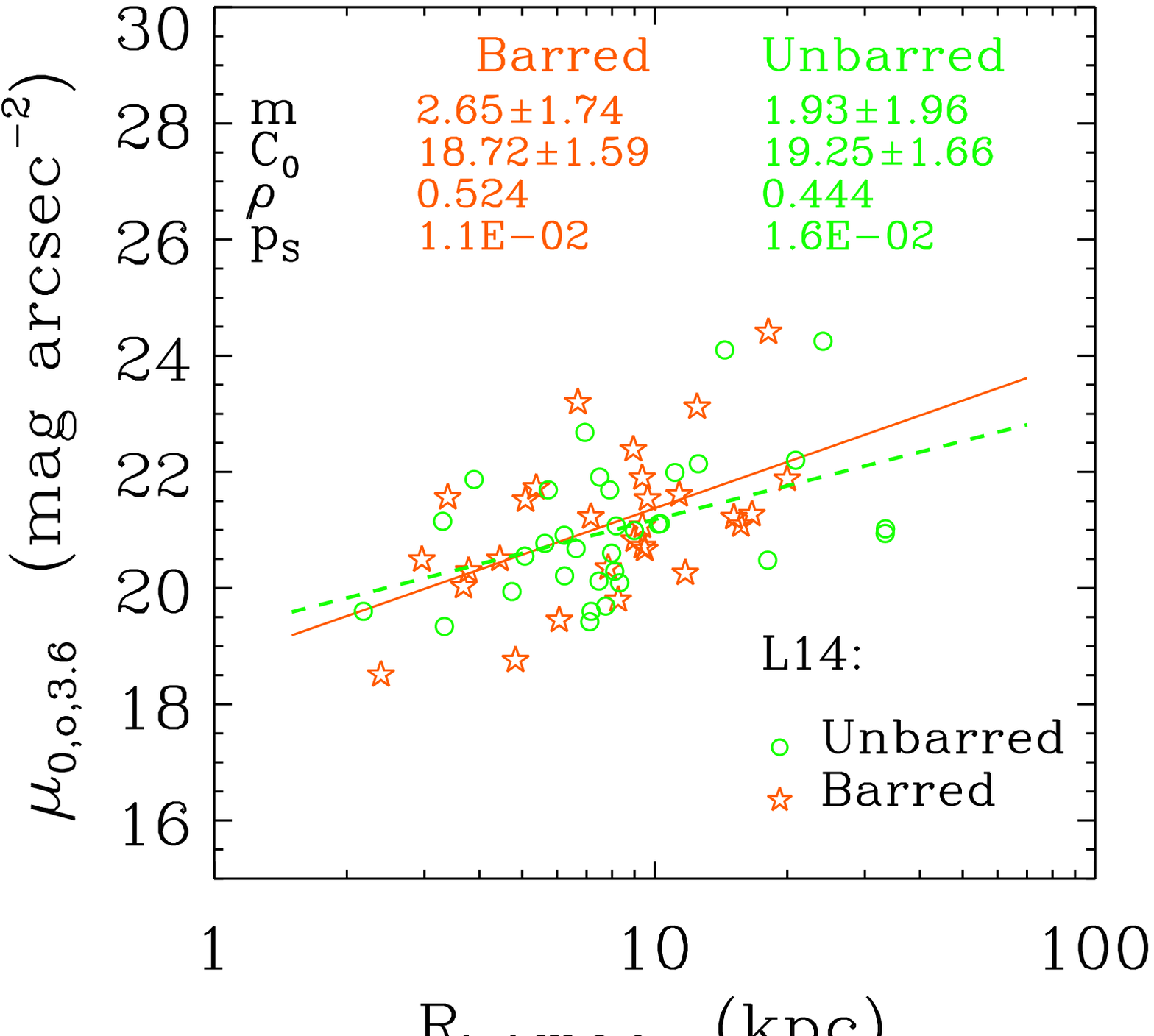}
   \includegraphics[width = 0.48\textwidth,bb=-35 -15 472 425, clip]{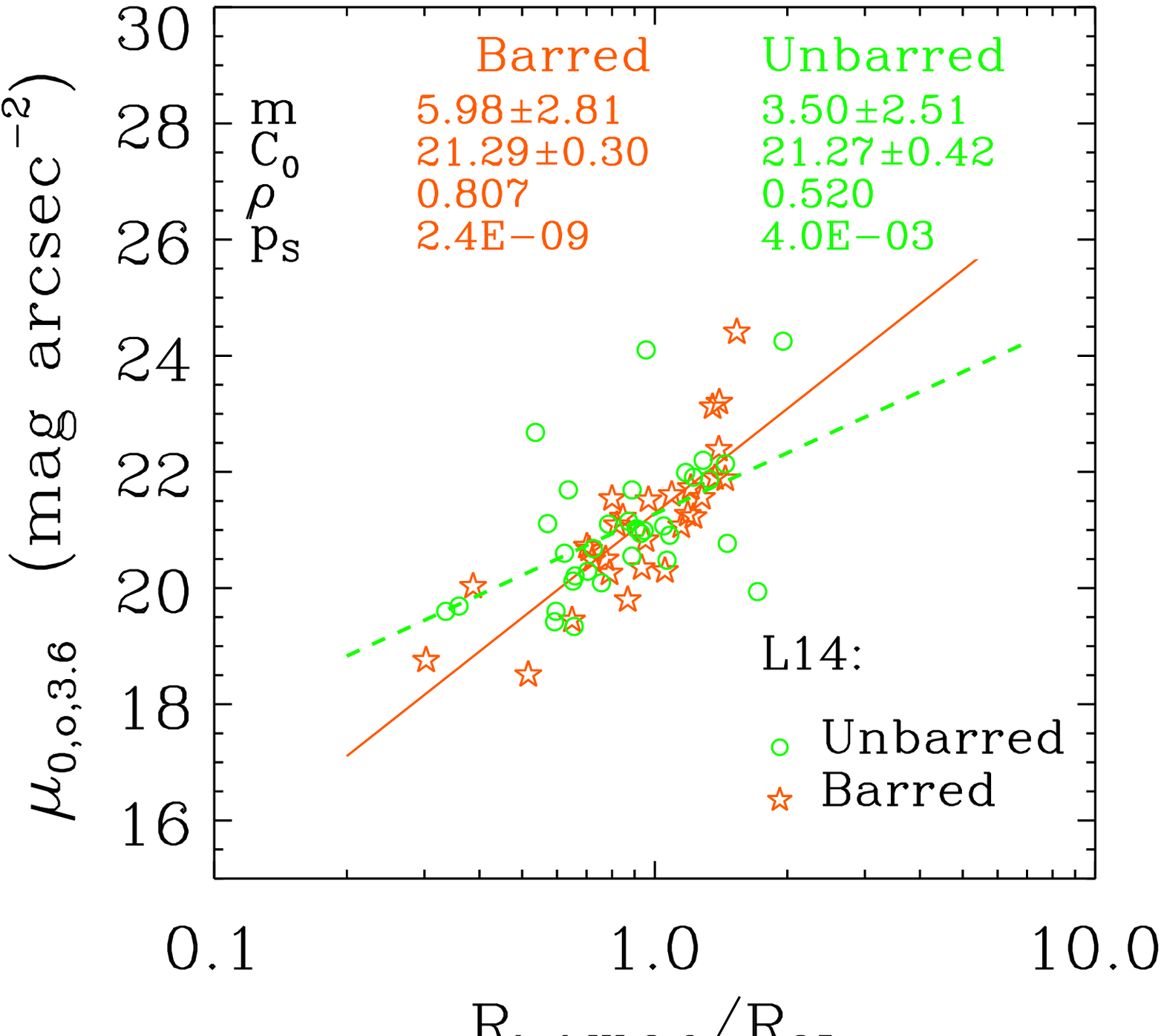}
   \includegraphics[width = 0.48\textwidth,bb=-35 -15 472 425, clip]{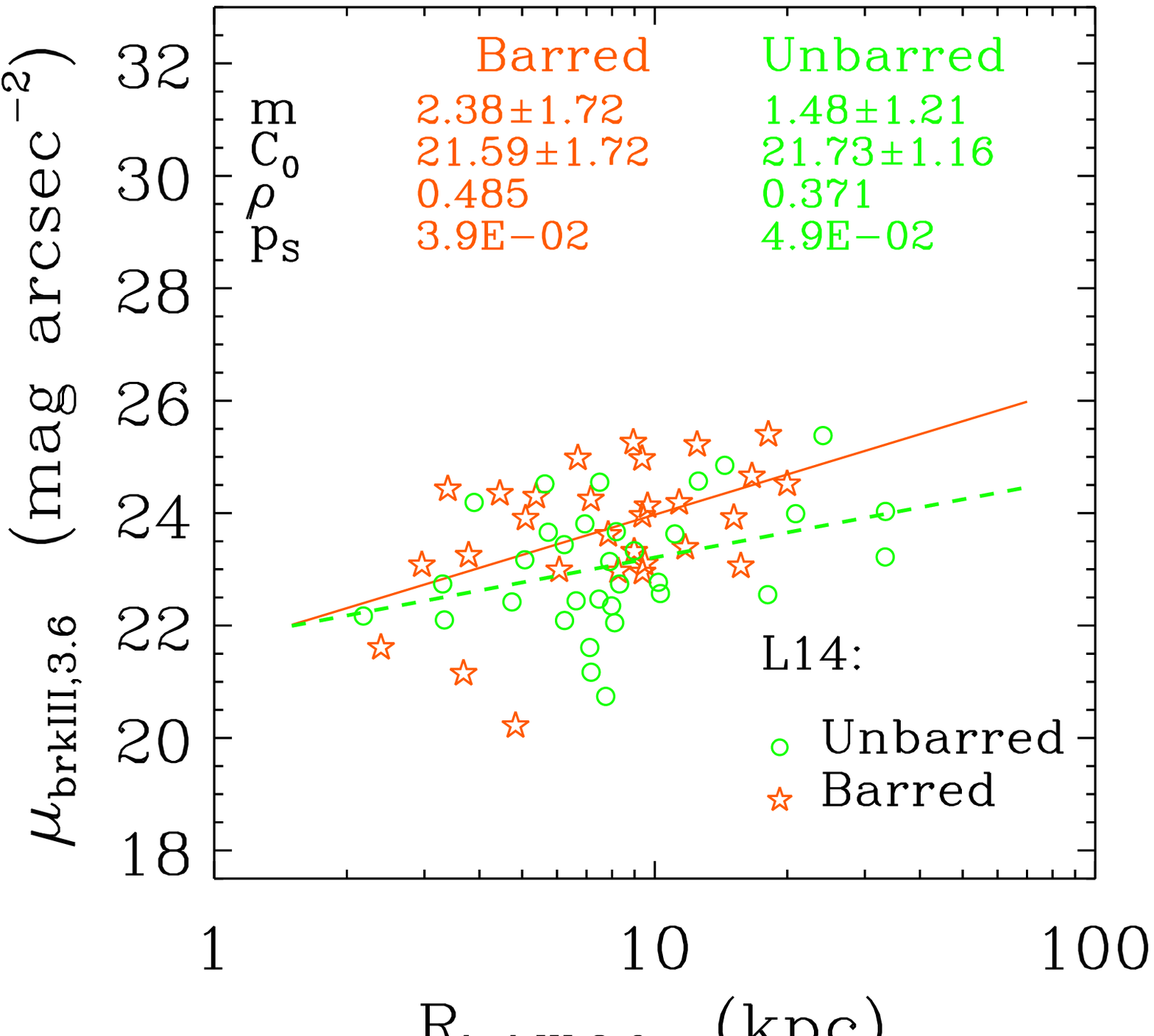}
   \includegraphics[width = 0.48\textwidth,bb=-35 -15 472 425, clip]{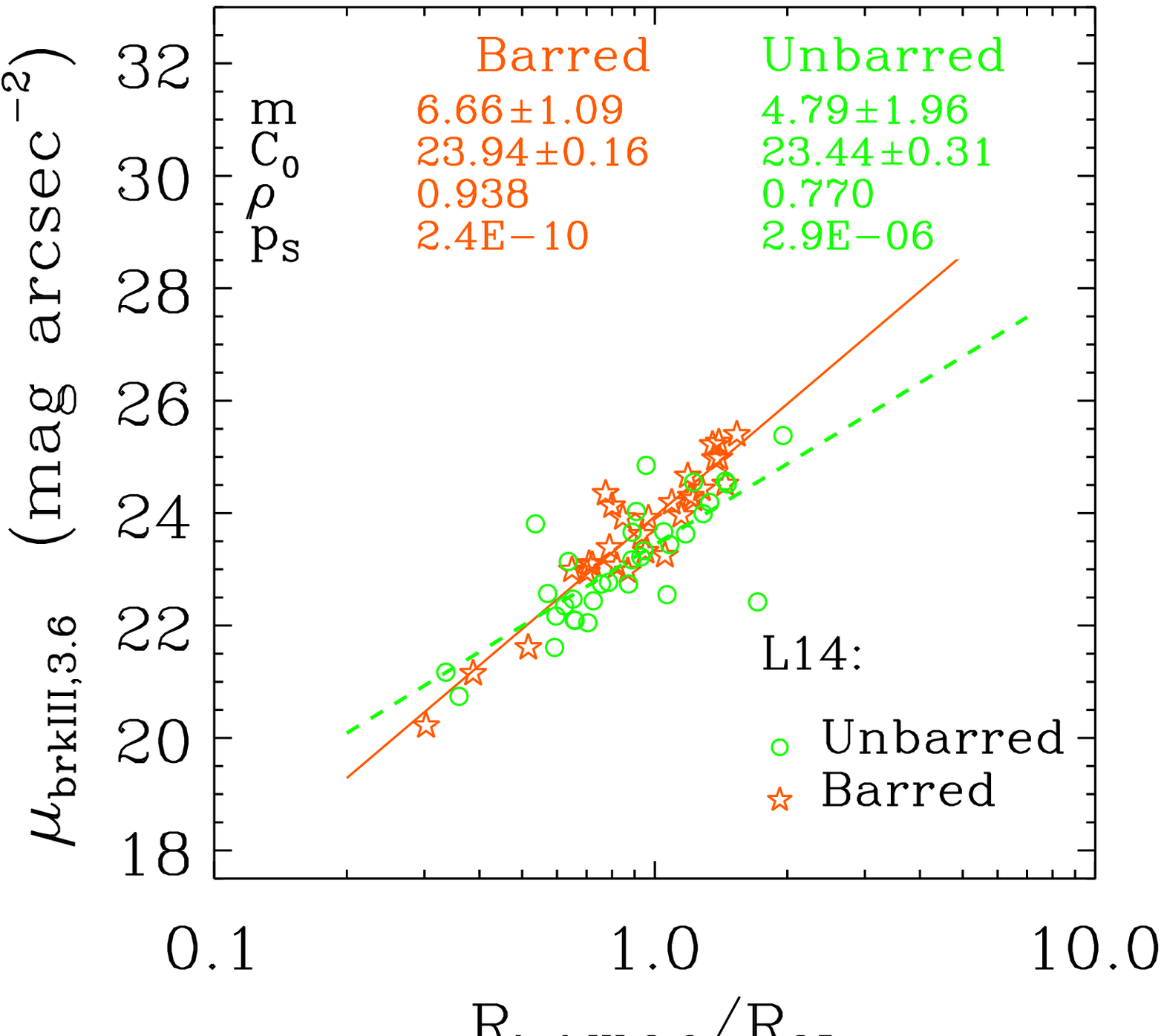}
\caption{The same as Fig.\,\ref{fig:withRbreak_3.6}, but for local antitruncated barred and unbarred galaxies in the \nir\ band (see Tables\,\ref{tab:hihorbreak}-\ref{tab:mubreakrbreak}). See the caption of Fig.\,\ref{fig:withRbreak_R_barred}.}
 \label{fig:withRbreak_3.6_barred}
 \end{minipage}
}
\end{tabular}
\end{figure*}

\section{Fits and correlation tests}
\label{sec:fits}

We  performed linear fits ($y = m\,x + C_0$) to the trends followed in each photometric plane by all galaxies, by spirals and S0s independently, by types of spirals (Sa--Sab, Sb--Sbc, Sc--Scd), as well as by barred and unbarred galaxies. The trends of the types in each sample have been fitted using ordinary least squares. The photometric parameters of the inner and outer discs characterized by E08, G11, and L14 had no errors assigned in their original papers, so no error weighting could be considered in the fits. In order to estimate realistic confidence intervals to the fitted regression coefficients, we adopted a bootstrapping method. We generated $n=10^5$ artificial data samples with the same size as the original one in each diagram with replacement. The final regression coefficients of each fit correspond to the median values of the probability distributions of each coefficient obtained with the $10^5$ results, in order to reduce the systematic bias introduced by outliers. The upper and lower errors considered for each coefficient are those enclosing 2.5\% and 97.5\% of the values in the corresponding probability distribution. The bootstrap distribution is closer to the real probability distribution of the coefficients than a simple Gaussian in general, so this method derives robust and conservative (asymmetric) confidence intervals for the regression coefficients, reducing the effects of possible outliers or high leverage points at the same time. 

We tested the significance of each photometric trend using the Spearman rank correlation test, which measures whether two variables are monotonically related and the level of correlation between them, and it has the advantage of being non parametric. Only those trends with an associated probability of random correlation below 5\% according to the test ($p_S<0.05$) are considered as statistically significant. The \nir\ dataset presents higher statistics than the $R$-band sample, but the Spearman rank correlation test accounts for the number of data pairs yielding the trend to derive $p_S$. Additionally, the Pearson coefficient  ($\rho$) has been used to determine the level of linear correlation of each trend. 

The slope ($m$) and $Y$-intercept ($C_0$) of the linear fits performed to the different trends analysed in each photometric plane, their asymmetric error intervals, as well as the values of $p_S$, $\rho$, and the number of available data pairs ($N_ \mathrm{pairs}$) for each trend, are listed in Tables\,\ref{tab:hihorbreak}-\ref{tab:r25}. 

\begin{figure*}[th!]
\begin{tabular}{cc} 
\framebox[0.48\textwidth][c]{Trends with \hi\ and \ho\  in $R$ (S0s/spirals)} &   \framebox[0.48\textwidth][c]{Trends with \hi\ and \ho\ in \nir\ (S0s/spirals)} \\
   \imagetop{
 \begin{minipage}{.48\textwidth}
\centering
   \includegraphics[width = 0.48\textwidth,bb=-30 -15 455 425, clip]{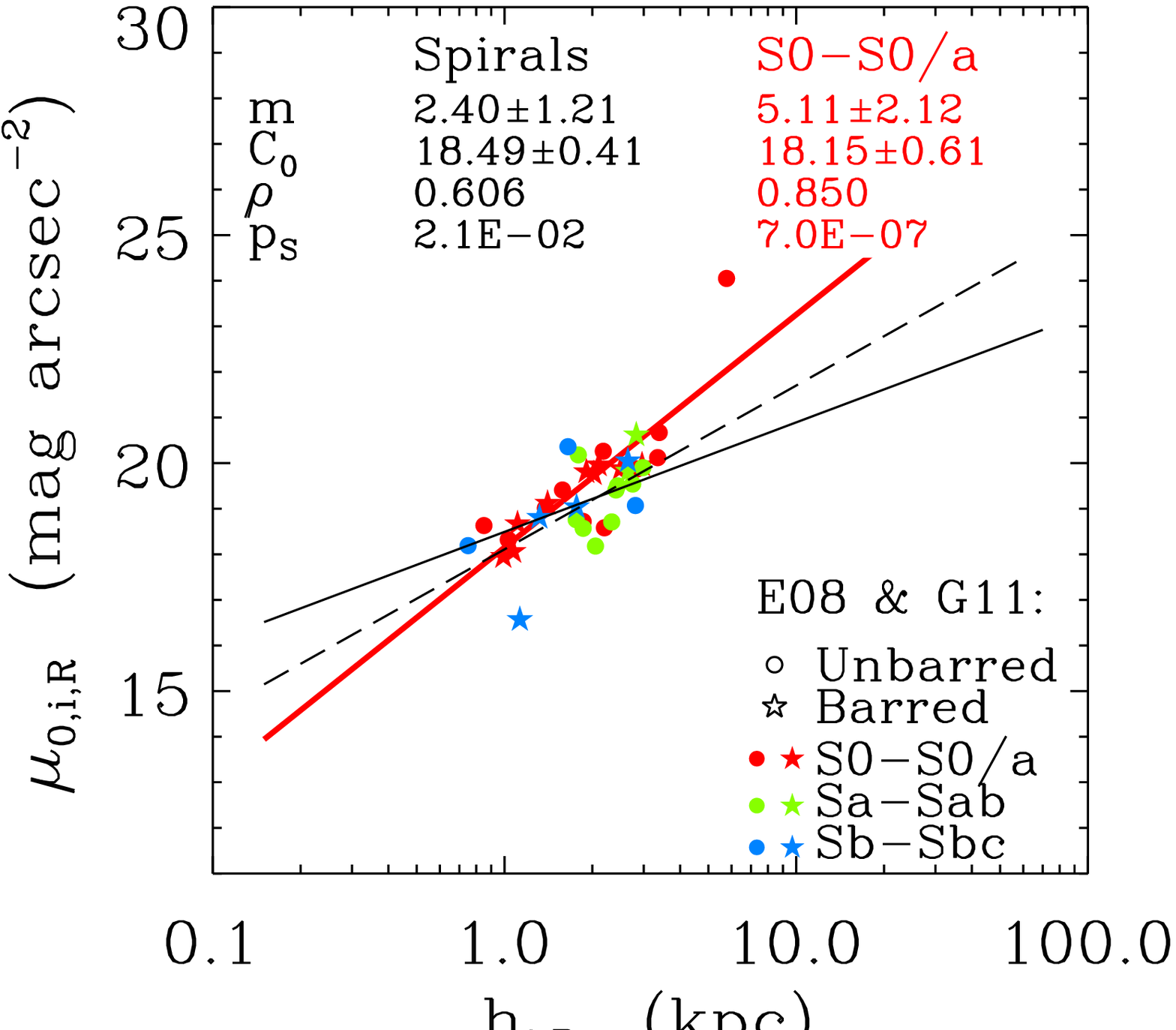}
   \includegraphics[width = 0.48\textwidth,bb=-30 -15 465 425, clip]{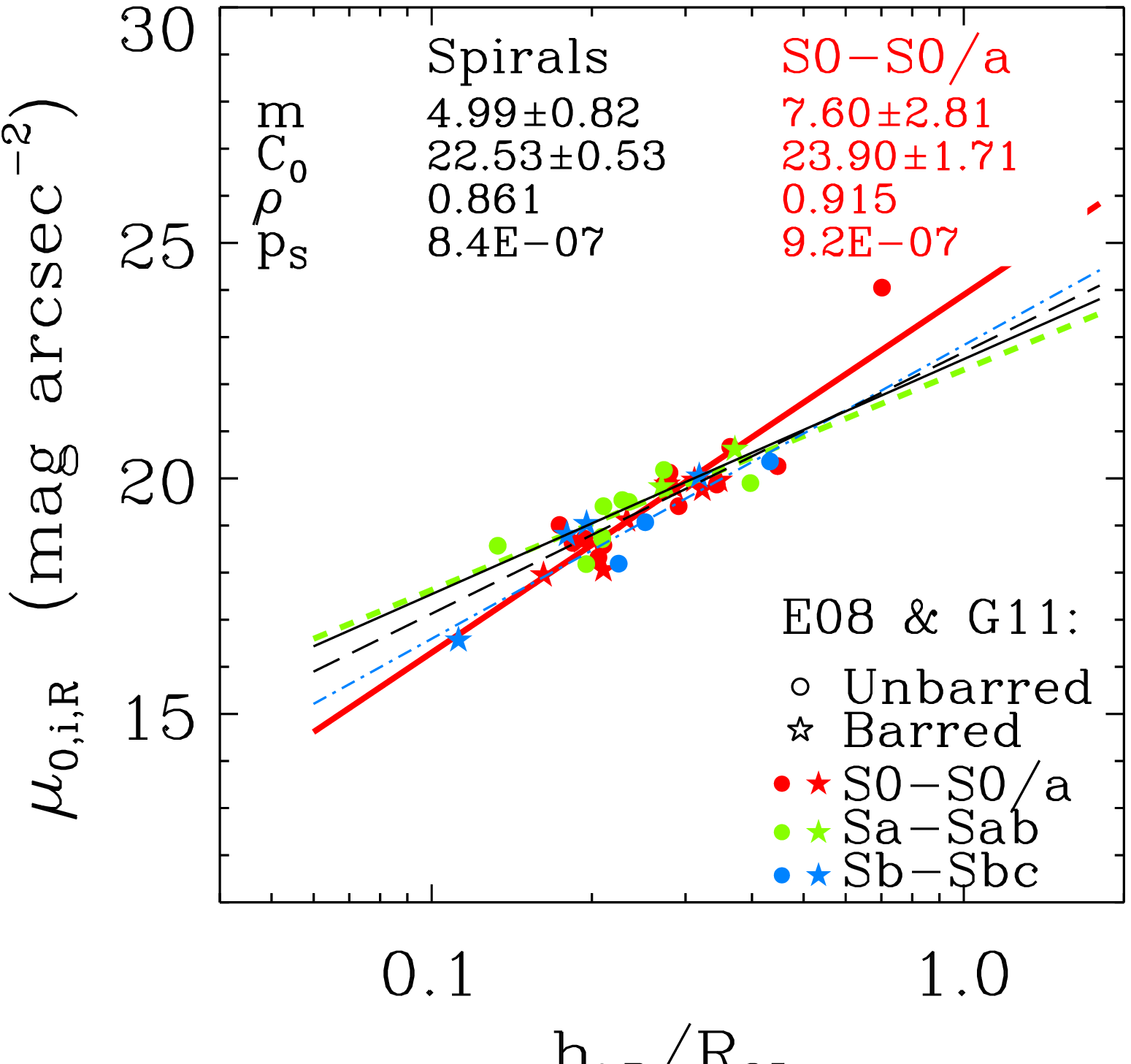}
   \includegraphics[width = 0.48\textwidth,bb=-30 -15 465 425, clip]{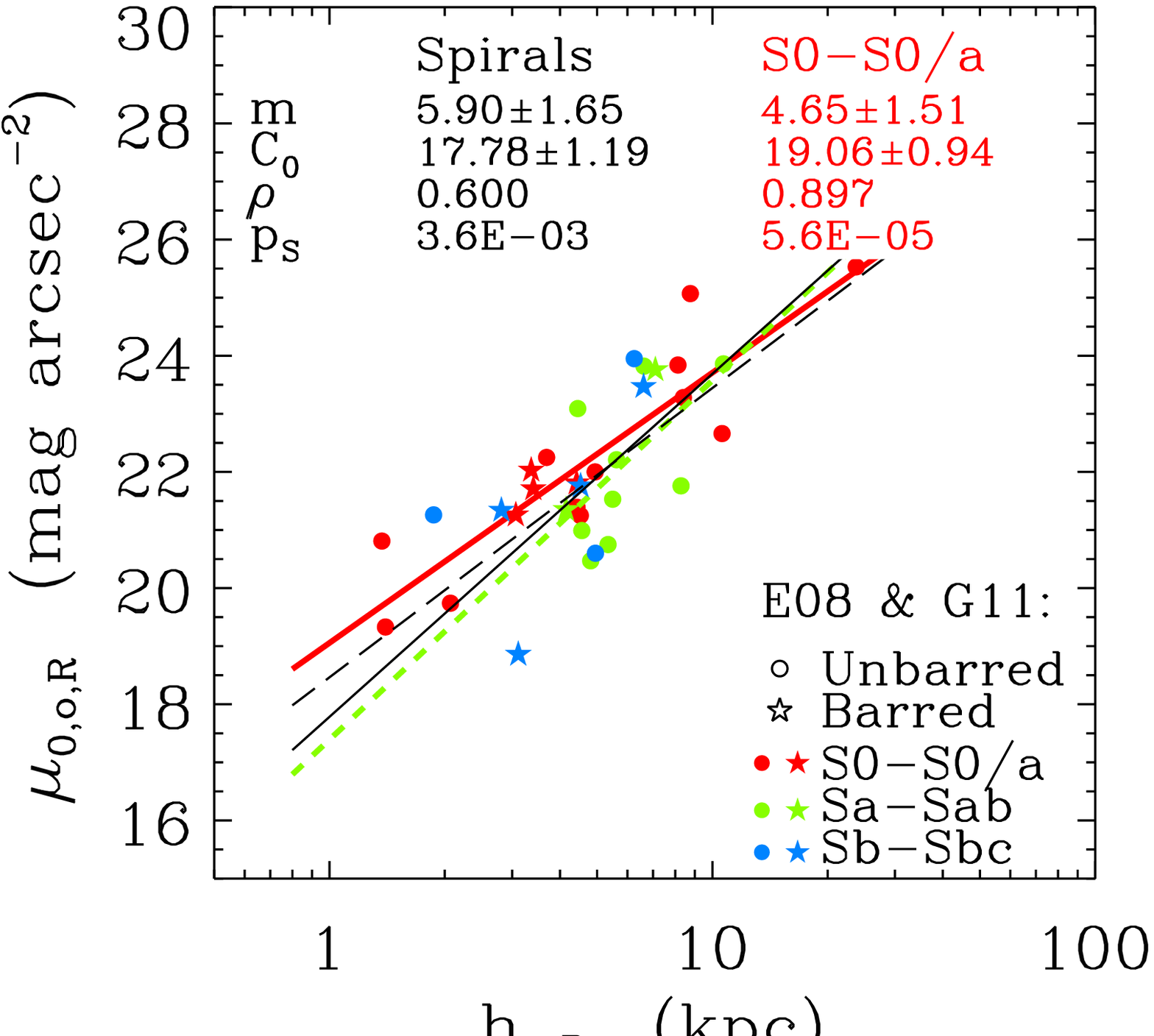}
   \includegraphics[width = 0.48\textwidth,bb=-30 -15 468 425, clip]{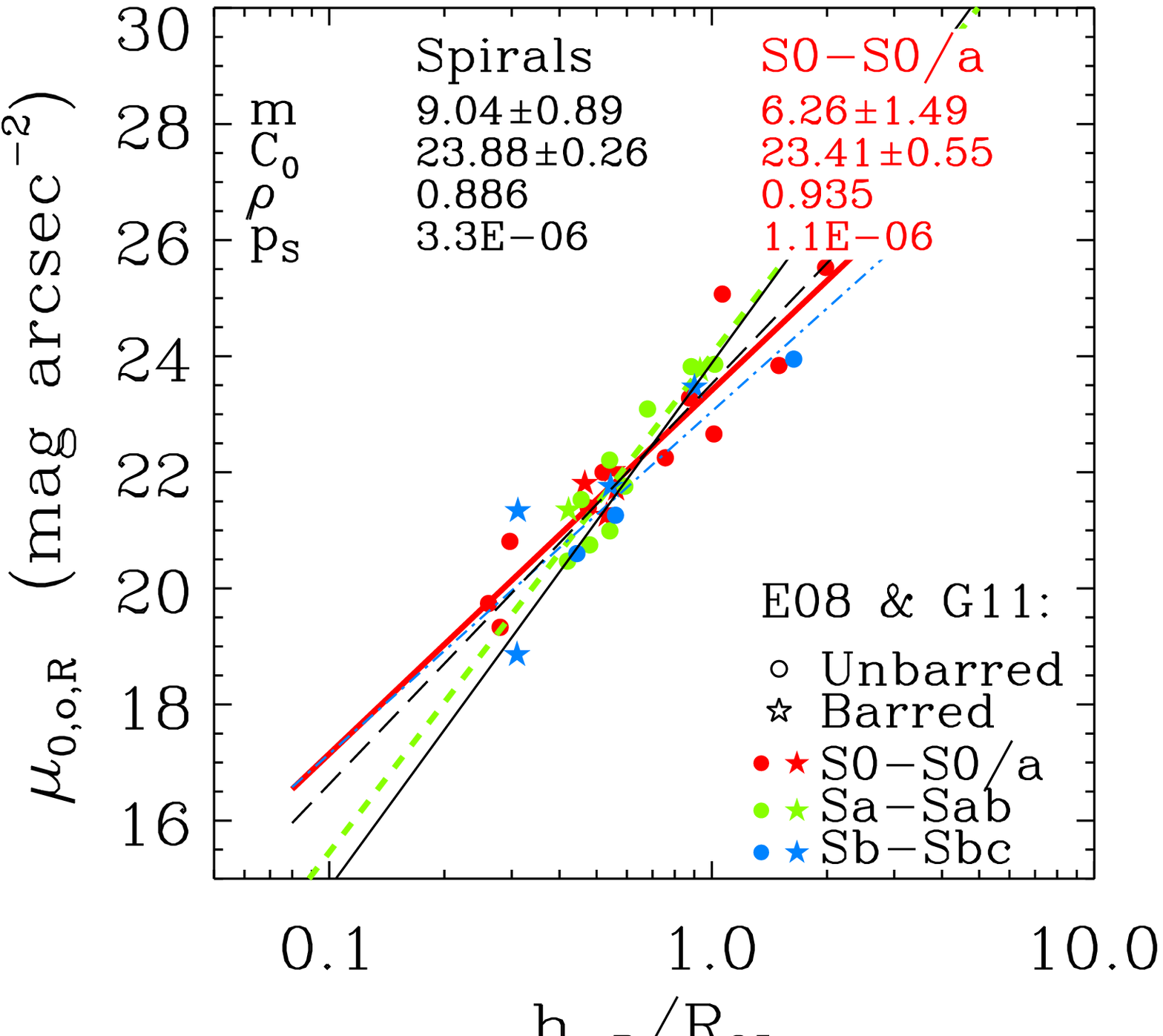}
   \includegraphics[width = 0.48\textwidth,bb=-30 -15 455 425, clip]{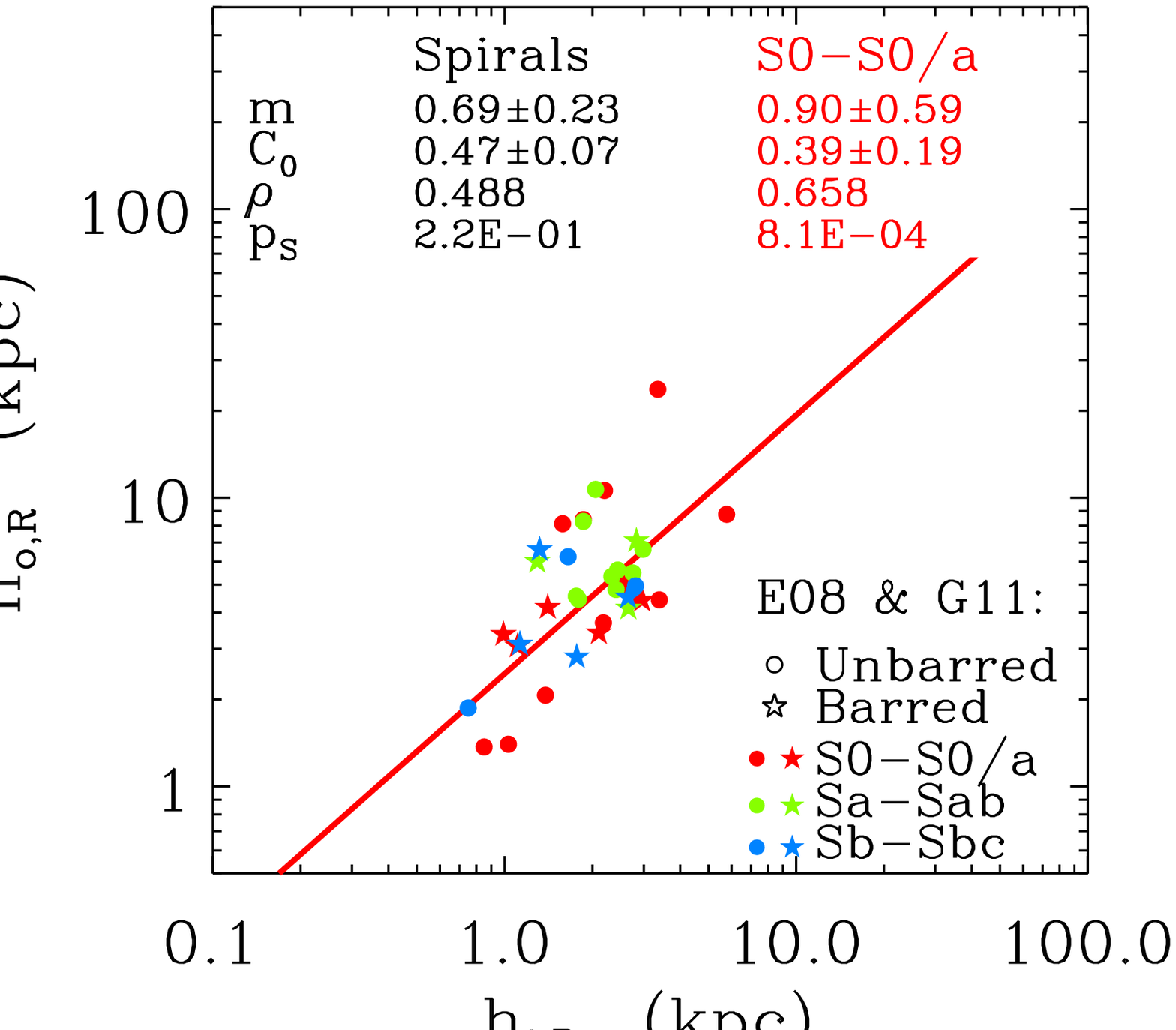}
   \includegraphics[width = 0.48\textwidth,bb=-30 -15 465 425, clip]{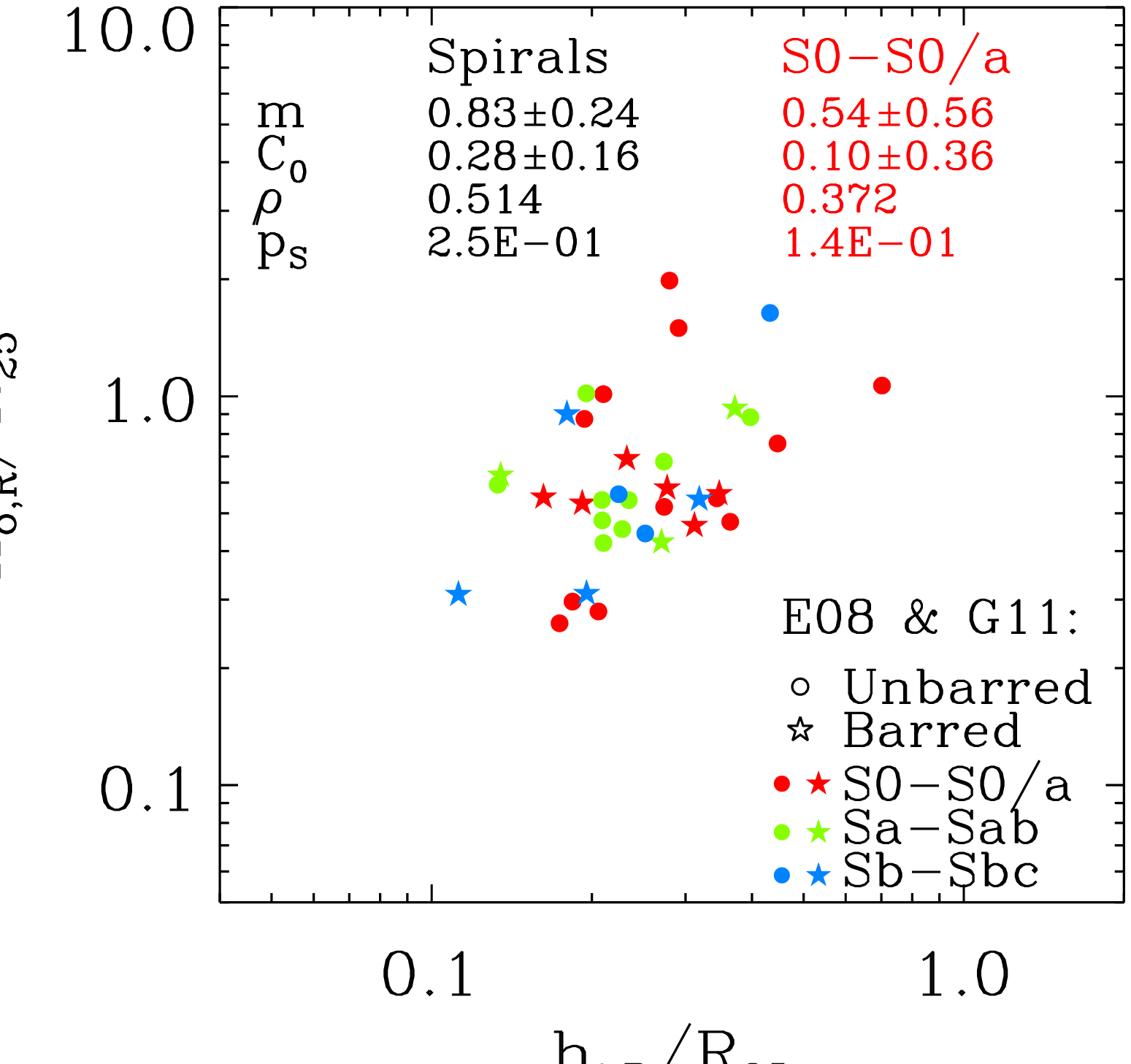}
\caption{Scaling relations between the parameters of the inner and outer discs of local antitruncated S0--Sbc galaxies in the $R$ band from E08 and G11 samples (these results are in Tables\,\ref{tab:muihimuoho} and \ref{tab:relhihorbreak}). See the caption of Fig.\,\ref{fig:withRbreak_R}.}
 \label{fig:withhiorho_R}
\end{minipage}
}
& 
\imagetop{
 \begin{minipage}{.48\textwidth}
\centering
   \includegraphics[width = 0.48\textwidth,bb=-30 -15 455 425, clip]{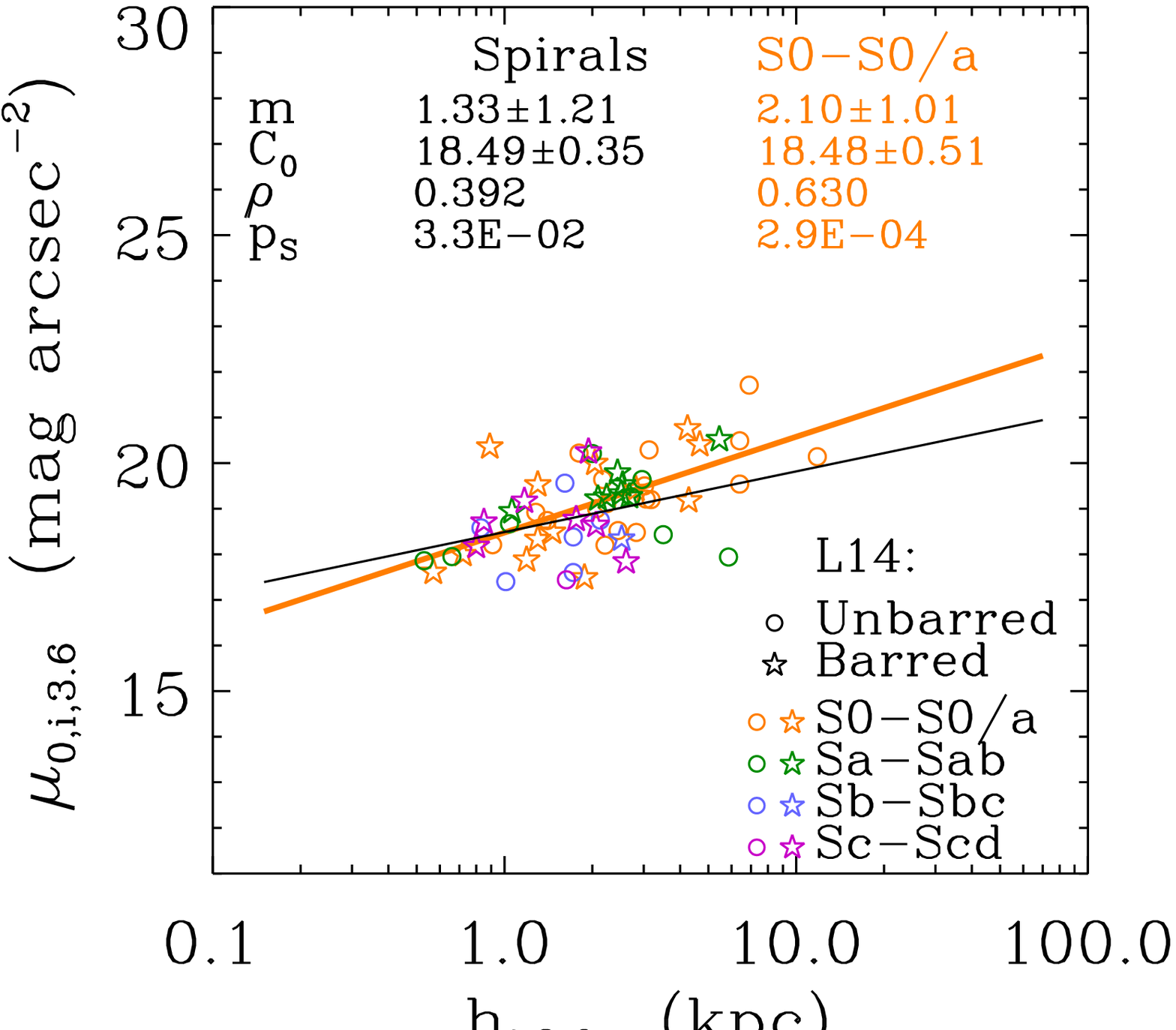}
   \includegraphics[width = 0.48\textwidth,bb=-30 -15 465 425, clip]{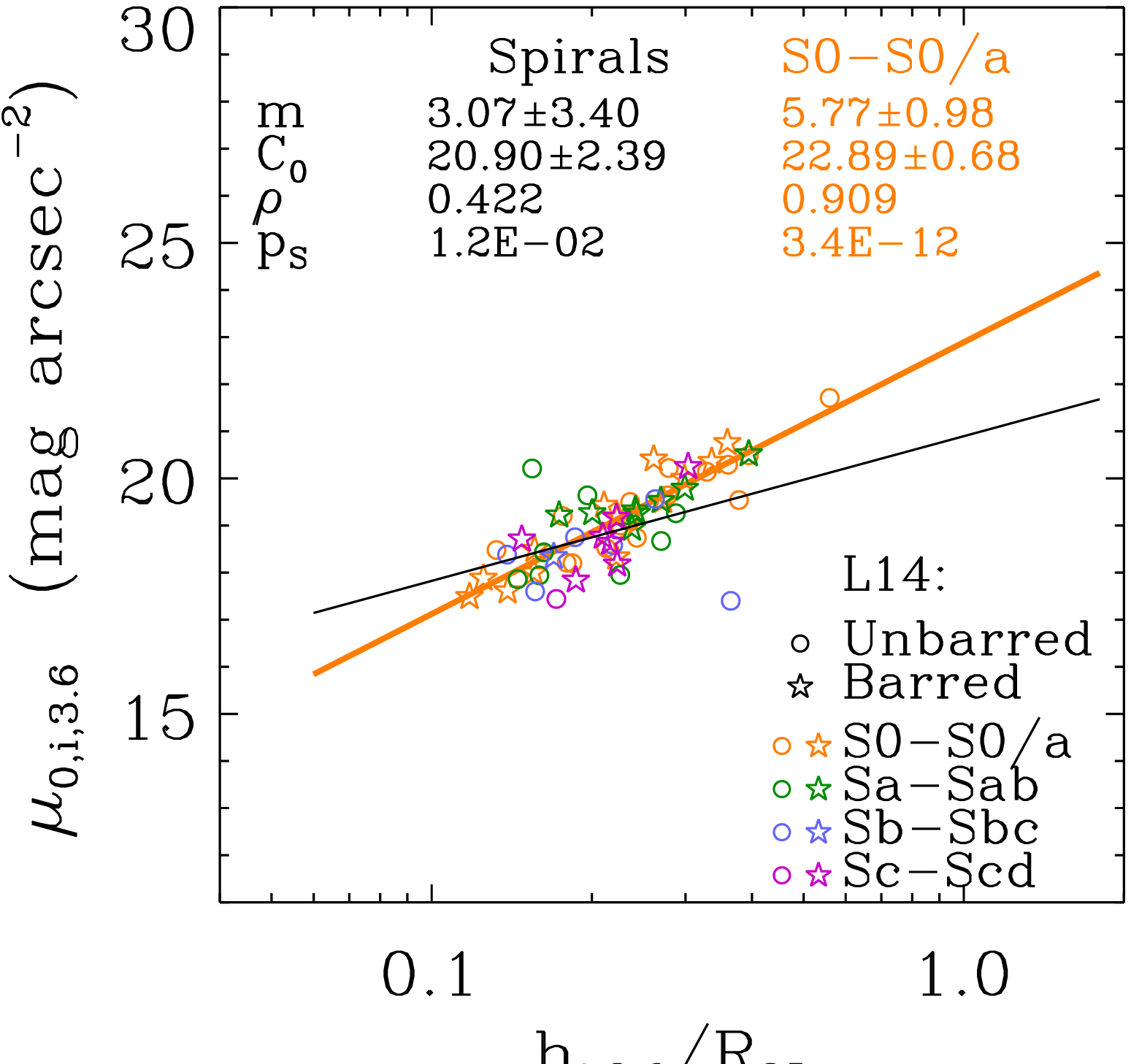}
   \includegraphics[width = 0.48\textwidth,bb=-30 -15 468 425, clip]{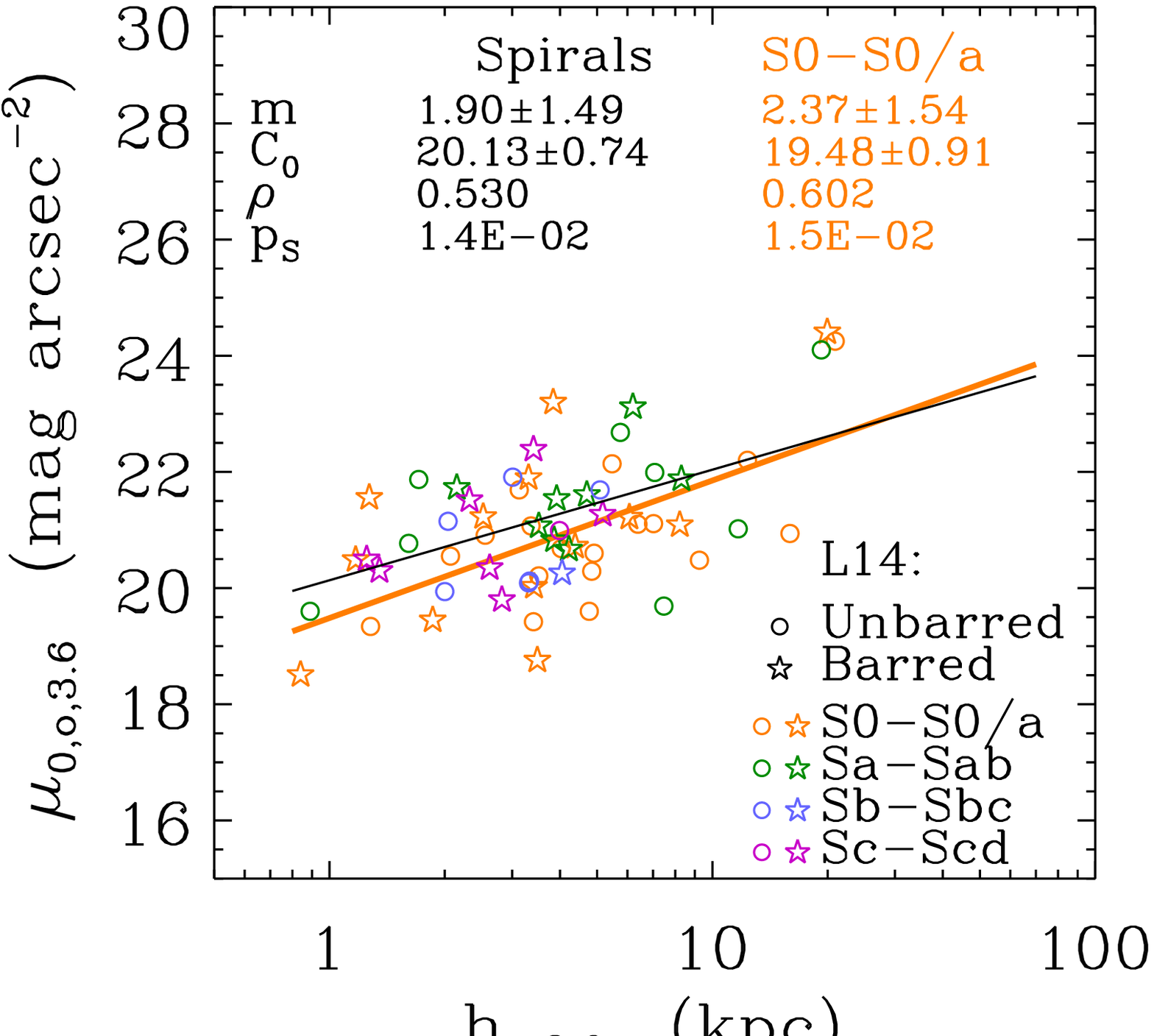}
   \includegraphics[width = 0.48\textwidth,bb=-30 -15 468 425, clip]{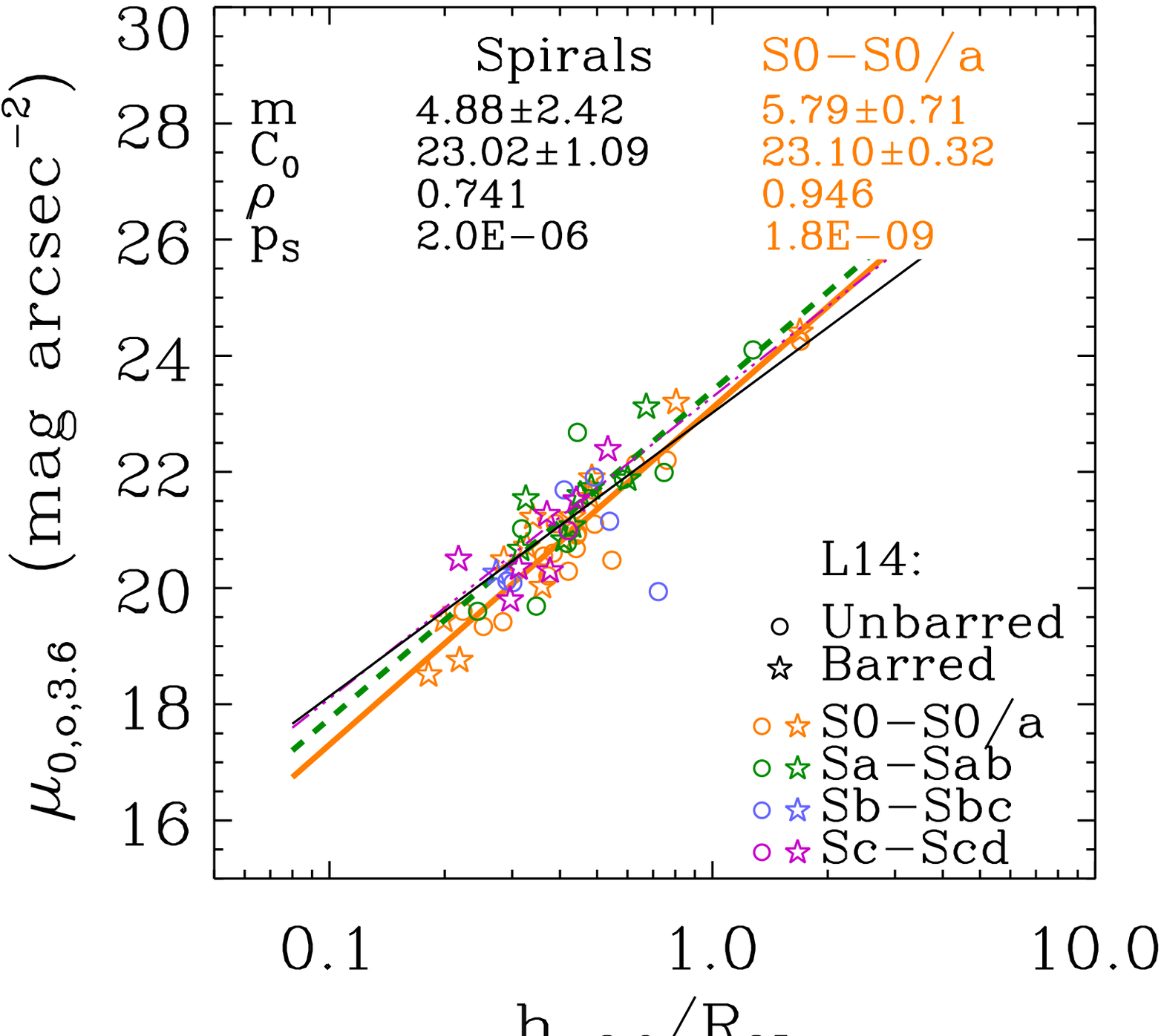}
   \includegraphics[width = 0.48\textwidth,bb=-30 -15 455 425, clip]{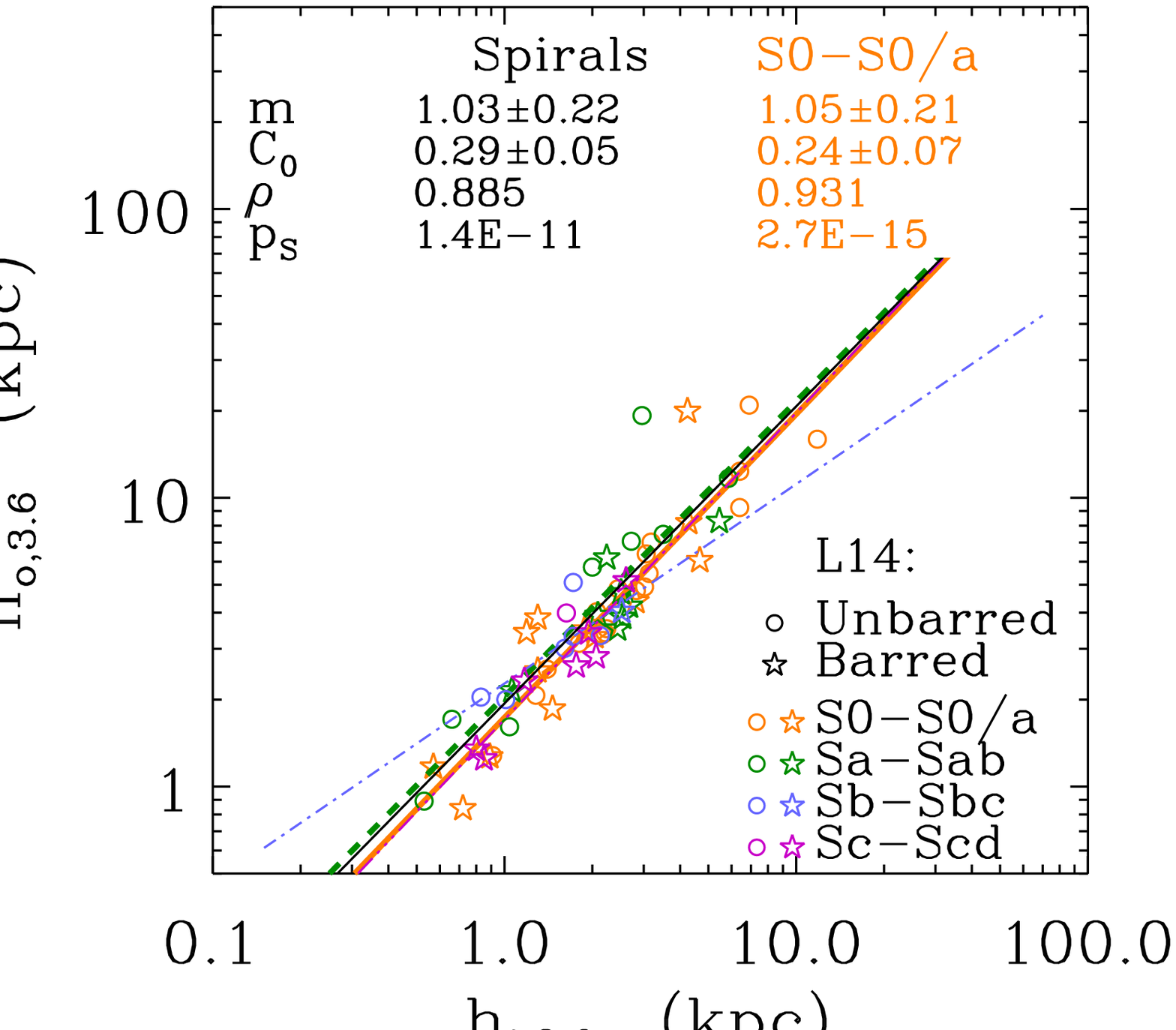}
   \includegraphics[width = 0.48\textwidth,bb=-30 -15 465 425, clip]{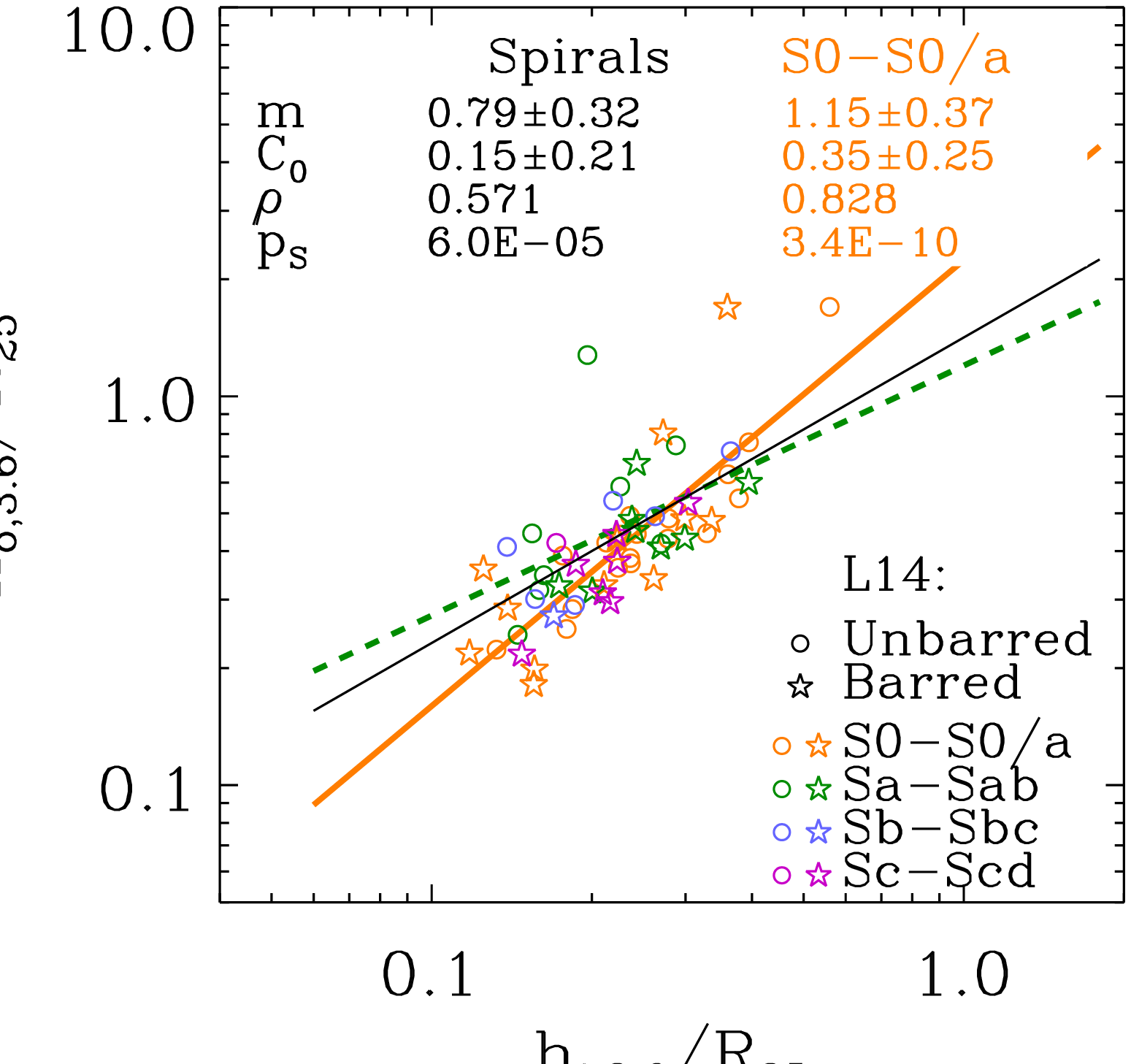}
\caption{The same as Fig.\,\ref{fig:withhiorho_R}, but for local antitruncated S0--Scd galaxies in the \nir\ band from the L14 sample  (the results are in Tables\,\ref{tab:muihimuoho} and \ref{tab:relhihorbreak}). See the caption of Fig.\,\ref{fig:withRbreak_3.6}.}
 \label{fig:withhiorho_3.6}
\end{minipage}
}
\end{tabular}
\end{figure*}

\begin{figure*}[!ht]
\begin{tabular}{cc}
\framebox[0.48\textwidth][c]{Trends with \hi\ and \ho\  in $R$ (barred/unbarred)} &   \framebox[0.48\textwidth][c]{Trends with \hi\ and \ho\ in \nir\ (barred/unbarred)} \\
   \imagetop{
 \begin{minipage}{.48\textwidth}
\centering
   \includegraphics[width = 0.48\textwidth,bb=-30 -15 455 425, clip]{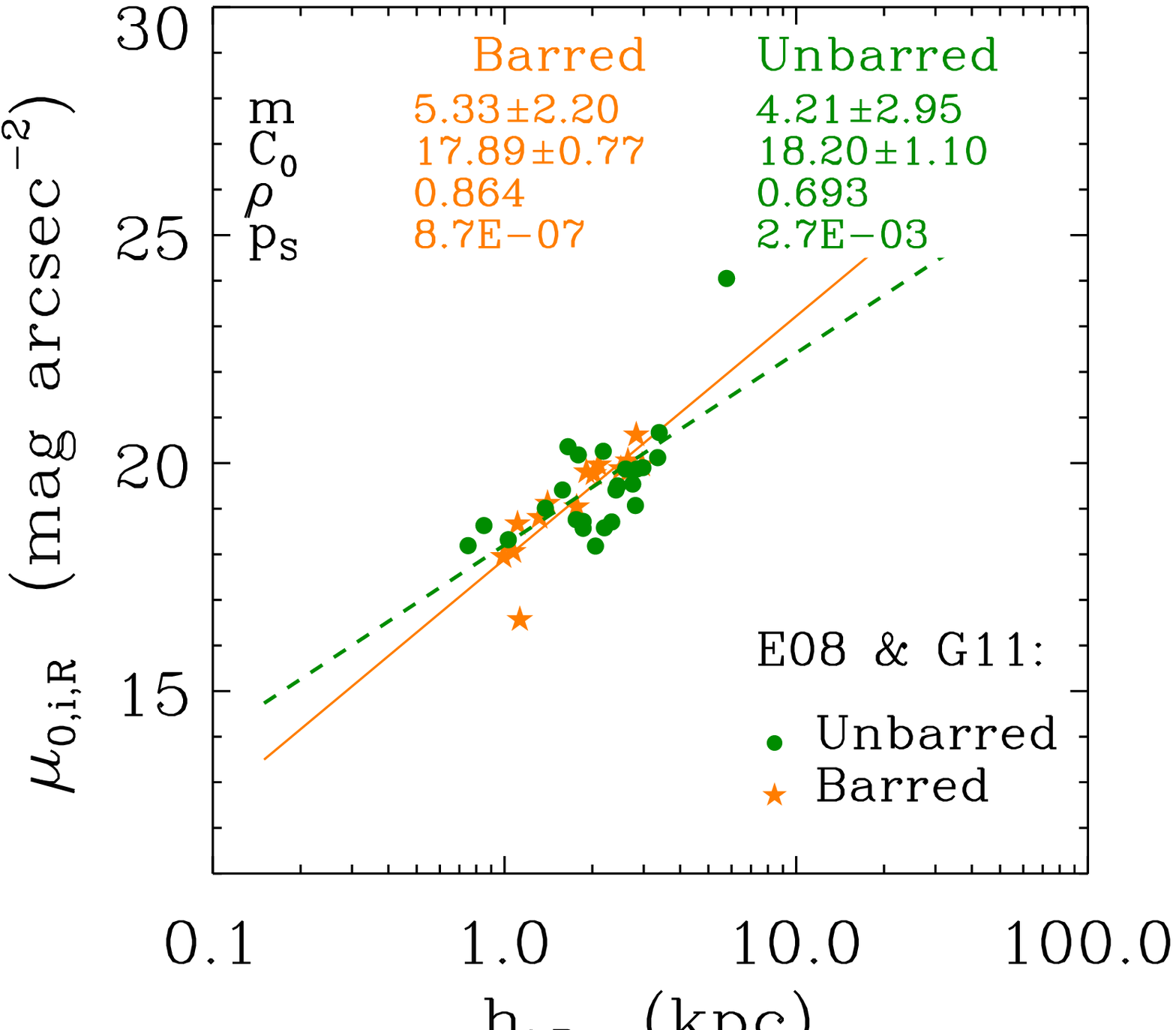}
   \includegraphics[width = 0.48\textwidth,bb=-30 -15 465 425, clip]{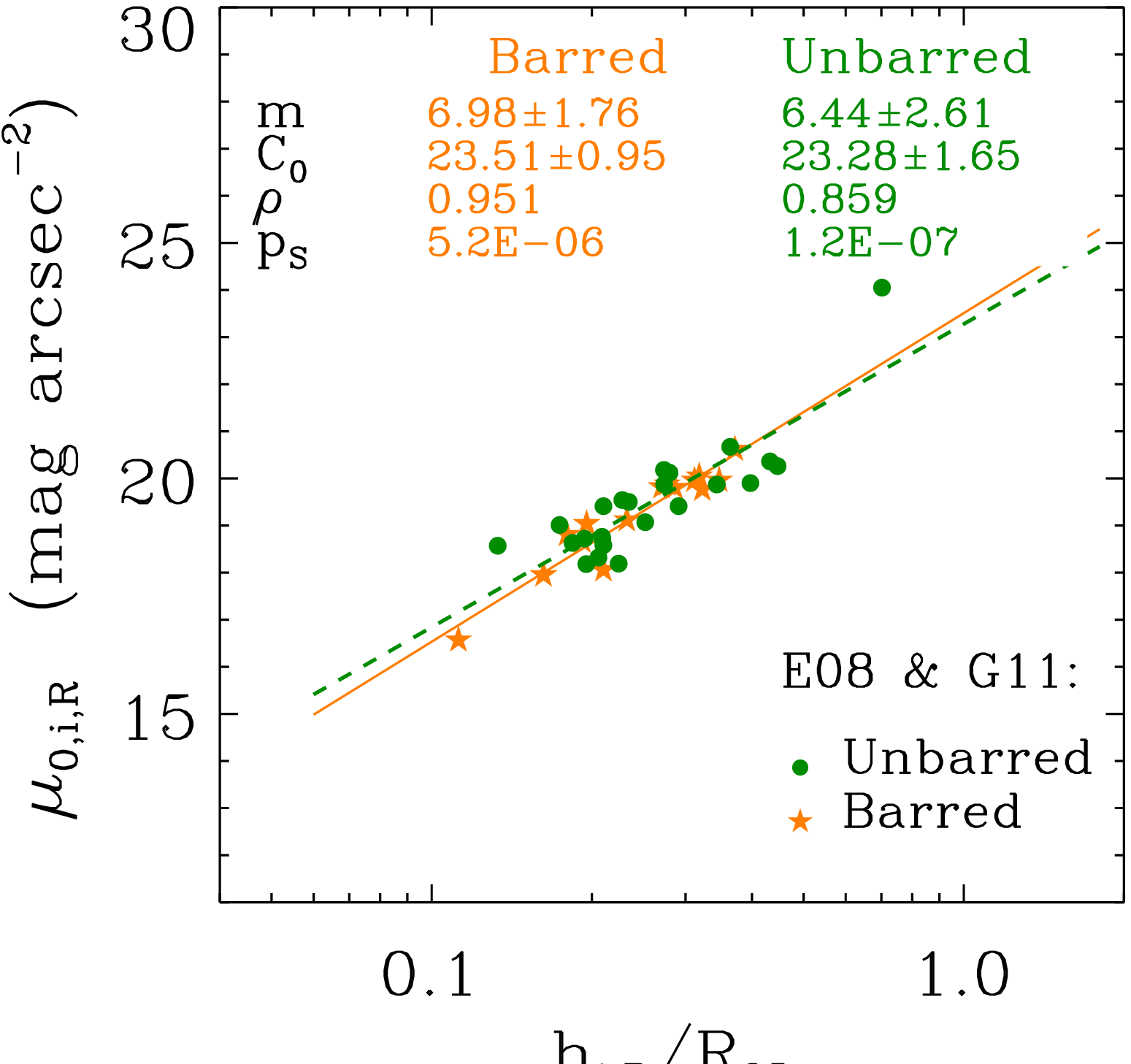}
   \includegraphics[width = 0.48\textwidth,bb=-30 -15 465 425, clip]{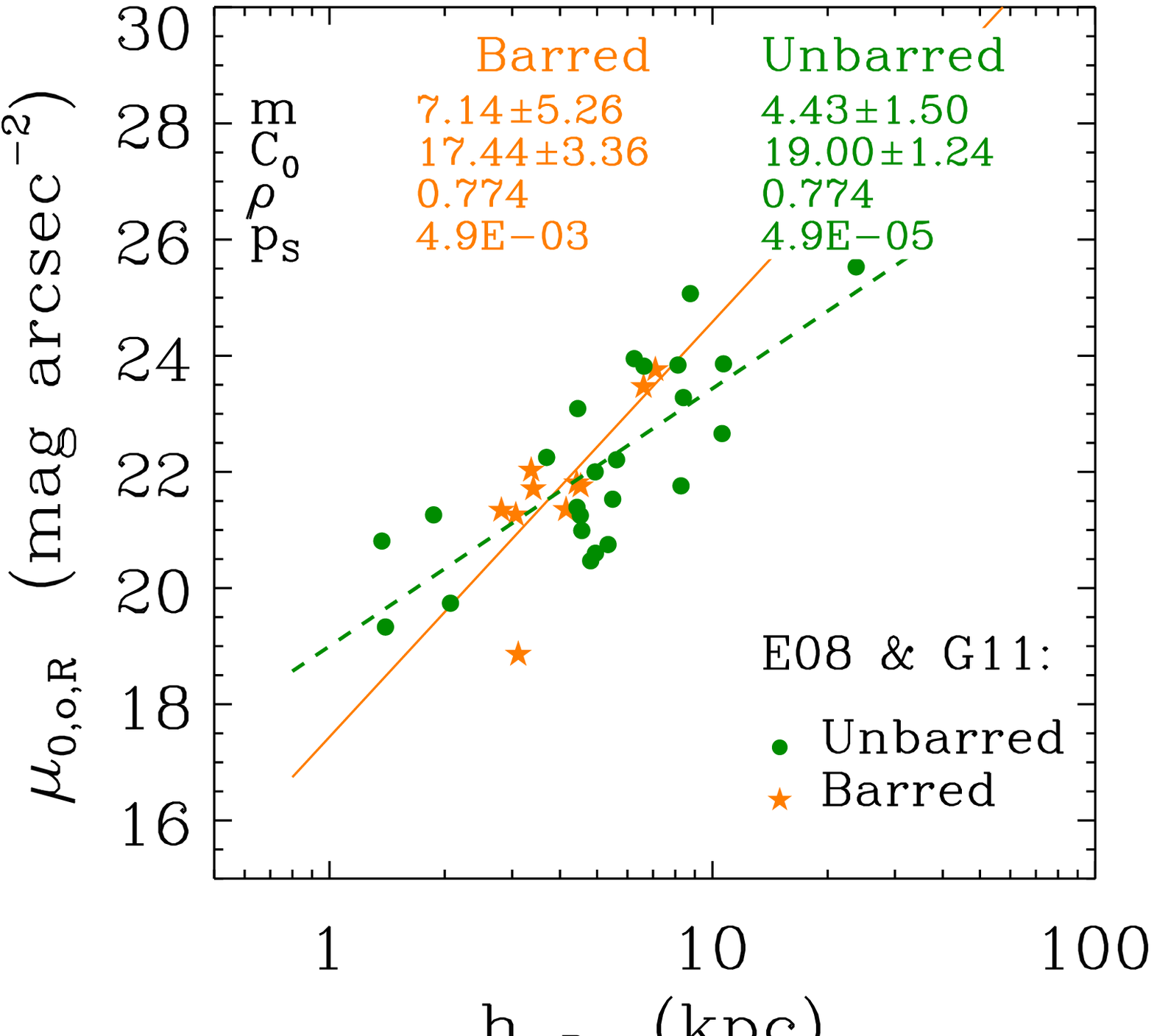}
   \includegraphics[width = 0.48\textwidth,bb=-30 -15 468 425, clip]{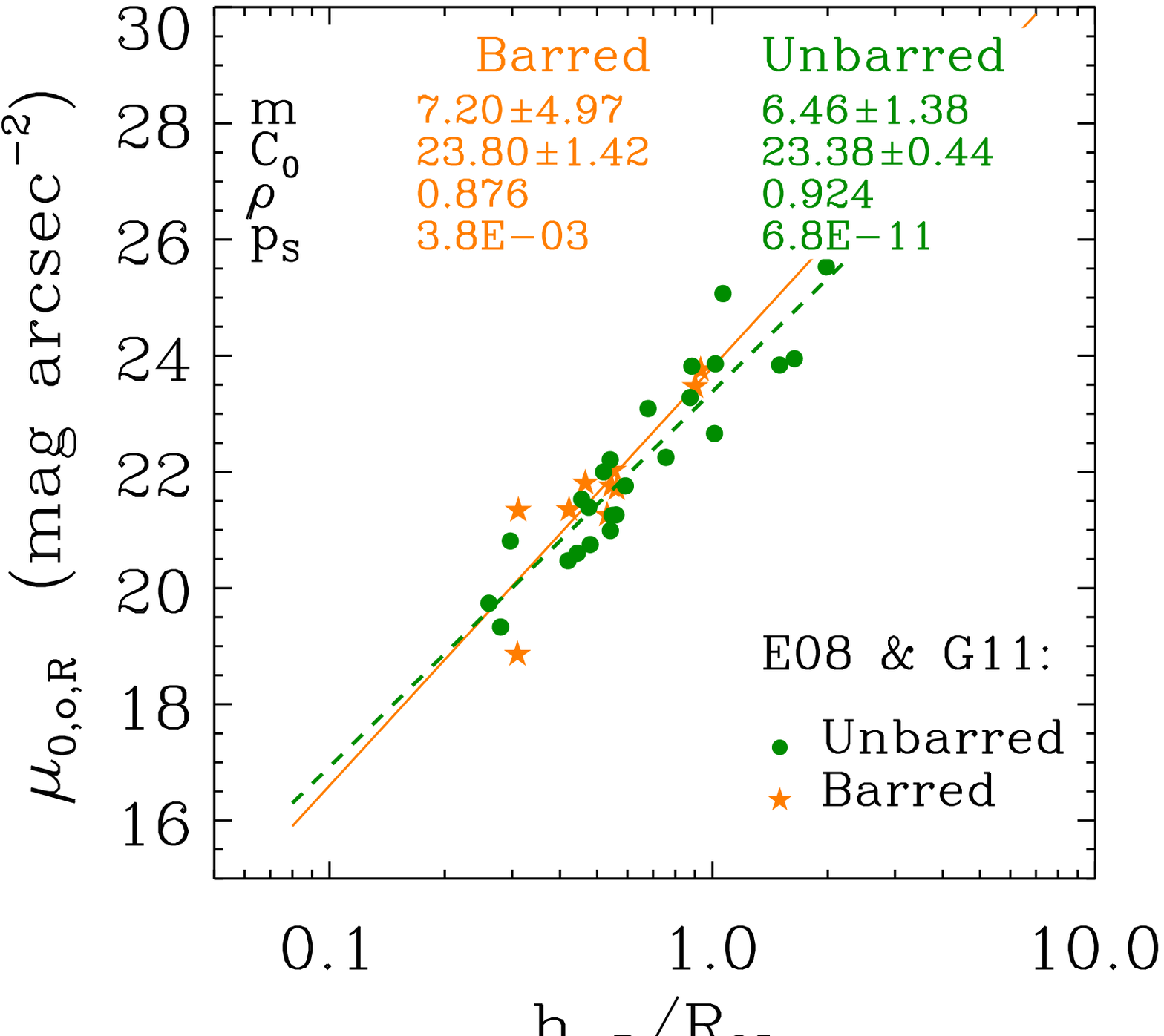}
   \includegraphics[width = 0.48\textwidth,bb=-30 -15 455 425, clip]{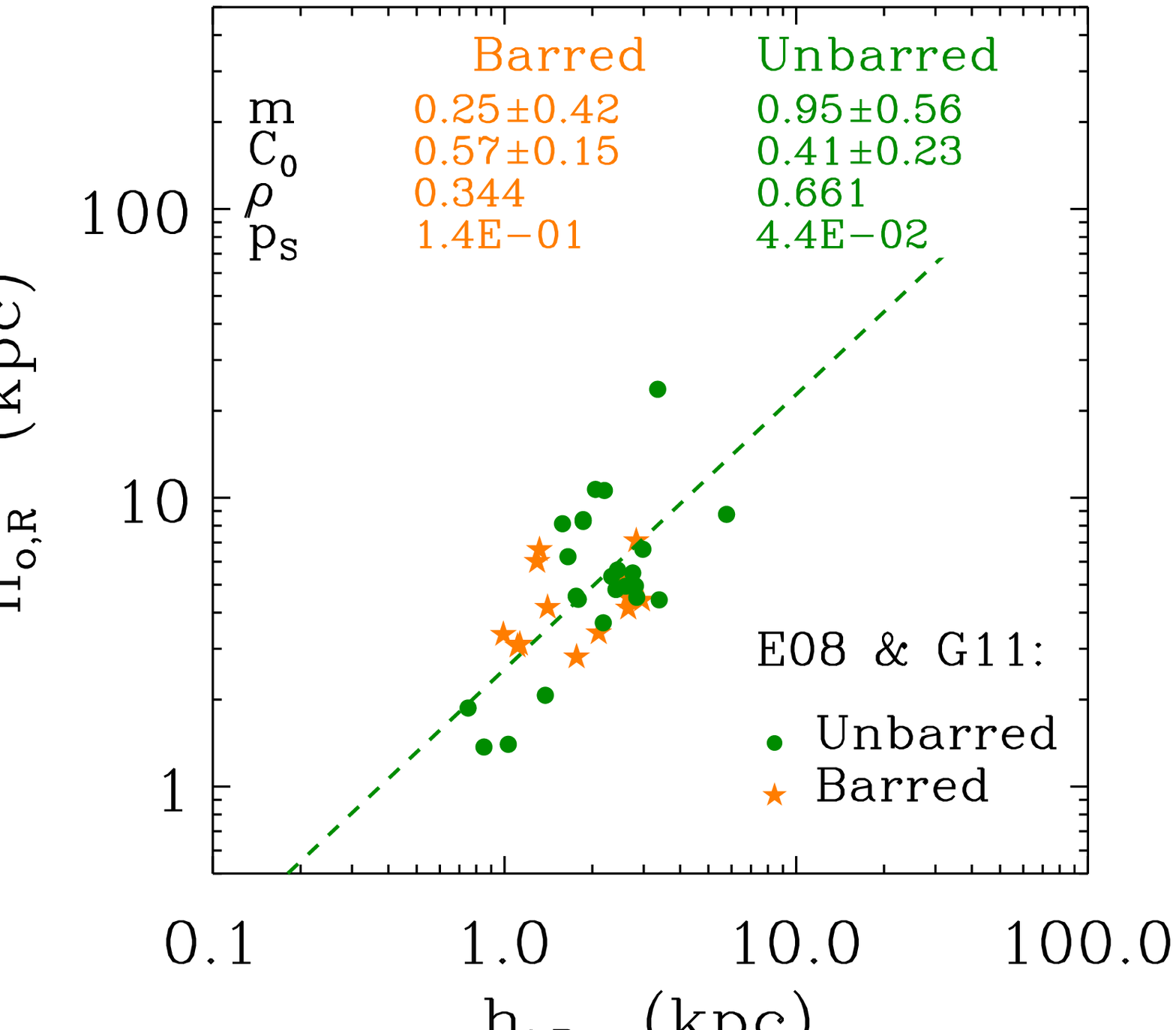}
   \includegraphics[width = 0.48\textwidth,bb=-30 -15 465 425, clip]{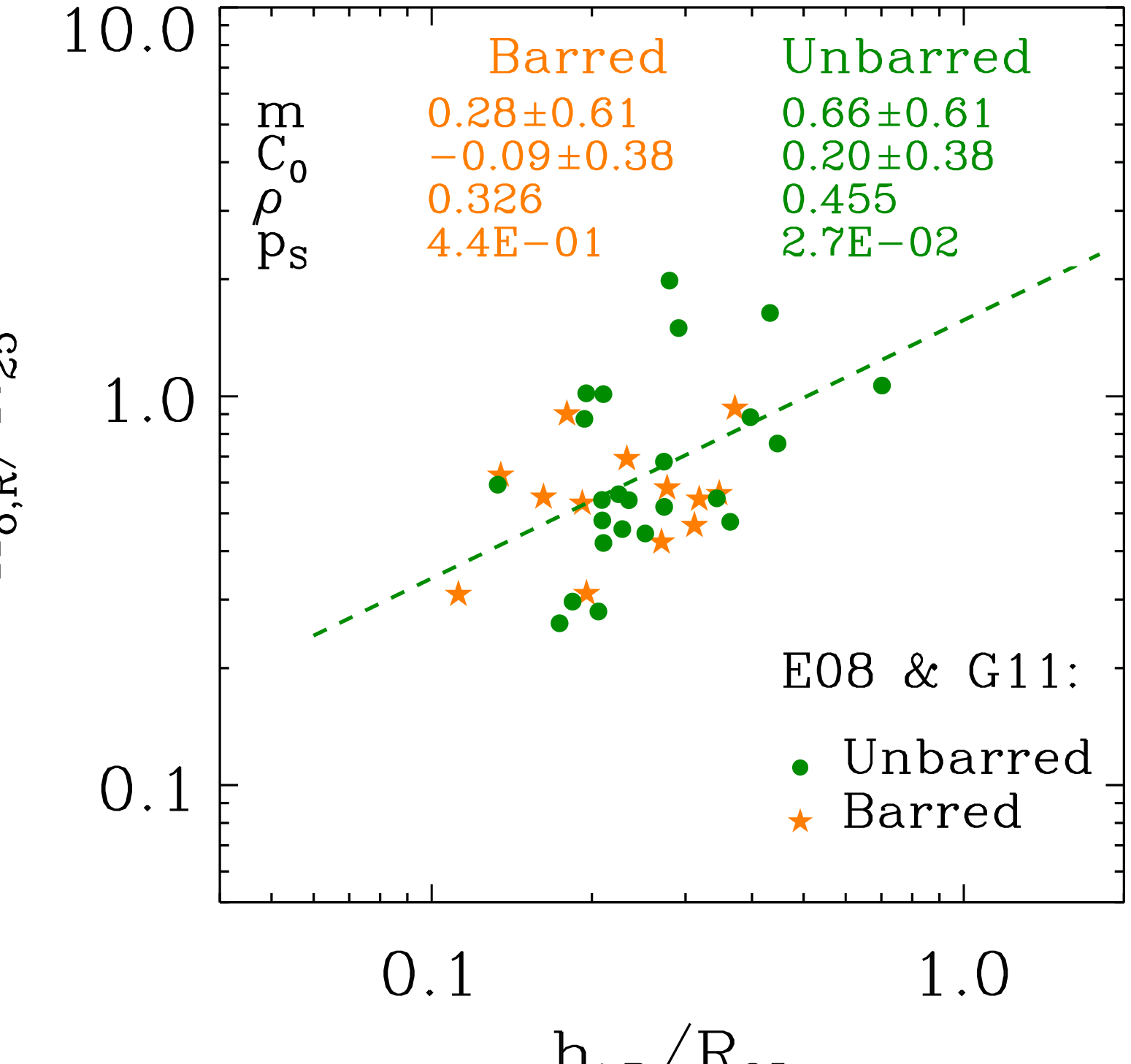}
\caption{The same as Fig.\,\ref{fig:withhiorho_R}, but for barred and unbarred galaxies in the $R$ sample  (the results are available in Tables\,\ref{tab:muihimuoho} and \ref{tab:relhihorbreak}). See the caption of Fig.\,\ref{fig:withRbreak_R_barred}.}
 \label{fig:withhiorho_R_barred}
\end{minipage}
}
& 
\imagetop{
 \begin{minipage}{.48\textwidth}
\centering
   \includegraphics[width = 0.48\textwidth,bb=-30 -15 455 425, clip]{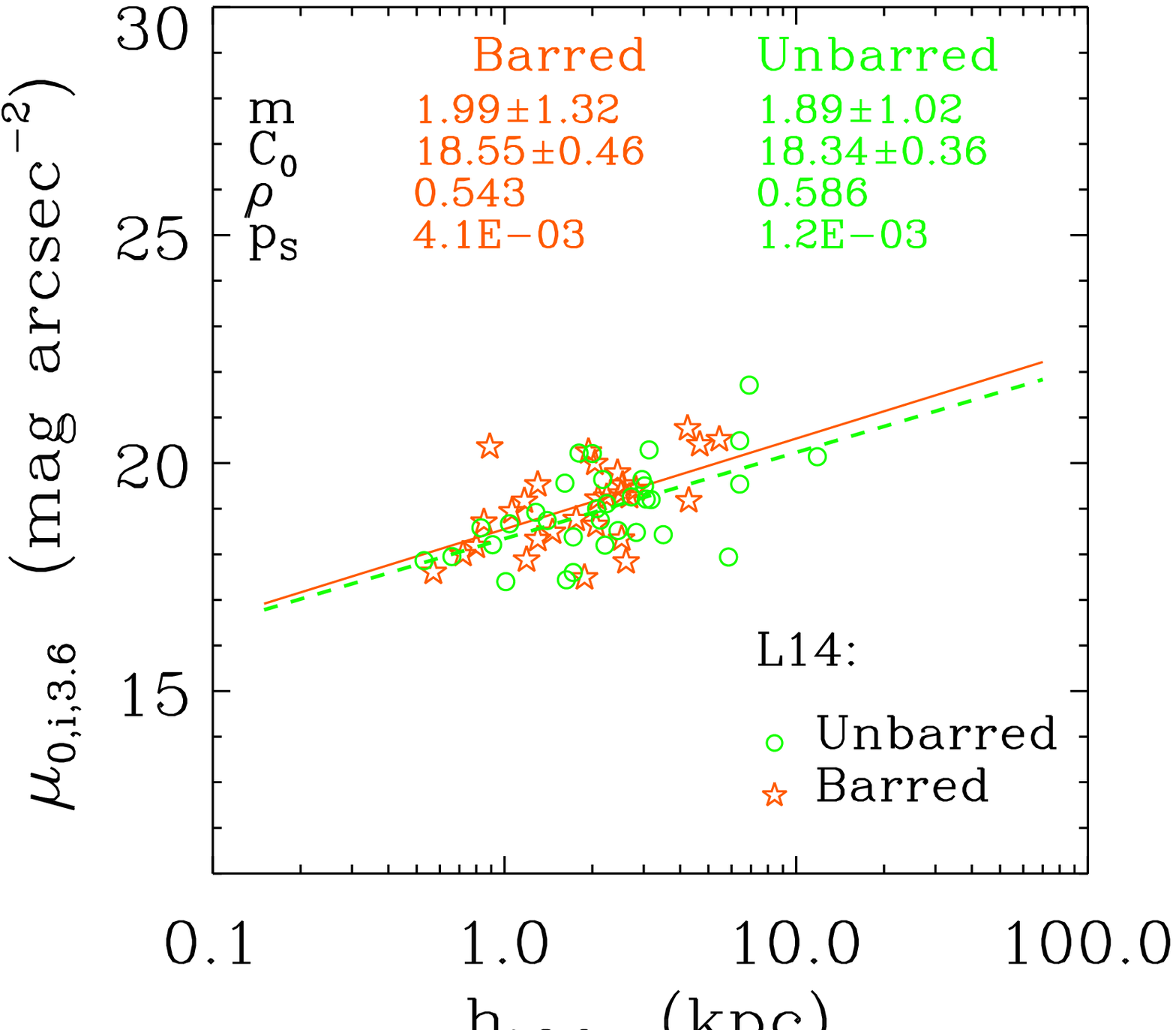}
   \includegraphics[width = 0.48\textwidth,bb=-30 -15 465 425, clip]{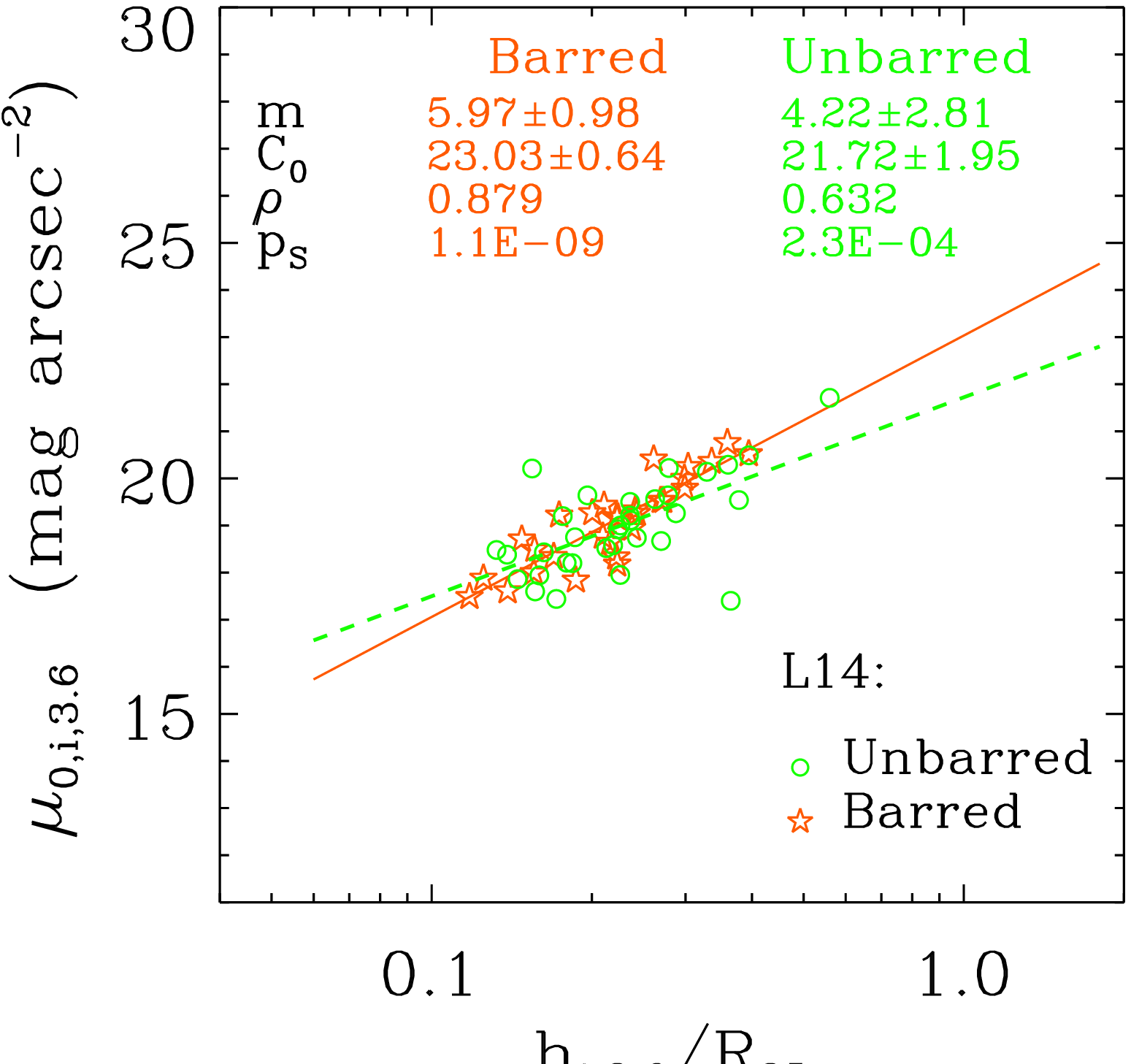}
   \includegraphics[width = 0.48\textwidth,bb=-30 -15 468 425, clip]{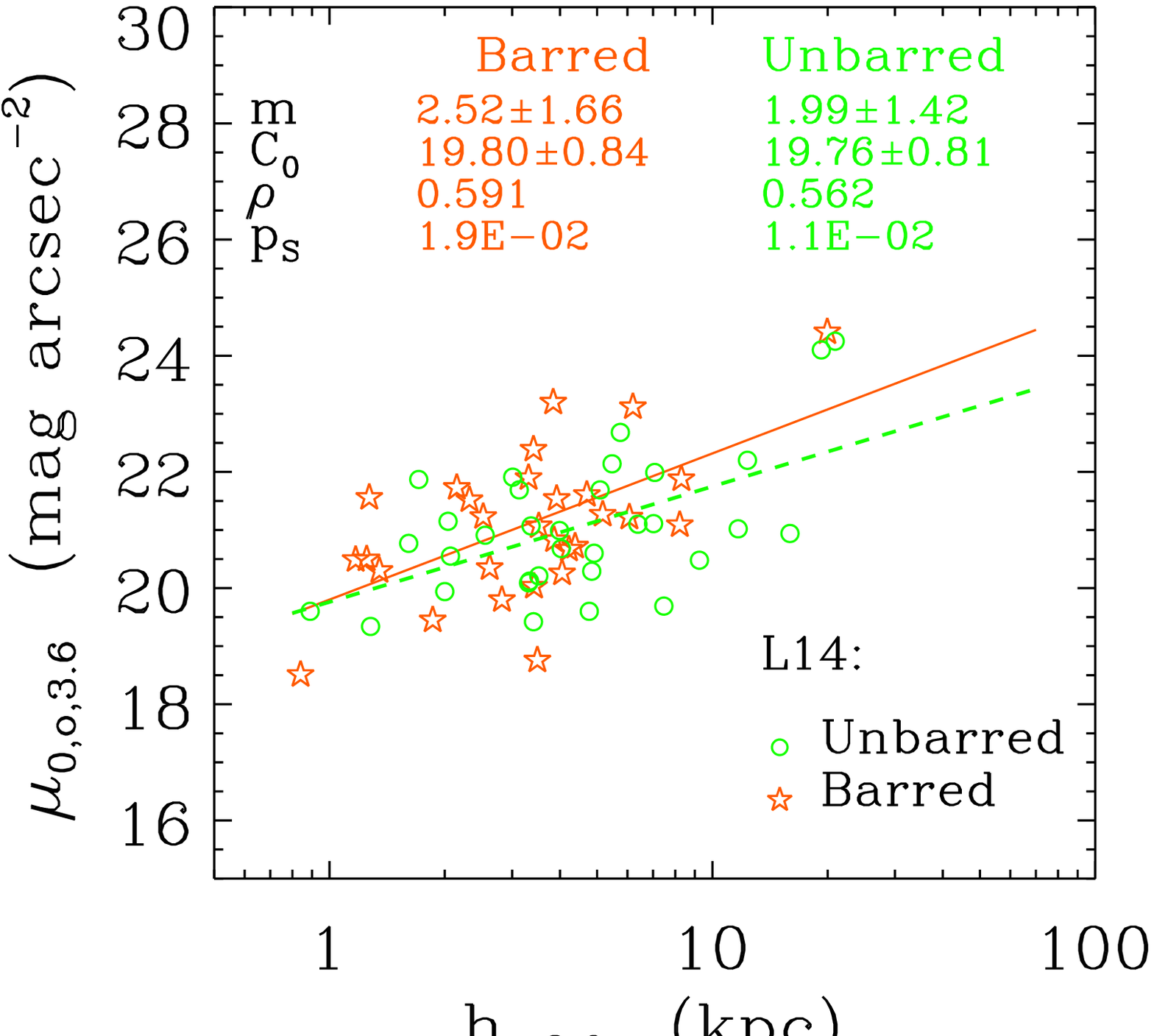}
   \includegraphics[width = 0.48\textwidth,bb=-30 -15 468 425, clip]{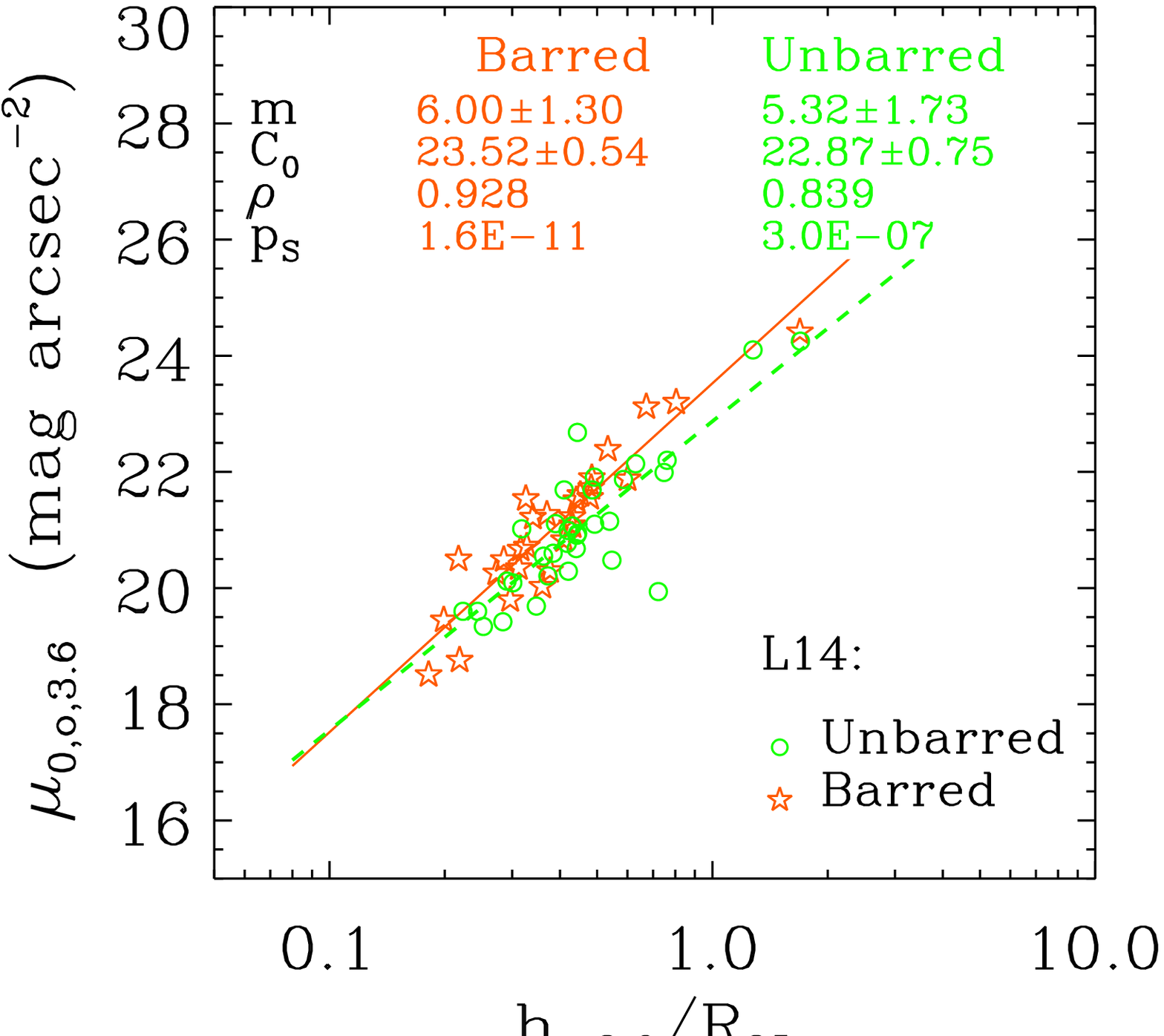}
   \includegraphics[width = 0.48\textwidth,bb=-30 -15 455 425, clip]{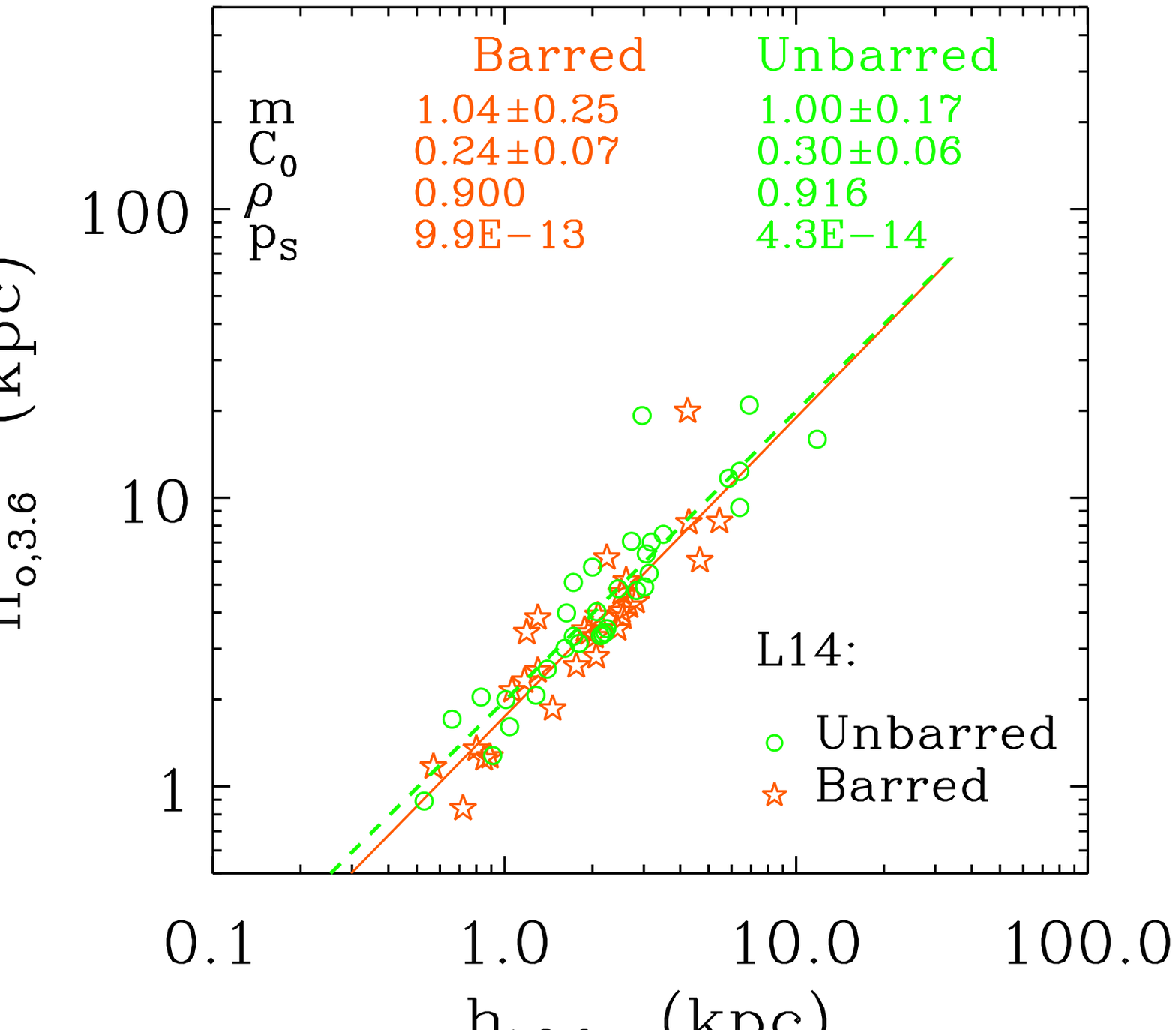}
   \includegraphics[width = 0.48\textwidth,bb=-30 -15 465 425, clip]{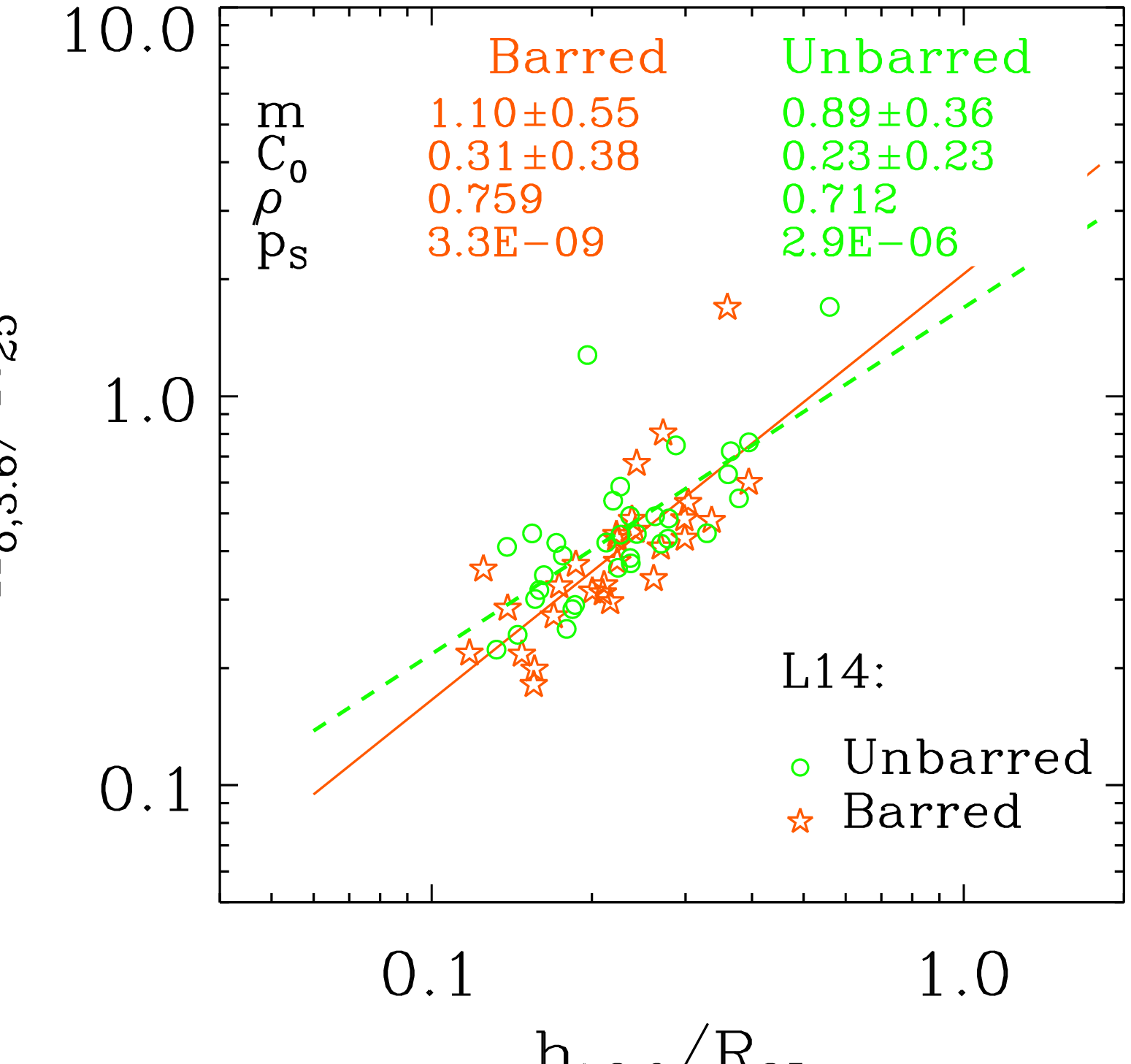}
\caption{The same as Fig.\,\ref{fig:withhiorho_3.6}, but for barred and unbarred galaxies in the \nir\ sample  (the results are in Tables\,\ref{tab:muihimuoho} and \ref{tab:relhihorbreak}). See the caption of Fig.\,\ref{fig:withRbreak_3.6_barred}.}
 \label{fig:withhiorho_3.6_barred}
\end{minipage}
}
\end{tabular}
\end{figure*}

Figures\,\ref{fig:withRbreak_R}-\ref{fig:r25} show the trends followed by Type-III galaxies in these photometric planes. We have overplotted the obtained linear fits \emph{only} when they fulfill the Spearman rank correlation test at 95\% of significance level, i.e., only if there is a significant correlation in the diagram. Note that, although there may be a significant correlation between two parameters according to the Spearman test, it does not have to be significantly linear. In fact, some trends are significant according to it ($p_S < 0.05$), but they exhibit low values of the Pearson coefficient $\rho$ (e.g., the \mubreak\ -- \rbreak\ trend in \nir\ in Fig.\,\ref{fig:withRbreak_3.6}). In these figures, we have written the results of the most relevant fits which are being compared at the top of each panel even when the correlations are not significant. For simplicity, we have symmetrized the error interval of $m$ and $C_0$ in the figures, but the asymmetrical upper and lower errors really obtained for the coefficients of the fits are available in Tables\,\ref{tab:hihorbreak}-\ref{tab:r25}. 

The characteristic scalelengths are plotted in logarithmic scales in all figures, because the correlations exhibit more defined linear trends in this way than using linear scales. In many photometric planes, we have normalized the characteristic scalelengths (\hi, \ho, \rbreak) to the optical radius of each galaxy. We have defined this following E08 and G11, i.e.,  as the radius of the isophote with $\mu = 25$\,mag arcsec$^{-2}$ in the $B$ band (\risoph). These authors provide \risoph\ for each galaxy in their samples, so we have used their tabulated values directly. The values of \risoph\ for the galaxies in the L14 sample have been obtained from HyperLeda\footnote{HyperLeda database is available at: http://leda.univ-lyon1.fr/}, and include a correction for Galactic extinction and inclination effects. 

\section{Results}
\label{sec:results}

In Section\,\ref{sec:trends}, we discuss the trends and correlations found in several photometric planes for the different morphological types and for barred and unbarred galaxies in the two datasets ($R$ and \nir). In Section\,\ref{sec:comparison}, we compare the slopes and $Y$-intercepts of the linear trends fitted in each photometric plane for S0s and spirals, as well as for barred and unbarred. The fits obtained for the $R$-band and \nir\ data are only compared in the photometric relations exclusively relating characteristic scalelengths.


\subsection{Trends and scaling relations}
\label{sec:trends}

\subsubsection{Trends with \rbreak}
\label{sec:rbreak}

Figures\,\ref{fig:withRbreak_R} and \ref{fig:withRbreak_3.6} show the trends of several photometric parameters of the inner and outer discs of Type-III galaxies with \rbreak\ and $\rbreak/\risoph$ in the $R$ and \nir\ bands, respectively. The two top rows of the figures display the trends of the inner and outer disc scalelengths with \rbreak\ by Hubble types in each band, in logarithmic scale. The distributions of spirals and S0s are similar in these planes and overlap.  

The main result is that \loghi, \logho, \mui, \muo, and \mubreak\ correlate strongly with \logrbreak\ in both spirals and S0s (all these trends have $p_S<0.05$), and furthermore, similar correlations are obeyed for the different spiral types surveyed by each sample (Sa--Sab and Sb--Sbc in both bands, and Sc--Scd in \nir) within the observational uncertainties. The dispersions around the fitted linear trends in S0s and spirals are similar in both bands, although the linear trends of \loghi\ and \logho\ with \logrbreak\ are better defined in \nir\ than in $R$ (i.e., they have higher values of the linear correlation coefficient $\rho$), whereas it is the opposite in the trends involving \mui, \muo, and \mubreak\ (this is noticeable just by visual inspection of the trends).

The only diagrams in which spirals (globally or by types) show no significant correlation according to the Spearman rank correlation test are $\ho$ -- $\rbreak$ in $R$ and $\mui$ or $\muo$ versus $\rbreak$ (or $\rbreakrisoph$) in \nir. The first may be only a question of the lower statistics of the $R$-band sample as compared to the \nir\ subsample, because the analogous plane in \nir\ shows significant linear correlations for all spirals and by their (sub-)types, and the data distributions in both planes are quite similar. Correspondingly, the distributions of S0s and spirals in the diagrams of $\mui$ (or $\muo$) -- \rbreak\ and $\mui$ (or $\muo$) -- $\rbreak/\risoph$ in \nir\ are also similar to the analogous distributions in the same diagrams of the $R$ band (compare the panels corresponding to \mui\ in both figures), so the lack of significance in the correlations in \nir\ might also be a question of small numbers.

The trends of these photometric parameters with $\logrbreakrisoph$ present similar or even higher values of linear correlation (as measured by $\rho$) than with \logrbreak. In general, the correlations of  \mui, \muo, and \mubreak\ improve when \rbreak\ is normalized to the optical size of the galaxy (in particular, compare the trends and the Pearson coefficients of \mubreak\ -- \rbreak\ and \mubreak\ -- \rbreakrisoph\ at the bottom panels of Figs.\,\ref{fig:withRbreak_R} and \ref{fig:withRbreak_3.6}). The values of \mui, \muo, and \mubreak\ in Type-III discs are fainter as the breaks are more external (see the corresponding panels in the figures). However, these values seem to be more closely linked to the relative location of \rbreak\ with respect to the outer radius of the galaxy (as measured by \risoph) than to \rbreak. 

In Figs.\,\ref{fig:withRbreak_R_barred} and \ref{fig:withRbreak_3.6_barred} we plot the same photometric planes as in Figs.\,\ref{fig:withRbreak_R} and \ref{fig:withRbreak_3.6}, but distinguishing between barred and unbarred galaxies. Again, the linear fits performed to the barred and unbarred galaxies have been overplotted only if they are significant. The majority of the photometric planes show significant scaling relations for both barred and unbarred galaxies in the two bands. The trends fitted to the barred galaxies look similar to those obtained for unbarred galaxies within the observational dispersion, as derived from the fact that the distributions for the two galaxy classes practically overlap in the diagrams. This suggests that bars seem to affect these scaling relations very little (at least, within the uncertainties implied by the data samples).

The only relations which are not significant in Figs.\,\ref{fig:withRbreak_R_barred} and \ref{fig:withRbreak_3.6_barred} are the trends involving \ho\ and \muo\ for barred galaxies in $R$ and the \mui\ -- \logrbreakrisoph\ trend in \nir\ for the unbarred galaxies. But again, the lack of correlation in each band may reflect the low statistics of the samples. 

The linear trends for barred and unbarred galaxies are very well defined in the \nir\ dataset in the planes involving \hi\ and \ho\, while those relating \mui, \muo, and \mubreak\ with \logrbreak\ have higher $\rho$ values in the $R$ band (as also happened in Figs.\,\ref{fig:withRbreak_R} and \ref{fig:withRbreak_3.6} for S0s and spirals). In any case, the trends in the photometric planes described by the $R$-band dataset look similar to their \nir\ analogs taking into account the data dispersion. Again, we find that the linear correlation coefficients of the trends relating \mui, \muo, and \mubreak\ with \logrbreak\ tend to improve if \rbreak\ is normalized to \risoph, for both barred and unbarred galaxies (see the three rows of panels at the bottom of Figs.\,\ref{fig:withRbreak_R_barred} and \ref{fig:withRbreak_3.6_barred}).

Summarizing, we have found that the inner and outer discs of antitruncated spirals obey tight photometric scaling relations with \rbreak, as \citet{2014A&A...570A.103B} discovered for Type-III S0 galaxies. The trends for each type look similar among different morphological types and among barred and unbarred galaxies within the dispersion of the data in the planes. This result suggests that antitruncations and bars are structurally independent phenomena in galaxies.

\subsubsection{Trends with \hi\ and \ho}
\label{sec:hoorhi}

In Figs.\,\ref{fig:withhiorho_R} and \ref{fig:withhiorho_3.6} we analyse the basic scaling relations obeyed by the inner and outer discs of Type-III galaxies in $R$ and \nir\ respectively. We show the photometric planes also normalizing \hi\ and \ho\ by \risoph. The different morphological types (S0s and spirals, as well as by Hubble types) yield significant linear relations in these photometric planes too, again similar among them within the observed data dispersion. The distribution in the planes of spirals and S0s overlap also in these diagrams.

L14 already reported that the two exponential sections of galaxy discs of Types II and III in their sample independently satisfied the basic scaling relation observed in pure exponential discs between their central surface brightness and their scalelengths, although they did not distinguish between different Hubble types in their Fig.\,11. The two top panels in the first column of Fig.\,\ref{fig:withhiorho_3.6} show that this result also applies for different morphological types (S0, Sa--Sab, Sb--Sbc, and Sc--Scd) and spirals in general, and that it can be extended to the $R$ band (see the corresponding panels in Fig.\,\ref{fig:withhiorho_R}). 

Again, the linear correlations involving \mui\ and \muo\ improve noticeably when the disc scalelengths are normalized to \risoph\ for the two main galaxy types being considered (compare the left panels with the right panels in the two figures). This is more striking in the \nir\ trends, where this improvement can be noticed by visual inspection: the dispersion around the fitted linear trends in the \mui\ -- \hi\ and \muo\ -- \ho\ plots is significantly reduced in Fig.\,\ref{fig:withhiorho_3.6} when \hi\ and \ho\ are normalized to \risoph. Moreover, $\rho$ increases significantly for both S0s and spirals in the two planes after this normalization. 

The bottom panels of Fig.\,\ref{fig:withhiorho_3.6} show that \loghi\ and \logho\ correlate linearly in both S0s and spiral galaxies in the \nir\ band (in fact, this applies independently for Sa--Sab, Sb--Sbc, and Sc--Scd types). In contrast, no significant trends are found in $R$, except for the S0s (see the same panel in Fig.\,\ref{fig:withhiorho_R}). Note that the \logho\ -- \loghi\ trends in \nir\ do not improve if the scalelengths are normalized to \risoph\ (compare the bottom panels of Fig.\,\ref{fig:withhiorho_3.6}).

We have plotted the same photometric relations in Figs.\,\ref{fig:withhiorho_R_barred} and \ref{fig:withhiorho_3.6_barred}, but now differentiating barred from unbarred galaxies. The linear fits obtained for each galaxy class (barred vs.\,unbarred) have been overplotted only if the correlations were significant according to the Spearman rank correlation test, as above. The figures show that barred and unbarred galaxies overlap in these diagrams and follow tight scaling relations in them, similar within the observational dispersion. Therefore, these scaling relations seem to be independent of the existence of a bar in the galaxy within the observational uncertainties, again suggesting that bars and antitruncations are structurally unrelated phenomena.

In conclusion, we have found that the inner and outer discs of Type-III spirals obey tight scaling relations too, as observed in Type-III S0 galaxies. Again, we find that the existence of bars in the galaxies affect negligibly to these scaling relations within the observational uncertainties and that the relations in the \mui\ -- \hi\ and \muo\ -- \ho\ planes significantly improve when the scalelengths are normalized by \risoph.


 \begin{figure}[!ht]
\centering
    \framebox[0.48\textwidth][c]{Trends of $\risoph$ in $R$ and \nir\ bands (S0s/spirals)} 
\includegraphics[width = 0.24\textwidth,bb=-15 -15 465 435, clip]{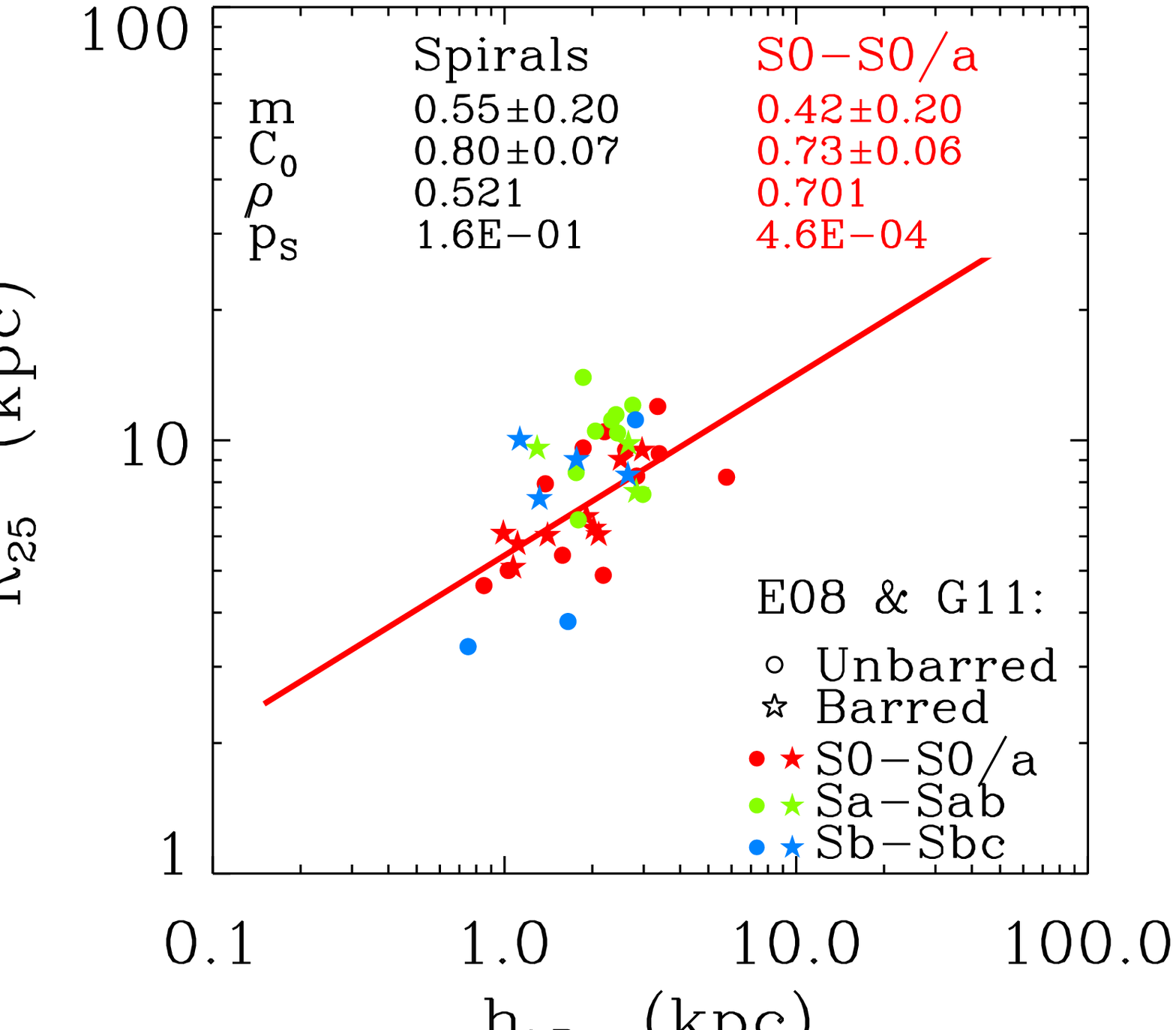}
   \includegraphics[width = 0.24\textwidth,bb=-15 -15 465 435, clip]{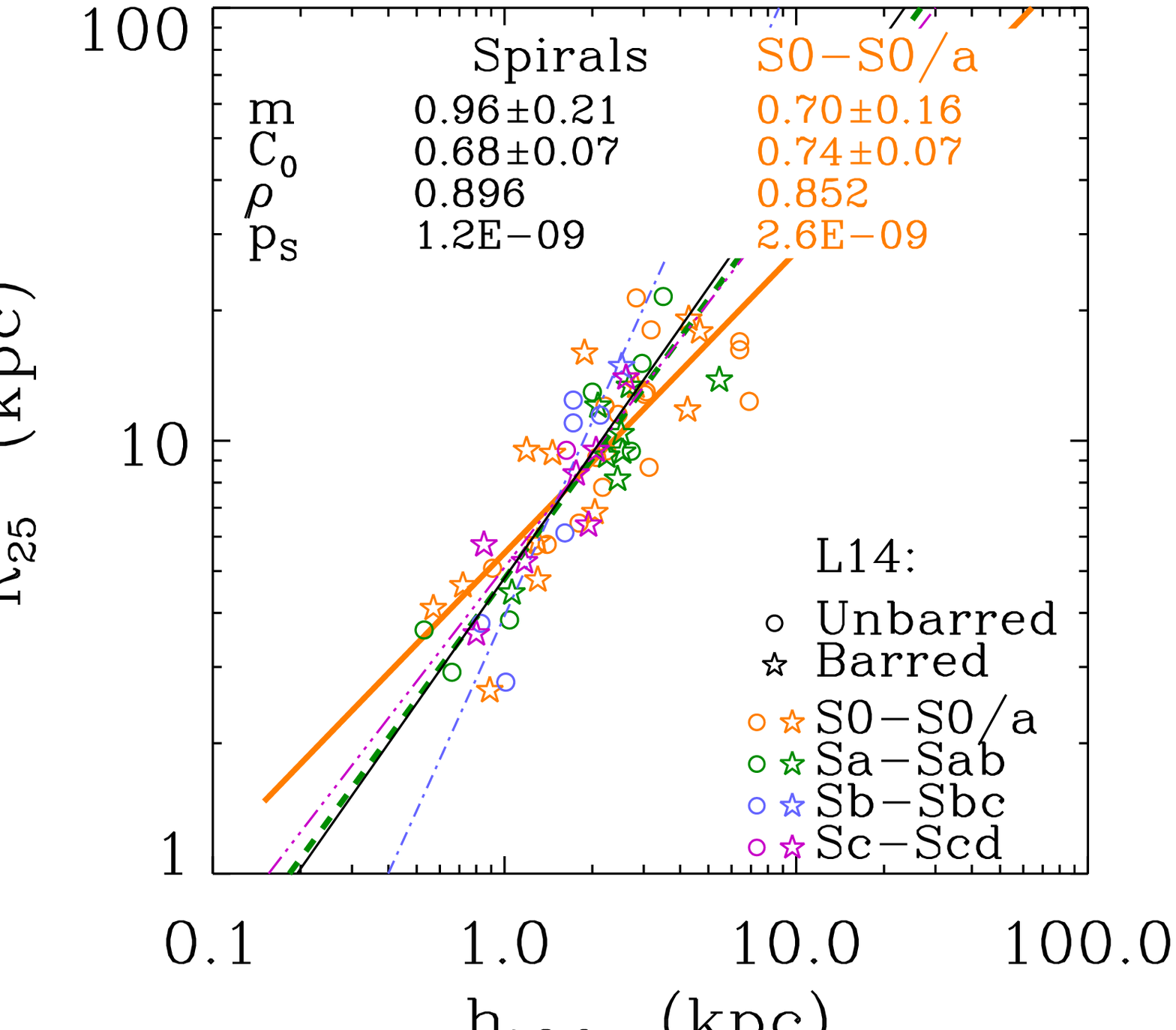}
  \includegraphics[width = 0.24\textwidth,bb=-15 -15 470 435, clip]{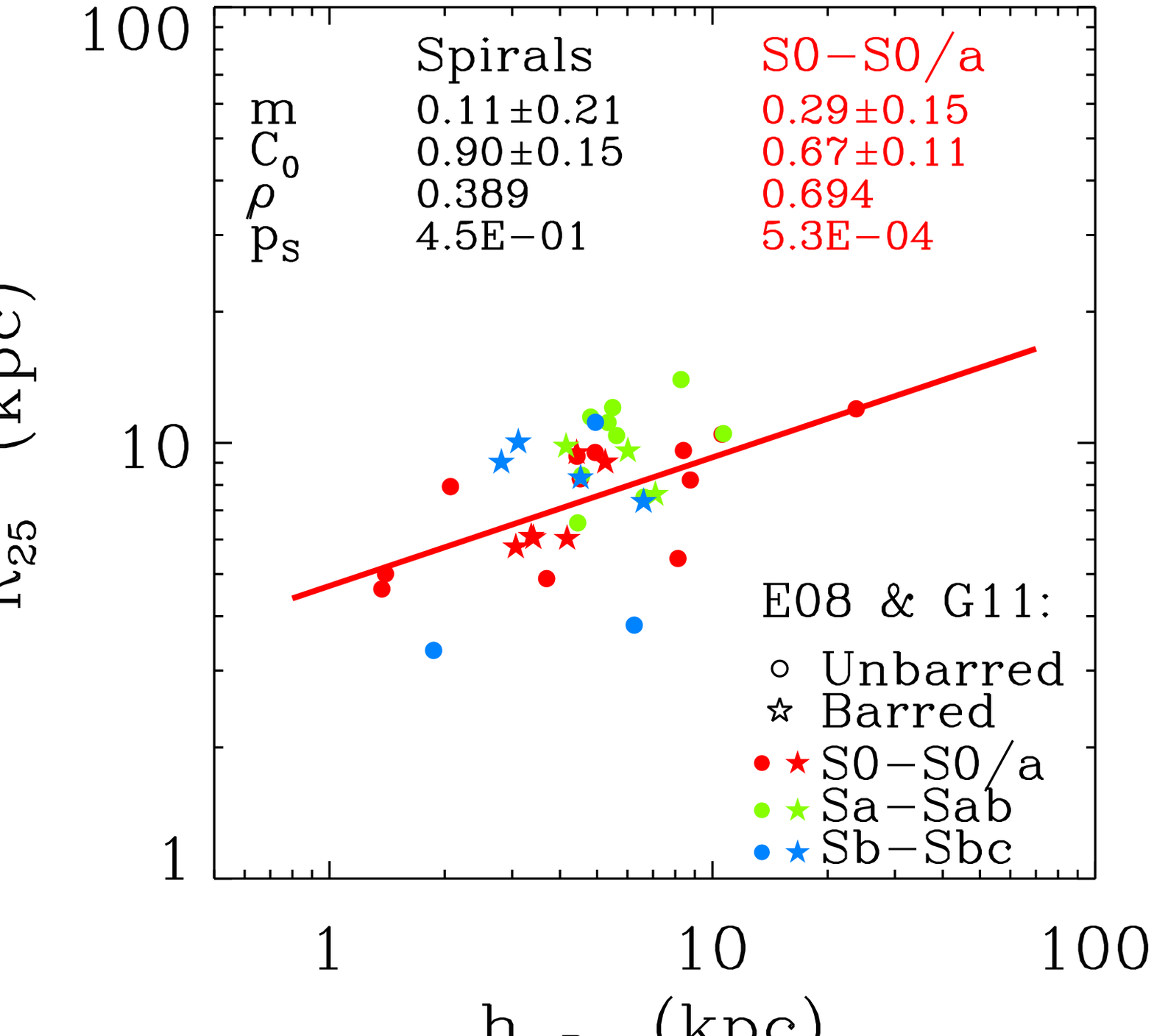}
   \includegraphics[width = 0.24\textwidth,bb=-15 -15 470 435, clip]{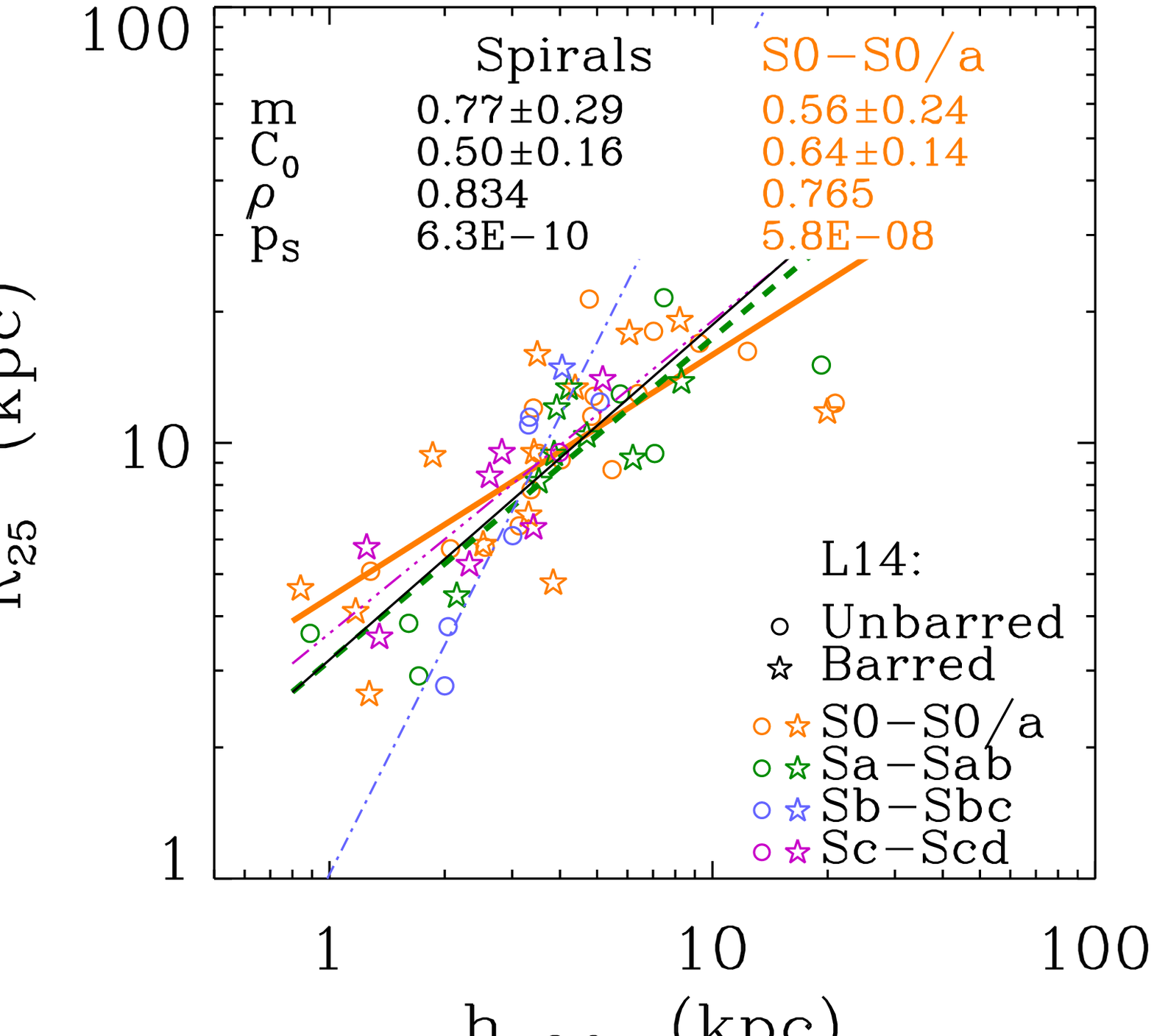}
  \includegraphics[width = 0.24\textwidth,bb=-15 -15 470 435, clip]{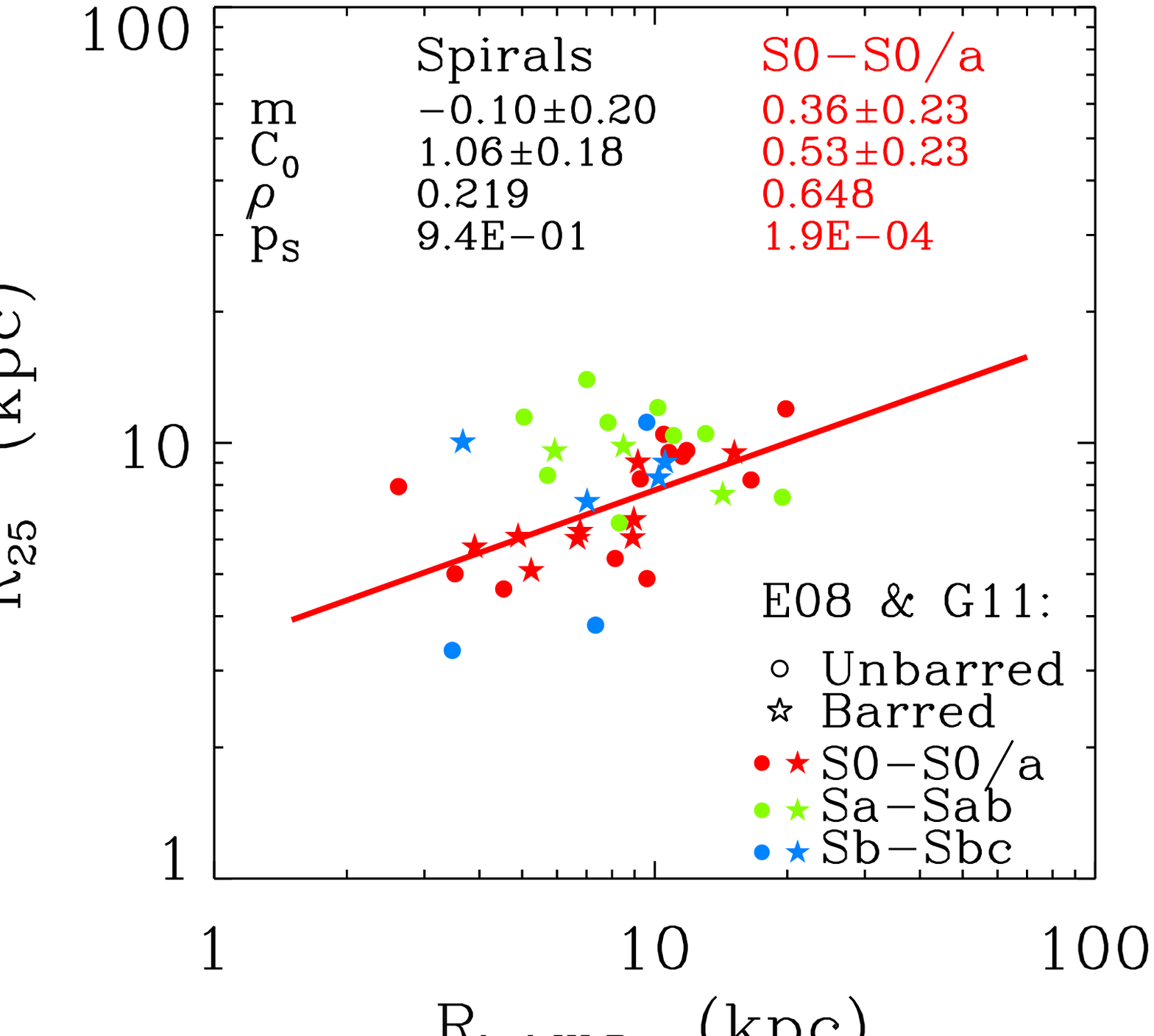}
   \includegraphics[width = 0.24\textwidth,bb=-15 -15 470 435, clip]{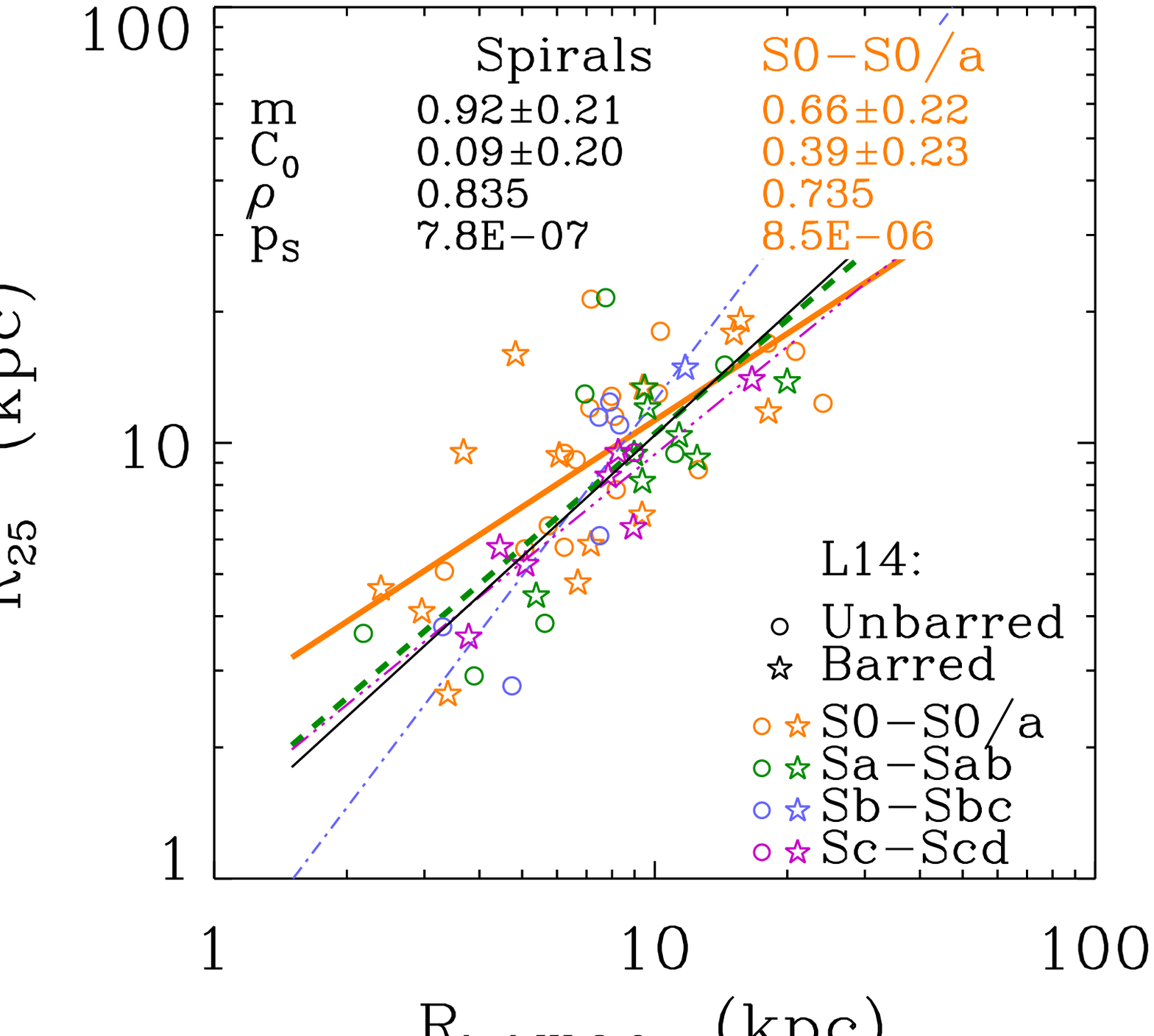}
 \caption{Trends of \risoph\ with \hi, \ho, and \rbreak\ in local antitruncated S0--Scd galaxies in the $R$ and \nir\ samples (left and right panels, respectively). The results are tabulated in Table\,\ref{tab:r25}. See the captions of Figs.\,\ref{fig:withRbreak_R} and \ref{fig:withRbreak_3.6}.}
 \label{fig:r25}
 \end{figure}  

\subsubsection{Trends with \risoph}
\label{sec:R25}

As shown above, the linear correlations between the characteristic surface brightness values and the scalelengths become better defined in many photometric planes after normalizing the relevant parameters to \risoph. We have analysed the trends between these characteristic scalelengths (\hi, \ho, and \rbreak) and \risoph\ in Fig.\,\ref{fig:r25} for several galaxy types. First, the values of \rbreak, \hi, and \ho\ span similar ranges for a given \risoph\ in both $R$ and \nir, implying that both bands must be sampling the same type of breaks, but in different wavelength ranges.

In $R$, only the S0s exhibit significant trends of \loghi, \logho, or \logrbreak\ with \logrisoph, despite the fact that S0s and spirals basically overlap in all planes (left panels in the figure). However, the \nir\ dataset shows well-defined linear correlations for S0s and spirals, as well as for spiral sub-types in these diagrams (right panels). The trends look similar among them, as  observed in the photometric parameters analysed previously. Again, the distributions of the $R$-band and \nir\ data are similar in the same photometric planes, so the lack of correlations for spirals in $R$ might be due to the small numbers in the sample, as commented above.

The scaling relations in the right panels of Fig.\,\ref{fig:r25} relate the size of the galaxy computed from an optical blue band ($B$) with the structure of the inner and outer discs observed in a NIR band (\nir), indicating that there is a clear size scaling in these galaxies, such that larger (Type-III) discs have larger inner and outer disc scalelengths, and hence larger break radii. These scaling relations may present higher dispersion because \risoph\ is measured in the $B$ band by definition, a band which is not a proxy of the galaxy stellar mass certainly, whereas \hi, \ho, and \rbreak\ have been derived using data in much redder bands. However, the effects of dust in the $B$ band must be very limited at radial locations near \risoph\ in the discs, so we can assume that in practice the $B$ band does trace the stellar mass similarly to the $R$ or \nir\ bands at these external radii. Moreover, the systematic improvement that we have found in many scaling relations after normalizing the scalelengths by \risoph\ implies that it must provide a robust estimate of the size of the stellar distribution, despite being computed in a blue band. 

\subsection{Comparison of the trends}
\label{sec:comparison}

In Sections\,\ref{sec:trends} we have seen that the scaling relations followed by S0s and spirals and by barred and unbarred galaxies look similar, both in the $R$ and \nir\ bands. Here we analyse if there is statistical significant evidence that these relations differ within the observational uncertainties. 

In order to do so, we have considered the relative differences between the fitted values of the slopes and $Y$-intercepts for each trend in the two pair of datasets compared in each case (S0 -- spirals, barred -- unbarred). We define the relative difference of the slopes $m$ obtained for one photometric relation $i$ between S0s and spirals in a given band as follows:

\begin{equation} \label{eq:delta}
\Delta(\mathrm{m, S0-Sp, band}) = \frac{m(\mathrm{Sp,band}) - m(\mathrm{S0,band})}{m(\mathrm{S0,band})},
\end{equation}

\noindent The errors in $\Delta(\mathrm{m, S0-Sp, band})$ correspond to the error propagation of the expression above, assuming as the error of each parameter the maximum between the absolute values of its upper and lower errors. Analogously, we have also defined the relative differences of the $Y$-intercepts ($C_0$) for the trends fulfilled by two datasets being compared, $\Delta (\mathrm{C_0, S0-Sp, band})$, and their associated errors. These $\Delta$ values for $m$ and $C_0$ have only been defined when the two data samples being compared exhibit statistically significant correlations in the photometric relation separately, according to the Spearman rank correlation test.

Even if $\Delta(m)$ and $\Delta(C_0)$ in a given trend were nearly zero, this does not ensure that the trends can be considered similar, because it depends on their errors. However, if $\Delta(m) \sim 0$ and $\Delta(C_0) \sim 0$ with errors below a given (low) percentage, the trends of the two samples  can be considered similar within these uncertainties. Obviously, we must keep in mind that deeper data can reveal differences in these trends that cannot be discriminated with the available datasets.

\begin{figure*}[!t]
\centering
\framebox[0.48\textwidth][c]{Comparison of $m$ and $C_0$ of spirals vs.~S0s} 
 \begin{minipage}{\textwidth}
\begin{tabular}{c c c}
   \imagetop{\includegraphics*[height=6.2cm,bb= 15 10 455 420, clip]{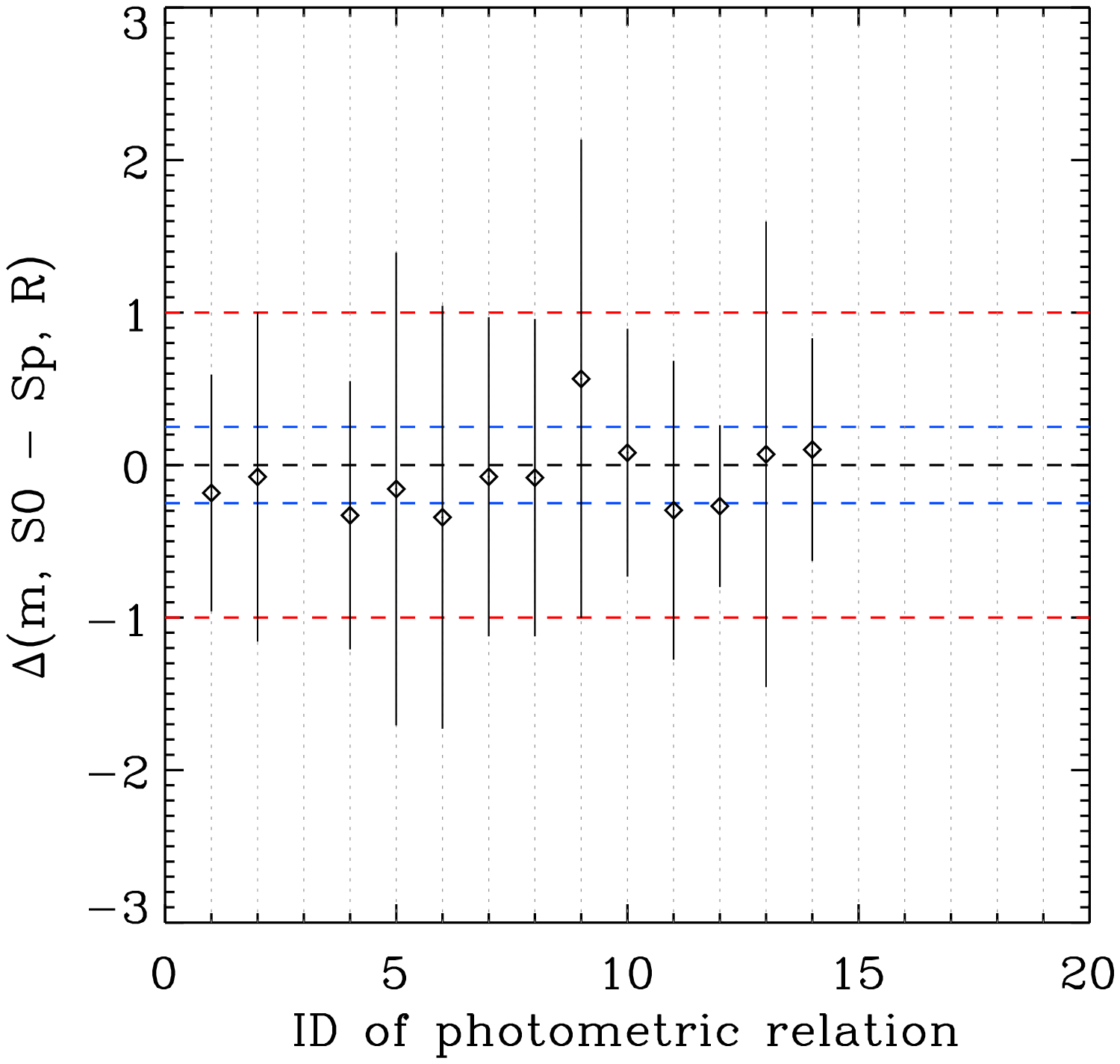}} &
   \imagetop{\includegraphics*[height=6.2cm,bb= 15 10 455 420, clip]{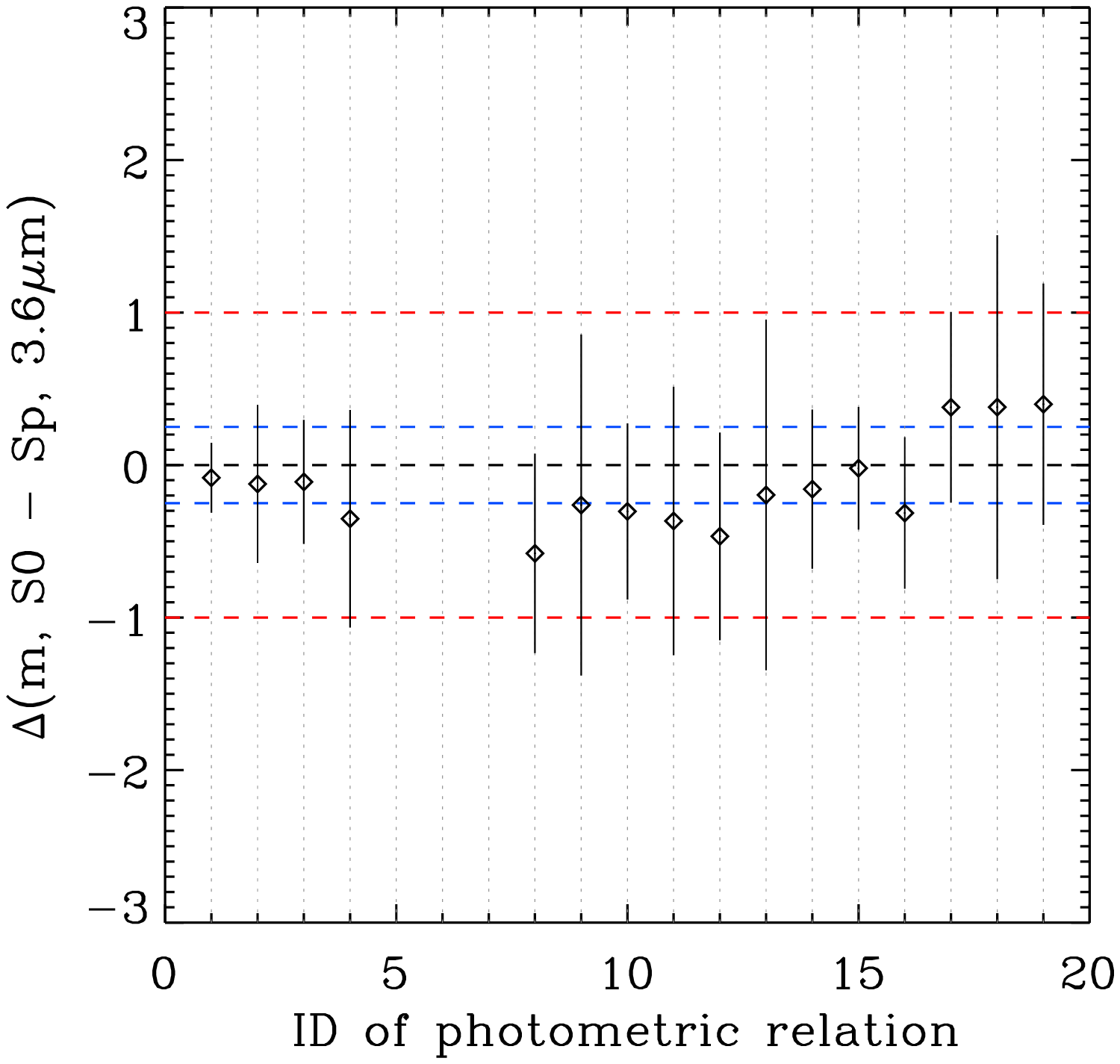}} &
   \imagetop{\includegraphics*[height=6cm,bb= 0 0 226 420, clip]{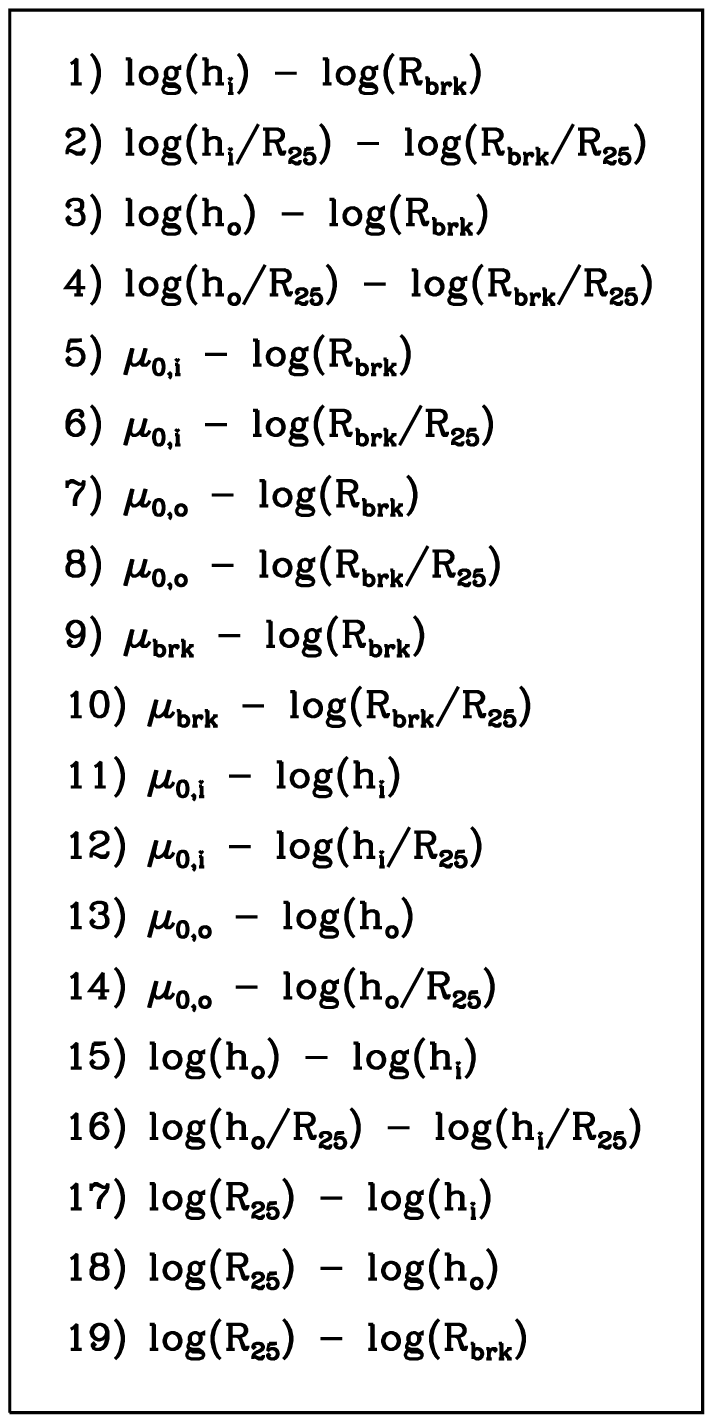}} \\
  \imagetop{\includegraphics*[height=6.2cm,bb= 15 10 455 420, clip]{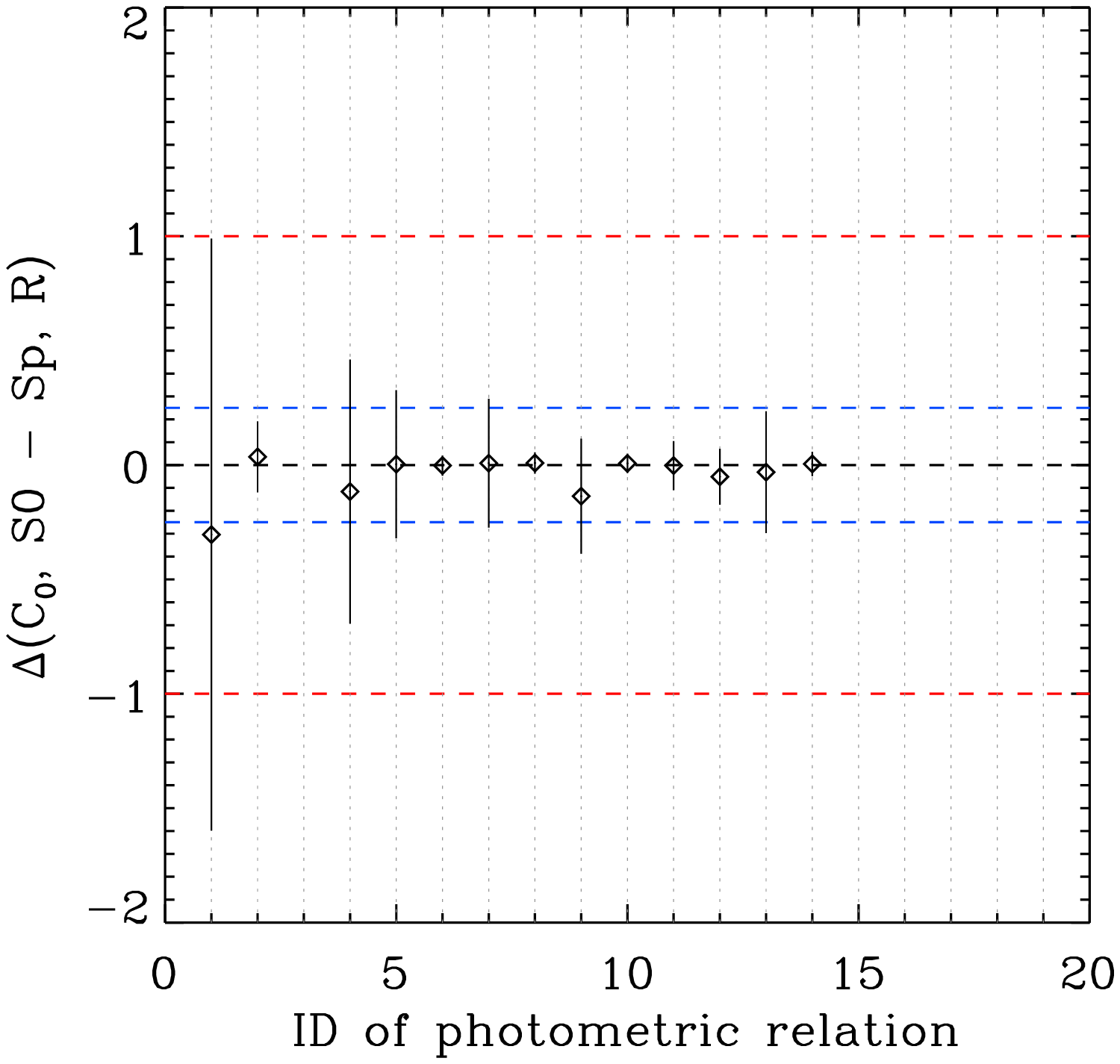}} &
   \imagetop{\includegraphics*[height=6.2cm,bb= 15 10 455 420, clip]{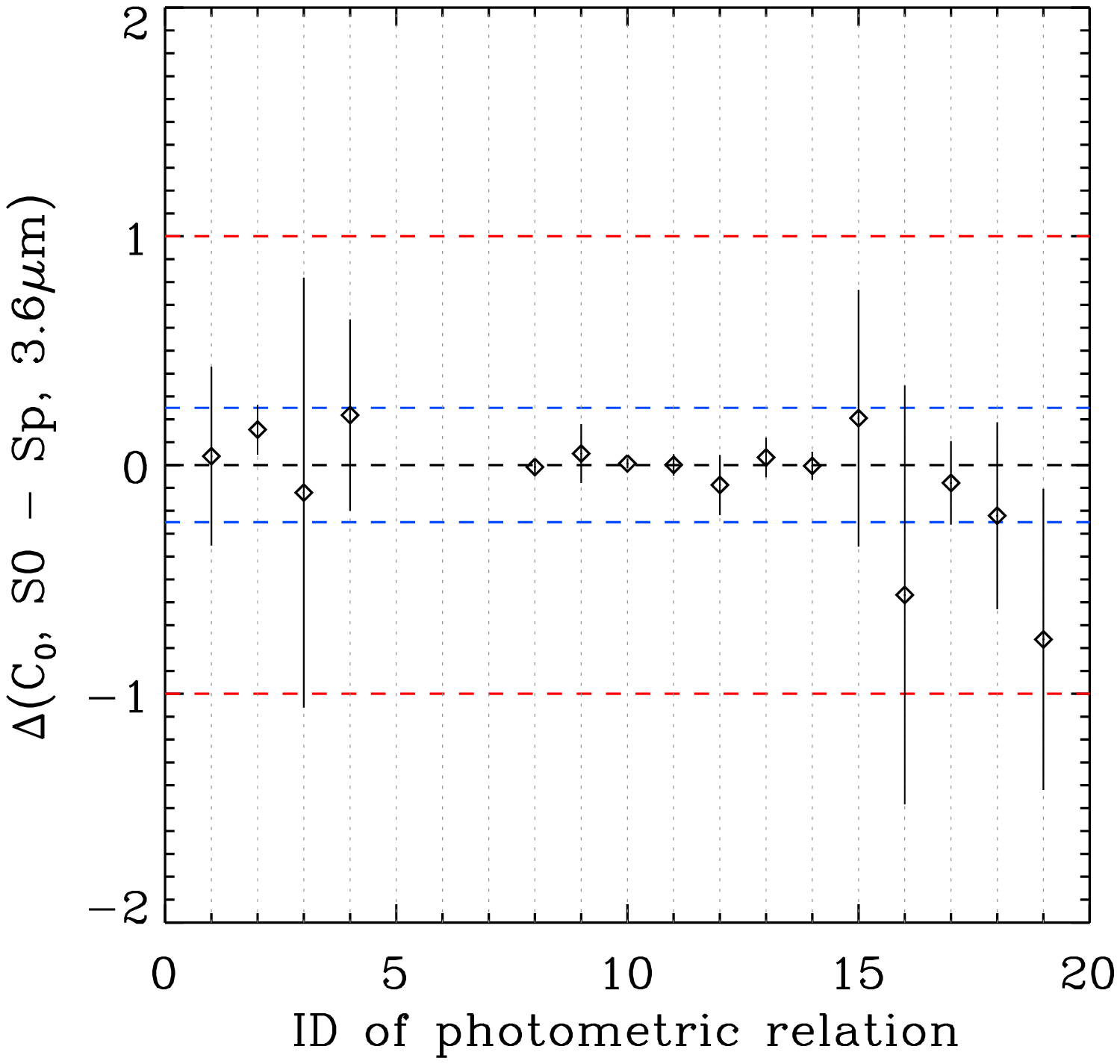}} &
   \imagetop{\includegraphics*[height=6cm,bb= 0 0 226 420, clip]{leyenda.ps}} \\
\end{tabular}
 \end{minipage}
 \caption{Comparison of the relative differences of slope ($m$) and $Y$-intercept ($C_0$) values obtained from the linear fits performed to Type-III spirals versus those performed to Type-III S0 galaxies in the photometric planes of Figs.\,\ref{fig:withRbreak_R} -- \ref{fig:r25}. \emph{Left panels}: $\Delta(m)$ and $\Delta(C_0)$ for the $R$-band data. \emph{Right panels}: $\Delta(m)$ and $\Delta(C_0)$ for the \nir\ data. Only the linear fits of the photometric trends which are significant according to the Spearman rank correlation test in the two datasets being compared are plotted. \emph{Blue and red horizontal dashed lines}: limiting values such that the differences in $m$ and $C_0$ fitted to the S0s and spirals reach 25\% and 100\% of their values respectively.  See the key on the right for the numbers in the $X$ axis representing each photometric relation. Their numerical identifiers are the same as those used in Tables\,\ref{tab:hihorbreak}--\ref{tab:r25}.}
 \label{fig:compareSpS0}
 \end{figure*}

In Fig.\,\ref{fig:compareSpS0}, we compare the relative differences of the slopes and $Y$-intercepts of the linear fits performed to the S0s and the spirals for each one of the 19 photometric relations analysed in Figs.\,\ref{fig:withRbreak_R}--\ref{fig:r25}, in $R$ and \nir\ (left and right panels, respectively). Let us assume that two fitted linear trends can be considered similar if the differences in $m$ and $C_0$ and their errors are below 25\%. The figure shows that, under this criterion, no linear trend followed by S0s can be considered similar to the analogous one in spirals, either in the $R$ band or \nir. Although $|\Delta(m)|$ and $|\Delta(C_0)|$  are lower than 0.25 in many trends in each band (i.e., their values are contained within the horizontal blue lines in the planes of Fig.\,\ref{fig:compareSpS0}), their errors (in one case and/or another) exceed this limit. The statistics of the samples are too small to conclude that the trends are similar within some reasonable uncertainty level. 

\begin{figure}[!]
\centering
\framebox[0.48\textwidth][c]{Comparison of $m$ and $C_0$ of barred  vs.~unbarred galaxies} 
   \includegraphics[width=0.24\textwidth,bb= 15 10 455 420, clip]{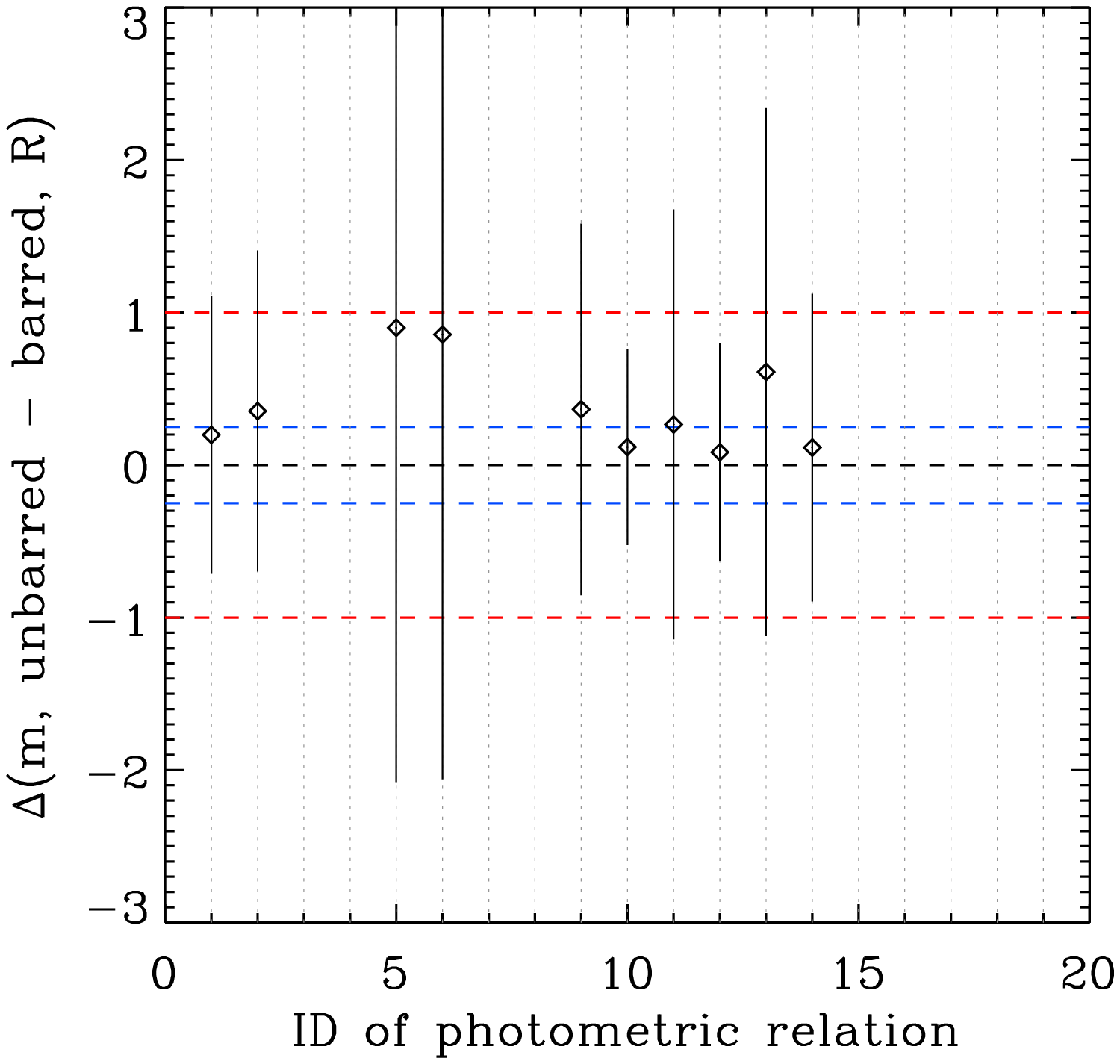} 
   \includegraphics[width=0.24\textwidth,bb= 15 10 455 420, clip]{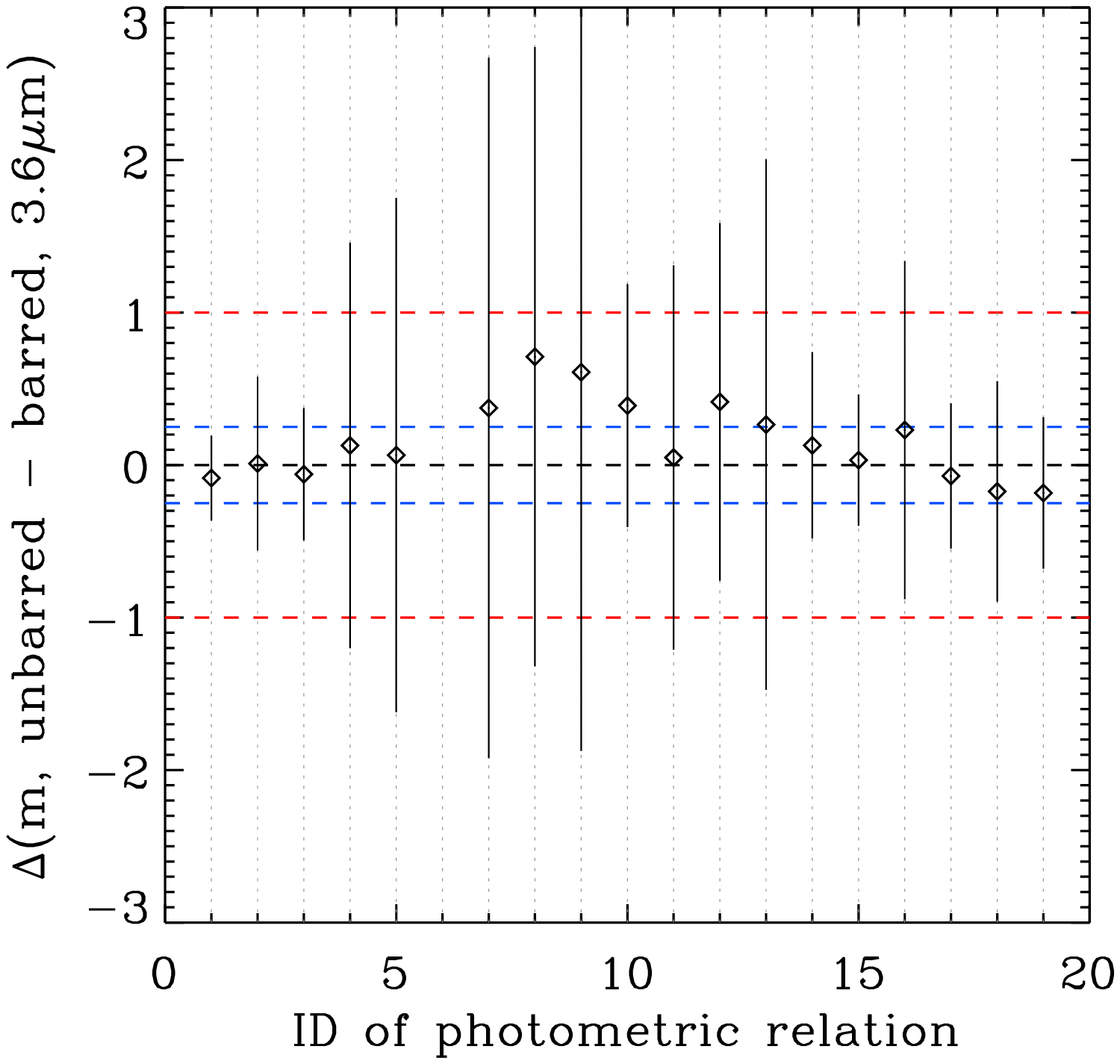} 
   \includegraphics[width=0.24\textwidth,bb= 15 10 455 420, clip]{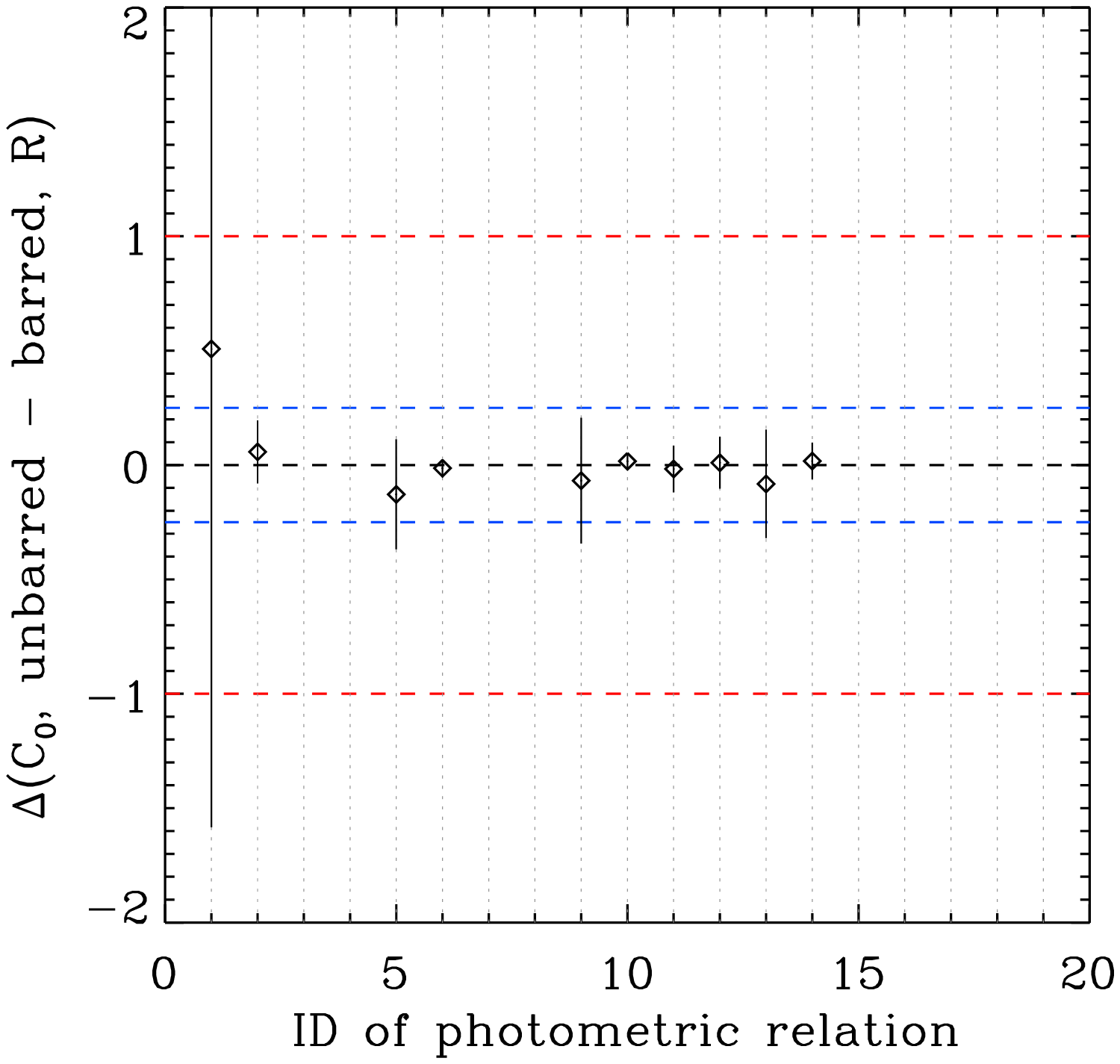}
   \includegraphics[width=0.24\textwidth,bb= 15 10 455 420, clip]{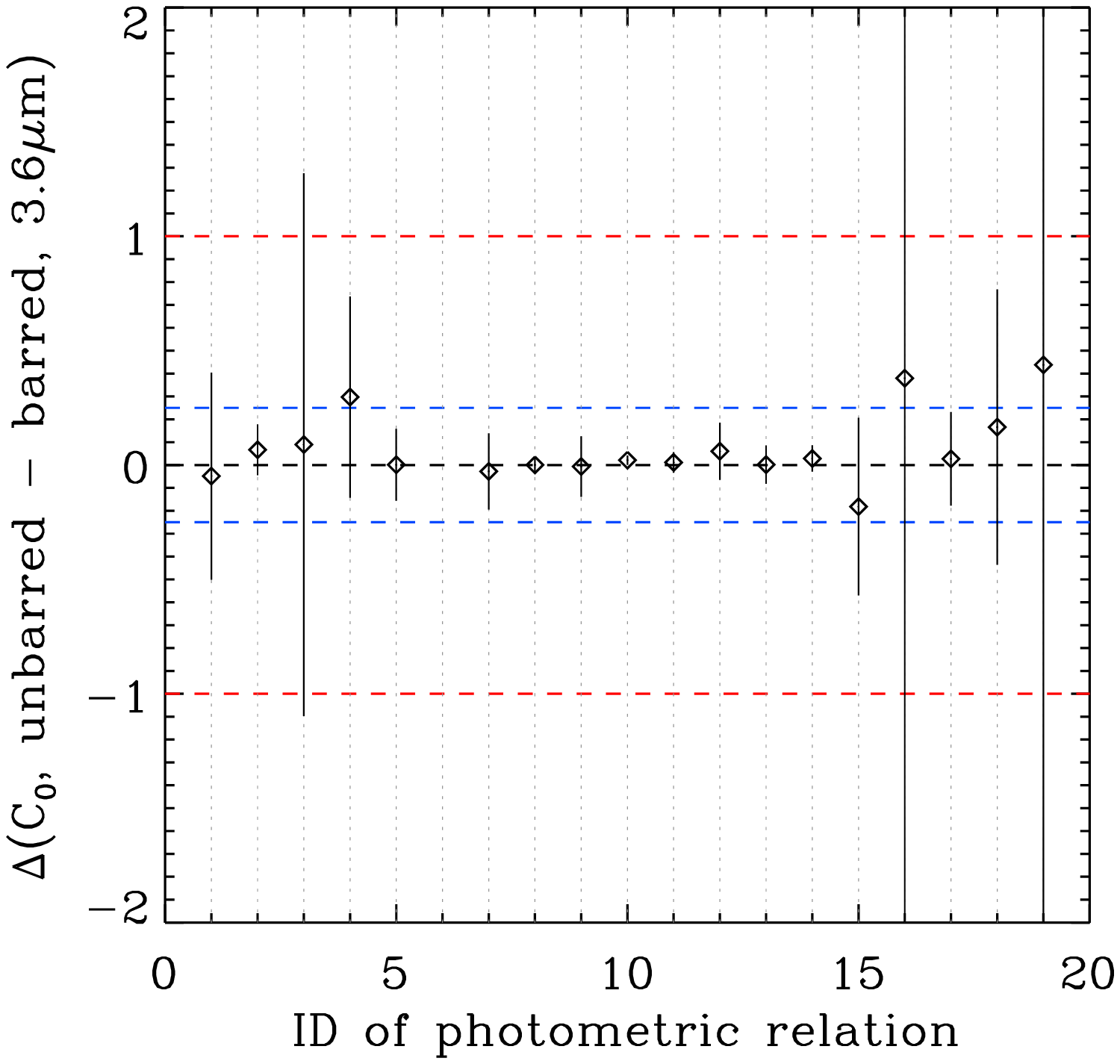}
 \caption{Comparison of the relative differences of slope ($m$) and $Y$-intercept ($C_0$) values obtained from the linear fits performed to the barred galaxies versus those performed to the unbarred galaxies in the photometric planes of Type-III galaxies being studied. \emph{Left}: trends for the $R$-band data by E08 and G11. \emph{Right}: trends for the \nir\ data by L14. See the caption and legend of Fig.\,\ref{fig:compareSpS0}.}
 \label{fig:comparebarredunbarred}
 \end{figure}

We have repeated the plot in Fig.\,\ref{fig:comparebarredunbarred}, but comparing $\Delta(m)$ and $\Delta(C_0)$ for the linear trends followed by barred and unbarred galaxies. Again, only those trends which are significant in both datasets are compared. The figure shows that the errors in these relative differences are too high again to conclude that the linear trends fitted to Type-III barred galaxies are similar to those of unbarred galaxies within uncertainties of 25\%, both in $R$ and \nir\ (left and right panels, respectively).

\begin{figure}[!]
\centering
\framebox[0.48\textwidth][c]{Comparison of $m$ of \nir\  vs.~$R$ data} 
   \includegraphics[width=0.24\textwidth,bb= 15 10 455 420, clip]{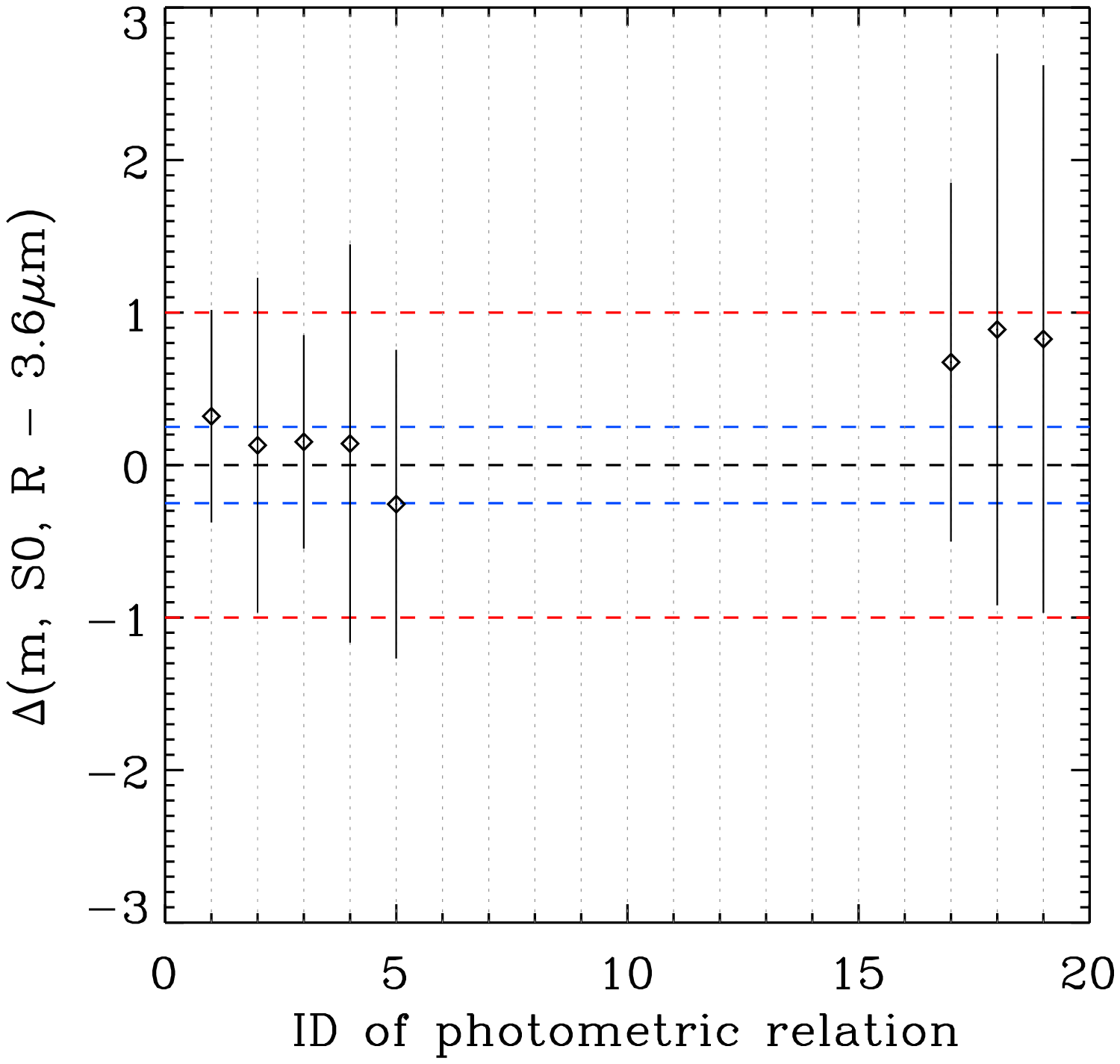} 
   \includegraphics[width=0.24\textwidth,bb= 15 10 455 420, clip]{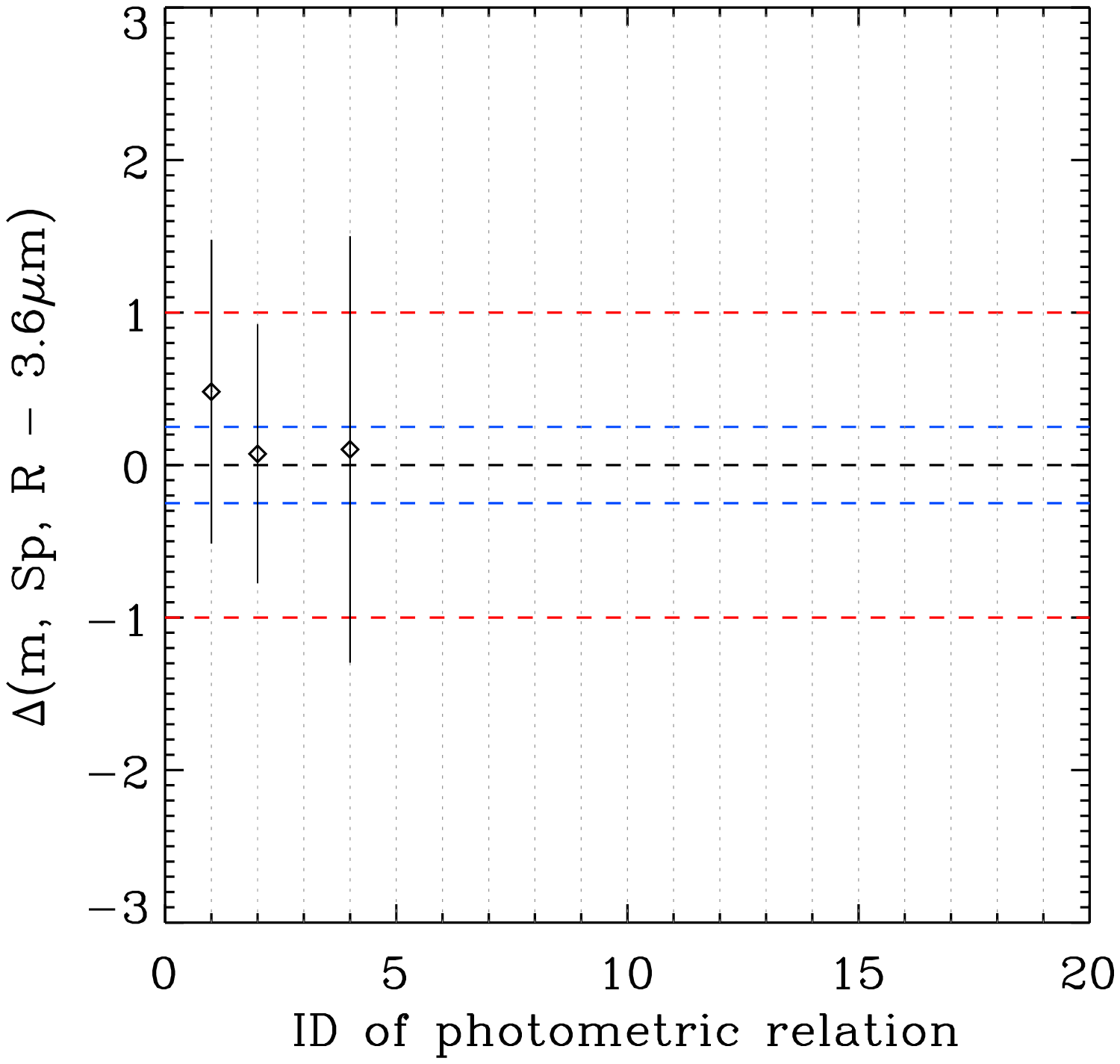} 
   \includegraphics[width=0.24\textwidth,bb= 15 10 455 420, clip]{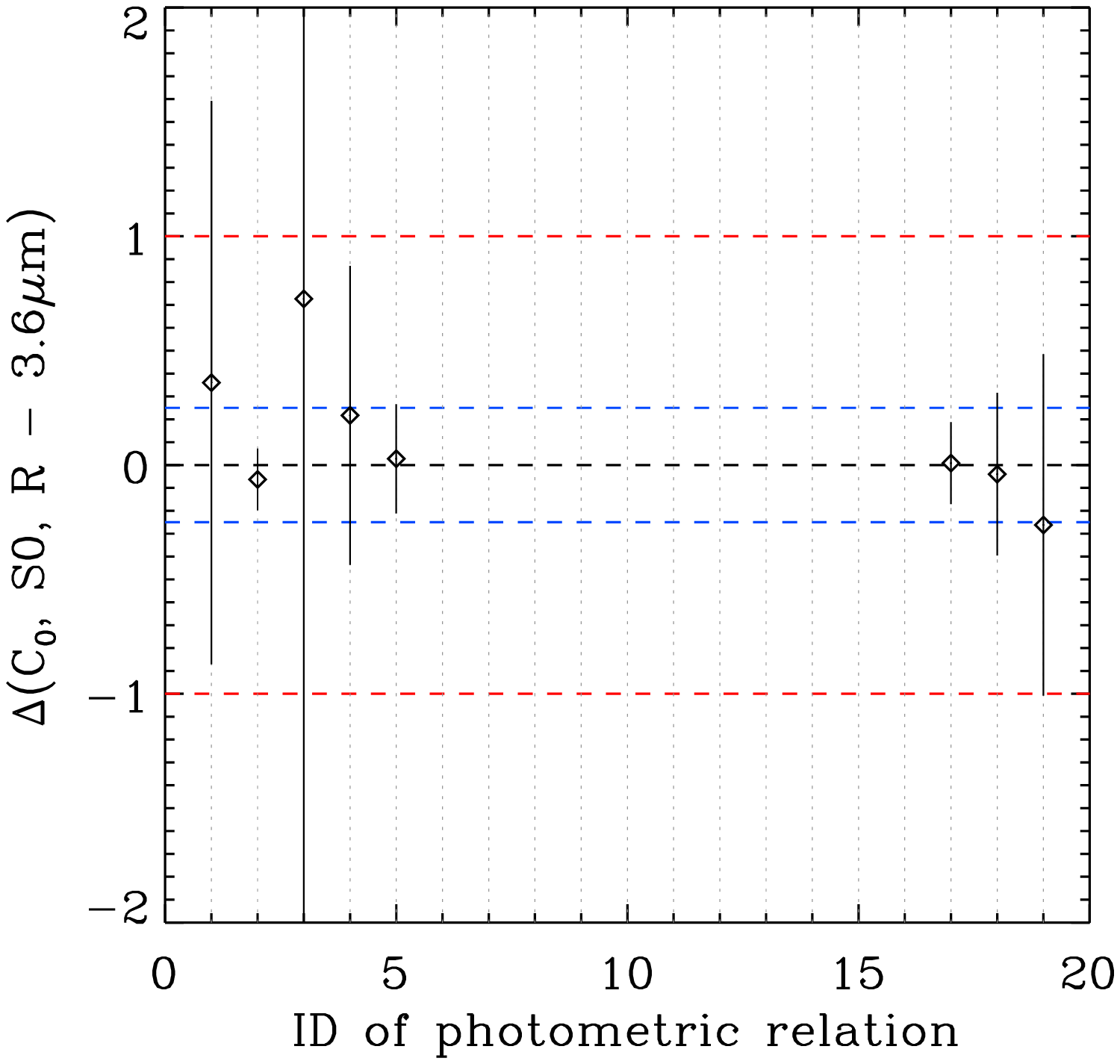}
   \includegraphics[width=0.24\textwidth,bb= 15 10 455 420, clip]{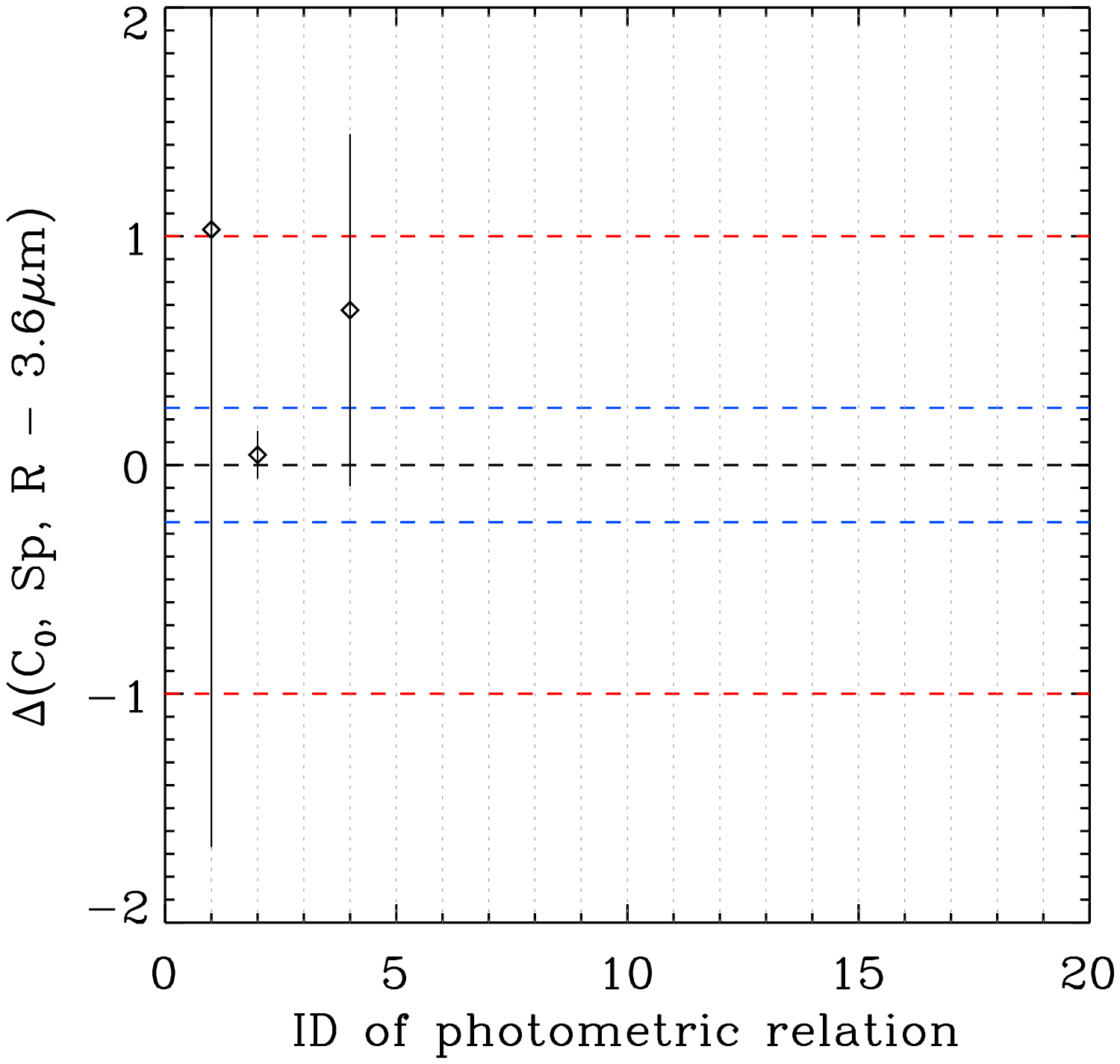}
 \caption{Comparison of the relative differences of slope ($m$) and $Y$-intercept ($C_0$) values obtained from the linear fits performed to the \nir\ data by L14 versus those obtained from the $R$-band data by E08 and G11 in the photometric planes relating physical scalelengths under study. \emph{Left}: trends for the S0 galaxies. \emph{Right}: trends for spirals. See the caption and legend of Fig.\,\ref{fig:compareSpS0}.}
 \label{fig:compareRNIR}
 \end{figure}

The comparison of the trends between distinct bands involving physical scalelengths is also reasonable, because if the breaks correspond to a change in the projected stellar density, they should be observed at a similar physical location in the disc in several bands. Therefore, we have compared the linear fits performed to the trends relating two characteristic scalelengths in the $R$ and \nir\ bands in Fig.\,\ref{fig:compareRNIR}. We have again considered only the trends which are significant according to the Spearman test in both bands. The left panels of the figure compare $\Delta(m)$ and $\Delta(C_0)$ for the linear trends fitted to the S0s in the two bands, while the right ones show the same for those fitted to spirals. Although the relative differences of $m$ and $C_0$ can be below 25\% for many trends, the uncertainties are too high to assess that these scaling relations observed in $R$ and \nir\ are similar.

In summary, Figs.\,\ref{fig:compareSpS0}--\ref{fig:compareRNIR} show that the observational dispersion is too high to robustly discern whether the analysed scaling relations are similar (or not) in both S0s and spirals, in barred and unbarred galaxies, and in the $R$ and \nir\ bands, although we do not find either statistical evidence of significant differences between the compared samples. Deeper data and larger samples are thus required to robustly confirm whether these scaling relations of antitruncated discs are really independent of the morphological type and the presence (or absence) of bars.


\section{Discussion}
\label{sec:discussion}

In Section\,\ref{sec:results}, we have shown that Type-III discs obey tight photometric scaling relations for galaxy types spanning the whole Hubble Sequence. We have found no statistical evidence of noticeable differences between the relations followed by S0s and spirals and barred and unbarred galaxies, a fact that suggests that the structure of antitruncated discs may be independent of the galaxy type and the existence, or not, of bars in the galaxy. Although the statistical evidence is not strong, considering the low numbers and the uncertainties of the available data samples, the structural independence of bars and antitruncations agrees pretty well with the similar relative frequency of Type-III profiles found in samples of barred and unbarred galaxies \citep[E08;][G11; L14]{2009IAUS..254..173S}. Our results thus support the idea that bars and antitruncations are decoupled structures in all morphological types. This does not imply that the two structures have formed independently, as some mechanisms are known to trigger the formation of both kind of features, such as mergers or flybys \citep[][]{1996ApJ...460..121W,2014arXiv1405.5832L}. However, this is indicative for that bars cannot have induced the formation of Type-III discs, as opposed to their tight structural link with Type-II discs \citep[see, e.g.,][]{2014ApJ...782...64K}. 

The scaling relations of Type-III discs found in the present study impose strong constraints on any formation scenarios proposed to explain the formation of antitruncations, independently of whether the relations really depend (or not) on the morphological type or the hosting of a bar. Accounting for the wide diversity of mechanisms proposed to explain the formation of antitruncations (see Section\,\ref{sec:introduction}), it is challenging to understand the physics underlying the scaling relations that we have found between Type-III discs across the whole Hubble Sequence. 

The dependence of these features on the environment becomes a key to discriminate between these mechanisms. Many studies have reported traces of recent or ongoing interactions in the outskirts of many Type-III discs, supporting a merging- or interaction-related nature \citep{2005ApJ...626L..81E}. L14 found a positive correlation between the scalelengths of Type-III discs and the tidal interaction strength, also pointing to external mechanisms. Coherently, flat and/or positive age gradients prevail in galaxies of the three profile types (in particular, of Type-III) in the Virgo cluster, contrary to the expectations of scenarios in which the formation of these discs were mostly driven by secular inside-out disc growth and/or stellar migrations \citep{2012ApJ...758...41R}. So, all these results suggest external processes as the main drivers of the formation of antitruncations, probably as a result of the gravitational response of the disc to a tidal interaction.

However, in this case we should also expect to find some dependence of the structural properties of Type-III discs on the local galaxy density. But, on the contrary, the inner and outer disc scalelengths and the break strength\footnote{The break strength of Type-II and Type-III profiles is defined as the logarithm of the outer-to-inner scalelengths ratio.} of Type-III discs show no trends with the environment, either in spiral or S0 galaxies \citep{2012MNRAS.419..669M,2015MNRAS.447.1506M}. Additionally, similar fractions of Type-III S0 galaxies are found in both cluster and field environments, suggesting that the environment hardly affects the outer structure of these galaxies \citep[][]{2012ApJ...744L..11E,2012ApJ...758...41R}.  How can all these results be reconciled?

External processes may induce the formation of antitruncations as the result of gravitational-driven instabilities in the disc or through gas/stars accretion in the galaxy outskirts. The tight scaling relations found here strongly support mechanisms related to the dynamical response of discs to tidal forces rather than other scenarios. In fact, a gravitational-driven mechanism would have three advantages. The first is that this can be induced through a wide diversity of processes (such as those commented in Section\,\ref{sec:introduction}). Secondly, it could provide a feasible explanation for the independence of these scaling relations of the Hubble type of the galaxy (although it must be confirmed more robustly, as explained above), since just a stellar disc and gravity are required to give rise to them. And finally, a gravitationally-induced mechanism could also explain some of the apparently contradictory results with the environment discussed previously. If the processes triggering antitruncations are mostly related to gravitational interactions, we expect to find them present in both groups and clusters and in similar fractions, because mergers and interactions can be equally relevant in both environments \citep{2007ApJ...671.1503M,2009ApJ...692..298W,2010AJ....139.2643P,2010AJ....140..612P,2011MNRAS.415.1783B,2013MNRAS.435.2713V}. Therefore, it is reasonable to find a weak dependence of their properties on the tidal interaction field (as reported by L14), but we do not expect to find significant trends of the break properties with the local density at the same time, because the antitruncation can have formed through an interaction not related with the current environment of the galaxy.

In any case, these speculative suggestions need to be confirmed through numerical simulations. At the moment, only \citet{2014A&A...570A.103B} have shown that major mergers are capable of producing antitruncated S0 galaxies that obey these scaling relations using N-body simulations. Nevertheless, it is obvious that the role of major mergers in the formation of late-type spirals must have been quite limited, so at least the Type-III discs in Sbc-Sd galaxies require different mechanisms, which also have to predict these scaling relations. In particular, satellite accretions are known to produce antitruncations \citep{2001MNRAS.324..685L,2006ApJ...650L..33P,2007ApJ...670..269Y}, inducing secular evolution in the disc that can couple the inner and outer galaxy structure at the same time \citep{2006A&A...457...91E,2011A&A...533A.104E,2012A&A...547A..48E,2013A&A...552A..67E}. This makes them good candidates to form antitruncated stellar discs. However, additional studies demonstrating the feasibility of this and other mechanisms in reproducing the scaling relations found here are required.


\section{Conclusions}
\label{sec:conclusions}

We have investigated whether the tight scaling relations recently observed by \citet{2014A&A...570A.103B} in Type-III S0s are satisfied by antitruncated galaxies of other Hubble types. We have used the samples of Type-III galaxies published by E08 and G11 in the $R$ band and by L14 in Spitzer \nir\ band, as well as the characterizations performed by these authors to the surface brightness profiles of these galaxies. The $R$-band dataset consists of 40 antitruncated galaxies with types spanning from S0 to Sbc, while the \nir\ sample has 62 galaxies of S0--Scd types. Nearly half of the galaxies in each sample are barred.

We have analysed the trends followed by S0s and spirals (all, Sa--Sab, Sb--Sbc, and Sc--Scd), as well as for barred and unbarred galaxies, in several planes relating the characteristic photometric parameters of the breaks (\mubreak, \rbreak) and of the inner and outer discs of these antitruncated galaxies (\mui, \hi, \muo, \ho), for the $R$ and \nir\ datasets separately. We have used the Spearman rank correlation test to select the correlations which are significant at 95\% of confidence level. Linear fits have been performed to the trends followed by each galaxy type in each photometric plane and the Pearson's coefficient has been used to measure the level of linear correlation. 

We have obtained the following results:

\begin{enumerate}
 \item The antitruncated discs of spirals (taking them all together, or dividing them into Hubble classes) obey tight photometric relations, like those observed in S0 galaxies, both in the $R$ and \nir\ bands. 

\item The antitruncated discs of barred and unbarred galaxies also follow tight photometric relations, again both in $R$ and \nir.

\item The majority of these correlations have high statistical significance despite the relatively low numbers of the available datasets, showing clear linear trends when \hi, \ho, \rbreak, and \risoph\ are plotted on a logarithmic scale. This implies the existence of strong scaling relations in the Type-III discs of all Hubble types between their characteristic parameters (\hi, \ho, \mui, \muo, \mubreak) and \rbreak, as well as between the parameters of the inner and outer discs (\mui\ -- \hi, \muo\ -- \ho, and \hi\ -- \ho). 

\item The correlations between \mui, \muo, or \mubreak\ with the logarithm of the characteristic scalelengths (\hi, \ho, or \rbreak)  improve significantly when the scalelengths are normalized to \risoph.

\item The logarithm of the characteristic scalelengths of antitruncated discs (\hi, \ho, and \rbreak) scale with $\log(\risoph)$ for all galaxy types in \nir. In $R$, the linear trends are less tight and lose significance in spiral types.  

\item The observational uncertainties of the data samples are too high to discern robustly whether the analysed scaling relations are similar (or not) in S0s and spirals, barred and unbarred galaxies, and in the $R$ and \nir\ bands. However, no statistical evidence is either found of significant differences between the relations followed by S0s and spirals and by barred and unbarred galaxies within errors. This result suggests that the scaling relations of antitruncated discs are independent of the morphological type and the presence or absence of bars. Deeper data and larger samples are required to confirm these results robustly.
\end{enumerate}

In conclusion, the tight scaling relations found in the present study for Type-III discs impose strong constraints on any formation scenarios proposed to explain the formation of antitruncations in stellar discs across the Hubble Sequence, independently of whether the relations really depend (or not) on the morphological type or the hosting of a bar.

\small  
%
\begin{acknowledgements}   
The authors thank to the anonymous referee for the provided input that helped to improve this publication significantly. We acknowledge the usage of the HyperLeda database (http://leda.univ-lyon1.fr). This research has made use of the NASA's Astrophysics Data System and NASA/IPAC Extragalactic Database (NED). Supported by the Ministerio de Econom\'{\i}a y Competitividad del Gobierno de Espa\~{n}a (MINECO) under project AYA2012-31277, the Instituto de Astrof\'{\i}sica de Canarias under project P3/86, and the Consejo Nacional de Ciencia y Tecnolog\'{\i}a de M\'{e}xico (CONACYT) under project 167236. JEB acknowledges financial support to the DAGAL network from the People Programme (Marie Curie Actions) of the European Union’s Seventh Framework Programme FP7/2007- 2013/ under REA grant agreement number PITN-GA-2011-289313.

\end{acknowledgements}

\vspace{-0.5cm}
\bibliography{antit.bib}



{\small

\begin{table*}
\begin{minipage}[t]{\textwidth}
\caption{Linear fits performed to the trends of Type-III galaxies in the photometric planes relating \hi\ and \ho\ with \rbreak}
\label{tab:hihorbreak}
\centering
{\fontsize{9}{9}\selectfont
\begin{tabular}{ccccccccc}
\hline\\\vspace{-0.5cm}\\
\multicolumn{9}{c}{1) $\log(\hi)$ vs.~$\log(\rbreak)$}\\\vspace{-0.1cm}\\
& \multicolumn{1}{c}{All} & \multicolumn{1}{c}{Spirals} & \multicolumn{1}{c}{S0} & \multicolumn{1}{c}{Sa--Sab} & \multicolumn{1}{c}{Sb--Sbc} & \multicolumn{1}{c}{Sc--Scd} & \multicolumn{1}{c}{Barred} & \multicolumn{1}{c}{Unbarred}
\\\vspace{-0.3cm}\\\hline\\\vspace{-0.5cm}\\
$m$ &  0.71 $^{\mathrm{+ 0.17 }}_{\mathrm{ -0.16 }}$ & 0.65 $^{\mathrm{+ 0.28 }}_{\mathrm{ -0.34 }}$ & 0.80 $^{\mathrm{+ 0.34 }}_{\mathrm{ -0.25 }}$ & 0.38 $^{\mathrm{+ 0.35 }}_{\mathrm{ -0.29 }}$ & 0.90 $^{\mathrm{+ 0.58 }}_{\mathrm{ -0.41 }}$ &  ... & 0.82 $^{\mathrm{+ 0.26 }}_{\mathrm{ -0.15 }}$ & 0.69 $^{\mathrm{+ 0.31 }}_{\mathrm{ -0.26 }}$ \\\vspace{-0.2cm}\\
 &  1.006 $^{\mathrm{+ 0.089 }}_{\mathrm{ -0.087 }}$ & 0.97 $^{\mathrm{+ 0.15 }}_{\mathrm{ -0.11 }}$ & 1.053 $^{\mathrm{+ 0.095 }}_{\mathrm{ -0.100 }}$ & 0.98 $^{\mathrm{+ 0.24 }}_{\mathrm{ -0.23 }}$ & 0.91 $^{\mathrm{+ 0.23 }}_{\mathrm{ -0.36 }}$ & 0.85 $^{\mathrm{+ 0.35 }}_{\mathrm{ -0.26 }}$ & 0.96 $^{\mathrm{+ 0.15 }}_{\mathrm{ -0.15 }}$ & 1.05 $^{\mathrm{+ 0.14 }}_{\mathrm{ -0.15 }}$ \\\vspace{-0.1cm}\\
$C_{0}$ &  -0.36 $^{\mathrm{+ 0.15 }}_{\mathrm{ -0.16 }}$ & -0.30 $^{\mathrm{+ 0.34 }}_{\mathrm{ -0.25 }}$ & -0.44 $^{\mathrm{+ 0.24 }}_{\mathrm{ -0.32 }}$ & -0.02 $^{\mathrm{+ 0.30 }}_{\mathrm{ -0.35 }}$ & -0.55 $^{\mathrm{+ 0.33 }}_{\mathrm{ -0.54 }}$ &  ... & -0.48 $^{\mathrm{+ 0.14 }}_{\mathrm{ -0.23 }}$ & -0.32 $^{\mathrm{+ 0.25 }}_{\mathrm{ -0.29 }}$  \\\vspace{-0.2cm}\\
 &  -0.613 $^{\mathrm{+ 0.081 }}_{\mathrm{ -0.081 }}$ & -0.62 $^{\mathrm{+ 0.11 }}_{\mathrm{ -0.13 }}$ & -0.59 $^{\mathrm{+ 0.10 }}_{\mathrm{ -0.10 }}$ & -0.62 $^{\mathrm{+ 0.26 }}_{\mathrm{ -0.22 }}$ & -0.57 $^{\mathrm{+ 0.36 }}_{\mathrm{ -0.17 }}$ & -0.57 $^{\mathrm{+ 0.27 }}_{\mathrm{ -0.25 }}$ & -0.59 $^{\mathrm{+ 0.14 }}_{\mathrm{ -0.14 }}$ & -0.62 $^{\mathrm{+ 0.14 }}_{\mathrm{ -0.13 }}$ \\\vspace{-0.1cm}\\
$\rho$ &  0.823  &  0.740  &  0.877  &  0.706  &  0.857  & ...  &  0.856  &  0.761 \\ \vspace{-0.3cm}\\
 & 0.896  &  0.869  &  0.939  &  0.791  &  0.775  &  0.833  &  0.932  &  0.861 \\\vspace{-0.2cm}\\
$p_{S}$ &  7.08e-11  &  2.89e-04  &  1.87e-07  &  1.02e-02  &  1.37e-02  & ...  &  2.34e-05  &  1.55e-05 \\\vspace{-0.3cm}\\ 
 & 7.96e-23  &  2.39e-10  &  5.24e-15  &  2.62e-04  &  4.08e-02  &  1.02e-02  &  1.90e-13  &  1.35e-10 \\\vspace{-0.2cm}\\
$N_\mathrm{pairs}$ &  40  &  19  &  21  &  12  &  7  & ... &  16  &  24 \\\vspace{-0.3cm}\\
 &  62  &  31  &  31  &  16  &  7  &  8  &  29  &  33 \\\vspace{-0.3cm}\\
 
\hline\\\vspace{-0.5cm}\\
\multicolumn{9}{c}{2) $\log(\hi/\risoph)$ vs.~$\log(\rbreak/\risoph)$}\\\vspace{-0.1cm} \\
 & \multicolumn{1}{c}{All} & \multicolumn{1}{c}{Spirals} & \multicolumn{1}{c}{S0} & \multicolumn{1}{c}{Sa--Sab} & \multicolumn{1}{c}{Sb--Sbc} & \multicolumn{1}{c}{Sc--Scd} & \multicolumn{1}{c}{Barred} & \multicolumn{1}{c}{Unbarred}
\\\vspace{-0.3cm}\\\hline\\\vspace{-0.5cm}\\
$m$ &  0.62 $^{\mathrm{+ 0.14 }}_{\mathrm{ -0.13 }}$ & 0.60 $^{\mathrm{+ 0.17 }}_{\mathrm{ -0.22 }}$ & 0.65 $^{\mathrm{+ 0.52 }}_{\mathrm{ -0.26 }}$ & 0.52 $^{\mathrm{+ 0.23 }}_{\mathrm{ -0.28 }}$ & 0.77 $^{\mathrm{+ 0.52 }}_{\mathrm{ -0.74 }}$ &  ... & 0.77 $^{\mathrm{+ 0.21 }}_{\mathrm{ -0.16 }}$ & 0.57 $^{\mathrm{+ 0.28 }}_{\mathrm{ -0.20 }}$ \\\vspace{-0.2cm}\\ 
 &  0.649 $^{\mathrm{+ 0.113 }}_{\mathrm{ -0.099 }}$ & 0.65 $^{\mathrm{+ 0.27 }}_{\mathrm{ -0.20 }}$ & 0.74 $^{\mathrm{+ 0.12 }}_{\mathrm{ -0.12 }}$ & 0.56 $^{\mathrm{+ 0.36 }}_{\mathrm{ -0.19 }}$ & 0.88 $^{\mathrm{+ 0.23 }}_{\mathrm{ -0.37 }}$ & 0.85 $^{\mathrm{+ 0.52 }}_{\mathrm{ -1.21 }}$ & 0.67 $^{\mathrm{+ 0.19 }}_{\mathrm{ -0.12 }}$ & 0.67 $^{\mathrm{+ 0.19 }}_{\mathrm{ -0.18 }}$ \\\vspace{-0.1cm}\\
$C_{0}$ &  -0.619 $^{\mathrm{+ 0.024 }}_{\mathrm{ -0.023 }}$ & -0.629 $^{\mathrm{+ 0.037 }}_{\mathrm{ -0.038 }}$ & -0.608 $^{\mathrm{+ 0.046 }}_{\mathrm{ -0.054 }}$ & -0.627 $^{\mathrm{+ 0.042 }}_{\mathrm{ -0.049 }}$ & -0.641 $^{\mathrm{+ 0.060 }}_{\mathrm{ -0.081 }}$ &  ... & -0.636 $^{\mathrm{+ 0.039 }}_{\mathrm{ -0.038 }}$ & -0.601 $^{\mathrm{+ 0.042 }}_{\mathrm{ -0.040 }}$ \\\vspace{-0.2cm}\\ 
 &  -0.625 $^{\mathrm{+ 0.018 }}_{\mathrm{ -0.018 }}$ & -0.657 $^{\mathrm{+ 0.026 }}_{\mathrm{ -0.026 }}$ & -0.569 $^{\mathrm{+ 0.030 }}_{\mathrm{ -0.031 }}$ & -0.644 $^{\mathrm{+ 0.040 }}_{\mathrm{ -0.038 }}$ & -0.651 $^{\mathrm{+ 0.037 }}_{\mathrm{ -0.037 }}$ & -0.686 $^{\mathrm{+ 0.042 }}_{\mathrm{ -0.047 }}$ & -0.637 $^{\mathrm{+ 0.026 }}_{\mathrm{ -0.027 }}$ & -0.597 $^{\mathrm{+ 0.035 }}_{\mathrm{ -0.037 }}$ \\\vspace{-0.1cm}\\
$\rho$ &  0.800  &  0.721  &  0.806  &  0.762  &  0.750  & ...  &  0.847  &  0.766  \\ \vspace{-0.3cm}\\
 & 0.801  &  0.860  &  0.871  &  0.782  &  0.893  &  0.619  &  0.859  &  0.770 \\ \vspace{-0.2cm}\\
$p_{S}$ &  5.79e-10  &  4.95e-04  &  1.01e-05  &  3.95e-03  &  5.22e-02  & ...  &  3.47e-05  &  1.28e-05  \\ \vspace{-0.3cm}\\
 & 5.42e-15  &  5.86e-10  &  1.87e-10  &  3.41e-04  &  6.81e-03  &  1.02e-01  &  2.45e-09  &  1.64e-07  \\ \vspace{-0.2cm}\\
$N_\mathrm{pairs}$ &  40  &  19  &  21  &  12  &  7  & ... &  16  &  24 \\ \vspace{-0.3cm}\\
 &  62  &  31  &  31  &  16  &  7  &  8  &  29  &  33 \\ \vspace{-0.3cm}\\

\hline\\\vspace{-0.5cm}\\
\multicolumn{9}{c}{3) $\log(\ho)$ vs.~$\log(\rbreak)$}\\\vspace{-0.1cm} \\
 & \multicolumn{1}{c}{All} & \multicolumn{1}{c}{Spirals} & \multicolumn{1}{c}{S0} & \multicolumn{1}{c}{Sa--Sab} & \multicolumn{1}{c}{Sb--Sbc} & \multicolumn{1}{c}{Sc--Scd} & \multicolumn{1}{c}{Barred} & \multicolumn{1}{c}{Unbarred}
\\\vspace{-0.3cm}\\\hline\\\vspace{-0.5cm}\\
$m$ &  0.72 $^{\mathrm{+ 0.24 }}_{\mathrm{ -0.24 }}$ & 0.49 $^{\mathrm{+ 0.35 }}_{\mathrm{ -0.47 }}$ & 1.00 $^{\mathrm{+ 0.44 }}_{\mathrm{ -0.40 }}$ & 0.31 $^{\mathrm{+ 0.40 }}_{\mathrm{ -0.31 }}$ & 0.51 $^{\mathrm{+ 0.70 }}_{\mathrm{ -2.28 }}$ &  ... & 0.27 $^{\mathrm{+ 0.28 }}_{\mathrm{ -0.45 }}$ & 0.99 $^{\mathrm{+ 0.36 }}_{\mathrm{ -0.37 }}$ \\\vspace{-0.2cm}\\ 
 &  1.08 $^{\mathrm{+ 0.14 }}_{\mathrm{ -0.12 }}$ & 1.03 $^{\mathrm{+ 0.30 }}_{\mathrm{ -0.18 }}$ & 1.16 $^{\mathrm{+ 0.18 }}_{\mathrm{ -0.19 }}$ & 1.05 $^{\mathrm{+ 0.45 }}_{\mathrm{ -0.29 }}$ & 0.67 $^{\mathrm{+ 0.66 }}_{\mathrm{ -0.44 }}$ & 0.98 $^{\mathrm{+ 0.42 }}_{\mathrm{ -0.29 }}$ & 1.05 $^{\mathrm{+ 0.26 }}_{\mathrm{ -0.26 }}$ & 1.12 $^{\mathrm{+ 0.24 }}_{\mathrm{ -0.17 }}$ \\\vspace{-0.1cm}\\
$C_{0}$ &  0.04 $^{\mathrm{+ 0.23 }}_{\mathrm{ -0.21 }}$ & 0.26 $^{\mathrm{+ 0.45 }}_{\mathrm{ -0.32 }}$ & -0.25 $^{\mathrm{+ 0.34 }}_{\mathrm{ -0.42 }}$ & 0.47 $^{\mathrm{+ 0.30 }}_{\mathrm{ -0.37 }}$ & 0.16 $^{\mathrm{+ 2.17 }}_{\mathrm{ -0.47 }}$ &  ... & 0.39 $^{\mathrm{+ 0.43 }}_{\mathrm{ -0.21 }}$ & -0.20 $^{\mathrm{+ 0.36 }}_{\mathrm{ -0.34 }}$ \\\vspace{-0.2cm}\\ 
 &  -0.39 $^{\mathrm{+ 0.11 }}_{\mathrm{ -0.12 }}$ & -0.38 $^{\mathrm{+ 0.17 }}_{\mathrm{ -0.24 }}$ & -0.43 $^{\mathrm{+ 0.19 }}_{\mathrm{ -0.16 }}$ & -0.37 $^{\mathrm{+ 0.33 }}_{\mathrm{ -0.37 }}$ & -0.07 $^{\mathrm{+ 0.43 }}_{\mathrm{ -0.52 }}$ & -0.44 $^{\mathrm{+ 0.31 }}_{\mathrm{ -0.35 }}$ & -0.42 $^{\mathrm{+ 0.25 }}_{\mathrm{ -0.22 }}$ & -0.39 $^{\mathrm{+ 0.16 }}_{\mathrm{ -0.20 }}$ \\\vspace{-0.1cm}\\
$\rho$ &  0.594  &  0.391  &  0.798  &  0.448  &  0.107  & ...  &  0.363  &  0.675  \\ \vspace{-0.3cm}\\
 & 0.868  &  0.839  &  0.900  &  0.824  &  0.714  &  0.976  &  0.918  &  0.882  \\ \vspace{-0.2cm}\\
$p_{S}$ &  1.07e-04  &  9.77e-02  &  7.30e-05  &  1.45e-01  &  8.19e-01  & ...  &  2.23e-01  &  2.98e-04  \\ \vspace{-0.3cm}\\
 & 7.07e-20  &  3.83e-09  &  5.97e-12  &  8.84e-05  &  7.13e-02  &  3.31e-05  &  2.45e-12  &  1.14e-11 \\ \vspace{-0.2cm}\\
$N_\mathrm{pairs}$ &  37  &  19  &  18  &  12  &  7  & ... &  13  &  24  \\ \vspace{-0.3cm}\\
 &  62  &  31  &  31  &  16  &  7  &  8  &  29  &  33  \\ \vspace{-0.3cm}\\

\hline\\\vspace{-0.5cm}\\
\multicolumn{9}{c}{4) $\log(\ho/\risoph)$ vs.~$\log(\rbreak/\risoph)$}\\\vspace{-0.1cm} \\
 & \multicolumn{1}{c}{All} & \multicolumn{1}{c}{Spirals} & \multicolumn{1}{c}{S0} & \multicolumn{1}{c}{Sa--Sab} & \multicolumn{1}{c}{Sb--Sbc} & \multicolumn{1}{c}{Sc--Scd} & \multicolumn{1}{c}{Barred} & \multicolumn{1}{c}{Unbarred}
\\\vspace{-0.3cm}\\\hline\\\vspace{-0.5cm}\\
$m$ &  0.60 $^{\mathrm{+ 0.18 }}_{\mathrm{ -0.17 }}$ & 0.54 $^{\mathrm{+ 0.30 }}_{\mathrm{ -0.31 }}$ & 0.80 $^{\mathrm{+ 0.60 }}_{\mathrm{ -0.39 }}$ & 0.42 $^{\mathrm{+ 0.24 }}_{\mathrm{ -0.23 }}$ & 0.8 $^{\mathrm{+ 1.1 }}_{\mathrm{ -2.5 }}$ &  ... & 0.31 $^{\mathrm{+ 0.29 }}_{\mathrm{ -0.61 }}$ & 0.76 $^{\mathrm{+ 0.38 }}_{\mathrm{ -0.26 }}$ \\\vspace{-0.2cm}\\ 
 &  0.73 $^{\mathrm{+ 0.22 }}_{\mathrm{ -0.18 }}$ & 0.59 $^{\mathrm{+ 0.42 }}_{\mathrm{ -0.25 }}$ & 0.92 $^{\mathrm{+ 0.37 }}_{\mathrm{ -0.32 }}$ & 0.49 $^{\mathrm{+ 0.55 }}_{\mathrm{ -0.27 }}$ & 0.81 $^{\mathrm{+ 0.48 }}_{\mathrm{ -0.70 }}$ & 1.2 $^{\mathrm{+ 1.2 }}_{\mathrm{ -1.0 }}$ & 0.85 $^{\mathrm{+ 0.64 }}_{\mathrm{ -0.34 }}$ & 0.75 $^{\mathrm{+ 0.32 }}_{\mathrm{ -0.28 }}$ \\\vspace{-0.1cm}\\
$C_{0}$ &  -0.220 $^{\mathrm{+ 0.041 }}_{\mathrm{ -0.040 }}$ & -0.216 $^{\mathrm{+ 0.066 }}_{\mathrm{ -0.068 }}$ & -0.245 $^{\mathrm{+ 0.080 }}_{\mathrm{ -0.082 }}$ & -0.206 $^{\mathrm{+ 0.056 }}_{\mathrm{ -0.058 }}$ & -0.25 $^{\mathrm{+ 0.14 }}_{\mathrm{ -0.17 }}$ &  ... & -0.259 $^{\mathrm{+ 0.072 }}_{\mathrm{ -0.076 }}$ & -0.208 $^{\mathrm{+ 0.066 }}_{\mathrm{ -0.066 }}$ \\\vspace{-0.2cm}\\ 
 &  -0.333 $^{\mathrm{+ 0.033 }}_{\mathrm{ -0.030 }}$ & -0.363 $^{\mathrm{+ 0.052 }}_{\mathrm{ -0.043 }}$ & -0.298 $^{\mathrm{+ 0.058 }}_{\mathrm{ -0.060 }}$ & -0.324 $^{\mathrm{+ 0.086 }}_{\mathrm{ -0.066 }}$ & -0.353 $^{\mathrm{+ 0.089 }}_{\mathrm{ -0.093 }}$ & -0.441 $^{\mathrm{+ 0.073 }}_{\mathrm{ -0.047 }}$ & -0.381 $^{\mathrm{+ 0.055 }}_{\mathrm{ -0.044 }}$ & -0.293 $^{\mathrm{+ 0.057 }}_{\mathrm{ -0.052 }}$ \\\vspace{-0.1cm}\\
$\rho$ &  0.614  &  0.619  &  0.581  &  0.608  &  0.571  & ...  &  0.264  &  0.753  \\ \vspace{-0.3cm}\\
 & 0.753  &  0.683  &  0.865  &  0.624  &  0.643  &  0.786  &  0.893  &  0.739  \\ \vspace{-0.2cm}\\
$p_{S}$ &  5.30e-05  &  4.69e-03  &  1.15e-02  &  3.58e-02  &  1.80e-01  & ...  &  3.84e-01  &  2.17e-05  \\ \vspace{-0.3cm}\\
 & 1.65e-12  &  2.29e-05  &  3.36e-10  &  9.86e-03  &  1.19e-01  &  2.08e-02  &  7.69e-11  &  9.21e-07  \\ \vspace{-0.2cm}\\
$N_\mathrm{pairs}$ &  37  &  19  &  18  &  12  &  7  & ... &  13  &  24  \\ \vspace{-0.3cm}\\
 &  62  &  31  &  31  &  16  &  7  &  8  &  29  &  33  \\ \vspace{-0.3cm}\\\hline\\ \vspace{-0.5cm}

\end{tabular}
}
\begin{minipage}[t]{0.9\textwidth}{\small
\emph{Comments}: For each photometric relation, we list the results obtained from the linear fits performed to the various galaxies subsamples in columns (all galaxies, spirals, S0s, Sa--Sab's, Sb--Sbc's, Sc--Scd's, barred galaxies, and unbarred ones). The first row of results in each parameter (for each relation and galaxy subsample considered) corresponds to the fits performed to the data in the $R$ band, while the second row corresponds to the results in the \nir\ band. In each linear fit, we provide the slope ($m$), the $Y$-intercept value ($C_0$), the Pearson coefficient of linear correlation ($\rho$), the Spearman rank probability of random correlation ($p_S$), and the number of data pairs available for each fit (see Section\,\ref{sec:fits} for more details). 
}
\end{minipage}

\end{minipage}
\end{table*}

\begin{table*}
\begin{minipage}[t]{\textwidth}
\caption{Linear fits performed to the trends of Type-III galaxies in the planes relating \mui\ and \muo\ with \rbreak}
\label{tab:muimuorbreak}
\centering
{\fontsize{9}{9}\selectfont
\begin{tabular}{ccccccccc}
\hline\\\vspace{-0.5cm}\\
\multicolumn{9}{c}{5) $\mui$ vs.~$\log(\rbreak)$}\\\vspace{-0.1cm}\\
& \multicolumn{1}{c}{All} & \multicolumn{1}{c}{Spirals} & \multicolumn{1}{c}{S0} & \multicolumn{1}{c}{Sa--Sab} & \multicolumn{1}{c}{Sb--Sbc} & \multicolumn{1}{c}{Sc--Scd} & \multicolumn{1}{c}{Barred} & \multicolumn{1}{c}{Unbarred}
\\\vspace{-0.3cm}\\\hline\\\vspace{-0.5cm}\\
$m$ &  3.1 $^{\mathrm{+ 1.7 }}_{\mathrm{ -1.3 }}$ & 2.9 $^{\mathrm{+ 2.4 }}_{\mathrm{ -2.7 }}$ & 3.5 $^{\mathrm{+ 3.2 }}_{\mathrm{ -1.9 }}$ & 1.5 $^{\mathrm{+ 2.7 }}_{\mathrm{ -3.1 }}$ & 4.5 $^{\mathrm{+ 3.4 }}_{\mathrm{ -7.6 }}$ &  ... & 4.8 $^{\mathrm{+ 2.3 }}_{\mathrm{ -2.2 }}$ & 2.5 $^{\mathrm{+ 2.8 }}_{\mathrm{ -1.8 }}$ \\\vspace{-0.2cm}\\ 
 &  1.72 $^{\mathrm{+ 0.91 }}_{\mathrm{ -0.91 }}$ & 1.0 $^{\mathrm{+ 1.5 }}_{\mathrm{ -1.3 }}$ & 2.6 $^{\mathrm{+ 1.1 }}_{\mathrm{ -1.1 }}$ & 1.1 $^{\mathrm{+ 1.9 }}_{\mathrm{ -2.3 }}$ & 0.4 $^{\mathrm{+ 4.7 }}_{\mathrm{ -5.5 }}$ & -0.9 $^{\mathrm{+ 4.5 }}_{\mathrm{ -2.3 }}$ & 2.0 $^{\mathrm{+ 1.3 }}_{\mathrm{ -1.6 }}$ & 1.9 $^{\mathrm{+ 1.4 }}_{\mathrm{ -1.5 }}$ \\\vspace{-0.1cm}\\
$C_{0}$ &  16.5 $^{\mathrm{+ 1.2 }}_{\mathrm{ -1.4 }}$ & 16.5 $^{\mathrm{+ 2.6 }}_{\mathrm{ -2.2 }}$ & 16.4 $^{\mathrm{+ 1.5 }}_{\mathrm{ -2.7 }}$ & 17.9 $^{\mathrm{+ 2.7 }}_{\mathrm{ -2.5 }}$ & 15.1 $^{\mathrm{+ 7.4 }}_{\mathrm{ -2.7 }}$ &  ... & 15.0 $^{\mathrm{+ 2.1 }}_{\mathrm{ -2.0 }}$ & 17.2 $^{\mathrm{+ 1.4 }}_{\mathrm{ -2.3 }}$ \\\vspace{-0.2cm}\\ 
 &  17.43 $^{\mathrm{+ 0.79 }}_{\mathrm{ -0.77 }}$ & 17.9 $^{\mathrm{+ 1.2 }}_{\mathrm{ -1.2 }}$ & 16.9 $^{\mathrm{+ 1.1 }}_{\mathrm{ -1.0 }}$ & 18.0 $^{\mathrm{+ 2.4 }}_{\mathrm{ -1.6 }}$ & 18.1 $^{\mathrm{+ 5.5 }}_{\mathrm{ -3.8 }}$ & 19.2 $^{\mathrm{+ 2.4 }}_{\mathrm{ -3.0 }}$ & 17.3 $^{\mathrm{+ 1.5 }}_{\mathrm{ -1.1 }}$ & 17.3 $^{\mathrm{+ 1.2 }}_{\mathrm{ -1.2 }}$ \\\vspace{-0.1cm}\\
$\rho$ &  0.595  &  0.484  &  0.703  &  0.436  &  0.643  & ...  &  0.829  &  0.447 \\\vspace{-0.3cm}\\ 
 & 0.508  &  0.292  &  0.687  &  0.394  &  -0.214  &  -0.310  &  0.499  &  0.439 \\\vspace{-0.2cm}\\
$p_{S}$ &  6.46e-05  &  4.18e-02  &  3.78e-04  &  1.80e-01  &  1.19e-01  & ...  &  1.32e-04  &  2.85e-02 \\\vspace{-0.3cm}\\ 
 & 2.51e-05  &  1.11e-01  &  1.96e-05  &  1.31e-01  &  6.45e-01  &  4.56e-01  &  5.82e-03  &  1.05e-02 \\\vspace{-0.2cm}\\
$N_\mathrm{pairs}$ &  39  &  18  &  21  &  11  &  7  & ... &  15  &  24 \\\vspace{-0.2cm}\\
 &  62  &  31  &  31  &  16  &  7  &  8  &  29  &  33 \\\vspace{-0.2cm}\\

\hline\\\vspace{-0.5cm}\\
\multicolumn{9}{c}{6) $\mui$ vs.~$\log(\rbreak/\risoph)$}\\\vspace{-0.1cm} \\
 & \multicolumn{1}{c}{All} & \multicolumn{1}{c}{Spirals} & \multicolumn{1}{c}{S0} & \multicolumn{1}{c}{Sa--Sab} & \multicolumn{1}{c}{Sb--Sbc} & \multicolumn{1}{c}{Sc--Scd} & \multicolumn{1}{c}{Barred} & \multicolumn{1}{c}{Unbarred}
\\\vspace{-0.3cm}\\\hline\\\vspace{-0.5cm}\\
$m$ &  3.3 $^{\mathrm{+ 1.7 }}_{\mathrm{ -1.3 }}$ & 2.9 $^{\mathrm{+ 1.9 }}_{\mathrm{ -1.9 }}$ & 4.4 $^{\mathrm{+ 6.3 }}_{\mathrm{ -2.8 }}$ & 1.7 $^{\mathrm{+ 2.0 }}_{\mathrm{ -2.0 }}$ & 5.3 $^{\mathrm{+ 2.7 }}_{\mathrm{ -2.4 }}$ &  ... & 5.3 $^{\mathrm{+ 1.1 }}_{\mathrm{ -2.2 }}$ & 2.9 $^{\mathrm{+ 3.3 }}_{\mathrm{ -1.8 }}$ \\\vspace{-0.2cm}\\ 
 &  2.0 $^{\mathrm{+ 1.2 }}_{\mathrm{ -1.3 }}$ & 0.7 $^{\mathrm{+ 2.6 }}_{\mathrm{ -2.4 }}$ & 4.0 $^{\mathrm{+ 1.2 }}_{\mathrm{ -1.3 }}$ & 0.8 $^{\mathrm{+ 2.7 }}_{\mathrm{ -3.3 }}$ & -0.9 $^{\mathrm{+ 7.1 }}_{\mathrm{ -2.5 }}$ & 3.8 $^{\mathrm{+ 7.4 }}_{\mathrm{ -10.7 }}$ & 3.5 $^{\mathrm{+ 1.3 }}_{\mathrm{ -1.3 }}$ & 1.6 $^{\mathrm{+ 2.4 }}_{\mathrm{ -2.4 }}$ \\\vspace{-0.1cm}\\
$C_{0}$ &  19.29 $^{\mathrm{+ 0.23 }}_{\mathrm{ -0.21 }}$ & 19.24 $^{\mathrm{+ 0.31 }}_{\mathrm{ -0.35 }}$ & 19.28 $^{\mathrm{+ 0.48 }}_{\mathrm{ -0.50 }}$ & 19.42 $^{\mathrm{+ 0.36 }}_{\mathrm{ -0.46 }}$ & 18.92 $^{\mathrm{+ 0.39 }}_{\mathrm{ -0.45 }}$ &  ... & 19.13 $^{\mathrm{+ 0.33 }}_{\mathrm{ -0.31 }}$ & 19.40 $^{\mathrm{+ 0.44 }}_{\mathrm{ -0.35 }}$ \\\vspace{-0.2cm}\\ 
 &  19.09 $^{\mathrm{+ 0.21 }}_{\mathrm{ -0.21 }}$ & 18.84 $^{\mathrm{+ 0.30 }}_{\mathrm{ -0.30 }}$ & 19.58 $^{\mathrm{+ 0.25 }}_{\mathrm{ -0.29 }}$ & 19.14 $^{\mathrm{+ 0.37 }}_{\mathrm{ -0.38 }}$ & 18.37 $^{\mathrm{+ 0.75 }}_{\mathrm{ -0.61 }}$ & 18.54 $^{\mathrm{+ 0.56 }}_{\mathrm{ -0.59 }}$ & 19.20 $^{\mathrm{+ 0.26 }}_{\mathrm{ -0.28 }}$ & 19.11 $^{\mathrm{+ 0.40 }}_{\mathrm{ -0.43 }}$ \\\vspace{-0.1cm}\\
$\rho$ &  0.702  &  0.616  &  0.788  &  0.627  &  0.750  & ...  &  0.781  &  0.617 \\\vspace{-0.3cm}\\ 
 & 0.431  &  0.194  &  0.747  &  0.113  &  -0.143  &  0.095  &  0.611  &  0.268 \\\vspace{-0.2cm}\\
$p_{S}$ &  6.52e-07  &  6.48e-03  &  2.20e-05  &  3.88e-02  &  5.22e-02  & ...  &  5.87e-04  &  1.33e-03 \\\vspace{-0.3cm}\\ 
 & 4.70e-04  &  2.95e-01  &  1.37e-06  &  6.76e-01  &  7.60e-01  &  8.23e-01  &  4.31e-04  &  1.32e-01 \\\vspace{-0.2cm}\\
$N_\mathrm{pairs}$ &  39  &  18  &  21  &  11  &  7  & ... &  15  &  24 \\\vspace{-0.2cm}\\
 &  62  &  31  &  31  &  16  &  7  &  8  &  29  &  33 \\\vspace{-0.2cm}\\

\hline\\\vspace{-0.5cm}\\
\multicolumn{9}{c}{7) $\muo$ vs.~$\log(\rbreak)$}\\\vspace{-0.1cm} \\
 & \multicolumn{1}{c}{All} & \multicolumn{1}{c}{Spirals} & \multicolumn{1}{c}{S0} & \multicolumn{1}{c}{Sa--Sab} & \multicolumn{1}{c}{Sb--Sbc} & \multicolumn{1}{c}{Sc--Scd} & \multicolumn{1}{c}{Barred} & \multicolumn{1}{c}{Unbarred}
\\\vspace{-0.3cm}\\\hline\\\vspace{-0.5cm}\\
$m$ &  4.8 $^{\mathrm{+ 1.3 }}_{\mathrm{ -1.4 }}$ & 4.7 $^{\mathrm{+ 2.3 }}_{\mathrm{ -2.9 }}$ & 5.0 $^{\mathrm{+ 1.9 }}_{\mathrm{ -2.6 }}$ & 6.2 $^{\mathrm{+ 2.2 }}_{\mathrm{ -1.9 }}$ & 3.2 $^{\mathrm{+ 6.8 }}_{\mathrm{ -20.0 }}$ &  ... & 3.2 $^{\mathrm{+ 3.7 }}_{\mathrm{ -4.7 }}$ & 5.4 $^{\mathrm{+ 1.7 }}_{\mathrm{ -1.8 }}$ \\\vspace{-0.2cm}\\ 
 &  2.06 $^{\mathrm{+ 1.10 }}_{\mathrm{ -0.94 }}$ & 1.5 $^{\mathrm{+ 1.7 }}_{\mathrm{ -1.3 }}$ & 2.8 $^{\mathrm{+ 1.9 }}_{\mathrm{ -1.7 }}$ & 1.5 $^{\mathrm{+ 2.6 }}_{\mathrm{ -2.0 }}$ & -0.5 $^{\mathrm{+ 5.5 }}_{\mathrm{ -6.3 }}$ & 1.2 $^{\mathrm{+ 2.6 }}_{\mathrm{ -3.0 }}$ & 2.7 $^{\mathrm{+ 1.7 }}_{\mathrm{ -1.7 }}$ & 1.9 $^{\mathrm{+ 2.0 }}_{\mathrm{ -1.4 }}$ \\\vspace{-0.1cm}\\
$C_{0}$ &  17.6 $^{\mathrm{+ 1.3 }}_{\mathrm{ -1.3 }}$ & 17.7 $^{\mathrm{+ 2.8 }}_{\mathrm{ -2.3 }}$ & 17.5 $^{\mathrm{+ 2.1 }}_{\mathrm{ -1.8 }}$ & 16.1 $^{\mathrm{+ 1.8 }}_{\mathrm{ -2.2 }}$ & 18.8 $^{\mathrm{+ 19.2 }}_{\mathrm{ -4.6 }}$ &  ... & 18.9 $^{\mathrm{+ 4.6 }}_{\mathrm{ -3.5 }}$ & 17.2 $^{\mathrm{+ 1.5 }}_{\mathrm{ -1.7 }}$ \\\vspace{-0.2cm}\\ 
 &  19.25 $^{\mathrm{+ 0.85 }}_{\mathrm{ -0.94 }}$ & 19.8 $^{\mathrm{+ 1.2 }}_{\mathrm{ -1.4 }}$ & 18.4 $^{\mathrm{+ 1.5 }}_{\mathrm{ -1.6 }}$ & 20.0 $^{\mathrm{+ 2.1 }}_{\mathrm{ -2.2 }}$ & 21.1 $^{\mathrm{+ 6.1 }}_{\mathrm{ -4.6 }}$ & 19.9 $^{\mathrm{+ 2.3 }}_{\mathrm{ -2.0 }}$ & 18.7 $^{\mathrm{+ 1.6 }}_{\mathrm{ -1.5 }}$ & 19.2 $^{\mathrm{+ 1.3 }}_{\mathrm{ -1.7 }}$ \\\vspace{-0.1cm}\\
$\rho$ &  0.644  &  0.561  &  0.712  &  0.827  &  0.250  & ...  &  0.467  &  0.703 \\\vspace{-0.3cm}\\ 
 & 0.452  &  0.339  &  0.569  &  0.409  &  0.071  &  0.405  &  0.465  &  0.416 \\\vspace{-0.2cm}\\
$p_{S}$ &  3.93e-05  &  1.55e-02  &  1.98e-03  &  1.68e-03  &  5.89e-01  & ...  &  1.74e-01  &  1.29e-04 \\\vspace{-0.3cm}\\ 
 & 2.28e-04  &  6.23e-02  &  8.47e-04  &  1.16e-01  &  8.79e-01  &  3.20e-01  &  1.10e-02  &  1.60e-02 \\\vspace{-0.2cm}\\
$N_\mathrm{pairs}$ &  34  &  18  &  16  &  11  &  7  & ... &  10  &  24 \\\vspace{-0.2cm}\\
 &  62  &  31  &  31  &  16  &  7  &  8  &  29  &  33 \\\vspace{-0.2cm}\\

\hline\\\vspace{-0.5cm}\\
\multicolumn{9}{c}{8) $\muo$ vs.~$\log(\rbreak/\risoph)$}\\\vspace{-0.1cm} \\
 & \multicolumn{1}{c}{All} & \multicolumn{1}{c}{Spirals} & \multicolumn{1}{c}{S0} & \multicolumn{1}{c}{Sa--Sab} & \multicolumn{1}{c}{Sb--Sbc} & \multicolumn{1}{c}{Sc--Scd} & \multicolumn{1}{c}{Barred} & \multicolumn{1}{c}{Unbarred}
\\\vspace{-0.3cm}\\\hline\\\vspace{-0.5cm}\\
$m$ &  5.5 $^{\mathrm{+ 1.2 }}_{\mathrm{ -1.1 }}$ & 5.4 $^{\mathrm{+ 1.6 }}_{\mathrm{ -1.7 }}$ & 5.9 $^{\mathrm{+ 4.9 }}_{\mathrm{ -2.8 }}$ & 4.9 $^{\mathrm{+ 2.5 }}_{\mathrm{ -1.4 }}$ & 6.5 $^{\mathrm{+ 3.6 }}_{\mathrm{ -10.3 }}$ &  ... & 4.6 $^{\mathrm{+ 2.1 }}_{\mathrm{ -5.8 }}$ & 5.8 $^{\mathrm{+ 2.7 }}_{\mathrm{ -1.7 }}$ \\\vspace{-0.2cm}\\ 
 &  4.0 $^{\mathrm{+ 1.4 }}_{\mathrm{ -1.5 }}$ & 2.5 $^{\mathrm{+ 2.8 }}_{\mathrm{ -3.0 }}$ & 5.9 $^{\mathrm{+ 2.0 }}_{\mathrm{ -1.7 }}$ & 2.9 $^{\mathrm{+ 2.9 }}_{\mathrm{ -4.5 }}$ & -0.4 $^{\mathrm{+ 8.1 }}_{\mathrm{ -5.4 }}$ & 7.6 $^{\mathrm{+ 4.4 }}_{\mathrm{ -7.2 }}$ & 6.0 $^{\mathrm{+ 2.8 }}_{\mathrm{ -1.7 }}$ & 3.5 $^{\mathrm{+ 2.3 }}_{\mathrm{ -2.5 }}$ \\\vspace{-0.1cm}\\
$C_{0}$ &  21.97 $^{\mathrm{+ 0.26 }}_{\mathrm{ -0.25 }}$ & 22.01 $^{\mathrm{+ 0.40 }}_{\mathrm{ -0.36 }}$ & 21.81 $^{\mathrm{+ 0.55 }}_{\mathrm{ -0.61 }}$ & 22.23 $^{\mathrm{+ 0.40 }}_{\mathrm{ -0.35 }}$ & 21.63 $^{\mathrm{+ 0.99 }}_{\mathrm{ -0.66 }}$ &  ... & 21.77 $^{\mathrm{+ 0.69 }}_{\mathrm{ -0.55 }}$ & 22.02 $^{\mathrm{+ 0.41 }}_{\mathrm{ -0.42 }}$ \\\vspace{-0.2cm}\\ 
 &  21.31 $^{\mathrm{+ 0.23 }}_{\mathrm{ -0.22 }}$ & 21.24 $^{\mathrm{+ 0.37 }}_{\mathrm{ -0.34 }}$ & 21.42 $^{\mathrm{+ 0.31 }}_{\mathrm{ -0.32 }}$ & 21.59 $^{\mathrm{+ 0.56 }}_{\mathrm{ -0.45 }}$ & 20.74 $^{\mathrm{+ 0.74 }}_{\mathrm{ -0.74 }}$ & 20.86 $^{\mathrm{+ 0.40 }}_{\mathrm{ -0.43 }}$ & 21.29 $^{\mathrm{+ 0.30 }}_{\mathrm{ -0.27 }}$ & 21.27 $^{\mathrm{+ 0.42 }}_{\mathrm{ -0.40 }}$ \\\vspace{-0.1cm}\\
$\rho$ &  0.774  &  0.819  &  0.709  &  0.836  &  0.786  & ...  &  0.612  &  0.834 \\\vspace{-0.3cm}\\ 
 & 0.667  &  0.435  &  0.844  &  0.409  &  -0.143  &  0.619  &  0.859  &  0.488 \\\vspace{-0.2cm}\\
$p_{S}$ &  7.76e-08  &  3.25e-05  &  2.11e-03  &  1.33e-03  &  3.62e-02  & ...  &  6.00e-02  &  4.15e-07 \\\vspace{-0.3cm}\\ 
 & 3.13e-09  &  1.44e-02  &  2.38e-09  &  1.16e-01  &  7.60e-01  &  1.02e-01  &  2.45e-09  &  4.00e-03 \\\vspace{-0.2cm}\\
$N_\mathrm{pairs}$ &  34  &  18  &  16  &  11  &  7  & ... &  10  &  24 \\\vspace{-0.2cm}\\
 &  62  &  31  &  31  &  16  &  7  &  8  &  29  &  33   \\ \vspace{-0.3cm}\\\hline\\ \vspace{-0.5cm}

\end{tabular}
}
\begin{minipage}[t]{0.9\textwidth}{\small
\emph{Comments}: See the notes of Table\,\ref{tab:hihorbreak}.}
\end{minipage}

\end{minipage}
\end{table*}

\begin{table*}
\begin{minipage}[t]{\textwidth}
\caption{Linear fits performed to the trends of Type-III galaxies in the photometric planes relating \mubreak\ with \rbreak}
\label{tab:mubreakrbreak}
\centering
{\fontsize{9}{9}\selectfont
\begin{tabular}{ccccccccc}
\hline\\\vspace{-0.5cm}\\
\multicolumn{9}{c}{9) $\mubreak$ vs.~$\log(\rbreak)$}\\\vspace{-0.1cm}\\
& \multicolumn{1}{c}{All} & \multicolumn{1}{c}{Spirals} & \multicolumn{1}{c}{S0} & \multicolumn{1}{c}{Sa--Sab} & \multicolumn{1}{c}{Sb--Sbc} & \multicolumn{1}{c}{Sc--Scd} & \multicolumn{1}{c}{Barred} & \multicolumn{1}{c}{Unbarred}
\\\vspace{-0.3cm}\\\hline\\\vspace{-0.5cm}\\
$m$ &  6.0 $^{\mathrm{+ 1.5 }}_{\mathrm{ -1.5 }}$ & 7.2 $^{\mathrm{+ 2.6 }}_{\mathrm{ -3.3 }}$ & 4.6 $^{\mathrm{+ 1.7 }}_{\mathrm{ -2.5 }}$ & 9.3 $^{\mathrm{+ 2.2 }}_{\mathrm{ -2.1 }}$ & 6.0 $^{\mathrm{+ 5.9 }}_{\mathrm{ -9.1 }}$ &  ... & 7.0 $^{\mathrm{+ 3.1 }}_{\mathrm{ -2.6 }}$ & 5.1 $^{\mathrm{+ 2.2 }}_{\mathrm{ -2.3 }}$ \\\vspace{-0.2cm}\\ 
 &  1.76 $^{\mathrm{+ 0.80 }}_{\mathrm{ -0.77 }}$ & 1.5 $^{\mathrm{+ 1.0 }}_{\mathrm{ -1.1 }}$ & 2.0 $^{\mathrm{+ 1.6 }}_{\mathrm{ -1.5 }}$ & 1.4 $^{\mathrm{+ 1.6 }}_{\mathrm{ -1.5 }}$ & 1.5 $^{\mathrm{+ 2.9 }}_{\mathrm{ -4.4 }}$ & 1.2 $^{\mathrm{+ 2.3 }}_{\mathrm{ -3.8 }}$ & 2.4 $^{\mathrm{+ 1.6 }}_{\mathrm{ -1.7 }}$ & 1.5 $^{\mathrm{+ 1.2 }}_{\mathrm{ -1.1 }}$ \\\vspace{-0.1cm}\\
$C_{0}$ &  18.2 $^{\mathrm{+ 1.4 }}_{\mathrm{ -1.5 }}$ & 17.0 $^{\mathrm{+ 3.1 }}_{\mathrm{ -2.6 }}$ & 19.7 $^{\mathrm{+ 2.1 }}_{\mathrm{ -1.4 }}$ & 14.7 $^{\mathrm{+ 2.1 }}_{\mathrm{ -2.5 }}$ & 18.5 $^{\mathrm{+ 8.5 }}_{\mathrm{ -4.8 }}$ &  ... & 17.7 $^{\mathrm{+ 2.4 }}_{\mathrm{ -3.1 }}$ & 19.0 $^{\mathrm{+ 2.0 }}_{\mathrm{ -2.3 }}$ \\\vspace{-0.2cm}\\ 
 &  21.83 $^{\mathrm{+ 0.76 }}_{\mathrm{ -0.79 }}$ & 22.3 $^{\mathrm{+ 1.1 }}_{\mathrm{ -1.0 }}$ & 21.2 $^{\mathrm{+ 1.5 }}_{\mathrm{ -1.6 }}$ & 22.4 $^{\mathrm{+ 1.7 }}_{\mathrm{ -1.8 }}$ & 21.8 $^{\mathrm{+ 4.5 }}_{\mathrm{ -2.3 }}$ & 22.9 $^{\mathrm{+ 2.9 }}_{\mathrm{ -1.9 }}$ & 21.6 $^{\mathrm{+ 1.7 }}_{\mathrm{ -1.7 }}$ & 21.7 $^{\mathrm{+ 1.1 }}_{\mathrm{ -1.2 }}$ \\\vspace{-0.2cm}\\
$\rho$ &  0.712  &  0.773  &  0.703  &  0.897  &  0.643  & ...  &  0.964  &  0.623 \\\vspace{-0.1cm}\\ 
 & 0.380  &  0.397  &  0.411  &  0.364  &  0.559  &  0.310  &  0.385  &  0.345 \\\vspace{-0.3cm}\\
$p_{S}$ &  2.39e-06  &  1.70e-04  &  2.38e-03  &  1.78e-04  &  1.19e-01  & ...  &  7.32e-06  &  1.15e-03 \\\vspace{-0.2cm}\\ 
 & 2.29e-03  &  2.68e-02  &  2.16e-02  &  1.66e-01  &  1.92e-01  &  4.56e-01  &  3.91e-02  &  4.93e-02 \\\vspace{-0.3cm}\\
$N_\mathrm{pairs}$ &  34  &  18  &  16  &  11  &  7  & ... &  10  &  24 \\\vspace{-0.2cm}\\
 &  62  &  31  &  31  &  16  &  7  &  8  &  29  &  33 \\\vspace{-0.2cm}\\

\hline\\\vspace{-0.5cm}\\
\multicolumn{9}{c}{10) $\mubreak$ vs.~$\log(\rbreak/\risoph)$}\\\vspace{-0.1cm} \\
 & \multicolumn{1}{c}{All} & \multicolumn{1}{c}{Spirals} & \multicolumn{1}{c}{S0} & \multicolumn{1}{c}{Sa--Sab} & \multicolumn{1}{c}{Sb--Sbc} & \multicolumn{1}{c}{Sc--Scd} & \multicolumn{1}{c}{Barred} & \multicolumn{1}{c}{Unbarred}
\\\vspace{-0.3cm}\\\hline\\\vspace{-0.5cm}\\
$m$ &  7.38 $^{\mathrm{+ 0.80 }}_{\mathrm{ -0.74 }}$ & 7.6 $^{\mathrm{+ 1.2 }}_{\mathrm{ -1.1 }}$ & 7.0 $^{\mathrm{+ 4.1 }}_{\mathrm{ -1.6 }}$ & 7.59 $^{\mathrm{+ 1.66 }}_{\mathrm{ -0.63 }}$ & 7.4 $^{\mathrm{+ 7.6 }}_{\mathrm{ -4.3 }}$ &  ... & 7.9 $^{\mathrm{+ 1.5 }}_{\mathrm{ -2.4 }}$ & 7.1 $^{\mathrm{+ 1.9 }}_{\mathrm{ -1.2 }}$ \\\vspace{-0.3cm}\\ 
 &  5.36 $^{\mathrm{+ 0.94 }}_{\mathrm{ -1.23 }}$ & 4.4 $^{\mathrm{+ 2.2 }}_{\mathrm{ -3.1 }}$ & 6.31 $^{\mathrm{+ 0.75 }}_{\mathrm{ -0.74 }}$ & 5.1 $^{\mathrm{+ 1.6 }}_{\mathrm{ -3.4 }}$ & 0.5 $^{\mathrm{+ 7.2 }}_{\mathrm{ -2.6 }}$ & 5.9 $^{\mathrm{+ 6.1 }}_{\mathrm{ -13.3 }}$ & 6.66 $^{\mathrm{+ 0.54 }}_{\mathrm{ -1.09 }}$ & 4.8 $^{\mathrm{+ 1.4 }}_{\mathrm{ -2.0 }}$ \\\vspace{-0.2cm}\\
$C_{0}$ &  23.69 $^{\mathrm{+ 0.18 }}_{\mathrm{ -0.21 }}$ & 23.75 $^{\mathrm{+ 0.28 }}_{\mathrm{ -0.26 }}$ & 23.58 $^{\mathrm{+ 0.42 }}_{\mathrm{ -0.71 }}$ & 23.86 $^{\mathrm{+ 0.20 }}_{\mathrm{ -0.24 }}$ & 23.61 $^{\mathrm{+ 0.77 }}_{\mathrm{ -0.54 }}$ &  ... & 23.95 $^{\mathrm{+ 0.38 }}_{\mathrm{ -0.30 }}$ & 23.56 $^{\mathrm{+ 0.31 }}_{\mathrm{ -0.40 }}$ \\\vspace{-0.3cm}\\ 
 &  23.69 $^{\mathrm{+ 0.15 }}_{\mathrm{ -0.17 }}$ & 23.79 $^{\mathrm{+ 0.25 }}_{\mathrm{ -0.29 }}$ & 23.62 $^{\mathrm{+ 0.16 }}_{\mathrm{ -0.18 }}$ & 23.94 $^{\mathrm{+ 0.29 }}_{\mathrm{ -0.27 }}$ & 23.12 $^{\mathrm{+ 0.84 }}_{\mathrm{ -0.52 }}$ & 23.83 $^{\mathrm{+ 0.47 }}_{\mathrm{ -0.54 }}$ & 23.94 $^{\mathrm{+ 0.16 }}_{\mathrm{ -0.15 }}$ & 23.44 $^{\mathrm{+ 0.25 }}_{\mathrm{ -0.31 }}$ \\\vspace{-0.2cm}\\
$\rho$ &  0.920  &  0.952  &  0.913  &  0.998  &  0.857  & ...  &  0.964  &  0.924 \\\vspace{-0.2cm}\\ 
 & 0.806  &  0.617  &  0.923  &  0.761  &  -0.072  &  0.476  &  0.883  &  0.715 \\\vspace{-0.1cm}\\
$p_{S}$ &  1.37e-14  &  1.13e-09  &  8.10e-07  &  7.46e-12  &  1.37e-02  & ...  &  7.32e-06  &  1.17e-10 \\\vspace{-0.2cm}\\ 
 & 2.66e-15  &  2.20e-04  &  1.57e-13  &  6.11e-04  &  8.78e-01  &  2.33e-01  &  2.37e-10  &  2.90e-06 \\\vspace{-0.1cm}\\
$N_\mathrm{pairs}$ &  34  &  18  &  16  &  11  &  7  & ... &  10  &  24 \\\vspace{-0.2cm}\\
 &  62  &  31  &  31  &  16  &  7  &  8  &  29  &  33 \\ \vspace{-0.3cm}\\\hline\\ \vspace{-0.5cm}

\end{tabular}
}
\begin{minipage}[t]{0.9\textwidth}{\small
\emph{Comments}: See the notes of Table\,\ref{tab:hihorbreak}.}
\end{minipage}

\end{minipage}
\end{table*}

\begin{table*}
\begin{minipage}[t]{\textwidth}
\caption{Linear fits performed to the trends of Type-III galaxies in the photometric planes relating \mui\ and \muo\ with \hi\ and \ho}
\label{tab:muihimuoho}
\centering
{\fontsize{9}{9}\selectfont
\begin{tabular}{ccccccccc}
\hline\\\vspace{-0.5cm}\\
\multicolumn{9}{c}{11) $\mui$ vs.~$\log(\hi)$}\\\vspace{-0.1cm}\\
& \multicolumn{1}{c}{All} & \multicolumn{1}{c}{Spirals} & \multicolumn{1}{c}{S0} & \multicolumn{1}{c}{Sa--Sab} & \multicolumn{1}{c}{Sb--Sbc} & \multicolumn{1}{c}{Sc--Scd} & \multicolumn{1}{c}{Barred} & \multicolumn{1}{c}{Unbarred}
\\\vspace{-0.3cm}\\\hline\\\vspace{-0.5cm}\\

 $m$ &  4.3 $^{\mathrm{+ 1.8 }}_{\mathrm{ -1.5 }}$ & 3.6 $^{\mathrm{+ 3.5 }}_{\mathrm{ -2.5 }}$ & 5.1 $^{\mathrm{+ 2.0 }}_{\mathrm{ -2.1 }}$ & 4.9 $^{\mathrm{+ 6.8 }}_{\mathrm{ -5.5 }}$ & 3.7 $^{\mathrm{+ 6.2 }}_{\mathrm{ -4.0 }}$ &  ... & 5.3 $^{\mathrm{+ 2.2 }}_{\mathrm{ -1.6 }}$ & 4.2 $^{\mathrm{+ 2.9 }}_{\mathrm{ -2.5 }}$ \\\vspace{-0.3cm}\\ 
 &  1.78 $^{\mathrm{+ 0.63 }}_{\mathrm{ -0.68 }}$ & 1.3 $^{\mathrm{+ 1.1 }}_{\mathrm{ -1.2 }}$ & 2.1 $^{\mathrm{+ 1.0 }}_{\mathrm{ -1.0 }}$ & 1.3 $^{\mathrm{+ 1.2 }}_{\mathrm{ -2.2 }}$ & 0.9 $^{\mathrm{+ 3.5 }}_{\mathrm{ -4.5 }}$ & -0.1 $^{\mathrm{+ 3.7 }}_{\mathrm{ -3.0 }}$ & 1.99 $^{\mathrm{+ 0.89 }}_{\mathrm{ -1.32 }}$ & 1.89 $^{\mathrm{+ 1.02 }}_{\mathrm{ -0.97 }}$ \\\vspace{-0.2cm}\\
$C_{0}$ &  18.06 $^{\mathrm{+ 0.44 }}_{\mathrm{ -0.58 }}$ & 18.11 $^{\mathrm{+ 0.99 }}_{\mathrm{ -1.33 }}$ & 18.15 $^{\mathrm{+ 0.46 }}_{\mathrm{ -0.61 }}$ & 17.6 $^{\mathrm{+ 2.2 }}_{\mathrm{ -2.8 }}$ & 18.1 $^{\mathrm{+ 1.5 }}_{\mathrm{ -1.7 }}$ &  ... & 17.89 $^{\mathrm{+ 0.57 }}_{\mathrm{ -0.77 }}$ & 18.20 $^{\mathrm{+ 0.63 }}_{\mathrm{ -1.10 }}$ \\\vspace{-0.3cm}\\ 
 &  18.46 $^{\mathrm{+ 0.23 }}_{\mathrm{ -0.21 }}$ & 18.49 $^{\mathrm{+ 0.35 }}_{\mathrm{ -0.31 }}$ & 18.48 $^{\mathrm{+ 0.51 }}_{\mathrm{ -0.43 }}$ & 18.69 $^{\mathrm{+ 0.88 }}_{\mathrm{ -0.24 }}$ & 18.21 $^{\mathrm{+ 1.50 }}_{\mathrm{ -0.91 }}$ & 18.62 $^{\mathrm{+ 0.81 }}_{\mathrm{ -0.53 }}$ & 18.55 $^{\mathrm{+ 0.46 }}_{\mathrm{ -0.37 }}$ & 18.34 $^{\mathrm{+ 0.34 }}_{\mathrm{ -0.36 }}$ \\\vspace{-0.2cm}\\
$\rho$ &  0.706  &  0.538  &  0.857  &  0.555  &  0.714  & ...  &  0.924  &  0.585 \\\vspace{-0.2cm}\\ 
 & 0.519  &  0.384  &  0.607  &  0.313  &  0.000  &  -0.048  &  0.516  &  0.538 \\\vspace{-0.1cm}\\
$p_{S}$ &  5.12e-07  &  2.14e-02  &  6.99e-07  &  7.67e-02  &  7.13e-02  & ...  &  8.66e-07  &  2.67e-03 \\\vspace{-0.2cm}\\ 
 & 1.57e-05  &  3.28e-02  &  2.92e-04  &  2.37e-01  &  1.00e+00  &  9.11e-01  &  4.14e-03  &  1.23e-03 \\\vspace{-0.1cm}\\
$N_\mathrm{pairs}$ &  39  &  18  &  21  &  11  &  7  & ... &  15  &  24 \\\vspace{-0.2cm}\\
 &  62  &  31  &  31  &  16  &  7  &  8  &  29  &  33 \\\vspace{-0.2cm}\\

\hline\\\vspace{-0.5cm}\\
\multicolumn{9}{c}{12) \mui\ vs.~$\log(\hi/\risoph)$}\\\vspace{-0.1cm} \\
 & \multicolumn{1}{c}{All} & \multicolumn{1}{c}{Spirals} & \multicolumn{1}{c}{S0} & \multicolumn{1}{c}{Sa--Sab} & \multicolumn{1}{c}{Sb--Sbc} & \multicolumn{1}{c}{Sc--Scd} & \multicolumn{1}{c}{Barred} & \multicolumn{1}{c}{Unbarred}
\\\vspace{-0.3cm}\\\hline\\\vspace{-0.5cm}\\

$m$ &  6.3 $^{\mathrm{+ 1.5 }}_{\mathrm{ -1.5 }}$ & 5.5 $^{\mathrm{+ 2.0 }}_{\mathrm{ -1.9 }}$ & 7.6 $^{\mathrm{+ 2.0 }}_{\mathrm{ -2.8 }}$ & 4.7 $^{\mathrm{+ 5.1 }}_{\mathrm{ -1.7 }}$ & 6.2 $^{\mathrm{+ 1.6 }}_{\mathrm{ -2.5 }}$ &  ... & 6.98 $^{\mathrm{+ 0.88 }}_{\mathrm{ -1.76 }}$ & 6.4 $^{\mathrm{+ 2.4 }}_{\mathrm{ -2.6 }}$ \\\vspace{-0.3cm}\\ 
 &  4.3 $^{\mathrm{+ 1.2 }}_{\mathrm{ -1.6 }}$ & 3.1 $^{\mathrm{+ 2.7 }}_{\mathrm{ -3.4 }}$ & 5.77 $^{\mathrm{+ 0.84 }}_{\mathrm{ -0.98 }}$ & 3.3 $^{\mathrm{+ 2.4 }}_{\mathrm{ -4.2 }}$ & -0.3 $^{\mathrm{+ 8.0 }}_{\mathrm{ -3.2 }}$ & 6.5 $^{\mathrm{+ 6.5 }}_{\mathrm{ -7.4 }}$ & 5.97 $^{\mathrm{+ 0.98 }}_{\mathrm{ -0.94 }}$ & 4.2 $^{\mathrm{+ 1.9 }}_{\mathrm{ -2.8 }}$ \\\vspace{-0.2cm}\\
$C_{0}$ &  23.16 $^{\mathrm{+ 0.95 }}_{\mathrm{ -0.94 }}$ & 22.7 $^{\mathrm{+ 1.3 }}_{\mathrm{ -1.1 }}$ & 23.9 $^{\mathrm{+ 1.2 }}_{\mathrm{ -1.7 }}$ & 22.3 $^{\mathrm{+ 3.2 }}_{\mathrm{ -1.1 }}$ & 22.8 $^{\mathrm{+ 1.3 }}_{\mathrm{ -1.3 }}$ &  ... & 23.51 $^{\mathrm{+ 0.56 }}_{\mathrm{ -0.95 }}$ & 23.3 $^{\mathrm{+ 1.5 }}_{\mathrm{ -1.7 }}$ \\\vspace{-0.3cm}\\ 
 &  21.84 $^{\mathrm{+ 0.81 }}_{\mathrm{ -1.10 }}$ & 20.9 $^{\mathrm{+ 1.8 }}_{\mathrm{ -2.4 }}$ & 22.89 $^{\mathrm{+ 0.52 }}_{\mathrm{ -0.68 }}$ & 21.3 $^{\mathrm{+ 1.4 }}_{\mathrm{ -2.7 }}$ & 18.2 $^{\mathrm{+ 5.8 }}_{\mathrm{ -2.2 }}$ & 23.2 $^{\mathrm{+ 4.2 }}_{\mathrm{ -5.5 }}$ & 23.03 $^{\mathrm{+ 0.61 }}_{\mathrm{ -0.64 }}$ & 21.7 $^{\mathrm{+ 1.2 }}_{\mathrm{ -2.0 }}$ \\\vspace{-0.2cm}\\
$\rho$ &  0.877  &  0.889  &  0.853  &  0.945  &  0.893  & ...  &  0.899  &  0.853 \\\vspace{-0.2cm}\\ 
 & 0.708  &  0.444  &  0.904  &  0.458  &  0.107  &  0.524  &  0.868  &  0.599 \\\vspace{-0.1cm}\\
$p_{S}$ &  2.55e-13  &  8.39e-07  &  9.24e-07  &  1.12e-05  &  6.81e-03  & ...  &  5.18e-06  &  1.19e-07 \\\vspace{-0.2cm}\\ 
 & 1.19e-10  &  1.24e-02  &  3.38e-12  &  7.46e-02  &  8.19e-01  &  1.83e-01  &  1.09e-09  &  2.33e-04 \\\vspace{-0.1cm}\\
$N_\mathrm{pairs}$ &  39  &  18  &  21  &  11  &  7  & ... &  15  &  24 \\\vspace{-0.2cm}\\
 &  62  &  31  &  31  &  16  &  7  &  8  &  29  &  33 \\\vspace{-0.2cm}\\

\hline\\\vspace{-0.5cm}\\
\multicolumn{9}{c}{13) \muo\ vs.~$\log(\ho)$}\\\vspace{-0.1cm} \\
 & \multicolumn{1}{c}{All} & \multicolumn{1}{c}{Spirals} & \multicolumn{1}{c}{S0} & \multicolumn{1}{c}{Sa--Sab} & \multicolumn{1}{c}{Sb--Sbc} & \multicolumn{1}{c}{Sc--Scd} & \multicolumn{1}{c}{Barred} & \multicolumn{1}{c}{Unbarred}
\\\vspace{-0.3cm}\\\hline\\\vspace{-0.5cm}\\

$m$ &  4.7 $^{\mathrm{+ 1.4 }}_{\mathrm{ -1.1 }}$ & 5.0 $^{\mathrm{+ 5.5 }}_{\mathrm{ -2.7 }}$ & 4.7 $^{\mathrm{+ 1.5 }}_{\mathrm{ -1.4 }}$ & 6.2 $^{\mathrm{+ 6.8 }}_{\mathrm{ -5.8 }}$ & 5.0 $^{\mathrm{+ 10.9 }}_{\mathrm{ -6.0 }}$ &  ... & 7.1 $^{\mathrm{+ 4.5 }}_{\mathrm{ -5.3 }}$ & 4.4 $^{\mathrm{+ 1.5 }}_{\mathrm{ -1.4 }}$ \\\vspace{-0.3cm}\\ 
 &  2.08 $^{\mathrm{+ 0.73 }}_{\mathrm{ -0.87 }}$ & 1.9 $^{\mathrm{+ 1.1 }}_{\mathrm{ -1.5 }}$ & 2.4 $^{\mathrm{+ 1.2 }}_{\mathrm{ -1.5 }}$ & 1.8 $^{\mathrm{+ 1.3 }}_{\mathrm{ -2.2 }}$ & 1.2 $^{\mathrm{+ 3.8 }}_{\mathrm{ -5.6 }}$ & 1.6 $^{\mathrm{+ 2.6 }}_{\mathrm{ -1.8 }}$ & 2.5 $^{\mathrm{+ 1.1 }}_{\mathrm{ -1.7 }}$ & 2.0 $^{\mathrm{+ 1.1 }}_{\mathrm{ -1.4 }}$ \\\vspace{-0.2cm}\\
$C_{0}$ &  18.77 $^{\mathrm{+ 0.79 }}_{\mathrm{ -1.07 }}$ & 18.5 $^{\mathrm{+ 1.8 }}_{\mathrm{ -4.2 }}$ & 19.06 $^{\mathrm{+ 0.88 }}_{\mathrm{ -0.94 }}$ & 17.4 $^{\mathrm{+ 4.3 }}_{\mathrm{ -5.0 }}$ & 18.7 $^{\mathrm{+ 2.7 }}_{\mathrm{ -7.8 }}$ &  ... & 17.4 $^{\mathrm{+ 3.1 }}_{\mathrm{ -3.4 }}$ & 19.00 $^{\mathrm{+ 0.92 }}_{\mathrm{ -1.24 }}$ \\\vspace{-0.3cm}\\ 
 &  19.90 $^{\mathrm{+ 0.48 }}_{\mathrm{ -0.44 }}$ & 20.13 $^{\mathrm{+ 0.74 }}_{\mathrm{ -0.63 }}$ & 19.48 $^{\mathrm{+ 0.91 }}_{\mathrm{ -0.83 }}$ & 20.37 $^{\mathrm{+ 1.32 }}_{\mathrm{ -0.84 }}$ & 20.2 $^{\mathrm{+ 2.9 }}_{\mathrm{ -2.2 }}$ & 20.21 $^{\mathrm{+ 0.77 }}_{\mathrm{ -0.97 }}$ & 19.80 $^{\mathrm{+ 0.84 }}_{\mathrm{ -0.76 }}$ & 19.76 $^{\mathrm{+ 0.81 }}_{\mathrm{ -0.74 }}$ \\\vspace{-0.2cm}\\
$\rho$ &  0.713  &  0.648  &  0.835  &  0.618  &  0.571  & ...  &  0.806  &  0.731 \\\vspace{-0.2cm}\\ 
 & 0.414  &  0.435  &  0.432  &  0.450  &  0.321  &  0.357  &  0.432  &  0.435 \\\vspace{-0.1cm}\\
$p_{S}$ &  2.24e-06  &  3.61e-03  &  5.64e-05  &  4.26e-02  &  1.80e-01  & ...  &  4.86e-03  &  4.91e-05 \\\vspace{-0.2cm}\\ 
 & 8.13e-04  &  1.44e-02  &  1.52e-02  &  8.03e-02  &  4.82e-01  &  3.85e-01  &  1.94e-02  &  1.14e-02 \\\vspace{-0.1cm}\\
$N_\mathrm{pairs}$ &  34  &  18  &  16  &  11  &  7  & ... &  10  &  24 \\\vspace{-0.2cm}\\
 &  62  &  31  &  31  &  16  &  7  &  8  &  29  &  33 \\\vspace{-0.2cm}\\

\hline\\\vspace{-0.5cm}\\
\multicolumn{9}{c}{14) \muo\ vs.~$\log(\ho/\risoph)$}\\\vspace{-0.1cm} \\
 & \multicolumn{1}{c}{All} & \multicolumn{1}{c}{Spirals} & \multicolumn{1}{c}{S0} & \multicolumn{1}{c}{Sa--Sab} & \multicolumn{1}{c}{Sb--Sbc} & \multicolumn{1}{c}{Sc--Scd} & \multicolumn{1}{c}{Barred} & \multicolumn{1}{c}{Unbarred}
\\\vspace{-0.3cm}\\\hline\\\vspace{-0.5cm}\\

$m$ &  6.60 $^{\mathrm{+ 1.20 }}_{\mathrm{ -0.98 }}$ & 6.9 $^{\mathrm{+ 2.9 }}_{\mathrm{ -1.9 }}$ & 6.3 $^{\mathrm{+ 1.5 }}_{\mathrm{ -1.4 }}$ & 8.6 $^{\mathrm{+ 2.0 }}_{\mathrm{ -1.6 }}$ & 5.9 $^{\mathrm{+ 4.1 }}_{\mathrm{ -2.7 }}$ &  ... & 7.2 $^{\mathrm{+ 3.2 }}_{\mathrm{ -5.0 }}$ & 6.5 $^{\mathrm{+ 1.4 }}_{\mathrm{ -1.1 }}$ \\\vspace{-0.3cm}\\ 
 &  5.33 $^{\mathrm{+ 0.69 }}_{\mathrm{ -0.89 }}$ & 4.9 $^{\mathrm{+ 1.6 }}_{\mathrm{ -2.4 }}$ & 5.79 $^{\mathrm{+ 0.71 }}_{\mathrm{ -0.59 }}$ & 5.6 $^{\mathrm{+ 1.6 }}_{\mathrm{ -1.9 }}$ & 1.3 $^{\mathrm{+ 7.3 }}_{\mathrm{ -4.0 }}$ & 5.2 $^{\mathrm{+ 5.2 }}_{\mathrm{ -3.8 }}$ & 6.00 $^{\mathrm{+ 1.30 }}_{\mathrm{ -0.90 }}$ & 5.32 $^{\mathrm{+ 0.89 }}_{\mathrm{ -1.73 }}$ \\\vspace{-0.2cm}\\
$C_{0}$ &  23.46 $^{\mathrm{+ 0.32 }}_{\mathrm{ -0.26 }}$ & 23.52 $^{\mathrm{+ 0.68 }}_{\mathrm{ -0.48 }}$ & 23.41 $^{\mathrm{+ 0.55 }}_{\mathrm{ -0.47 }}$ & 24.03 $^{\mathrm{+ 0.44 }}_{\mathrm{ -0.36 }}$ & 23.05 $^{\mathrm{+ 1.02 }}_{\mathrm{ -0.91 }}$ &  ... & 23.80 $^{\mathrm{+ 0.72 }}_{\mathrm{ -1.42 }}$ & 23.38 $^{\mathrm{+ 0.44 }}_{\mathrm{ -0.35 }}$ \\\vspace{-0.3cm}\\ 
 &  23.08 $^{\mathrm{+ 0.32 }}_{\mathrm{ -0.41 }}$ & 23.02 $^{\mathrm{+ 0.71 }}_{\mathrm{ -1.09 }}$ & 23.10 $^{\mathrm{+ 0.32 }}_{\mathrm{ -0.27 }}$ & 23.40 $^{\mathrm{+ 0.63 }}_{\mathrm{ -0.68 }}$ & 21.2 $^{\mathrm{+ 3.5 }}_{\mathrm{ -1.4 }}$ & 23.3 $^{\mathrm{+ 1.9 }}_{\mathrm{ -2.1 }}$ & 23.52 $^{\mathrm{+ 0.54 }}_{\mathrm{ -0.33 }}$ & 22.87 $^{\mathrm{+ 0.42 }}_{\mathrm{ -0.75 }}$ \\\vspace{-0.2cm}\\
$\rho$ &  0.888  &  0.866  &  0.909  &  0.873  &  0.821  & ...  &  0.818  &  0.928 \\\vspace{-0.2cm}\\ 
 & 0.804  &  0.739  &  0.848  &  0.882  &  0.000  &  0.714  &  0.905  &  0.759 \\\vspace{-0.1cm}\\
$p_{S}$ &  2.55e-12  &  3.35e-06  &  1.10e-06  &  4.55e-04  &  2.34e-02  & ...  &  3.81e-03  &  6.81e-11 \\\vspace{-0.2cm}\\ 
 & 3.47e-15  &  2.04e-06  &  1.79e-09  &  6.10e-06  &  1.00e+00  &  4.65e-02  &  1.60e-11  &  2.98e-07 \\\vspace{-0.1cm}\\
$N_\mathrm{pairs}$ &  34  &  18  &  16  &  11  &  7  & ... &  10  &  24 \\\vspace{-0.2cm}\\
 &  62  &  31  &  31  &  16  &  7  &  8  &  29  &  33  \\ \vspace{-0.3cm}\\\hline\\ \vspace{-0.5cm}

\end{tabular}
}
\begin{minipage}[t]{0.9\textwidth}{\small
\emph{Comments}: See the notes of Table\,\ref{tab:hihorbreak}.}
\end{minipage}

\end{minipage}
\end{table*}

\begin{table*}
\begin{minipage}[t]{\textwidth}
\caption{Linear fits performed to the trends of Type-III galaxies in the photometric planes relating \hi\ and \ho}
\label{tab:relhihorbreak}
\centering
{\fontsize{9}{9}\selectfont
\begin{tabular}{ccccccccc}
\hline\\\vspace{-0.5cm}\\
\multicolumn{9}{c}{15) $\log(\ho)$ vs.~$\log(\hi)$}\\\vspace{-0.1cm}\\
& \multicolumn{1}{c}{All} & \multicolumn{1}{c}{Spirals} & \multicolumn{1}{c}{S0} & \multicolumn{1}{c}{Sa--Sab} & \multicolumn{1}{c}{Sb--Sbc} & \multicolumn{1}{c}{Sc--Scd} & \multicolumn{1}{c}{Barred} & \multicolumn{1}{c}{Unbarred}
\\\vspace{-0.3cm}\\\hline\\\vspace{-0.5cm}\\
$m$ &  0.70 $^{\mathrm{+ 0.27 }}_{\mathrm{ -0.30 }}$ & 0.51 $^{\mathrm{+ 0.34 }}_{\mathrm{ -0.72 }}$ & 0.90 $^{\mathrm{+ 0.59 }}_{\mathrm{ -0.47 }}$ & -0.03 $^{\mathrm{+ 0.61 }}_{\mathrm{ -0.86 }}$ & 0.56 $^{\mathrm{+ 0.63 }}_{\mathrm{ -1.05 }}$ &  ... & 0.25 $^{\mathrm{+ 0.33 }}_{\mathrm{ -0.42 }}$ & 0.95 $^{\mathrm{+ 0.51 }}_{\mathrm{ -0.56 }}$ \\\vspace{-0.3cm}\\ 
 &  1.018 $^{\mathrm{+ 0.115 }}_{\mathrm{ -0.095 }}$ & 1.03 $^{\mathrm{+ 0.22 }}_{\mathrm{ -0.16 }}$ & 1.05 $^{\mathrm{+ 0.21 }}_{\mathrm{ -0.16 }}$ & 1.02 $^{\mathrm{+ 0.28 }}_{\mathrm{ -0.20 }}$ & 0.69 $^{\mathrm{+ 0.46 }}_{\mathrm{ -0.73 }}$ & 1.06 $^{\mathrm{+ 0.25 }}_{\mathrm{ -0.45 }}$ & 1.04 $^{\mathrm{+ 0.25 }}_{\mathrm{ -0.20 }}$ & 1.00 $^{\mathrm{+ 0.17 }}_{\mathrm{ -0.13 }}$ \\\vspace{-0.2cm}\\
$C_{0}$ &  0.488 $^{\mathrm{+ 0.113 }}_{\mathrm{ -0.089 }}$ & 0.56 $^{\mathrm{+ 0.26 }}_{\mathrm{ -0.13 }}$ & 0.39 $^{\mathrm{+ 0.19 }}_{\mathrm{ -0.15 }}$ & 0.78 $^{\mathrm{+ 0.34 }}_{\mathrm{ -0.26 }}$ & 0.49 $^{\mathrm{+ 0.37 }}_{\mathrm{ -0.13 }}$ &  ... & 0.57 $^{\mathrm{+ 0.15 }}_{\mathrm{ -0.11 }}$ & 0.41 $^{\mathrm{+ 0.23 }}_{\mathrm{ -0.14 }}$ \\\vspace{-0.3cm}\\ 
 &  0.282 $^{\mathrm{+ 0.035 }}_{\mathrm{ -0.035 }}$ & 0.288 $^{\mathrm{+ 0.049 }}_{\mathrm{ -0.046 }}$ & 0.239 $^{\mathrm{+ 0.070 }}_{\mathrm{ -0.068 }}$ & 0.312 $^{\mathrm{+ 0.097 }}_{\mathrm{ -0.074 }}$ & 0.357 $^{\mathrm{+ 0.239 }}_{\mathrm{ -0.056 }}$ & 0.232 $^{\mathrm{+ 0.120 }}_{\mathrm{ -0.053 }}$ & 0.242 $^{\mathrm{+ 0.066 }}_{\mathrm{ -0.063 }}$ & 0.296 $^{\mathrm{+ 0.060 }}_{\mathrm{ -0.056 }}$ \\\vspace{-0.2cm}\\
$\rho$ &  0.463  &  0.296  &  0.717  &  0.126  &  0.321  & ...  &  0.434  &  0.415 \\\vspace{-0.2cm}\\ 
 & 0.923  &  0.893  &  0.942  &  0.868  &  0.811  &  0.810  &  0.923  &  0.919 \\\vspace{-0.1cm}\\
$p_{S}$ &  3.87e-03  &  2.18e-01  &  8.07e-04  &  6.97e-01  &  4.82e-01  & ...  &  1.38e-01  &  4.36e-02 \\\vspace{-0.2cm}\\ 
 & 1.22e-26  &  1.39e-11  &  2.67e-15  &  1.33e-05  &  2.69e-02  &  1.49e-02  &  9.91e-13  &  4.31e-14 \\\vspace{-0.1cm}\\
$N_\mathrm{pairs}$ &  37  &  19  &  18  &  12  &  7  & ... &  13  &  24 \\\vspace{-0.2cm}\\
 &  62  &  31  &  31  &  16  &  7  &  8  &  29  &  33 \\\vspace{-0.2cm}\\

\hline\\\vspace{-0.5cm}\\
\multicolumn{9}{c}{16) $\log(\ho/\risoph)$ vs.~$\log(\hi/\risoph)$}\\\vspace{-0.1cm} \\
 & \multicolumn{1}{c}{All} & \multicolumn{1}{c}{Spirals} & \multicolumn{1}{c}{S0} & \multicolumn{1}{c}{Sa--Sab} & \multicolumn{1}{c}{Sb--Sbc} & \multicolumn{1}{c}{Sc--Scd} & \multicolumn{1}{c}{Barred} & \multicolumn{1}{c}{Unbarred}
\\\vspace{-0.3cm}\\\hline\\\vspace{-0.5cm}\\

$m$ &  0.56 $^{\mathrm{+ 0.28 }}_{\mathrm{ -0.27 }}$ & 0.58 $^{\mathrm{+ 0.49 }}_{\mathrm{ -0.63 }}$ & 0.54 $^{\mathrm{+ 0.56 }}_{\mathrm{ -0.55 }}$ & 0.28 $^{\mathrm{+ 0.67 }}_{\mathrm{ -0.75 }}$ & 0.9 $^{\mathrm{+ 1.0 }}_{\mathrm{ -1.8 }}$ &  ... & 0.28 $^{\mathrm{+ 0.44 }}_{\mathrm{ -0.61 }}$ & 0.66 $^{\mathrm{+ 0.61 }}_{\mathrm{ -0.42 }}$ \\\vspace{-0.3cm}\\ 
 &  0.93 $^{\mathrm{+ 0.22 }}_{\mathrm{ -0.22 }}$ & 0.79 $^{\mathrm{+ 0.32 }}_{\mathrm{ -0.31 }}$ & 1.15 $^{\mathrm{+ 0.35 }}_{\mathrm{ -0.37 }}$ & 0.64 $^{\mathrm{+ 0.52 }}_{\mathrm{ -0.63 }}$ & 0.87 $^{\mathrm{+ 0.66 }}_{\mathrm{ -0.78 }}$ & 0.94 $^{\mathrm{+ 0.50 }}_{\mathrm{ -1.56 }}$ & 1.10 $^{\mathrm{+ 0.55 }}_{\mathrm{ -0.40 }}$ & 0.89 $^{\mathrm{+ 0.29 }}_{\mathrm{ -0.36 }}$ \\\vspace{-0.2cm}\\
$C_{0}$ &  0.13 $^{\mathrm{+ 0.17 }}_{\mathrm{ -0.18 }}$ & 0.15 $^{\mathrm{+ 0.30 }}_{\mathrm{ -0.43 }}$ & 0.10 $^{\mathrm{+ 0.36 }}_{\mathrm{ -0.30 }}$ & -0.03 $^{\mathrm{+ 0.35 }}_{\mathrm{ -0.57 }}$ & 0.37 $^{\mathrm{+ 0.53 }}_{\mathrm{ -1.10 }}$ &  ... & -0.09 $^{\mathrm{+ 0.28 }}_{\mathrm{ -0.38 }}$ & 0.20 $^{\mathrm{+ 0.38 }}_{\mathrm{ -0.25 }}$ \\\vspace{-0.3cm}\\ 
 &  0.24 $^{\mathrm{+ 0.15 }}_{\mathrm{ -0.15 }}$ & 0.15 $^{\mathrm{+ 0.21 }}_{\mathrm{ -0.19 }}$ & 0.35 $^{\mathrm{+ 0.23 }}_{\mathrm{ -0.25 }}$ & 0.08 $^{\mathrm{+ 0.37 }}_{\mathrm{ -0.35 }}$ & 0.21 $^{\mathrm{+ 0.44 }}_{\mathrm{ -0.57 }}$ & 0.20 $^{\mathrm{+ 0.33 }}_{\mathrm{ -1.08 }}$ & 0.31 $^{\mathrm{+ 0.38 }}_{\mathrm{ -0.27 }}$ & 0.23 $^{\mathrm{+ 0.19 }}_{\mathrm{ -0.23 }}$ \\\vspace{-0.2cm}\\
$\rho$ &  0.367  &  0.279  &  0.362  &  0.084  &  0.536  & ...  &  0.236  &  0.451 \\\vspace{-0.2cm}\\ 
 & 0.775  &  0.657  &  0.865  &  0.506  &  0.643  &  0.619  &  0.856  &  0.716 \\\vspace{-0.1cm}\\
$p_{S}$ &  2.54e-02  &  2.47e-01  &  1.40e-01  &  7.95e-01  &  2.15e-01  & ...  &  4.37e-01  &  2.69e-02 \\\vspace{-0.2cm}\\ 
 & 1.52e-13  &  5.98e-05  &  3.36e-10  &  4.56e-02  &  1.19e-01  &  1.02e-01  &  3.32e-09  &  2.86e-06 \\\vspace{-0.1cm}\\
$N_\mathrm{pairs}$ &  37  &  19  &  18  &  12  &  7  & ... &  13  &  24 \\\vspace{-0.2cm}\\
 &  62  &  31  &  31  &  16  &  7  &  8  &  29  &  33 \\\vspace{-0.2cm}\\

\hline\\\vspace{-0.5cm}\\

\end{tabular}
}
\begin{minipage}[t]{0.9\textwidth}{\small
\emph{Comments}: See the notes of Table\,\ref{tab:hihorbreak}.}
\end{minipage}

\end{minipage}
\end{table*}

\begin{table*}
\begin{minipage}[t]{\textwidth}
\caption{Linear fits performed to the trends of Type-III galaxies in the photometric planes relating \risoph\ with \hi, \ho, and \rbreak}
\label{tab:r25}
\centering
{\fontsize{9}{9}\selectfont
\begin{tabular}{ccccccccc}
\hline\\\vspace{-0.5cm}\\
\multicolumn{9}{c}{17) $\log(\risoph)$ vs.~$\log(\hi)$}\\\vspace{-0.1cm} \\
 & \multicolumn{1}{c}{All} & \multicolumn{1}{c}{Spirals} & \multicolumn{1}{c}{S0} & \multicolumn{1}{c}{Sa--Sab} & \multicolumn{1}{c}{Sb--Sbc} & \multicolumn{1}{c}{Sc--Scd} & \multicolumn{1}{c}{Barred} & \multicolumn{1}{c}{Unbarred}
\\\vspace{-0.3cm}\\\hline\\\vspace{-0.5cm}\\

$m$ &  0.47 $^{\mathrm{+ 0.19 }}_{\mathrm{ -0.21 }}$ & 0.50 $^{\mathrm{+ 0.37 }}_{\mathrm{ -0.55 }}$ & 0.42 $^{\mathrm{+ 0.20 }}_{\mathrm{ -0.16 }}$ & 0.00 $^{\mathrm{+ 0.64 }}_{\mathrm{ -0.83 }}$ & 0.60 $^{\mathrm{+ 0.52 }}_{\mathrm{ -0.90 }}$ &  ... & 0.25 $^{\mathrm{+ 0.23 }}_{\mathrm{ -0.30 }}$ & 0.56 $^{\mathrm{+ 0.29 }}_{\mathrm{ -0.33 }}$ \\\vspace{-0.3cm}\\ 
 &  0.84 $^{\mathrm{+ 0.12 }}_{\mathrm{ -0.11 }}$ & 0.96 $^{\mathrm{+ 0.21 }}_{\mathrm{ -0.20 }}$ & 0.70 $^{\mathrm{+ 0.16 }}_{\mathrm{ -0.15 }}$ & 0.92 $^{\mathrm{+ 0.29 }}_{\mathrm{ -0.24 }}$ & 1.49 $^{\mathrm{+ 0.73 }}_{\mathrm{ -0.54 }}$ & 0.87 $^{\mathrm{+ 0.31 }}_{\mathrm{ -0.46 }}$ & 0.76 $^{\mathrm{+ 0.20 }}_{\mathrm{ -0.15 }}$ & 0.82 $^{\mathrm{+ 0.20 }}_{\mathrm{ -0.17 }}$ \\\vspace{-0.2cm}\\
$C_{0}$ &  0.771 $^{\mathrm{+ 0.079 }}_{\mathrm{ -0.065 }}$ & 0.79 $^{\mathrm{+ 0.20 }}_{\mathrm{ -0.14 }}$ & 0.734 $^{\mathrm{+ 0.057 }}_{\mathrm{ -0.048 }}$ & 0.99 $^{\mathrm{+ 0.34 }}_{\mathrm{ -0.23 }}$ & 0.73 $^{\mathrm{+ 0.26 }}_{\mathrm{ -0.19 }}$ &  ... & 0.812 $^{\mathrm{+ 0.100 }}_{\mathrm{ -0.083 }}$ & 0.726 $^{\mathrm{+ 0.140 }}_{\mathrm{ -0.087 }}$ \\\vspace{-0.3cm}\\ 
 &  0.705 $^{\mathrm{+ 0.040 }}_{\mathrm{ -0.041 }}$ & 0.682 $^{\mathrm{+ 0.061 }}_{\mathrm{ -0.067 }}$ & 0.740 $^{\mathrm{+ 0.073 }}_{\mathrm{ -0.074 }}$ & 0.681 $^{\mathrm{+ 0.083 }}_{\mathrm{ -0.099 }}$ & 0.59 $^{\mathrm{+ 0.19 }}_{\mathrm{ -0.17 }}$ & 0.707 $^{\mathrm{+ 0.093 }}_{\mathrm{ -0.082 }}$ & 0.725 $^{\mathrm{+ 0.067 }}_{\mathrm{ -0.072 }}$ & 0.706 $^{\mathrm{+ 0.071 }}_{\mathrm{ -0.065 }}$ \\\vspace{-0.2cm}\\
$\rho$ &  0.530  &  0.335  &  0.696  &  -0.014  &  0.571  & ...  &  0.418  &  0.528 \\\vspace{-0.2cm}\\ 
 & 0.859  &  0.852  &  0.843  &  0.859  &  0.901  &  0.881  &  0.849  &  0.855 \\\vspace{-0.1cm}\\
$p_{S}$ &  4.41e-04  &  1.61e-01  &  4.57e-04  &  9.66e-01  &  1.80e-01  & ...  &  1.07e-01  &  7.95e-03 \\\vspace{-0.2cm}\\ 
 & 4.07e-19  &  1.22e-09  &  2.64e-09  &  2.05e-05  &  5.62e-03  &  3.85e-03  &  5.87e-09  &  2.28e-10 \\\vspace{-0.1cm}\\
$N_\mathrm{pairs}$ &  40  &  19  &  21  &  12  &  7  & ... &  16  &  24 \\\vspace{-0.2cm}\\
 &  62  &  31  &  31  &  16  &  7  &  8  &  29  &  33 \\\vspace{-0.2cm}\\

\hline\\\vspace{-0.5cm}\\
\multicolumn{9}{c}{18) $\log(\risoph)$ vs.~$\log(\ho)$}\\\vspace{-0.1cm} \\
 & \multicolumn{1}{c}{All} & \multicolumn{1}{c}{Spirals} & \multicolumn{1}{c}{S0} & \multicolumn{1}{c}{Sa--Sab} & \multicolumn{1}{c}{Sb--Sbc} & \multicolumn{1}{c}{Sc--Scd} & \multicolumn{1}{c}{Barred} & \multicolumn{1}{c}{Unbarred}
\\\vspace{-0.3cm}\\\hline\\\vspace{-0.5cm}\\

$m$ &  0.33 $^{\mathrm{+ 0.16 }}_{\mathrm{ -0.17 }}$ & 0.34 $^{\mathrm{+ 0.40 }}_{\mathrm{ -0.65 }}$ & 0.29 $^{\mathrm{+ 0.11 }}_{\mathrm{ -0.15 }}$ & 0.19 $^{\mathrm{+ 0.55 }}_{\mathrm{ -0.59 }}$ & 0.17 $^{\mathrm{+ 0.97 }}_{\mathrm{ -1.40 }}$ &  ... & 0.14 $^{\mathrm{+ 0.46 }}_{\mathrm{ -0.32 }}$ & 0.34 $^{\mathrm{+ 0.18 }}_{\mathrm{ -0.18 }}$ \\\vspace{-0.3cm}\\ 
 &  0.68 $^{\mathrm{+ 0.16 }}_{\mathrm{ -0.13 }}$ & 0.77 $^{\mathrm{+ 0.29 }}_{\mathrm{ -0.21 }}$ & 0.56 $^{\mathrm{+ 0.24 }}_{\mathrm{ -0.19 }}$ & 0.74 $^{\mathrm{+ 0.36 }}_{\mathrm{ -0.25 }}$ & 1.75 $^{\mathrm{+ 0.88 }}_{\mathrm{ -0.98 }}$ & 0.72 $^{\mathrm{+ 0.36 }}_{\mathrm{ -0.43 }}$ & 0.58 $^{\mathrm{+ 0.30 }}_{\mathrm{ -0.21 }}$ & 0.71 $^{\mathrm{+ 0.26 }}_{\mathrm{ -0.20 }}$ \\\vspace{-0.2cm}\\
$C_{0}$ &  0.69 $^{\mathrm{+ 0.13 }}_{\mathrm{ -0.11 }}$ & 0.69 $^{\mathrm{+ 0.47 }}_{\mathrm{ -0.28 }}$ & 0.672 $^{\mathrm{+ 0.108 }}_{\mathrm{ -0.076 }}$ & 0.84 $^{\mathrm{+ 0.44 }}_{\mathrm{ -0.40 }}$ & 0.71 $^{\mathrm{+ 0.96 }}_{\mathrm{ -0.43 }}$ &  ... & 0.81 $^{\mathrm{+ 0.24 }}_{\mathrm{ -0.30 }}$ & 0.66 $^{\mathrm{+ 0.15 }}_{\mathrm{ -0.13 }}$ \\\vspace{-0.3cm}\\ 
 &  0.554 $^{\mathrm{+ 0.080 }}_{\mathrm{ -0.088 }}$ & 0.50 $^{\mathrm{+ 0.13 }}_{\mathrm{ -0.16 }}$ & 0.64 $^{\mathrm{+ 0.13 }}_{\mathrm{ -0.14 }}$ & 0.50 $^{\mathrm{+ 0.17 }}_{\mathrm{ -0.21 }}$ & 0.01 $^{\mathrm{+ 0.58 }}_{\mathrm{ -0.37 }}$ & 0.56 $^{\mathrm{+ 0.17 }}_{\mathrm{ -0.19 }}$ & 0.62 $^{\mathrm{+ 0.13 }}_{\mathrm{ -0.15 }}$ & 0.53 $^{\mathrm{+ 0.13 }}_{\mathrm{ -0.15 }}$ \\\vspace{-0.2cm}\\
$\rho$ &  0.479  &  0.182  &  0.734  &  0.203  &  0.036  & ...  &  0.115  &  0.536 \\\vspace{-0.2cm}\\ 
 & 0.841  &  0.859  &  0.802  &  0.876  &  0.964  &  0.810  &  0.796  &  0.864 \\\vspace{-0.1cm}\\
$p_{S}$ &  2.73e-03  &  4.55e-01  &  5.28e-04  &  5.27e-01  &  9.39e-01  & ...  &  7.07e-01  &  6.98e-03 \\\vspace{-0.2cm}\\ 
 & 1.23e-17  &  6.34e-10  &  5.82e-08  &  8.44e-06  &  4.54e-04  &  1.49e-02  &  2.38e-07  &  9.44e-11 \\\vspace{-0.1cm}\\
$N_\mathrm{pairs}$ &  37  &  19  &  18  &  12  &  7  & ... &  13  &  24 \\\vspace{-0.2cm}\\
 &  62  &  31  &  31  &  16  &  7  &  8  &  29  &  33 \\\vspace{-0.2cm}\\

\hline\\\vspace{-0.5cm}\\
\multicolumn{9}{c}{19) $\log(\risoph)$ vs.~$\log(\rbreak)$}\\\vspace{-0.1cm}\\
& \multicolumn{1}{c}{All} & \multicolumn{1}{c}{Spirals} & \multicolumn{1}{c}{S0} & \multicolumn{1}{c}{Sa--Sab} & \multicolumn{1}{c}{Sb--Sbc} & \multicolumn{1}{c}{Sc--Scd} & \multicolumn{1}{c}{Barred} & \multicolumn{1}{c}{Unbarred}
\\\vspace{-0.3cm}\\\hline\\\vspace{-0.5cm}\\

$m$ &  0.26 $^{\mathrm{+ 0.20 }}_{\mathrm{ -0.18 }}$ & 0.17 $^{\mathrm{+ 0.51 }}_{\mathrm{ -0.38 }}$ & 0.36 $^{\mathrm{+ 0.23 }}_{\mathrm{ -0.22 }}$ & -0.19 $^{\mathrm{+ 0.34 }}_{\mathrm{ -0.23 }}$ & 0.45 $^{\mathrm{+ 1.53 }}_{\mathrm{ -0.80 }}$ &  ... & 0.19 $^{\mathrm{+ 0.30 }}_{\mathrm{ -0.29 }}$ & 0.33 $^{\mathrm{+ 0.31 }}_{\mathrm{ -0.27 }}$ \\\vspace{-0.3cm}\\ 
 &  0.82 $^{\mathrm{+ 0.12 }}_{\mathrm{ -0.13 }}$ & 0.92 $^{\mathrm{+ 0.21 }}_{\mathrm{ -0.18 }}$ & 0.66 $^{\mathrm{+ 0.20 }}_{\mathrm{ -0.22 }}$ & 0.87 $^{\mathrm{+ 0.30 }}_{\mathrm{ -0.30 }}$ & 1.34 $^{\mathrm{+ 1.38 }}_{\mathrm{ -0.47 }}$ & 0.82 $^{\mathrm{+ 0.30 }}_{\mathrm{ -0.32 }}$ & 0.68 $^{\mathrm{+ 0.25 }}_{\mathrm{ -0.23 }}$ & 0.84 $^{\mathrm{+ 0.19 }}_{\mathrm{ -0.21 }}$ \\\vspace{-0.2cm}\\
$C_{0}$ &  0.67 $^{\mathrm{+ 0.19 }}_{\mathrm{ -0.20 }}$ & 0.78 $^{\mathrm{+ 0.39 }}_{\mathrm{ -0.48 }}$ & 0.53 $^{\mathrm{+ 0.20 }}_{\mathrm{ -0.23 }}$ & 1.16 $^{\mathrm{+ 0.26 }}_{\mathrm{ -0.31 }}$ & 0.48 $^{\mathrm{+ 0.65 }}_{\mathrm{ -1.46 }}$ &  ... & 0.71 $^{\mathrm{+ 0.28 }}_{\mathrm{ -0.29 }}$ & 0.60 $^{\mathrm{+ 0.29 }}_{\mathrm{ -0.31 }}$ \\\vspace{-0.3cm}\\ 
 &  0.22 $^{\mathrm{+ 0.13 }}_{\mathrm{ -0.13 }}$ & 0.09 $^{\mathrm{+ 0.18 }}_{\mathrm{ -0.20 }}$ & 0.39 $^{\mathrm{+ 0.23 }}_{\mathrm{ -0.20 }}$ & 0.15 $^{\mathrm{+ 0.34 }}_{\mathrm{ -0.33 }}$ & -0.24 $^{\mathrm{+ 0.46 }}_{\mathrm{ -1.16 }}$ & 0.15 $^{\mathrm{+ 0.27 }}_{\mathrm{ -0.23 }}$ & 0.32 $^{\mathrm{+ 0.24 }}_{\mathrm{ -0.25 }}$ & 0.22 $^{\mathrm{+ 0.21 }}_{\mathrm{ -0.19 }}$ \\\vspace{-0.2cm}\\
$\rho$ &  0.404  &  0.018  &  0.726  &  -0.315  &  0.429  & ...  &  0.294  &  0.352 \\\vspace{-0.2cm}\\ 
 & 0.735  &  0.758  &  0.708  &  0.691  &  0.786  &  0.857  &  0.742  &  0.752 \\\vspace{-0.1cm}\\
$p_{S}$ &  9.67e-03  &  9.43e-01  &  1.95e-04  &  3.19e-01  &  3.37e-01  & ...  &  2.69e-01  &  9.15e-02 \\\vspace{-0.2cm}\\ 
 & 1.09e-11  &  7.84e-07  &  8.50e-06  &  3.02e-03  &  3.62e-02  &  6.53e-03  &  4.14e-06  &  4.61e-07 \\\vspace{-0.1cm}\\
$N_\mathrm{pairs}$ &  40  &  19  &  21  &  12  &  7  & ... &  16  &  24 \\\vspace{-0.2cm}\\
 &  62  &  31  &  31  &  16  &  7  &  8  &  29  &  33 \\ \vspace{-0.3cm}\\\hline\\ \vspace{-0.5cm}

\end{tabular}
}
\begin{minipage}[t]{0.9\textwidth}{\small
\emph{Comments}: See the notes of Table\,\ref{tab:hihorbreak}.}
\end{minipage}

\end{minipage}
\end{table*}

}  

\end{document}